\documentclass[10pt,onehalfspaced,normalmargins]{ut-thesis}

\degree{Doctor of Philosophy}
\department{Physics}
\gradyear{2020}
\author{Wenyuan Wang}
\title{Adaptive Techniques in Practical Quantum Key Distribution}

\usepackage{graphicx}
\usepackage{dcolumn}
\usepackage{bm}
\usepackage{amsthm}
\usepackage{amsmath}
\usepackage{mathtools}
\usepackage{amssymb}
\usepackage{graphicx}
\usepackage{url}
\usepackage{float}
\usepackage{bm}
\usepackage{ifthen}
\usepackage[usenames,dvipsnames]{color}
\usepackage{mathrsfs}
\usepackage[colorlinks=true,citecolor=blue,urlcolor=black]{hyperref}
\usepackage{float}
\usepackage{subfigure}
\usepackage{enumitem}
\usepackage{color}
\usepackage{setspace}
\usepackage{braket}
\usepackage{lipsum}
\usepackage{comment}
\usepackage{tablefootnote}
\usepackage{braket}
\usepackage{multirow}
\usepackage{bm}

\begin{document}

\begin{preliminary}

\maketitle


\begin{abstract}
Quantum Key Distribution (QKD) can provide information-theoretically secure communications and is a strong candidate for the next generation of cryptography. However, in practice, the performance of QKD is limited by ``practical imperfections" in realistic sources, channels, and detectors (such as multi-photon components or imperfect encoding from the sources, losses and misalignment in the channels, or dark counts in detectors). Addressing such practical imperfections is a crucial part of implementing QKD protocols with good performance in reality. There are two highly important future directions for QKD: (1) QKD over free space, which can allow secure communications between mobile platforms such as handheld systems, drones, planes, and even satellites, and (2) fibre-based QKD networks, which can simultaneously provide QKD service to numerous users at arbitrary locations. These directions are both highly promising, but so far they are limited by practical imperfections in the channels and devices, which pose huge challenges and limit their performance. In this thesis, we develop adaptive techniques with innovative protocol and algorithm design, as well as novel techniques such as machine learning, to address some of these key challenges, including (a) atmospheric turbulence in channels for free-space QKD, (b) asymmetric losses in channels for QKD network, and (c) efficient parameter optimization in real time, which is important for both free-space QKD and QKD networks. We believe that this work will pave the way to important implementations of free-space QKD and fibre-based QKD networks in the future.
\end{abstract}





\begin{acknowledgements}
	
	First, I would like to sincerely thank my PhD supervisor, Prof. Hoi-Kwong Lo, who has led me into the world of academic research. His kind guidance, deep knowledge in the field, and tremendous patience were all indispensable to my being able to complete this program so smoothly and take my first step in a research career. I am deeply grateful for the very kind help he has provided over the many years, and for making the PhD program a very fruitful and pleasant journey for me.
	
	I would also like to thank Prof. Li Qian and Prof. Daniel James for kindly being on my PhD supervisory committee and helping me over the years. I have benefited greatly from Prof. Qian's countless helpful suggestions and comments on many projects, and have also very much enjoyed Prof. James' two graduate courses. I would also like to thank Prof. Aephraim Steinberg and Prof. John Sipe for being on my SGS defense committee, and my special thanks to Dr. Zhilian Yuan for kindly being my external examiner.
	
	I would also like to take this opportunity to thank the many people who have helped and supported me in my academic life and made many projects possible. Specifically, I would like to thank Prof. Feihu Xu, who has offered me much guidance and support when I started out in this field, come up with many brilliant ideas and suggestions, and collaborated with us closely over the many years, and Prof. Marcos Curty, for the kind help and instruction during our many pleasant collaboration/fruitful discussions. I would like to thank Dr. Chen Feng who has offered much help and guidance when I first visited the University of Toronto in 2014, when he patiently taught me about Error-Correction Codes when I was still a clueless beginner. I would like to thank Dr. Zhiyuan Tang for kindly introducing fellow students and me to the laboratory devices. I would also like to thank Prof. Zidan Wang and Prof. Francis Chin, for being the supervisors for my undergraduate research at the University of Hong Kong in 2014 and 2015 and first introducing me to hands-on research. Moreover, I would like to thank Dr. Bing Qi, Prof. George Siopsis, Prof. Norbert Lutkenhaus, Prof. Thomas Jennewein, Dr. Charles Lim, Dr. Koji Azuma, Prof. Kiyoshi Tamaki, Dr. Marco Lucamarini, Dr. Zhiliang Yuan,  Dr. Cunlu Zhou, Prof. Xiongfeng Ma, Prof. Paul Kwiat for the help and insights over the collaborations and/or academic discussions. 
	
	Also, I have benefited greatly from the many helpful discussions with and support from fellow students, including current and past fellow group members Olinka Bedroya, Xiaoqing Zhong, Chenyang Li, Eli Bourassa, Ilan Tzitrin, Michael Tisi, Reem Mandil, Yongtao Zhan, Amita Gnanapandithan, Thomas Van Himbeeck, and Anqi Mou, as well as friends from other research groups and other fields, including Chi Yan, Zihan Wang, Jie Luo, Jie Lin, Ian George, and Alistair Duff. Special thanks to Reem and Yongtao for their kind suggestions on my draft thesis.
	
	I would like to thank the support and funding of the University of Toronto Physics Department and thank the Natural Sciences and Engineering Research Council of Canada (NSERC), National Research Council of Canada (NRC), and U.S. Office of Naval Research (ONR) for funding several projects. I would like to thank Dr. Kevin McBryde and Dr. Stephen Hammel from U.S. SPAWAR Systems Center Pacific for kindly providing real-world atmospheric data and helpful discussions for the ONR project of QKD through turbulence. I'd also like to thank Nvidia and SciNet/Compute Canada for providing the computing equipment and services that supported many of the projects.
	
	Lastly, I would like to thank my family for all the love, support, and encouragement. I could not have been here without them. My parents' cultivation of my interest in science from a very young age, their selfless support, encouragement and guidance in my life, all make it possible for me to pursue my dream in the world of scientific research.
	
	This work is dedicated to the family I love, to the mentor I respect and look up to, and to the many people who have supported me on this path.

\end{acknowledgements}

\tableofcontents




\end{preliminary}


\chapter{Introduction}

\section{Quantum Key Distribution (QKD)}


Information privacy is a crucial part of our daily lives. From financial transactions, personal identification, to banking and even national defense, cryptography is an indispensable part of our modern society. Quantum computing threatens the security of conventional public-key cryptography, which is widely used for key agreement and digital signature, and plays a key role in internet security. Many public-key cryptographic algorithms, such as the popular RSA, depend on an assumed intractability of difficult mathematics problems, such as large number factorization. However, an algorithm proposed for a universal quantum computer, the Shor's algorithm \cite{Shor}, can efficiently solve the large number factorization problem, thus breaking down the security of RSA, which would bring huge risks to modern security applications in fields such as finance and national defense. 

Quantum computer uses quantum bits (``qubits"), which are two-level systems possessing quantum mechanical properties (such as superposition and entanglement), to encode and process information. With the inherent parallelism from the superposed quantum states of qubits, it is believed that quantum computers can provide computing power that exponentially exceeds that of the classical computers we use today. Many applications have been proposed (and being developed) for quantum computers, such as acceleration of optimization and searching problems, as well as simulation of quantum systems, but perhaps the most influential application of quantum computers so far is their aforementioned ability to break cryptographic systems.

The trouble is - this seemingly impending doom of cryptography brought by quantum computers might not be that far away. Though a scalable, universal quantum computer has yet to be built (and there have been only rudimentary demonstrations of Shor's algorithm), many companies such as Google, IBM, Intel have already started building quantum computers based on superconducting qubits, and there are also implementations or proposals with linear optics, ion traps, or nuclear magnetic resonance. Importantly, though we are still rather far from a large-scale error-corrected universal quantum computer, there have already been demonstrations where the current-technology quantum computers solve problems that are intractable on classical computers (an advantage commonly called ``quantum supremacy"), such as in Ref. \cite{GoogleQS} for quantum circuits and Ref. \cite{bosonsampling} for boson sampling.

Moreover, even if it might be years before a universal quantum computer capable of running Shor's algorithm at a practical scale is built, we also need to consider a \textit{retroactive} threat to security. That is, an eavesdropper Eve can simply intercept and save for decades the transcript of any classical communication whose secure keys are generated from e.g. RSA, and wait for the invention of new algorithms or construction of new computers such as quantum computers to crack the key. This means that important information from the present day might be leaked in the future once quantum computers become practical. This further calls for a cryptographic scheme that can resist attacks from quantum computers as soon as possible.

To address such an increasing threat from quantum computing, one strong candidate for the next generation of cryptography is quantum key distribution (QKD). QKD uses qubits encoded in photons to transfer information, and allows two parties (traditionally called Alice and Bob) to share a pair of random bit strings (called ``key") with information-theoretic, i.e. provable, security. QKD can protect users from the attacks of even quantum computers. Because of this, QKD is considered as one of the strong candidates for the next-generation technology for secure communications. \footnote{There are also other candidates such as ``post-quantum cryptography", for instance lattice-based cryptography (such as ``learning with error" \cite{LWE} and ``ring learning with error" \cite{RLWE}), which can resist the Shor's algorithm. The advantage of such schemes is that they are software-based and can be implemented on classical computers. However, in a way they are only temporary solutions since they only resist the attacks from \textit{known} quantum algorithms, and their security might be compromised if new quantum algorithms targeting specific post-quantum cryptographic schemes arise in the future. On the other hand, the security of QKD is independent of whichever types of equipment or algorithms the eavesdropper might have. In this thesis, we will focus our discussion on QKD only.}

The security of QKD fundamentally comes from the ``\textit{no-cloning theorem}" \cite{nocloning}, that is, a qubit cannot be perfectly duplicated with perfect fidelity. In other words, information that is encoded in qubits and sent through a public quantum channel cannot be retrieved by an eavesdropper without inevitably disturbing the quantum state. Once a qubit is measured, the measurement will disturb the state. Such disturbance can subsequently be detected by Alice and Bob, quantifiable by a ``quantum-bit-error-rate" (QBER). If Alice and Bob detect such error, they will either reduce their key length and thus remove information leaked to an eavesdropper, or abort the communication if too much disturbance is present. This ensures an information-theoretically quantifiable level of security, independent of Eve's computing capabilities (even in the scenario that she possesses a quantum computer).

Once a pair of secure keys are distributed between Alice and Bob via QKD, Alice can use the ``One-Time-Pad" protocol (which encrypts a message by performing an XOR operation between the secure random key and the actual message she wants to send) to send the encrypted message to Bob, who then decrypts the message by performing XOR again using his secure key. The One-Time-Pad \cite{onetimepad} protocol is proven to be secure as long as the keys are secure and not reused. Therefore, QKD can enable secure communications between Alice and Bob, protected from even the attacks a quantum computer.

The problem is, while QKD is theoretically secure, side channels still exist in a system built with practical components, such as in sources and detectors. There have been multiple quantum hacking attacks, e.g. Refs. \cite{blinding,timeshift,phaseremapping,fieldhacking,detectordeadtime,devicecalibration} that target the practical weaknesses in QKD systems. Therefore, an important question in quantum cryptography is to determine how secure a system really is in practice. 

Ideally, QKD requires single-photon sources, which are difficult to implement in reality. To be able to use ``imperfect" realistic photon sources (which contain both single photons and multi-photon contributions), the decoy-state method \cite{decoystate_LMC,decoystate_Hwang,decoystate_Wang} is proposed, which uses multiple intensities to estimate the single-photon contributions, and allows the secure use of the relatively easily attainable weak coherent pulse (WCP) sources in QKD systems, drastically improving the practical usefulness of QKD and increasing its key rate.

Aside from multi-photon contributions from WCP sources, sources might also have other imperfections such as fluctuating intensities, which requires characterization of the intensity fluctuation and corresponding reduction to the secure key rate. Also, sources might have imperfect encoding, which is addressed by loss-tolerant QKD protocols.

Among the components of a QKD system, detectors are especially susceptible to attacks (and a majority of hacking attempts target the detectors), making them the Achilles' Heel of QKD systems. The measurement-device-independent (MDI) QKD~\cite{mdiqkd} protocol allows an untrusted third-party to make measurements, thus avoiding all security breaches from detector side channels. Since its proposal, MDI-QKD has attracted worldwide interest, and there have been hundreds of follow-up theory and experimental papers. For example, some notable experimental implementations have been reported in Refs. \cite{mdiexp1,mdiexp2,mdiexperiment,mdi200km,mdiexp3,mdi404km,mdiPOP}.

A recently proposed protocol, Twin-Field (TF) QKD protocol \cite{TFQKD}, maintains a similar measurement-device-independence as MDI-QKD, but can significantly extend the maximum distance and overcome the maximum key rate versus distance trade-off for repeaterless QKD (such as the TGW and PLOB bounds \cite{TGW,PLOB}, which state that the maximum key rate of QKD scales linearly with the transmittance of the channel, without the help quantum repeater). This groundbreaking advantage of TF-QKD has generated much interest in the community, both theoretically \cite{TFQKD01,TFQKD02,TFQKD03,TFQKD04,simpleTFQKD} and experimentally \cite{TFexperiment01,TFexperiment02,TFexperiment03,TFexperiment04}.

More details on QKD protocols and common techniques to address practical imperfections in QKD systems are included in Chapter 2.

Two promising future directions of QKD are to implement it over free-space, and implement it as a multi-user fibre-based quantum network. However, though highly attractive if successfully implemented, there are several key challenges involved in free-space QKD and fibre-based quantum networks. 

This thesis aims at addressing some of these practical challenges, and paving the way for more robust implementations of free-space QKD and QKD fibre networks. 

\section{Motivation}

The introductions to the motivations of each project here are based on the background sections of our papers Refs. \cite{this_BB84,this_asymMDI,this_ML}.

\subsection{Free-Space QKD}

There has been increasing interest in implementing QKD through free-space channels. A major attraction for free-space QKD is that, when performed efficiently, it could potentially be applied to airborne or maritime quantum communications where participating parties are on mobile platforms. Furthermore, it could even enable applications for ground to satellite quantum communications, and eventually, a global quantum communication network.

Free-space quantum communication has seen great advances over the past 25 years. The first demonstration of free-space QKD was published by Bennett et al. from IBM research in 1992 \cite{freespace_IBM} over 32cm of free-space channel, which was also the first successful demonstration of experimental QKD. Over the next two decades, numerous demonstrations for free-space QKD have been made. In 1998, Buttler and Hughes et al. \cite{freespace_Hughes} performed QKD over 1km of free-space channel outdoors at night-time. In 2005, Peng et al. \cite{freespace_2005_Peng} performed distribution of entangled photons over 13km. In 2007, two successful experimental ground-to-ground free-space QKD experiments based on BB84 and E91 protocol \cite{free2007,freespace_144km_E91} were implemented over a 144km link between the Canary Islands of La Palma and Tenerife. Ling et al. \cite{freespace_urban_E91} performed another entanglement-based QKD Experiment with modified E91 protocol over 1.4km in urban area in 2008. In 2012, Yin et al. and Ma et al. \cite{freespace_2012_100km, freespace_2012_143km} respectively performed quantum teleportation over 100km and 143 km.

In recent years, free-space QKD has also seen much development over rapidly moving platforms, with an air-to-ground experiment in 2013 by Nauerth et al. \cite{freespace_plane} over a plane 20km from ground, a demonstration of QKD with devices on moving turntable and floating hot balloon over 20km and 40km by J-Y Wang et al. \cite{freespace_2013_satellite} in 2013, a very recent report on drone-to-drone QKD experiment in 2017 by D. Hill et al. \cite{freespace_drone}, and notably, satellite-based quantum communication experiments in 2017 \cite{freespace_satellite1, freespace_satellite2, freespace_satellite3}, including a QKD experiment from a quantum satellite to the ground over a 1200km channel. Meanwhile, there is a lot of interest in doing QKD in a maritime environment either over sea water\cite{freespace_maritime_data} or through seawater \cite{freespace_underwater}. A study on quantum communication performance over a turbulent free-space maritime channel using real atmospheric data can be found in Ref.\cite{freespace_maritime_data}.

A major characteristic of a free-space channel is its time-dependent transmittance. This is due to the temporal fluctuations of the local refractive index in the free-space channel, i.e. \textit{atmospheric turbulence}, which causes scintillation and beam wandering \cite{freespacethesis}, and results in fluctuations in the channel transmittance, which in turn affect QKD performance. 

Therefore, addressing turbulence is a major challenge for QKD over free-space. In Chapters 3 and 4, we will discuss a real-time selection technique, that can drastically improve the maximum tolerated transmission distance (channel loss), and greatly improve the performance of free-space QKD when turbulence is present in the channels.

\subsection{Fibre-based Quantum Network}


As we now live in an era of the Internet of Things (IoT) that interconnects many users and devices, for QKD to be widely deployed in the future, an important step is to study it in a network setting, i.e. designing \textit{quantum networks} that can connect and provide service to numerous users, who may freely join or leave a network.

Up till now, multiple field implementations of point-to-point QKD networks have been reported in e.g., Refs.\cite{quantumnetwork1,quantumnetwork2,QKDbackbonenetwork}. However, they all relied on \textit{trusted} relays (where the information stops being quantum at the relays), which are undesirable for security. MDI-QKD protocol (which will be explained in more detail in chapter 2) solves this problem by allowing \textit{untrusted} relays in a quantum network, which is a huge advantage over previous point-to-point QKD networks, making MDI-QKD a powerful candidate for future quantum networks. For instance, a first three-user star-shaped MDI-QKD network experiment in a metropolitan setting has been reported in Ref. \cite{mdinetwork}.


However, a major limitation of MDI-QKD is that it \textit{requires all users to have near-identical (i.e. symmetric) channel losses to the untrusted relay} for the protocol to work well \cite{mdipractical,HOM}, and the key rate will degrade very quickly with an increased level of asymmetry between channel losses. As standard fibre has a constant loss per distance of $0.2dB/km$, the above limitation means that the users Alice and Bob are limited to similar fibre lengths, or \textit{geographical distances} between their sites and the relay in a realistic fibre-based network. Because of this limitation, previous experiments of MDI-QKD either were performed in the laboratory over symmetric fibre spools~\cite{mdiexp1,mdiexp2,mdiexperiment,mdi200km,mdiexp3,mdi404km}, or had to deliberately add a tailored length of fibre to the shorter channel (to introduce additional loss) in exchange for better symmetry~\cite{mdiPOP}.  Adding additional fibres not only is cumbersome as it requires halting the system (and not practical when there are many pairs of connections in a quantum network, or when channel loss is changing) but also results in suboptimal key rate when channels are asymmetric.

In a realistic setup, a quantum network will very likely have asymmetric channels due to different geographical locations of sites (or, if one intends to implement MDI-QKD over free-space, due to moving platforms and changing channels). The requirement on symmetric channels significantly limits the key rate of previous MDI-QKD protocols in a general quantum network, thus seriously hindering the widespread deployment of MDI-QKD.

In Chapter 5, we demonstrated a technique of compensating for channel loss with different laser intensities from users. We show that we can lift this limitation on channel symmetry, and allow completely arbitrary levels of asymmetry between any two channels. This makes it possible to enable a scalable high-rate MDI-QKD network.


Furthermore, as TF-QKD allows a similar measurement-device-independence and can be a candidate for a quantum network with untrusted relays too, in Chapter 6 we extend our results and show that it works well for TF-QKD with asymmetric channels in quantum networks too.

\subsection{Machine Learning in QKD}

Machine learning is a type of algorithm that extracts the implicit relationships between input and output data, and use this learnt mapping to analyze new data. Machine learning techniques, especially neural networks, have been widely used in computer vision, language recognition, and automated control. There is increasing interest in the field in applying machine learning to improve the performance of quantum communication. For instance, there is recent literature that applies machine learning to continuous-variable (CV) QKD to improve the noise-filtering \cite{CVQKD1} and the prediction/compensation of intensity evolution of light over time \cite{CVQKD2}, respectively. 

In Chapter 7 of this thesis, we will discuss the application of machine learning to a specific problem: parameter optimization for QKD.

In reality, a QKD experiment always has a limited transmission time. Therefore, the total number of signals is finite. This means that, when estimating the single-photon contributions with decoy-state analysis, one would need to take into consideration the statistical fluctuations of the observables: the Gain and Quantum Bit Error Rate (QBER). This is called the finite-key analysis of QKD. When considering finite-size effects, the choice of intensities and probabilities of sending these intensities is crucial to getting the optimal rate. Therefore, we would need to perform optimizations for the search of parameters. 

Traditionally, the optimization of parameters is implemented as either a brute-force global search for smaller numbers of parameters, or local search algorithms for larger numbers of parameters. For instance, in several papers studying MDI-QKD protocols in symmetric \cite{mdiparameter} and asymmetric channels \cite{this_asymMDI}, a local search method called coordinate descent algorithm is used to find the optimal set of intensity and probabilities.

However, optimization of parameters often requires significant computational power. This means that a QKD system has to either wait for an optimization off-line (and suffer from delay) or use sub-optimal or even unoptimized parameters in real time. Moreover, due to the amount of computing resources required, parameter optimization is usually limited to relatively powerful devices such as a desktop PC.

There is increasing interest in implementing QKD in free-space on mobile platforms, such as drones, handheld systems, and even satellites. Such devices (e.g. single-board computers and mobile System-on-Chips) are usually limited in computational power. As low-latency is important in such free-space applications, fast and accurate parameter optimization based on a changing environment in real time is a difficult task on such low-power platforms.

Moreover, with the advent of the internet of things (IoT), a highly attractive future direction of QKD is a quantum network that connects multiple devices, each of which could be portable and mobile, and numerous connections are present at the same time. This will present a great computational challenge for the controller of a quantum network with many pairs of users (where real-time optimization might simply be infeasible for even a moderate number of connections).

With the development of machine learning technologies based on neural networks in recent years, and with more and more low-power devices implementing on-board acceleration chips for neural networks, we present a new method of using neural networks to help predict optimal parameters efficiently on low-power devices. Such a method makes it possible to support \textit{real-time} parameter optimization for free-space QKD systems, or large-scale QKD networks with thousands of connections.

\section{Structure of Thesis}


\begin{table}[t]

\caption{An outline of how the chapters 3-7 (corresponding to results from 5 of our papers) are organized in this thesis, including the protocols (BB84, MDI-QKD, TF-QKD), the topics (turbulence, channel asymmetry, machine learning) and the applications (free-space QKD, and fibre-based QKD network). Due to limited spacing, we have used abbreviations for some terms. Chapters 3 and 4 discuss turbulence in free-space QKD for BB84 and MDI-QKD. Chapters 5 and 6 discuss channel asymmetry for MDI-QKD and TF-QKD in a network setting. Chapter 7 discusses using machine learning (in parameter optimization) for all three protocols, which can be useful for both free-space QKD and fibre-based QKD networks.} 
\begin{center}
	\begin{tabular}{c|ccc}		
		\hline \hline
		Protocol & Turbulence & Asymmetry & Machine Learning \\
		\hline
		BB84 & Chapter 3 \cite{this_BB84}& - & Chapter 7 \cite{this_ML}\\
		MDI-QKD & Chapter 4 \cite{this_MDI}& Chapter 5 \cite{this_asymMDI}& Chapter 7 \cite{this_ML}\\
		TF-QKD & - & Chapter 6 \cite{this_asymTF}& Chapter 7 \cite{this_ML}\\
		\hline \hline
	\end{tabular}
	
\end{center}

\end{table}

\begin{itemize}
\item In Chapter 2, we review the QKD protocols including BB84, MDI-QKD and TF-QKD, and common techniques such as decoy-state method, finite-size analysis, and parameter optimization to address practical imperfections.
\end{itemize}

Our new results are included in Chapters 3-7. \\

Chapters 3 and 4 correspond to our results for free-space QKD.

\begin{itemize}
\item In Chapter 3, we discuss a real-time selection technique that greatly improves the performance of free-space BB84 when turbulence is present in the channels. This chapter corresponds to our paper Ref. \cite{this_BB84}
\item In Chapter 4, we extend the real-time selection technique to MDI-QKD and show similar effectiveness when turbulence is present. This chapter corresponds to our paper Ref. \cite{this_MDI}
\end{itemize}

Chapters 5 and 6 correspond to our results mainly for fibre-based QKD networks.

\begin{itemize}
\item In Chapter 5, we demonstrated a technique of compensating for asymmetric channel loss with different laser intensities from users, allowing MDI-QKD to be performed over asymmetric channels, and therefore enabling a scalable MDI-QKD network. This chapter corresponds to our paper Ref. \cite{this_asymMDI}
\item In Chapter 6, we extend the asymmetric-intensity method to TF-QKD, and demonstrate that TF-QKD can be performed efficiently over asymmetric channels too. This chapter corresponds to our paper Ref. \cite{this_asymTF}
\end{itemize}

Chapter 7 corresponds to our results for machine learning in QKD.

\begin{itemize}
\item In Chapter 7, we apply machine learning to a specific problem: parameter optimization for QKD, and show that it can greatly improve the efficiency of parameter optimization and allow it to be performed in real-time on low power devices, making it a highly useful tool for both free-space QKD and QKD networks. This chapter corresponds to our paper Ref. \cite{this_ML}.
\end{itemize}

We conclude the thesis in Chapter 8.

\begin{itemize}
	\item In Chapter 8, we conclude the thesis, and also discuss some unsolved problems and future research directions, in terms of short-term projects and long-term goals.
\end{itemize}

\section{List of Publications and Presentations}

\subsection{Publications Included in this Thesis}

\begin{itemize}
	
\item Wenyuan Wang, Feihu Xu, and Hoi-Kwong Lo. ``Prefixed-threshold Real-Time Selection for Free-Space Measurement-Device-Independent Quantum Key Distribution." arXiv:1910.10137 (2019). \cite{this_MDI}
	
\item Wenyuan Wang, and Hoi-Kwong Lo. ``Simple Method for Asymmetric Twin-Field Quantum Key Distribution" A Simple Method for Asymmetric Twin-Field QKD”, New Journal of Physics 22.1: 013020 (2020). \cite{this_asymTF}

\item Wenyuan Wang, and Hoi-Kwong Lo. ``Machine Learning for Optimal Parameter Prediction in Quantum Key Distribution." Physical Review A 100, 062334 (2019). \cite{this_ML}

\item Wenyuan Wang, Feihu Xu, and Hoi-Kwong Lo. ``Asymmetric Protocols for Scalable High-Rate Measurement-Device-Independent Quantum Key Distribution Networks." Physical Review X 9, 041012 (2019). \cite{this_asymMDI}

\item Wenyuan Wang, Feihu Xu, and Hoi-Kwong Lo. ``Prefixed-threshold real-time selection method in free-space quantum key distribution." Physical Review A 97.3: 032337 (2018). \cite{this_BB84}
\end{itemize}

\subsection{Publications not Included in this Thesis}

\begin{itemize}
\item Xiaoqing Zhong, Wenyuan Wang, Li Qian, Hoi-Kwong Lo. ``Proof-of-Principle Experimental Demonstration of Twin-Field QKD over Optical Channels with Asymmetric Losses", arXiv: 2001.10599 (2020).

\item Hui Liu*, Wenyuan Wang*, Kejin Wei, Xiao-Tian Fang, Li Li, Nai-Le Liu, Hao Liang, Si-Jie Zhang, Weijun Zhang, Hao Li, Lixing You, Zhen Wang, Hoi-Kwong Lo, Teng-Yun Chen, Feihu Xu, Jian-Wei Pan. ``Experimental demonstration of high-rate measurement-device-independent quantum key distribution over asymmetric channels." Physical Review Letters 122.16: 160501 (2019). 
(*these authors contributed equally to this work)
	
\item Kiyoshi Tamaki, Hoi-Kwong Lo, Wenyuan Wang, Marco Lucamarini. ``Information theoretic security of quantum key distribution overcoming the repeaterless secret key capacity bound." arXiv preprint arXiv:1805.05511 (2018).

\item Álvaro Navarrete, Wenyuan Wang, Feihu Xu, and Marcos Curty. ``Characterizing multi-photon quantum interference with practical light sources and threshold single-photon detectors." New Journal of Physics 20.4: 043018 (2018).

\item Feihu Xu, Juan Arrazola, Kejin Wei, Wenyuan Wang, Pablo Palacios, Chen Feng, Shihan Sajeed, Norbert Lütkenhaus, and Hoi-Kwong Lo. ``Experimental quantum fingerprinting with weak coherent pulses." Nature Communications 6 (2015).

\end{itemize}

\subsection{Presentations}

\begin{itemize}
	
	\item Poster presentation and rump session talk at the 9th International Conference on Quantum Cryptography (QCrypt 2019), Montreal, on ``Machine Learning for Optimal Parameter Prediction in Quantum Key Distribution" (Poster), ``Prefixed-threshold Real-Time Selection for Free-Space Measurement-Device-Independent Quantum Key Distribution" (Poster), and ``Simple Method for Asymmetric Twin-Field QKD" (Rump Session Talk)
	
	\item Oral presentation at the QKD Security Proof Workshop at the University of Toronto, Toronto (2019) on ``Machine Learning for Optimal Parameter Prediction in Quantum Key Distribution and ``Simple Method for Asymmetric Twin-Field Quantum Key Distribution".
	
	\item Invited talk at the IQC-China conference on Quantum Technologies, Waterloo (2018) on ``Enabling a scalable high-rate measurement-device-independent quantum key distribution network".
	
	\item Oral presentation at the 8th International Conference on Quantum Cryptography, Shanghai (QCrypt 2018), on ``Enabling a scalable high-rate measurement-device-independent quantum key distribution network", \textbf{Best Student Paper Award}.
	
	\item Oral presentation and poster presentation at the QKD Security Proof Workshop at IQC, Waterloo (2018) on ``Enabling a scalable high-rate measurement-device-independent quantum key distribution network" (talk) and ``Practical Real Time Selection Method for Quantum Communication against Atmospheric Turbulence" (poster).
	
	\item Poster presentation and rump session talk at the 7th International Conference on Quantum Cryptography (QCrypt 2017), Cambridge, on ``Practical Real Time Selection Method for Quantum Communication against Atmospheric Turbulence".
		
\end{itemize}

\chapter{Background on QKD Protocols}

In this chapter, we will review some of the key concepts in QKD protocols and practical techniques in previous literature, which lay the foundations of the new proposals in this thesis. The content of this chapter is a recapitulation of several previous important papers on the subject, especially Refs. \cite{bb84,Preskill,GLLP,decoystate_LMC,decoypractical,mdiqkd,mdipractical,mdiparameter}. We also refer to Ref. \cite{reviewQKD2019_arxiv, reviewQKD2019, QKD_review1, QKD_review2, MDI_review} for an overview of topics such as QKD, hacking attacks, and MDI-QKD. One can refer to Ref. \cite{reviewQKD2019_arxiv} for a comprehensive review of QKD with realistic devices.

\section{List of Abbreviations}

In this section, we list some of the frequently used abbreviations in this thesis:\\

Theoretical concepts and protocols:

\begin{itemize}
	\item QKD: Quantum Key Distribution
	\item RSA: Rivest–Shamir–Adleman (cryptosystem)
	\item BB84: Bennett-Brassard-84 (protocol)
	\item MDI-QKD: Measurement-Device-Independent QKD
	\item TF-QKD: Twin-Field QKD
	\item CV-QKD: Continuous-Variable QKD
	\item GLLP: Gottesman-Lo-Lutkenhaus-Preskill (proof)
	\item PLOB: Pirandola-Laurenza-Ottaviani-Banchi (bound)
	\item HOM: Hong-Ou-Mandel (interference)
	\item QBER: Quantum Bit Error Rate
	\item XOR: Exclusive-OR (gate)
	\item ECC: Error Correction Code
	\item BSM: Bell State Measurement
	\item PDTC: Probability Distribution of Transmission Coefficient
	\item ARTS: Adaptive Real Time Selection
	\item P-RTS: Prefixed-threshold Real Time Selection
	\item LS: Local Search
	\item CD: Coordinate Descent
	\item NN: Neural Network
	\item ReLU: Rectified Linear Unit
	\item IoT: Internet of Things
\end{itemize}

Experimental devices and hardware components:

\begin{itemize}
	\item WCP: Weak Coherent Pulse (source)
	\item SPDC: Spontaneous Parametric Down-Conversion (source)
	\item SPAD: Single-Photon Avalanche Detector
	\item SNSPD: Superconducting Nanowire Single-Photon Detector
	\item BS: Beam Splitter
	\item PM: Phase Modulator
	\item Pol-M: Polarization Modulator
	\item IM: Intensity Modulator
	\item CPU: Central Processing Unit
	\item GPU: Graphical Processing Unit
\end{itemize}

\section{BB84}

The contents of this section are based on Refs. \cite{bb84,Preskill}.

In 1984, Bennett and Brassard published the paper ``Quantum cryptography: Public key distribution and coin tossing" \cite{bb84}, which later became known as the BB84 protocol, and marked the first QKD protocol proposed.

\begin{figure}[h]
	\includegraphics[scale=0.35]{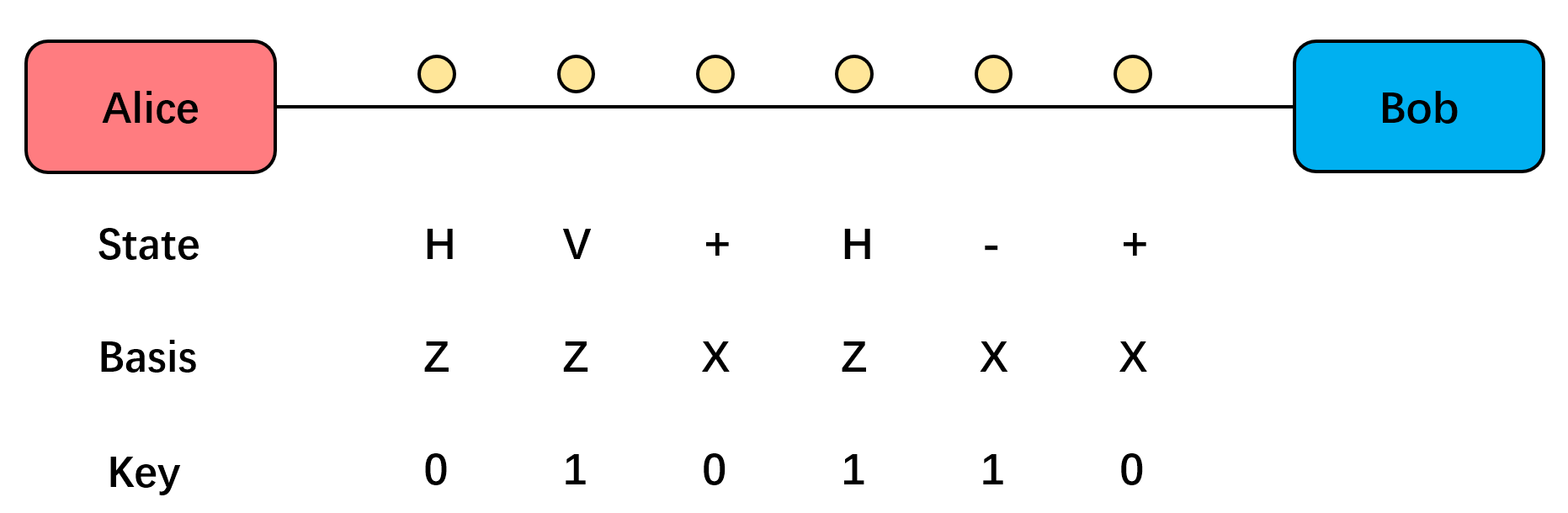}
	\caption{An illustration of the states sent in the BB84 protocol. Alice sends states $\ket{H},\ket{V}$ in Z basis, and $\ket{+},\ket{-}$ in X basis. The states correspond to bit values 0 or 1 in each basis. Bob measures the bits randomly in X or Z basis, and later announce the basis choice where he and Alice chose the same basis (sifting). They then randomly choose and  publicly announce a portion of the data to sample error rate, which they subsequently use to reduce the key length in order to reduce an eavesdropper's information to near zero (privacy amplification). Figure reproduced with modifications based on Table 1 from Ref. \cite{bb84}.}
	\label{fig:intro_BB84}
\end{figure}

The key steps of BB84 are:

\begin{itemize}
	\item Each of Alice and Bob, the two users who would like to share a secret key, respectively \textit{prepares and measures} in two non-orthogonal bases, which we can denote as the diagonal (X) and rectilinear (Z) bases. In the Z (rectilinear) basis, Alice (Bob) prepares (measures) states in $\ket{H},\ket{V}$ corresponding to 0 and 1 bit, while in the X (diagonal) basis, Alice (Bob) prepares (measures) states in $\ket{+}={1\over\sqrt{2}}(\ket{H}+\ket{V}),\ket{-}={1\over\sqrt{2}}(\ket{H}-\ket{V})$, also corresponding to 0 and 1 bit. An illustration of this can be found in Fig. \ref{fig:intro_BB84}.
	\item After the transmission of signals is complete, Alice and Bob then announce the bases they chose, and keep only events where they chose the same basis (which is a process called \textit{sifting}). After the sifting, Alice and Bob acquire a pair of raw keys (random bits).
	\item Alice and Bob perform \textit{random sampling} on part of the data, and test the quantum-bit-error-rate (QBER). Alice and Bob abort if the QBER exceeds a certain threshold.
	\item They then perform \textit{error correction} on the data (raw key) to ensure that they have the same random string with a high probability. In this process, some additional information is leaked to an eavesdropper.
	\item Lastly, they perform \textit{privacy amplification} on the error-corrected keys (which usually takes the form of a universal hashing function), that discards part of the keys, to reduce the information an eavesdropper can hold to close to zero. The amount of information an eavesdropper has (and subsequently the amount of hashing necessary) is estimated by the aforementioned random sampling of part of the data and the estimated quantum-bit-error-rate (QBER) among the data. 
\end{itemize} 

After the above process, Alice and Bob acquire a pair of random bits (keys) that contains no information that can possibly be obtained by an eavesdropper (i.e. information-theoretically secure). They can then use this pair of random key for e.g. a one-time-pad \cite{onetimepad}, where the sender uses the key to perform a logical XOR operation on any message to encode it, and the receiver uses the same key to perform another XOR operation, decoding the original message. This process is proven to be secure if the random key itself is secure.



\begin{figure}[h]
	\includegraphics[scale=0.35]{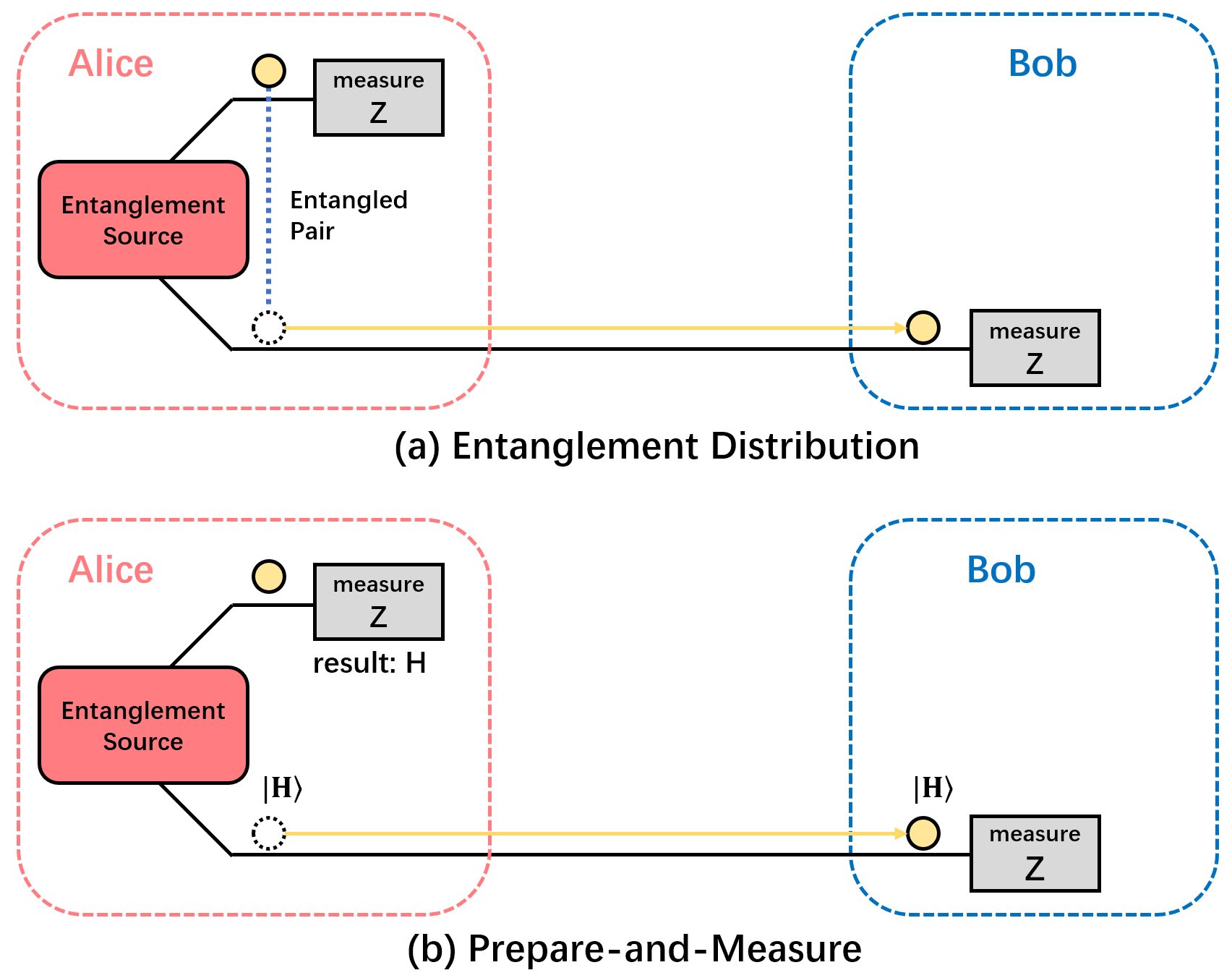}
	\caption{An illustration of the idea behind the security proof of BB84. (a) Alice has an entanglement source that distributes entanglement pairs between Alice and Bob. The pairs are then measured in Z axis or X axis (the latter not shown in the figure). A random sample from each basis can help them quantify the bit error and the phase error rate used for error-correction and privacy amplification. (b) Alice first measures in Z basis (or X basis - not shown in the figure), and then send the qubit - now deterministically in Z basis after the measurement - to Bob who measures it in Z (X) basis. The two cases are in fact equivalent because the measurements are commutable with Eve's potential disturbance of the entanglement pairs as well as Alice and Bob's error correction. Therefore, with a prepare-and-measure scheme like BB84 combined with classical error-correction and hashing, we can achieve the same security as the corresponding entanglement distillation protocol with quantum error correction and privacy amplification. Figure reproduced with modifications based on Fig .1 from Ref. \cite{GLLP}.}
	\label{fig:intro_BB84security}
\end{figure}

There have been subsequent security proofs that rigorously showed the unconditional security of the BB84 protocol against eavesdropping, importantly Refs. \cite{LoChau,Preskill}. The key idea of these proofs is to link a prepare-and-measure scheme (as described above) with an entanglement distribution scheme. Instead of sending out the four states $\ket{H},\ket{V},\ket{+},\ket{-}$, let us imagine that Alice possesses an entanglement source that generates entangled EPR pairs \cite{EPR} of qubits. One of the qubits is kept locally (say, with quantum memory) at Alice's lab, and another qubit is sent to Bob. After Bob receives the other half of the EPR pair, Alice and Bob then ``distill" pure entanglement pairs by performing quantum error correction by measuring the ``bit error rate" (X basis QBER) and the ``phase error rate" (Z basis QBER) and performing correction operations. Once pure entanglement pairs are obtained, they can then measure either both in X basis or Z basis (or in different bases, in which case the event is discarded during sifting) to obtain the key. Since pure entanglement pairs do not carry basis nor bit information before measurement, such a scheme gives them unconditionally secure key pairs.


The Lo-Chau proof \cite{LoChau}, which is built on \cite{Deusch}, proposed the above idea of entanglement distillation, and showed the unconditional security of QKD when Alice and Bob possess quantum computers and are able to perform quantum error correction. Furthermore, the Shor-Preskill proof relaxed the requirement by allowing Alice and Bob to perform \textit{classical} error-correction and hashing (privacy amplification), to gain the same unconditional security. A key observation is that the above entanglement distribution scheme is equivalent to the case where Alice \textit{first measures} in either Z or X basis, and sends the (now no longer in entanglement) qubit to Bob, who does another measurement on the qubit in the same basis, i.e. the entanglement distribution scheme is equivalent to the prepare-and-measure scheme. This is because the local measurement at Alice is commutable with the error corrections or Eve's potential perturbations on the qubit sent to Bob. The quantum error correction is replaced with classical error correction and privacy amplification, based on the Z basis and X basis QBER, which are still respectively called the bit error rate and the phase error rate (because they correspond to the quantum error correction operations, in the imaginary entanglement distribution scenario).

In the Shor-Preskill Proof \cite{Preskill}, it is shown that the secure key rate can be bounded by:

\begin{equation}
	R = 1 - h_2(e_X) - h_2(e_Z)
\end{equation}

\noindent where $h_2(x)= -xlog_2(x)-(1-x)log_2(1-x)$ is the binary entropy function, and $e_X,e_Z$ are respectively the bit error rate and the phase error rate.

An illustration of the above security proof is shown in Fig. \ref{fig:intro_BB84security}.



\section{Decoy-State Method}

The contents of this section are based on Refs. \cite{GLLP,decoystate_LMC,decoypractical}.

The Shor-Preskill rate proves unconditional security for BB84 and proposes a lower bound for secret key rate. However, the protocol assumes a perfect single-photon source. In practice, there are two types of commonly used sources for QKD: (1) entanglement source based on spontaneous parametric down-conversion (SPDC) (which can either be used for entanglement-based QKD protocols such as Ekert 91 \cite{E91}, or for BB84 and its variants by allowing the sender to measure one of the two entangled photons generated, creating a \textit{heralded} single photon source). However, entanglement sources have a low generation rate of single photons, and another type of source (2) weak coherent pulse (WCP) source is commonly used due to its ease of implementation and high repetition rate. A WCP source is simply a heavily attenuated laser, which outputs a coherent state 

\begin{equation}
	\ket{\alpha} = \sum_n e^{-|\alpha|^2/2} {{\alpha^n}\over{\sqrt{n!}}} \ket{n}
\end{equation}

\noindent which is a superposition of photon number states (Fock states) $\ket{n}$. To use a WCP source, Alice needs to make the coherent state \textit{phase randomized}, in order to minimize the phase correlation between adjacent pulses \cite{decoypractical}. In some cases, Alice actively randomizes the phase of her signals (e.g. with a phase modulator) \cite{phase_randomization}. In some other experiments, phase randomization is achieved by switching a laser above and below its lasing threshold (i.e. gain-switched laser) during each signal generation, such as demonstrated in \cite{phase_randomization2}. When we perform a phase randomization on the state $\ket{\alpha}$ (i.e. integrate over $\theta$ in $\ket{\alpha e^{-i\theta}}$) we can obtain

\begin{equation}
\ket{\alpha}\bra{\alpha} = \sum_n e^{-|\alpha|^2} {{|\alpha|^{2n}}\over{{n!}}} \ket{n}\bra{n} = \sum_n e^{-\mu} {{\mu^{n}}\over{{n!}}} \ket{n}\bra{n}
\end{equation}

\noindent where the intensity $\mu=|\alpha|^2$. This means that a phase-randomized WCP source can be viewed of as a \textit{classical} mixture of pulses containing $n$ photons, with a classical Poissonian probabilistic distribution described by $\mu$ and $n$. The security of discrete phase randomization (where the phase randomly takes several discrete values instead of sampling continuously between $[0,2\pi)$) for WCP sources has also been proven \cite{discretephase}.

While a WCP source with a reasonably low intensity $\mu$ has a fair amount of probability to send single photons, it also has some probabilities of sending a vacuum state or multiple photons, the latter of which would pose a security risk to QKD, due to an eavesdropper being able to perform a photon-number splitting attack \cite{PNS}. In such an attack, since all photons in a multi-photon pulse undergo the same e.g. polarization or phase modulation processes, the eavesdropper can simply split of one of these multiphotons, and send the remaining photon (or photons) to Bob. In this way, Eve has a perfect copy of the signal sent to Bob, resulting in a security breach. Note that this doesn't violate the no-cloning theorem, as multiple copies of the same information are prepared in the first place, while the theorem only states that unknown quantum information on a single qubit (i.e. single-photon) cannot be perfectly cloned. Therefore, we can see that the multiphoton proportion in a WCP source poses a security risk.


The paper by Gottesman-Lo-Lutkenhaus-Preskill (GLLP) \cite{GLLP} proves that, even for imperfect sources with multiphoton components, it is possible to modify the upper bound for secure key rate by considering \text{only the single photon components}, in the form of (from Eq. (50) in \cite{GLLP}):

\begin{equation}
 R = max\left\{ (1-\Delta)\left[1-h_2\left({{\delta}\over{1-\Delta}}\right)\right] - h_2(\delta) ,0\right\}
\end{equation}

\noindent where $\Delta={p_M \over p_D}$ is the proportion of multiphoton contribution among detected photons, and $\delta$ is the observed bit error rate.

To accurately estimate the proportion of single photons and the phase-error rate among single photons, in Refs. \cite{decoystate_Hwang,decoystate_LMC,decoystate_Wang}, a ``decoy-state" method is proposed. The key idea is that, instead of constantly using one intensity $\mu$ for the WCP source, Alice can randomly switch between multiple intensity settings $\{\mu,\nu,\omega, \dots\}$. The different intensities allow different Poissonian photon number distributions. Combining the data from multiple intensity settings (decoy states), we can obtain linear constraints from the observables, gain $Q$ and error-gain $QE$ (the probability of getting a count or an error, respectively, for each signal sent), for each intensity setting. From the gains we can obtain:

\begin{equation}
\begin{aligned}
Q_\mu &= \sum_{n} e^{-\mu} {{\mu^n}\over{{n!}}} Y_n = \sum_{n}P_{n}^\mu Y_n\\
Q_\nu &= \sum_{n} e^{-\nu} {{\nu^n}\over{{n!}}} Y_n = \sum_{n}P_{n}^\nu Y_n\\
Q_\omega &= \sum_{n} e^{-\omega} {{\omega^2}\over{{n!}}} Y_n = \sum_{n}P_{n}^\omega Y_n\\
&\dots
\end{aligned}
\end{equation}

\noindent and from the error-gains:

\begin{equation}
\begin{aligned}
Q_\mu E_\mu &= \sum_{n} e^{-\mu} {{\mu^n}\over{{n!}}} e_n Y_n = \sum_{n}P_{n}^\mu e_n Y_n\\
Q_\nu E_\nu &= \sum_{n} e^{-\nu} {{\nu^n}\over{{n!}}} e_n Y_n = \sum_{n}P_{n}^\nu e_n Y_n\\
Q_\omega E_\omega &= \sum_{n} e^{-\omega} {{\omega^2}\over{{n!}}} e_n Y_n = \sum_{n}P_{n}^\omega e_n Y_n\\
&\dots
\end{aligned}
\end{equation}

Note that, importantly, a key assumption is made here, that  

\begin{equation}
\begin{aligned}
Y_n (\mu) &= Y_n (\nu) = Y_n (\omega) = ... \\
e_n (\mu) &= e_n (\nu) = e_n (\omega) = ...
\end{aligned}
\end{equation}

\noindent which physically means that the eavesdropper does not know which intensity setting a given pulse comes from (as she only has the information of the photon number $n$ in the pulse, not the intensity $\{\mu,\nu,\omega, \dots\}$). \footnote{Note that, by ``intensity", we actually mean the intensity setting of a laser source - which determines the probability distribution of the photon numbers in a pulse being sent. Of course, a phase-randomized WCP source sends a classical mixture of photon number states, and we assume that Eve can make e.g. a quantum nondemolition measurement on the photon number of a pulse, and obtain the EM field intensity of the pulse. However, she cannot obtain information on the Poissonian probability distribution of photon numbers in e.g. Alice's laser source to obtain values of $\{\mu,\nu,\omega, \dots\}$, which is the cornerstone for the security of decoy-state analysis. By convention, throughout the rest of the text, in the context of WCP sources we will use the phrases ``intensity setting" and ``intensity" interchangeably.}This is because the process of generating a pulse with photon number $n$ from a probability distribution $P(n)$ is a \textit{Markov process}, which means that the process is memoryless, and the pulse itself with $n$ photons has no information of the probability distribution $P(n)$ that generated it. This is the keystone to the security of decoy-state method. 

The above equations constitute two sets of linear programs, with $Y_n$ and $e_nY_n$ as variables. If Alice is allowed to use an infinite amount of intensity settings (commonly denoted as the ``asymptotic" case), she will be able to perfectly estimate all $Y_n$ and $e_nY_n$. Using the GLLP key rate formula along with $Y_1,e_1$ estimated from the linear program, we can obtain the key rate for decoy-state BB84 with WCP source, as in Ref. \cite{decoypractical} Eq. 42:

\begin{equation}
	R = q(P_1^\mu Y_1 (1-h_2(e_1)) - f_e Q_\mu h_2(E_\mu))
\end{equation}

\noindent where we assume that $\mu$ is the signal state, $P_1^\mu=\mu e^{-\mu}$, $q$ is the basis probability (1/2 for standard BB84 and $\approx 1$ for efficient BB84), $h_2$ is the binary entropy function, $Y_1,e_1$ are the single-photon contributions estimated from the linear program, and $f_e$ is the error-correction efficiency (a constant larger than 1, the smaller the better), which depends on the error-correction code used.

In reality, Alice can only use a finite number of decoy states. In such a case, the bounds obtained from the linear programs will not be perfect, and Alice and Bob would only obtain lower and upper bounds on the variables $Y_n$ and $e_nY_n$. The single-photon contributions $Y_1$ and $e_1$ in the key rate expression need to be replaced by $Y_1^L$ and $e_1^U$.  

In practice, a 3-decoy protocol with $\{\mu,\nu,\omega, \dots\}$ settings is commonly used - and in fact it is shown in \cite{decoypractical} that using these three decoy states already generate a key rate very close to the infinite-decoy asymptotic key rate, and using more decoys will bring little further benefit. For a 3-decoy protocol, in Ref. \cite{decoypractical} Eqs. 34, 37, analytical forms for $Y_1^L$ and $e_1^U$ have been obtained:

\begin{equation}
\begin{aligned}
Y_1^L &= {{\mu}\over{\mu\nu - \nu^2}}(Q_\nu e^\nu - Q_\mu e^\mu {{\nu^2}\over{\mu^2}} - {{\mu^2-\nu^2}\over{\mu^2}}Y_0)\\
e_1^U &= {{Q_\nu E_\nu e^\nu - e_0 Y_0}\over{Y_1^L \nu}}
\end{aligned}
\end{equation}

\noindent where $Y_0$ equals the dark count rate from detectors, and $e_0={1\over 2}$ is the error for vacuum state.

\section{Measurement-Device-Independent (MDI) QKD}

\begin{figure}[h]
	\includegraphics[scale=0.32]{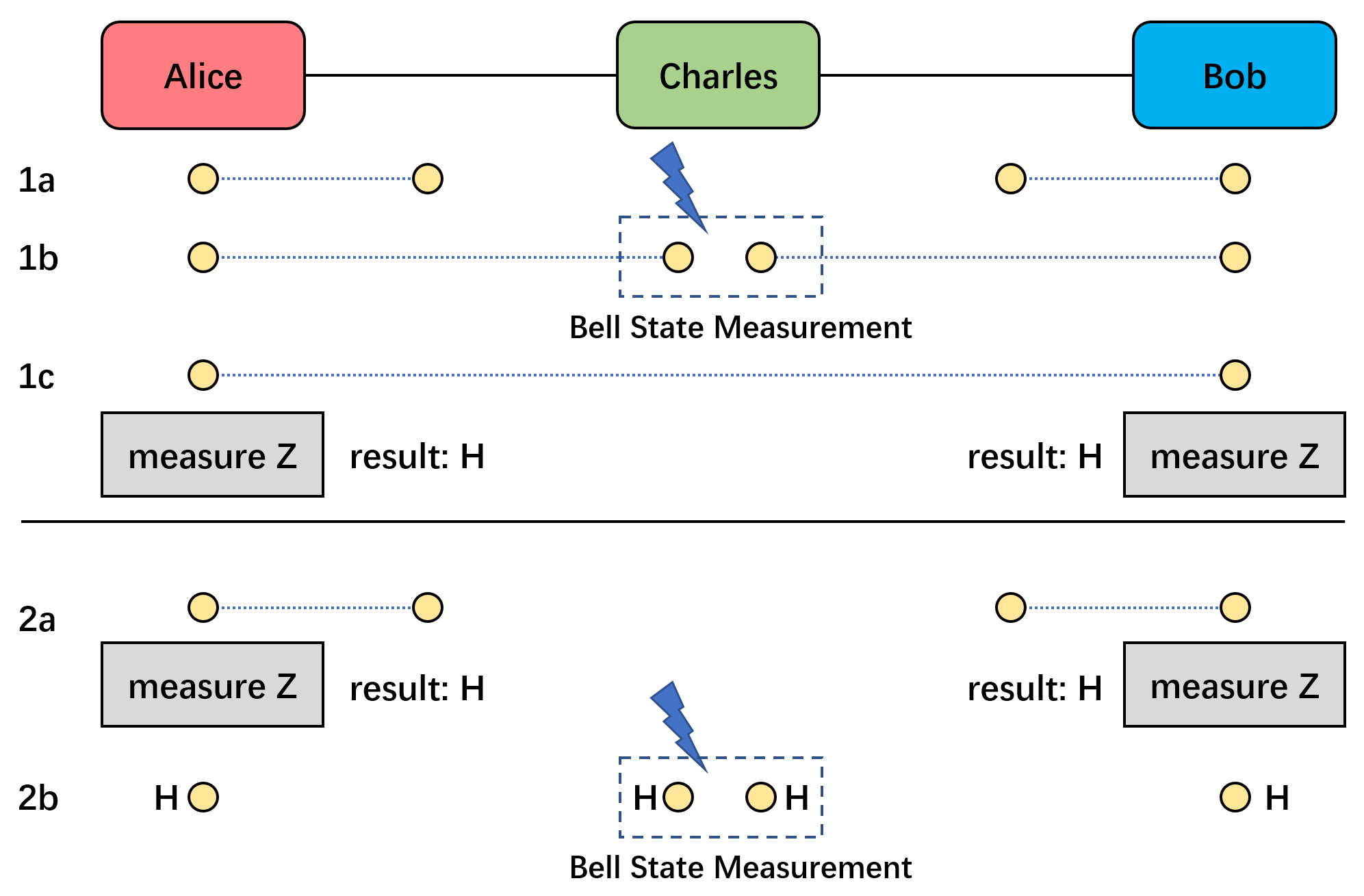}
	\caption{An illustration of the setup of MDI-QKD and how its security is proven. The protocol is equivalent to an entanglement distribution protocol (in 1a-1c) where Alice and Bob each keep half an entangled pair locally and send the other half to Charles, who performs a Bell State Measurement. By entanglement swapping, Alice's and Bob's local qubits are now entangled, and they can measure locally in X or Z basis (here in the illustration it's the Z basis) to obtain bit (Z) and phase (X) error rate. However, in practice, the Bell State Measurement and Alice and Bob's local measurements can be time-reversed, making the protocol a prepare-and-measure setup (2a-2b) where Alice and Bob can simply send signals encoded in corresponding $\ket{H},\ket{V}$ and $\ket{+},\ket{-}$ states instead of part of an entangled pair, but the security of the protocol is unchanged. Figure reproduced with modifications based on Fig. 3.1 from Ref. \cite{feihu_thesis}, while the original setup is based on Fig .1 from Ref. \cite{time_reverse_EPR1}.}
	\label{fig:intro_MDIQKD}
\end{figure}

This section is based on Refs. \cite{mdiqkd,mdipractical,mdifourintensity}.

\subsection{Protocol}

As described in Chapter 1, while QKD is theoretically proven to be secure, practical components in QKD systems allow for the existence of side channels, and there have been many hacking attempts \cite{blinding,timeshift,phaseremapping,fieldhacking,detectordeadtime,devicecalibration} that attack practical QKD systems. 

An ultimate solution to this is to perform Device-Independent (DI) QKD \cite{diqkd}, which is based on the testing of Bell's inequality and does not make any assumptions on the devices. However, a Bell test is difficult to implement in reality. Hence, up so far DI-QKD remains mostly of theoretical interest only. Also, DI-QKD is vulnerable to memory attacks \cite{memoryattack} and covert channels \cite{covert}. 

On the other hand, a practical solution that is feasible with off-the-shelf components and high repetition-rate sources is Measurement-Device-Independent (MDI) QKD \cite{mdiqkd}. The motivation of MDI-QKD is that, among the many types of hacking attempts, a large number of attacks target the side channels in detectors, which become the Achilles' Heel of practical QKD systems. MDI-QKD is a scheme that eliminates all \textit{detector side channels} (a review on hacking attacks, particularly ones targeting detectors, e.g. time-shift attack, can be found in e.g. Refs. \cite{MDI_review,reviewQKD2019_arxiv}), hence protecting a QKD system against a majority of the types of attacks, while still maintaining a satisfactory level of key rate and requiring only practical off-the-shelf components. This makes MDI-QKD a good balance point between the security in practice and ease of implementation.

The key idea of MDI-QKD is that, instead of Alice sending signals to Bob, they can both send signals to a third party, Charles, who performs a Bell measurement that tests the parity (and not the bits) of the incoming signals. More specifically, the procedures of the protocol can be described as follows:

\begin{itemize}
\item Alice and Bob each randomly switches between two bases X and Z to send signals (following Ref. \cite{mdiqkd}, let us suppose Alice and Bob encodes bits in polarization), sending $\ket{H},\ket{V}$ in the Z basis and $\ket{+}={1 \over \sqrt{2}}(\ket{H}+\ket{V}),\ket{-}={1 \over \sqrt{2}}(\ket{H}-\ket{V})$ in the X basis.
\item Charles sets up a beam-splitter (BS), and two pairs of detectors $c_H,c_V,d_H,d_V$, each pair behind a polarizing beam-splitter (PBS) along the Z basis $H,V$, and measures the incoming signals from Alice and Bob.
\item Charles announces the Bell test results publicly. Only events where exactly one $H$ and one $V$ detector clicks simultaneously (behind the same PBS or each from one PBS) are announced as a successful event. The cases $c_Hd_V$ and $c_Vd_H$ correspond to the Bell state $\ket{\psi^-}={1\over \sqrt{2}}(\ket{HV}-\ket{VH})={1\over \sqrt{2}}(\ket{-+}-\ket{+-})$, while the cases $c_Hc_V$ and $d_Hd_V$ correspond to the Bell state $\ket{\psi^+}={1\over \sqrt{2}}(\ket{HV}+\ket{VH})={1\over \sqrt{2}}(\ket{++}-\ket{--})$. Data whose patterns of clicked detectors are not among the above four cases, and data where fewer (or more) than two detectors click are discarded. Note that, in the MDI-QKD scheme, only two states $\ket{\psi^+},\ket{\psi^-}$ out of the four Bell states can be distinguished.
\item Alice and Bob announce their bases choices and sift the results. Based on Charles's announced results, they can determine the parity of their bits and generate key. The details of how detection results generate key can be seen in Table 2.1.
\item As in BB84, they can then perform error-correction, and sample part of the results for the gain and QBER data and perform privacy amplification to get the final key.
\end{itemize} 

A more detailed illustration of the MDI-QKD experimental setup is shown in Chapter 5, Fig. \ref{fig:mdiqkd}.


\begin{table}[t]
	
	\caption{Key generation for MDI-QKD. This is an expanded version based on Table I in Ref. \cite{mdiqkd}. According to the protocol, Alice and Bob need to flip one of their bits before generating the raw key, unless they sent in X basis and $\ket{\psi^+}$ is detected. This table describes whether Alice and Bob can generate the correct key based on their sent signals and Charles' announced event $\ket{\psi^-}$ and $\ket{\psi^+}$ (other detector events are discarded). The ``-" signs are impossible combinations for sent signals versus detector events in the ideal case (although if there is misalignment in the system, or if dark count rate is non-negligible, events might be detected in these cases, which would contribute to the QBER).} 
	\begin{center}
		\begin{tabular}{cccc}		
			\hline \hline
			Basis & Alice \& Bob & $\ket{\psi^-}$ & $\ket{\psi^+}$ \\
			\hline
			Z & HV or VH & key (flip) & key (flip) \\
			Z & HH or VV & - & - \\
			\hline
			X & +- or -+ & key (flip) & - \\
			X & ++ or -- & - & key (no flip)\\
			\hline \hline
		\end{tabular}
	\end{center}
	
\end{table}

Security-wise, MDI-QKD is equivalent to an entanglement-distribution scenario (illustrated in Fig. \ref{fig:intro_MDIQKD}): Assume that Alice and Bob each possess an entanglement pair. They can each send half of their entanglement pair to Charles, who performs a Bell measurement on the two incoming qubits. This constitutes an \textit{entanglement swapping} operation, where the two qubits sent to Charles are mapped onto a Bell state, and the two remaining qubits held by Alice and Bob are now entangled. Since Alice and Bob now share an entanglement pair, they can measure the pair in X or Z basis to obtain a pair of bits with perfect security. 

A key point used in Ref. \cite{mdiqkd} is that, since Charles measurement (on the qubits sent to Charles) and Alice and Bob's measurements (on the local qubits remaining in their labs) commute, the above scenario is equivalent to the case where Alice and Bob first measure their local qubits in X or Z basis, and send the other (now no longer in entanglement) qubit to Charles, i.e. a \textit{time-reversed scheme}.

Moreover, Alice and Bob do not even need to possess entanglement pairs to begin with. They can simply prepare random qubits in X or Z basis (among the four states $\ket{H},\ket{V},\ket{+},\ket{-}$) and send them to Charles for a Bell state measurement. The qubits that are sent out are in the same states as the scenario above (where they start out with entanglement pairs but measure their local qubits before sending). This means that, MDI-QKD can be reduced to a prepare-and-measure scheme, which is much easier to implement.

The idea of a time-reversed EPR scheme was first proposed in Ref. \cite{time_reverse_EPR1}, and Ref. \cite{time_reverse_EPR2} provided a security proof. However, such a scheme was not practical when first proposed, and had rather low key rate; thus, it received little attention in the community. It was not until over a decade later when Ref. \cite{mdiqkd} added the ingredient of decoy-state technique (which will be discussed in the next subsection), and made use of the important known fact that a Bell-state measurement (BSM) can be easier performed by standard linear optics components (such as beam-splitters and polarizing beam-splitters) and threshold single-photon detectors, that the scheme was made highly practical. Ref. \cite{mdiqkd} performed explicit calculation and showed that MDI-QKD can have very high key rate, and the MDI-QKD protocol subsequently gained widespread attention in the QKD community.

\subsection{Decoy-State MDI-QKD}

Practically, similar to decoy-state BB84, MDI-QKD is also compatible with decoy-state analysis, which enables the use of WCP sources instead of single photon sources. The difference here is that, since both Alice and Bob send signals now, each of them needs to choose between different levels of intensities $\mu_i,\mu_j$, and the linear equations contain variables of $Y_{mn}, Y_{mn}e_{mn}$ (instead of $Y_n,Y_ne_n$), because one needs to consider the cases where $m,n$ photons are respectively sent from Alice and Bob to Charles through the two channels.

\begin{equation}
\begin{aligned}
Q_{\mu_i\mu_j} &= \sum_{m,n} (e^{-\mu_i} {{{\mu_i}^m}\over{{m!}}})(e^{-\mu_j} {{{\mu_j}^m}\over{{n!}}}) Y_{mn} = \sum_{m,n}P_{m}^{\mu_i} P_{n}^{\mu_j} Y_{mn}\\
&\dots
\end{aligned}
\end{equation}

\noindent and from the error-gains:

\begin{equation}
\begin{aligned}
Q_{\mu_i\mu_j} E_{\mu_i\mu_j} &= \sum_{m,n} (e^{-\mu_i} {{{\mu_i}^m}\over{{m!}}})(e^{-\mu_j} {{{\mu_j}^m}\over{{n!}}}) e_{mn}Y_{mn} = \sum_{m,n}P_{m}^{\mu_i} P_{n}^{\mu_j} e_{mn}Y_{mn}\\
&\dots
\end{aligned}
\end{equation}

Again, a key assumption is made here, that all $Y_{m,n} (\mu_i,\mu_j)$ (and all $e_{m,n} (\mu_i,\mu_j)$) are values independent of the choice of $\mu_i,\mu_j$

Similar to BB84, one can apply again the GLLP key rate formula and incorporate the single-photon contributions,

\begin{equation}
R = P_Z^2((P_1^{\mu})^2 Y_{11}^Z (1-h_2(e_{11}^X)) - f_e Q_{\mu\mu}^Z h_2(E_{\mu\mu}^Z))
\end{equation}

\noindent where we assume Alice and Bob both use intensity $\mu$ as the signal state and Z as the encoding basis. $P_1^\mu = \mu e^{-\mu}$ is the probability of sending single photons, $Y_{11}^Z,e_{11}^X$ are the single-photon (both Alice and Bob sent single-photon) yield and QBER, $f_e$ is again the error-correction probability, and $P_Z$ is the basis choice probability ($1/2$ for equal probability of choosing $X,Z$ bases, and $\approx 1$ for ``efficient" case where $Z$ basis is chosen with close to 1 probability).

Again, in practice, Alice and Bob can only choose a finite number of decoy states. This results in an imperfect (but close to asymptotic) bound on the single-photon contributions, $Y_{11}^{Z,L},e_{11}^{X,U}$. A good choice is again where Alice and Bob each choose 3-decoys, $\{\mu,\nu,\omega\}$. The analytical forms for $Y_{11}^{Z,L},e_{11}^{X,U}$ have been obtained in Refs. \cite{mdipractical,mdifourintensity}:

\begin{equation}
\begin{aligned}
Y_{11}^{Z,L} &= {{1}\over{\mu - \nu}}({{\mu}\over{\nu^2}}Q_{\nu\nu}^{Z,M1}  - {{\nu}\over{\mu^2}}Q_{\nu\nu}^{Z,M2})\\
Y_{11}^{X,L} &= {{1}\over{\mu - \nu}}({{\mu}\over{\nu^2}}Q_{\nu\nu}^{X,M1}  - {{\nu}\over{\mu^2}}Q_{\nu\nu}^{X,M2})\\
e_{11}^{X,U} &= {1\over {\nu^2 Y_{11}^{X,L}}}(e^{2\nu}Q_{\nu\nu}^XE_{\nu\nu}^X-e^{\nu}Q_{\nu\omega}^XE_{\nu\omega}^X -e^{\nu}Q_{\omega\nu}^XE_{\omega\nu}^X+Q_{\omega\omega}^XE_{\omega\omega}^X)
\end{aligned}
\end{equation}

\noindent where

\begin{equation}
\begin{aligned}
Q_{\nu\nu}^{Z,M1} &= e^{2\nu}Q_{\nu\nu}^Z - e^{\nu}Q_{\nu\omega}^Z - e^{\nu}Q_{\omega\nu}^Z + Q_{\omega\omega}^Z\\
Q_{\mu\mu}^{Z,M2} &= e^{2\mu}Q_{\mu\mu}^Z - e^{\mu}Q_{\mu\omega}^Z - e^{\mu}Q_{\omega\mu}^Z + Q_{\omega\omega}^Z\\
Q_{\nu\nu}^{X,M1} &= e^{2\nu}Q_{\nu\nu}^X - e^{\nu}Q_{\nu\omega}^X - e^{\nu}Q_{\omega\nu}^X + Q_{\omega\omega}^X\\
Q_{\mu\mu}^{X,M2} &= e^{2\mu}Q_{\mu\mu}^X - e^{\mu}Q_{\mu\omega}^X - e^{\mu}Q_{\omega\mu}^X + Q_{\omega\omega}^X\\
\end{aligned}
\end{equation}

A later proposal in Ref. \cite{mdifourintensity} made the observation that in fact $Y_{11}^{Z,L}=Y_{11}^{X,L}$ (since the pair of single photons are in Bell states and basis-independent). This means that Alice and Bob in fact only need to perform decoy-state analysis in one basis (X basis) to obtain $Y_{11}^{X,L},e_{11}^{X,U}$, and the entire Z basis can be used for key generation, and only one signal intensity $s$ is required. Furthermore, this signal intensity $s$ can be different from $\mu$ as Z basis is decoupled from the X basis. This constitutes a ``four-intensity" MDI-QKD protocol, which (as we will describe in Section 2.5) not only improves finite-size analysis, but also is a useful construction for scenarios where channels are asymmetric. We will further analyze this point and discuss our findings in Chapter 5.

\section{Twin-Field (TF) QKD}

This section is based on Refs. \cite{TFQKD, simpleTFQKD}. Parts of the overview of different proposals of TF-QKD is also based on the review paper Ref. \cite{reviewQKD2019}.

\begin{figure}[h]
	\includegraphics[scale=0.31]{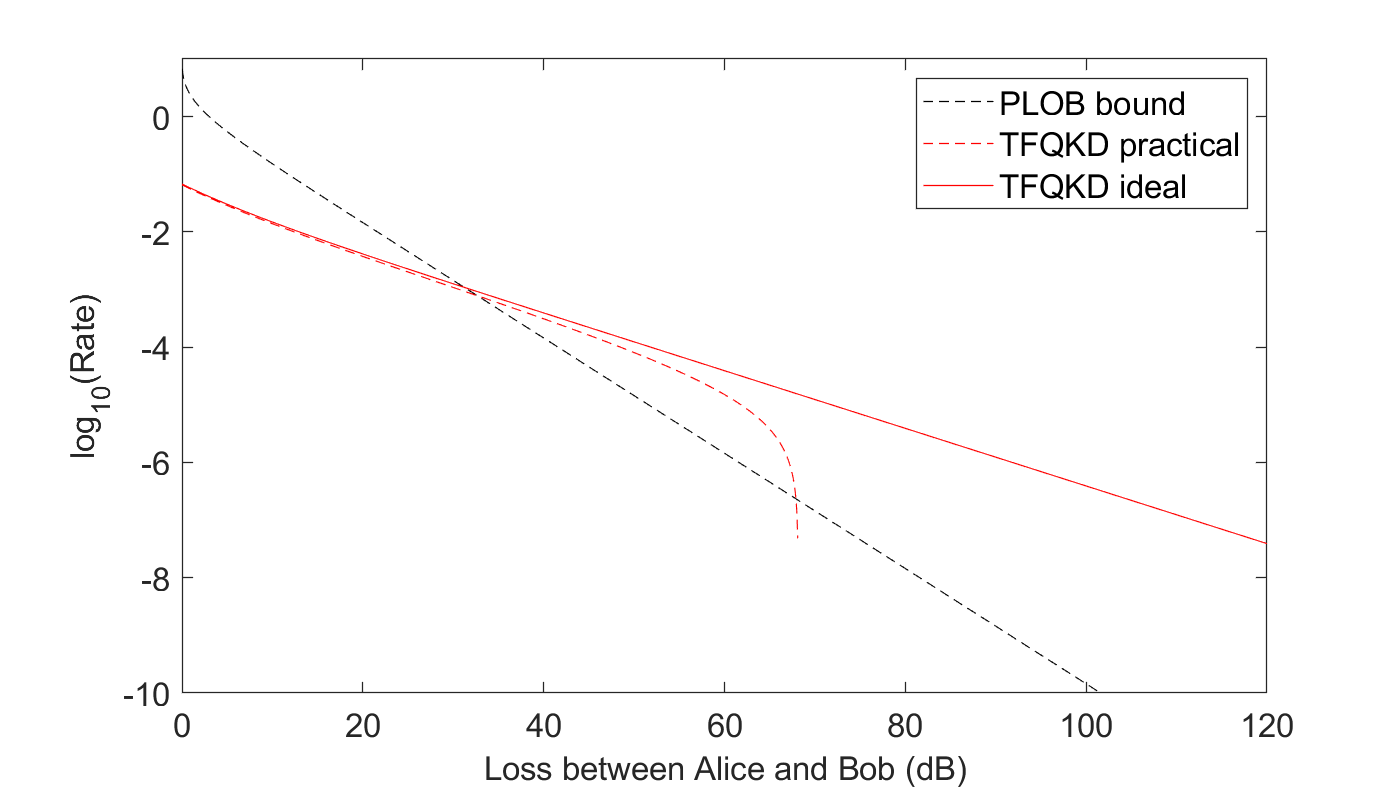}
	\caption{A comparison of the PLOB bound and the key rates of TFQKD. Here we use the simple TF-QKD protocol \cite{simpleTFQKD}, and simulate the key rate for an ideal case and a case with practical parameters. For the ideal case, we assume no dark count or misalignment error, and the only error comes from imperfect estimation of phase error rate due to multiphoton contributions in WCP sources (we also assume an infinite number of decoy states and infinite data). For the practical case, we set a misalignment of $0.09$ rad, and a dark count rate of $7\times 10^{-7}$. In both cases, we fix the signal intensity as $\mu=0.05$. As can be seen, TF-QKD is able to beat the PLOB bound in both ideal and practical scenarios.}
	\label{fig:intro_TFQKD}
\end{figure}

\subsection{Motivation}

One main challenge of QKD is the maximum distance over which the quantum signals can be sent in order to establish secure communications. Because fundamentally, QKD relies on the sending and receiving of single-photons (and that it cannot be cloned), its secure key rate is limited by the transmittance of the quantum channel (i.e. the probability of the photon being able to pass through the channel and reach a receiver). There are papers that study the fundamental upper bound of the distance versus key rate trade-offs for QKD, such as the TGW \cite{TGW} bound and PLOB \cite{PLOB} bound (also called ``linear bounds"), which state that the maximum QKD key rate scales linearly with transmittance in the channel (here we show the PLOB \cite{PLOB} bound):

\begin{equation}
	R \leq -log_2(1-\eta_{AB}) = O(\eta_{AB})
\end{equation}

\noindent where $\eta_{AB}$ is the transmittance between Alice and Bob. To put things into perspective, each 50km of standard fibre (10dB loss in transmittance) will cause the key rate to drop by one order-of-magnitude.

One solution to this problem is to use classical, trusted relays. That is, Alice performs QKD and generates keys between Alice-Relay, and Bob performs QKD and generates keys between Relay-Bob. The relay can then perform an XOR on the keys and announce the combined keys, which can allow Alice and Bob to recover a mutual key. (Effectively, this can be viewed as the relay creating a secure channel between itself and Bob with One-Time-Pad using the Bob-Relay key from QKD, and securely sending Alice's key to Bob over this secure channel.) This operation can be repeated over many relays, indefinitely extending the distance of secure communications. There have been several demonstrations of quantum networks using trusted relays, some even over thousands of kilometres \cite{QKDbackbonenetwork}. However, a major problem of such an approach is that the signals stop being quantum at the relays, therefore requiring the crucial assumption that all relays must be trusted, where one compromised link could lead to a breach of security.

An alternative is to use quantum repeaters, which are able to generate and store entanglement pairs. Alice and Bob can use entanglement pairs to perform teleportation of signals (which doesn't require the quantum signals to physically pass through the channel, given previously stored entanglement pairs). Multiple relays can use entanglement swapping to extend the range of entanglement pairs, which extends the maximum distance over which Alice and Bob can securely establish communication. However, quantum repeaters require quantum memories, which are still at a stage of infancy and are not practical with current technology. There are also proposals that aim at avoiding the use of quantum memory, such as ``all-photonic quantum repeater" \cite{all_photonic}, which uses a pre-generated highly entangled photon state (called a cluster state), Bell measurements and post-selection to establish connections between itself and Alice and Bob. However, cluster states are experimentally difficult to implement too, and an all-photon quantum repeater has yet to be experimentally demonstrated.

MDI-QKD allows untrusted relays. Here Alice and Bob both act as senders and a relay is set up in the middle. This eliminates detector side channels and provides better practical security for QKD. However, MDI-QKD does not improve the fundamental rate-distance scaling properties of QKD, as the key rate of MDI-QKD still scales with the total transmittance between Alice and Bob (i.e. the transmittances in the channels Alice-Charles and Charles-Bob are multiplied in the scaling of key rate), because MDI-QKD generates keys from coincidences, where both signals in the two channels have to pass through the channel successfully. Note that, however, in principle MDI-QKD can provide longer maximum distance than say BB84, because it uses coincidences between detectors to generate key, and the dark count rate of the detectors (which ultimately determines the maximum cutoff distance of QKD - keys cannot be generated at the point where the level of signal is so weak as to be comparable to the level of noise from dark counts) has less of an effect on the signal-to-noise ratio, since only coincidences of dark counts contribute to QBER for MDI-QKD. Nonetheless, at such maximum distances the key rate will be extremely low.\\

\subsection{Proposal of TF-QKD}

\begin{figure}[h]
	\includegraphics[scale=0.3]{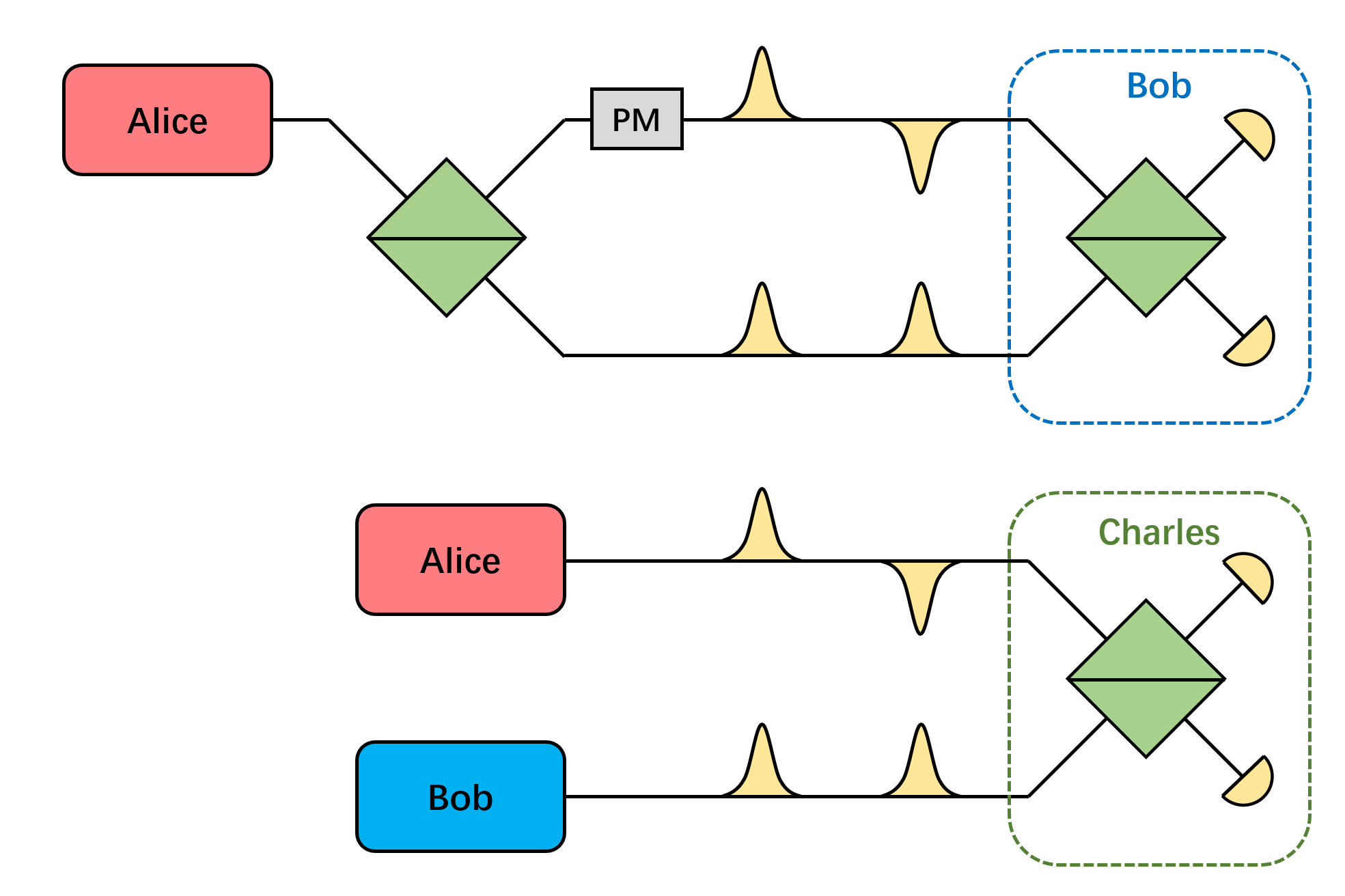}
	\caption{An illustration of the setup of TF-QKD. As can be seen, intuitively, TF-QKD can be thought of as a modified phase-encoding QKD as in (a), but instead of splitting Alice's pulse into signal and reference pulses, here in TF-QKD as in (b), Alice and Bob independently send coherent pulses (based on a shared phase reference), and a third party Charles performs a swap test, which tests the difference between the phases in Alice's and Bob's signals (i.e. just like MDI-QKD, only the parity between Alice's and Bob's bits are announced, but not the values itself, giving TF-QKD a similar kind of measurement-device-independence). Figure reproduced with modifications based on Fig. 2 from Ref. \cite{TFQKD}.}
	\label{fig:intro_TFQKD2}
\end{figure}

Recently, there is a new protocol proposed by Lucamarini et al., called ``Twin-Field" (TF) QKD \cite{TFQKD}, that can surpass the fundamental distance-key rate trade-off described by the linear bound. Furthermore, TF-QKD also retains the measurement-device-independence properties similar to that of MDI-QKD. Because of these striking properties, TF-QKD has gained worldwide attention in the QKD community since its proposal.

Before we describe the TF-QKD protocol, let us consider first a phase-encoding QKD scheme between Alice and Bob. Alice's signal (suppose its a WCP source) is first split into two pulses on different paths by a beam-splitter (usually called a ``signal" pulse and a ``reference" pulse). Alice encodes her bit string by applying phase modulation on the signal pulse (with a phase shift of $0, \pi$ for X basis and ${1 \over 2}\pi, {3 \over 2}\pi$ for Z basis). After the two pulses reach the receiver Bob\footnote{By applying a delay to one of the pulses, the two pulses can be recombined by another beam-splitter to travel in the same fibre channel, but this is equivalent to a time-multiplexing process, and logically the pulses are still in different channels.}, he can again apply a phase modulation to the incoming signal pulse, and combine the pulses with a beam-splitter connected to two detectors. The idea is, if Alice and Bob choose the same basis, depending on whether their phase modulations differ or coincide, the two pulses will have a $0$ or $\pi$ phase difference, and exit the beam splitter at different ports, triggering a different detector. Physically, Alice and Bob together form a Mach-Zehnder interferometer for the incoming signal.

The simple but important observation in TF-QKD is that, instead of splitting off an incoming signal at Alice's beam-splitter, it's possible for \textit{two independent sources} to send signals with matching or un-matching phases $0, \pi$ or ${1 \over 2}\pi, {3 \over 2}\pi$ (provided that they have a common global phase reference), and the two signals will interfere at the receiver, reaching one of the two detectors depending on whether their phases are the same or opposite. In this way, physically the setup is still similar to the above phase-encoding QKD scheme, but the two sources could be two different users, Alice and Bob, sending to a receiver Charles. Since the two sources Alice and Bob don't need to be located together geographically, this protocol effectively doubles the distance between Alice and Bob, with the same key rate scaling as if Alice performs a phase-encoding QKD with the receiver Charles in the middle (over only half the distance between Alice and Bob). That is, the key rate of TF-QKD

\begin{equation}
	R_{TF-QKD} = O(\sqrt{\eta_{AB}})
\end{equation}

Fundamentally, TF-QKD exhibits this advantageous scaling, because it relies on single-photon interference (rather than two-photon Hong-Ou-Mandel interference as in MDI-QKD) to generate the key. For each key bit generated, on average only one signal has passed through either Alice's channel or Bob's channel (and the relay Charles cannot tell which one it came from), but not through both channels. Therefore, only the loss in one of the channels is considered at any time.

One important note is that, in order to establish a global phase reference, in practice TF-QKD requires phase stability between the two quantum signals, one from each of Alice and Bob. This can be experimentally challenging. In contrast, MDI-QKD does not require such phase stability.

\subsection{Variants of TF-QKD and Security Proofs}

However, though promising, there are still some important caveats to such a proposal. The original proposal of TF-QKD \cite{TFQKD} has not provided a rigorous security proof. The main problem is that decoy-state analysis (which is necessary in order to enable the use of practical WCP sources) requires \textit{global phase randomization}. This is not a problem for phase-encoding QKD, as the signal and reference pulses share the same global random phase. However, for TF-QKD, Alice and Bob are at two separate locations. This means that they need to independently perform phase randomization, and later announce the global phase to perform post-selection on cases where they share a close enough global phase. However, the phase and the photon number in a pulse are \textit{incompatible observables}, which cannot be simultaneously measured with certainty. This means that, decoy-state analysis is not compatible with the public announcement of global phase.


Since the original proposal of TF-QKD, there have been several papers that aim at designing a modified TF-QKD protocol for which a complete security proof can be given. One underlying idea for several of these proofs is to divide the signals into two parts - an ``encoding" part that publicly announces the global phase and generates the key (and samples the bit error rate), and another ``testing" part that randomizes the phase, and tries to obtain data that can upper-bound the phase error rate for the encoding part. It is assumed that an eavesdropper cannot tell the encoding part from the testing part, which means that data in the testing part can be used to estimate some \textit{invariant quantity}, that can be used to bound the phase error rate for encoding part. Once the phase error is bounded, combining it with the gain and error of the encoding part, the key rate still takes the form of Shor-Preskill key rate:

\begin{equation}
R = 1 - h_2(e_{bit}) - h_2(e_{phase})
\end{equation}

\noindent where $e_{bit},e_{phase}$ are the bit error and phase error, and the key rate in practice will be multiplied by the gain in the signal state.

There are numerous variants for the TF-QKD protocol, that differ in their choices of ``encoding" and ``testing" phases, as well as the respective usage/absence of phase randomization in the two bases. We give a brief introduction to some representative TF-QKD type protocols in Appendix A.1.

In this thesis, we will focus on the ``simple TF-QKD protocol" \cite{simpleTFQKD}. In Chapter 6 of this thesis, we discuss the security and performance of the simple TF-QKD protocol when channels have asymmetric levels of loss. By making TF-QKD resistant to asymmetric channels, we enable applications of TF-QKD in a fibre network setting, just like for asymmetric MDI-QKD in Chapter 5.





\section{Finite-Size Analysis}

In the above discussions on security proof and decoy-state analysis, we have assumed that the observables (gain, QBER) Alice and Bob obtain in the experiment are always the same as their expected values. However, this is only true when there is an infinite number of signals being sent (the ``asymptotic scenario"). In practice, as the sending of each signal can be considered as a random event (whether the signal reaches the receiver, whether the signal reaches the wrong detector, etc.), statistical fluctuation might cause the overall sum of the random events (i.e. the total counted number of detected signals, and the total error counts) to deviate from the expected value. This might cause an overestimation of the key rate if we directly assume that the expected values such as gain and QBER are equal to these observed values, which will result in a security breach.

A simple model is using a ``standard error analysis" (e.g. in Ref. \cite{decoypractical,MDIAnalytical,mdiparameter,mdifourintensity}): We assume that the random events of each signal being detected (counting towards the gain) and being incorrectly detected (counting towards the QBER) are independently and identically distributed (i.e. ``i.i.d."). From the central limit theorem, the sum of these events (the total counts and error counts) will follow a normal distribution.

If a random variable $n$ follows a normal distribution and has an expected value $\langle n \rangle$, we can bound the probability that the variable $n$ takes a value within a confidence interval within $\gamma$ numbers of standard deviation from the expected value:

\begin{equation}
	P(n<\langle n \rangle-\gamma \langle n \rangle)=P(n>\langle n \rangle+\gamma \langle n \rangle)={1\over 2}[1-erf(\gamma/\sqrt{2})]
\end{equation}

\noindent where we have used the cumulative probability distribution of a normal distribution, and $erf$ is the error-function. Inversely, if we know the observed value $n$, we have the same probability of $\epsilon=erf(\gamma/\sqrt{2})$ that the expected value $\langle n \rangle$ will fall within the confidence interval of $\gamma$ standard deviations near the observable $n$:

\begin{equation}
\begin{aligned}
\underline{n} = n - \gamma \sqrt{n} \leq \langle n \rangle \leq  n + \gamma \sqrt{n} = \overline{n}
\end{aligned}
\end{equation} 

\noindent here we take an estimated value of $\sqrt{n}$ for the standard deviation of variable n.

With the above method of bounding the expected value from observed value, we can apply this to our decoy-state analysis (which calculates the key rate from the \textit{expected values} of the gain and QBER). Let us take MDI-QKD as an example. Assuming that Alice and Bob perform decoy-state analysis and choose intensities $\mu_i,\mu_j$ in the X basis with probabilities $P_{\mu_i},P_{\mu_j}$, with a total number of $N$ signals. We can then obtain the gain $Q_{\mu_i,\mu_j}^X$, error-gain $T_{\mu_i,\mu_j}^X$, and QBER $E_{\mu_i,\mu_j}^X$ as:

\begin{equation}
\begin{aligned}
Q_{\mu_i,\mu_j}^X &= {{n_{\mu_i,\mu_j}^X} \over {NP_{\mu_i}P_{\mu_j}}} \\
T_{\mu_i,\mu_j}^X &= {{m_{\mu_i,\mu_j}^X} \over {NP_{\mu_i}P_{\mu_j}}} \\
E_{\mu_i,\mu_j}^X &= {{T_{\mu_i,\mu_j}^X} \over {Q_{\mu_i,\mu_j}^X}} \\
\end{aligned}
\end{equation}

\noindent where $n_{\mu_i,\mu_j}^X,m_{\mu_i,\mu_j}^X$ are the actually observed counts and error counts. Here the error-gain is defined as:

\begin{equation}
T_{\mu_i,\mu_j}^X = Q_{\mu_i,\mu_j}^X E_{\mu_i,\mu_j}^X
\end{equation}

Now, if we would like to consider finite-size effects by applying standard error analysis, we can upper and lower bound the gain and error gain by applying the confidence interval:

\begin{equation}
\begin{aligned}
\overline{Q_{\mu_i\mu_j}^X}=Q_{\mu_i\mu_j}^X + \gamma \sqrt{Q_{\mu_i\mu_j}^X \over {N P_{\mu_i} P_{\mu_j}}} \\
\underline{Q_{\mu_i\mu_j}^X}=Q_{\mu_i\mu_j}^X - \gamma \sqrt{Q_{\mu_i\mu_j}^X \over {N P_{\mu_i} P_{\mu_j}}} \\
\overline{T_{\mu_i\mu_j}^X}=T_{\mu_i\mu_j}^X + \gamma \sqrt{T_{\mu_i\mu_j}^X \over {N P_{\mu_i} P_{\mu_j}}} \\
\underline{T_{\mu_i\mu_j}^X}=T_{\mu_i\mu_j}^X - \gamma \sqrt{T_{\mu_i\mu_j}^X \over {N P_{\mu_i} P_{\mu_j}}} \\
\end{aligned}
\end{equation}

This loosens the bound in the linear constraints as shown in Section 2.4.2.

\begin{equation}
\begin{aligned}
\underline{Q_{\mu_i\mu_j}^X} &\leq= \sum_{m,n}P_{m}^{\mu_i} P_{n}^{\mu_j} Y_{mn} \leq \overline{Q_{\mu_i\mu_j}^X}\\
\underline{T_{\mu_i\mu_j}^X} &\leq \sum_{m,n}P_{m}^{\mu_i} P_{n}^{\mu_j} e_{mn}Y_{mn} \leq \overline{T_{\mu_i\mu_j}^X}\\
\end{aligned}
\end{equation}

\noindent which will result in a worse estimation for the single-photon statistics (i.e. higher QBER and lower yield) and consequently a lower key rate, in order to secure the protocol when finite-size effects are considered.

Note that, such an analysis only protects the protocol against ``individual attacks" from Eve (i.e. assuming that she performs the same attacks on each signal independently), since it assumes an i.i.d. distribution. In practice, Eve could theoretically store all signals and perform a ``joint attack" on all signals (making the distributions different for each signal). In this case, more general finite-size analysis techniques are required, that relax the aforementioned i.i.d. assumptions. For instance, the Chernoff bound only requires \textit{independent} variables (without having to be identically distributed), and it has been applied to finite-size security analysis for BB84 \cite{finitebb84}, MDI-QKD \cite{mdiChernoff}, as well as TF-QKD \cite{TFfinite1}. Furthermore, a ``composable" security analysis (such as in Ref. \cite{mdiChernoff}) quantifies the success rate of each one of the security estimation process (e.g. estimation of yields and QBER for different photon numbers, the privacy amplification process, etc.) as well as the success rate of error-correction, to provide an overall bound to the probability for correctness and security. 

For simplicity, in most parts of this thesis, if not specified, we will use the standard error analysis described above (secure against individual attacks) when discussing finite-size effects. 

\section{Parameter Optimization}

In this section we will discuss the motivation for parameter optimization in QKD and the algorithms commonly used. This section provides the background knowledge mainly for Chapter 7 (and is also included in the technical details in Appendix C corresponding Chapter 5). If preferred, one may first skip ahead and come back later to this section for references when reading Chapter 7.

In the above sections, we have described the important role decoy-state method plays in various protocols such as BB84, MDI-QKD, and TF-QKD. The choice of intensity values for the decoy states greatly affects the key rate of QKD - their optimal values are determined by various factors: the channel loss and the background noise or detector dark counts (which together determine the signal-to-noise ratio), the basis misalignment, and the asymmetry between channels for MDI-QKD and TF-QKD with two channels (which we will discuss more in Chapters 5 and 6). A good choice of the set of intensities can provide a good signal gain (hence key generation rate) while also ensuring low values for the observed bit error rate and the estimated phase error rate.

Moreover, as described in Section 2.6, the finite-size effect is also an important limiting factor for the key rate of QKD in practice. When considering finite-size effects, one must carefully choose the probabilities of sending each signal (e.g. too little decoy-state data will cause statistical fluctuation to increase and increase the estimated phase error rate, while too little signal data will cause a low key generation rate, thus creating a trade-off between different states). Such a choice of probabilities for intensity settings \footnote{Additionally, sometimes one also needs to optimize the basis choice probability. For instance, for BB84, if one assumes infinite data size, it is possible to set almost \textit{all data} to Z basis and a fraction of data to X basis, as in efficient BB84, while in the finite-size scenario, e.g. Ref. \cite{finitebb84}, probability of basis choice is one of the parameters to be optimized. The exceptions are some protocols where intensity choice implies basis choice, such as for MDI-QKD in Ref.\cite{mdifourintensity,this_asymMDI} (for TF-QKD in Ref.\cite{simpleTFQKD,this_asymTF}), Z (X) basis uses a different signal intensity setting from the decoy states in X (Z) basis, so choosing signal intensity automatically means choosing Z (X) basis.} is very important in the finite-data scenario, where a non-optimal set of probabilities usually results in low or even zero key rate. This is especially important when data size is small, e.g. $10^{12}$ or fewer total pulses sent, which is a common data size for systems running at a clock speed near 100MHz, over the course of a few hours. Free-space systems usually have even less data (such as $10^{10}$) due to the very limited communication time over e.g. minutes.

Therefore, generally for QKD using decoy-state method and considering finite-size effects, users would need to optimize over both the values of intensity settings, and the probabilities of using each setting. Of course, there are other user parameters that can be optimized in QKD, such as the number of decoy states (which are discussed in Refs. \cite{QKD1,QKD2,QKD3,DecoyChau} for BB84. We also briefly discuss this for MDI-QKD in Appendix C and in Chapter 5, which corresponds to our paper Ref. \cite{this_asymMDI}), but in the context of this thesis, unless specified, by ``parameter optimization" we will be denoting such an optimization of intensity and probability settings.

Parameter optimization can be considered a problem of searching for a global maximum point over a given parameter space that maximizes a given function (key rate versus parameters). It is usually a rather computationally intensive task, due to the large size of the search space. For instance, for MDI-QKD (suppose Alice and Bob use the same parameter settings and use the 4-intensity protocol in Ref. \cite{mdifourintensity}), there are six parameters that need optimizing:

\begin{equation}
	\{s,\mu,\nu,P_s,P_\mu,P_\nu\}
\end{equation}

\noindent and the key rate is a function of these parameters:

\begin{equation}
	R(s,\mu,\nu,P_s,P_\mu,P_\nu)
\end{equation}

\noindent If Alice and Bob are allowed to use different parameters (which we discuss in Chapters 5-6), there can be as many as 12 parameters to optimize.

When the number of parameters is relatively small (say, 1-5 parameters), a brute-force search may be applied, which searches all combinations of parameters and finds the global maximum. When there are more parameters (e.g. $\geq 6$ parameters \footnote{Of course, this number depends on the speed of the computer, actual algorithm implementation, user's tolerance of computing time, and whether one can accept smaller search range and lower resolution to reduce brute-force search time.}), a brute-force search takes too much time or simply becomes infeasible, and a local-search approach is commonly applied. 

For instance, Ref. \cite{mdiparameter} proposes to use the ``coordinate descent" algorithm for MDI-QKD (with finite-size analysis), which iteratively optimizes the function in each one dimension while fixing all other dimensions, and moves on to the next dimension when the maximum on this 1-D region is found. After all dimensions are optimized, the algorithm either enters a new cycle that iterates over the dimensions again, or stops when multiple cycles provide similar results (suggesting that the current point is a local maximum), or maximum cycle count is reached. For instance, to illustrate this, for MDI-QKD mentioned above, based on a given set of current parameters $\{s^i,\mu^i,\nu^i,P_s^i,P_\mu^i,P_\nu^i\}$, we can search $s$ over $(s^{min},s^{max})$ to find the value that maximizes the current key rate function:

\begin{equation}
	\begin{aligned}
	R^{i+1}&=\max_{s\in (s^{min},s^{max})} R(s^i,\mu^i,\nu^i,P_s^i,P_\mu^i,P_\nu^i)\\
	&= R(s^{i+1},\mu^i,\nu^i,P_s^i,P_\mu^i,P_\nu^i)
	\end{aligned}
\end{equation}

\noindent where after the iteration, $s^i$ is replaced by the new optimal value $s^{i+1}$, and $R^{i}$ is replaced by $R^{i+1}$. The algorithm then continues to the next iteration over $\mu$, and then over $\nu$, and so on (and repeats the iteration from $s$ again in a new cycle when all six parameters are optimized, until the local maximum is found or the maximum cycle count is reached). Such an algorithm is shown to be able to find the same local maximum as gradient descent \cite{mdiparameter}. 

Overall, these local search algorithms are based on the assumption that the key rate versus parameters function for QKD is a convex function where a local maximum is equivalent to a global maximum. This assumption is generally true in practice (although not proven). There are some cases that deviate from this assumption. For instance, for asymmetric MDI-QKD, the function does not have a continuous first-order gradient (which we will show a method to circumvent this problem in Chapter 5). For QKD protocols whose key rate estimation involves linear programs, the linear solvers usually introduce some level of non-convexity (with multiple maxima or discontinuities in the function). This can be alleviated by applying certain global search techniques, such as starting local searches from multiple starting points, or applying algorithms such as evolution algorithm. However, neither an analytical explanation for such non-convexities in linear programs nor a method to completely remove them has been found yet, and it could be a subject for future studies.

\chapter{BB84 over a Turbulent Free-Space Channel}

This chapter is largely reproduced from our paper Ref. \cite{this_BB84} (with some minor modifications to keep the consistency with other chapters of the thesis).

\section{Background}

As we have introduced in Chapter 2, there has been increasing interest in implementing QKD through free-space channels, which can enable QKD over moving platforms, such as airborne or maritime QKD, or even ground to satellite quantum communications. Free-space QKD holds the potential of enabling a global quantum communication network.

A major characteristic of a free-space channel is its time-dependent transmittance, which is caused by the temporal fluctuations of the local refractive index in the free-space channel, i.e. \textit{atmospheric turbulence}. Turbulence causes effects such as scintillation and beam wandering \cite{freespacethesis}, which results in fluctuations in the channel transmittance that, in turn, affect QKD performance. Therefore, addressing turbulence is a major challenge for QKD over free-space. This fluctuation due to turbulence can be modelled as a probability distribution, called the Probability Distribution of Transmission Coefficient (PDTC), i.e. the real-time transmittance $\eta$ is a random time-dependent variable that can be described by the PDTC. 

\begin{table*}[t]
	\caption{Comparison of transmittance post-selection methods in QKD through turbulence channel}
	\begin{center}
		\begin{tabular}{cccc}            
			Method & Threshold choice & Model of signals & Sampling of transmittance\\
			\hline
			ARTS \cite{probetest} & post-determined & single-photon & secondary probe laser\\
			SNRF \cite{SNRF} & post-determined & single-photon & detector count (coincidence) rate \\
			P-RTS & pre-determined & general & general
			
		\end{tabular}
	\end{center}
\end{table*}

As free-space channels have time-varying transmittance due to turbulence, the QBER (and hence the secure key rate) for QKD changes with time. In previous literature discussing free-space QKD, such as \cite{free2007,Hughes}, the time variance of the channel is ignored, i.e. the secure key rate is calculated based on the time-average of channel transmittance. Having knowledge of the PDTC, however, Vallone et al. proposed a method named Adaptive Real-Time Selection (ARTS)\cite{probetest} that acquires information about real-time transmittance fluctuation due to turbulence, and makes use of this information to perform post-selection and improve the key rate of QKD.

However, ARTS method needs to ``adaptively" choose an optimal threshold by performing numerical optimization after collecting all the data. A similar proposal by Erven et al. \cite{SNRF} called ``signal-to-noise-ratio-filter (SNRF)" also discusses the idea of using a threshold to post-select high-transmittance periods, but uses the quantum data itself rather than a secondary classical channel. However, it needs to numerically optimize the threshold after collecting all experiment data, too. 


Here we ask the question, is scanning through all acquired data after the experiment and finding such an ``adaptive" threshold really necessary? The answer is in fact no. In this chapter, we propose a new method called ``pre-fixed threshold real-time selection (P-RTS)", and show the important observation that, for post-selection based on transmittance in a turbulent channel, the optimal post-selection threshold is independent of the channel, and can be directly calculated from experimental parameters of the devices beforehand - thus simplifying the implementation and enabling post-selection of signals in real time, which can also reduce the data storage requirements and computational resources in Bob's system. This is because, instead of having to wait until all data is collected to optimize the threshold, Bob can immediately discard the data that are obtained below the pre-fixed threshold and doesn't need to store all collected data. Moreover, he doesn't need to have a model for the PDTC of the channel, and no longer need to run a numerical optimization to find the optimal threshold. Thus we can additionally save software development effort and computing resource for Bob, too.

Furthermore, both ARTS and SNRF are limited to a single photon model only, while decoy state must be used for QKD with a practical weak coherent pulse (WCP) source. Here we also propose a general framework for QKD through a channel with fluctuating transmittance, for not only single-photon BB84, but also practical decoy-state BB84 with WCP source, and decoy-state BB84 with finite-size effects (both of which we are the first to apply threshold post-selection to), thus greatly improving its usefulness in practice. We also propose a model to estimate the maximum improvement in key rate from using threshold post-selection, and show that with P-RTS method we can achieve a key rate very close to the maximum performance with an optimal threshold. In this chapter, for simplicity, we focus on the BB84 protocol (with a single-photon source or with decoy states). Nonetheless, our idea of P-RTS is rather general and can, in principle, be applied to not only other QKD protocols but also other quantum communication protocols.


A comparison of P-RTS with post-selection methods in previous literature can be seen in Table 3.1. As shown here, P-RTS has the great advantage of being able to predict the optimal threshold independently of the channel. This means that one no longer needs to store all the data after experiment and optimize the threshold, but can perform real-time selection with a single threshold, regardless of the actual channel turbulence and loss condition. Moreover, our result is valid not only for BB84 with single photons, but for any general protocol that has a fluctuating transmittance. It is also not restricted to transmittance sampling with a secondary laser as in ARTS, but for instance can also use observed photon count rates in a given time interval as in SNRF.

Here a point worth noting is that, for P-RTS, as well as previous post-selection schemes including ARTS and SNRF, the post-selection depends on things such as the transmittance of the classical channel or the quantum channel. It might first appear that this would raise concerns about the security of such schemes, as it is possible for Eve to control the transmittance (or just the classical signal, if one is used), thus manipulating the post-selection. But we emphasize here that the channel transmittance information is classical, and by default, such information is already considered public in QKD (and can be controlled by Eve), thus using it would not undermine the security of the protocol. Eve is indeed able to control the classical signal, but by doing so she is only selecting an arbitrary (yet still uniform) sample among the quantum signals sent by Alice, and this is no more than a denial-of-service attack in the worst case (e.g. when Eve chooses to deliberately attenuate the classical signal below the threshold, all quantum signals get discarded by Bob). Moreover, for decoy state, despite Eve's information on the transmittance of the classical channel, given an n-photon state prepared by Alice, there is no way for Eve to distinguish a signal state from a decoy state. This is because the density matrices in the two cases are exactly the same. For this reason, decoy-state QKD applies to P-RTS.

Lastly, we have performed a computer simulation to show the actual advantage of using P-RTS in practical decoy-state BB84, with up to 170\% improvement in decoy-state BB84 key rate for certain scenarios, or 5.1dB increase in the maximum tolerant loss at $R=10^{-7}$, under medium-level turbulence. We also include a numerical demonstration for applying P-RTS to BB84 with finite-size effects, which still shows a significant increase in rate even when the total number of signals is limited, e.g. maximum tolerant loss at $R=10^{-7}$ gains an increase of 1.4dB to 5.2dB, for $N=10^{11}-10^{13}$ under high turbulence.

The organization of the chapter is listed as follows: in section 3.2 we first present a brief recapitulation of ARTS method, and proceed to propose a general framework for QKD key rate in turbulent channel. We then propose P-RTS method, and discuss how and why an optimal threshold can be pre-fixed, and show an upper bound for the rate of P-RTS. We also present numerical results from simulations to show how P-RTS behaves compared to no post-selection. Lastly, we discuss P-RTS in decoy-state BB84, for the asymptotic case in Section 3.3 and for finite-size regime in Section 3.4, and show with simulation results that P-RTS works effectively for both of them. Further details including the validity of our assumption of the strong correlation between the transmittance of the classical channel and that of the quantum channel can be found in Appendix B. 

\section{Methods}
\subsection{The ARTS Method}

In Ref. \cite{probetheory}, Capraro et al. performed an experiment to study the impact of turbulence on a quantum laser transmitted through a 143km channel on the Canary Islands, and proposed the idea of improving SNR with a threshold at the expense of the number of signals. Subsequently, in Ref. \cite{probetest}, Vallone et al. from the same group performed an experiment of free-space single-photon B92 QKD through the same channel, and showed the effectiveness of using real-time transmittance information in a turbulent channel to improve secure key rate, by performing post-selection on signals with a threshold, hence naming the method adaptive real-time selection (ARTS). 

\begin{figure}[b]
	\includegraphics[scale=0.3]{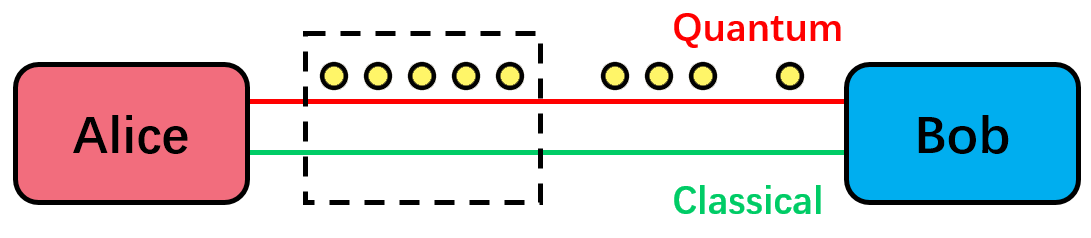}
	\caption{The ARTS setup by Vallone et al., where Alice and Bob are linked by a quantum channel and a classical channel. One can post-select quantum signals passing through the channel with high-transmittance, using a threshold on the corresponding classical channel signal. Reproduced from \cite{this_BB84} @2018 APS.}
	\label{fig:ARTS}
\end{figure}

This is realized by using a classical probe signal (a strong laser beam) alongside the quantum channel. In the quantum channel, the bits are polarization-encoded into quantum signals, which are detected by single-photon avalanche diodes (SPADs) that return click events. Meanwhile, the laser passing through the classical channel is detected by an avalanche photodetector that returns a voltage proportional to received light intensity, which is also proportional to the channel transmittance at that moment. An illustration of the setup can be seen in Fig. \ref{fig:ARTS}.

The key idea is that the transmittance of the classical channel will correspond to that of the quantum channel. Therefore, by reading the voltage from the classical detector (defined as $V$), one can gain information about the periods of high transmittance. Combined with a threshold on the classical signal (defined as $V_T$), this information can be used to post-select only those quantum signals received by Bob during high transmittance periods (only when $V \geq V_T$), thus increasing the overall average transmittance, at the expense of a smaller number of signals due to post-selection. 

This post-selection increases the signal-to-noise ratio among the selected signals, and hence reduces the QBER, which subsequently increases the key rate. However, post-selection also reduces the total number of signals. Therefore, this becomes an optimization problem, and the choice of threshold value becomes critical. By numerically choosing an optimal threshold that maximizes the rate, it is possible to acquire a secure key rate much higher than before applying post-selection. This, as defined in Ref. \cite{probetest}, is called the adaptive real time selection (ARTS) method.

\subsection{General Framework for QKD Key Rate in a Turbulent Channel}

In this section, we will expand upon this threshold post-selection idea from ARTS, and apply it to a general framework of post-selection upon transmittance. We will then discuss the effects of threshold post-selection based on transmittance on the secure key rate. Our following discussions will be based on the channel transmittance $\eta$ only, and they are not limited to the secondary-laser transmittance sampling as in ARTS, but can be applied to any sampling method of transmittance, including photon count rate such as in SNRF.

\begin{figure}[b!]
	\includegraphics[scale=0.4]{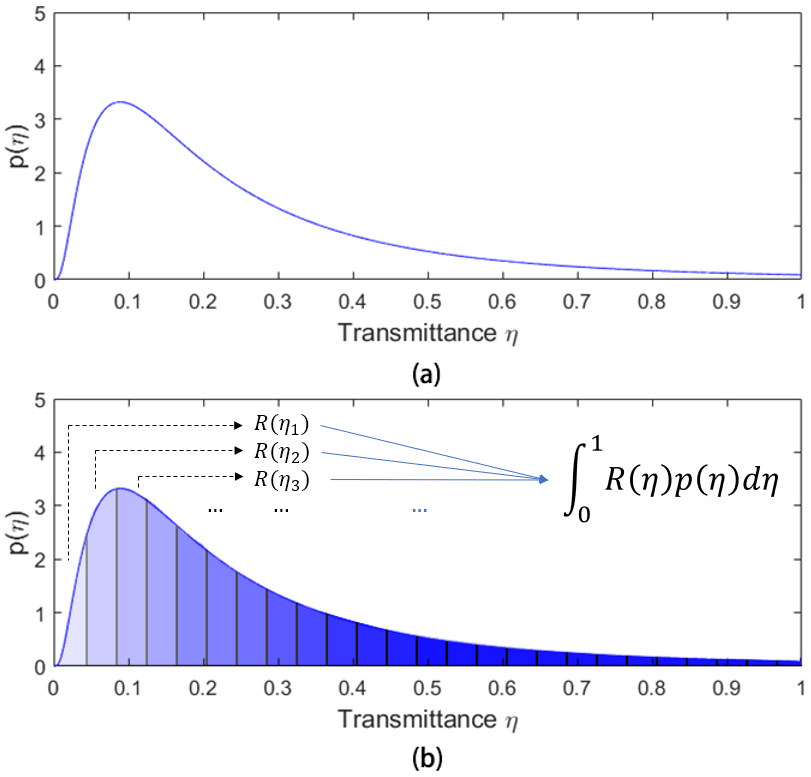}
	\caption{(a) The Probability Distribution of Transmittance Coefficient (PDTC), where $\eta$ is the transmittance, taking a value between [0,1], while $p(\eta)$ is the probability density function of $\eta$. Here we are showing a plot generated from the log-normal model of the PDTC, with $\eta_0=0.3, \sigma=0.9$; (b) Dividing signals into bins according to their respective $\eta$, in the rate-wise integration model. Reproduced from \cite{this_BB84} @2018 APS.}
	\label{fig:PDTC}
\end{figure}

As mentioned in Section 3.1, an important characteristic of a turbulent channel is the time-dependent transmittance, which follows a probability distribution called the PDTC. There have been multiple efforts to accurately characterize the PDTC, and a widely accepted model is the log-normal distribution \cite{distribution,laser} (a plot of which is shown in Fig. \ref{fig:PDTC} (a)):
\begin{equation}
p(\eta)_{\eta_0,\sigma}={1 \over {\sqrt{2\pi}}\sigma\eta}e^{-{{[ln({\eta \over \eta_0})+{1\over 2}{\sigma}^2]^2}\over{2\sigma^2}}}
\end{equation}

\noindent where p is the probability density, $\eta$ the transmittance, and $\eta_0$ and $\sigma$ the mean and variance. The distribution is solely determined by the two parameters $\eta_0$ and $\sigma$, which are inherent to the channel itself. $\eta_0$ is the expected atmospheric transmittance, with a typical value of $10^{-3}$ to $10^{-4}$ (corresponding to 30-40 dB of loss) for a 100km channel, while $\sigma$, typically taking a value between 0 and 1, is determined by the amount of turbulence - the larger the amount of turbulence, the larger the variance. The pair $(\eta_0,\sigma)$ hence contains all information of the PDTC.

Now, we make an important observation: For any given protocol implementation (say, single-photon BB84, or decoy-state BB84), if all experimental parameters in the system except $\eta$ are fixed - i.e. the device parameters including background and dark count rate, detector efficiency, laser intensities, and optical misalignment are all fixed - then the key rate solely depends upon the transmittance $\eta$, and can be written as a single-variable function of $\eta$, i.e. $R(\eta)$. 

To estimate secure key rate of QKD through turbulent channel, the question therefore becomes studying how the function $R(\eta)$ changes, when $\eta$ is a random variable following a probability distribution that we know, the PDTC. 

Here, we will propose two models for $R(\eta)$ under turbulence:

\begin{enumerate}
	\item \textbf{Rate-wise integration model}, $R^{\text{Rate-wise}}$, which is the case where we integrate the rate over PDTC, thus making use of all information of the PDTC. This rate only depends on the rate function and the PDTC, and is independent of what actual threshold we choose.
	\item \textbf{Simplified model}, $R^{\text{Simplified}}(\eta_T)$, which estimates the performance of decoy-state QKD with P-RTS, using post-selection with a threshold $\eta_T$ on channel transmittance. It is a function of the threshold $\eta_T$ that one uses.
\end{enumerate}

Let us first start with the rate-wise integration model. We can begin by considering an ideal case, where we assume that we have complete knowledge of the channel transmittance $\eta$ when each single signal passes through the channel. Moreover, here we discuss the asymptotic case where there is an infinite number of signals sent. Then, it is possible to order all signals from low to high transmittance $\eta$ when they pass through the channel, and divide the signals into bins of $[\eta,\eta+\Delta \eta)$ (which ranges from 0 to 1), as shown in Fig. \ref{fig:PDTC} (b).

Therefore, within the bin $[\eta,\eta+\Delta \eta)$, we can assume that all signals pass through the channel with the same transmittance $\eta$, given that the bin is sufficiently small, i.e. $\Delta \eta \rightarrow 0$. That is, the signals in the same bin can be considered as in a ``static channel", and enjoy the same rate formula $R(\eta)$ and security analysis as if $\eta$ is a static constant.

Then, we can calculate the number of secure key bits from each bin, according to their respective $\eta$, and add all bins together. In the limit of $\Delta \eta \rightarrow 0$, this is an integration of $R(\eta)$ over $\eta$, with $p_{\eta_0,\sigma}(\eta)$ being the weight (i.e. the proportion of signals in each bin). We call this model the \textit{``rate-wise integration model"}. Its rate $R^{\text{Rate-wise}}$ satisfies:

\begin{equation}
R^{Rate-wise}=\int_{0}^{1}R(\eta)p_{\eta_0,\sigma}(\eta)d\eta
\end{equation}

$R^{\text{Rate-wise}}$ makes use of all PDTC information from turbulence. Since all bins have either zero or positive rate, using a threshold $\eta_T$ in the rate-wise integration model will always result in either the same or lower rate. i.e.  

\begin{equation}
\begin{aligned}
R^{Rate-wise}(0)&=\int_{0}^{1}R(\eta)p_{\eta_0,\sigma}(\eta)d\eta \\
&\geq \int_{\eta_T}^{1}R(\eta)p_{\eta_0,\sigma}(\eta)d\eta \\
& = R^{\text{Rate-wise}}(\eta_T)
\end{aligned}
\end{equation}

\noindent Hence, from here on if $\eta_T$ is not specified, by $R^{\text{Rate-wise}}$ we will always mean $R^{\text{Rate-wise}}(0)$, which is a constant-value ``max possible performance" of the key rate that is only dependent on the PDTC of the channel and the experimental device parameters.\\

Now, let us consider applying post-selection to free-space QKD. Using a similar model as in ARTS method (instead of using classical detector voltage V, here we will directly use $\eta$, which is proportional to V). We can set a threshold $\eta_T$ and perform post-selection: we select quantum signals received when transmittance $\eta\geq \eta_T$, and discard all signals received when $\eta<\eta_T$.

Unlike the ideal case of the rate-wise integration model, in reality we do not have an infinite resolution from the classical detector, nor do we have an infinite number of signals. In practice, we are post-selecting signals with only two statuses: ``pass" or ``fail". To make use of this ``pass/fail" information, here we propose a practical model that estimates the rate with only the mean transmittance of the post-selected signals. We name it the \textit{``simplified model"}. First, with no post-selection applied, the rate is:

\begin{equation}
R^{\text{Simplified}}(0)=R(\eta_0)
\end{equation}

\noindent which means that we simply use the mean value of transmittance $\eta_0$ in the channel for all calculations and assume a ``static channel", using the same rate formula for a static channel, too. This is, in fact, what has been done in most literature for free-space QKD that don't consider fluctuations due to turbulence, such as in \cite{free2007,Hughes}.

Now, when a threshold is used and post-selection is performed, $R^{\text{Simplified}}$ is written as:

\begin{equation}
R^{\text{Simplified}}(\eta_T)=\int_{\eta_T}^{1}p_{\eta_0,\sigma}(\eta)d\eta \times R(\langle \eta \rangle)
\end{equation}

\noindent here we again treat all post-selected signals as having passed through a ``static channel", and use the same rate expression for static case. But the difference is that we use the new mean transmittance among only the post-selected signals, denoted as $\langle \eta \rangle$, as the transmittance of the channel. $\langle \eta \rangle$ satisfies (using expected value formula for a truncated distribution):

\begin{equation}
\langle \eta \rangle={{\int_{\eta_T}^{1}\eta p_{\eta_0,\sigma}(\eta)d\eta}\over{\int_{\eta_T}^{1}p_{\eta_0,\sigma}(\eta)d\eta}}
\end{equation}

When we apply post-selection (like the case with ARTS), in the rate formula for $R^{\text{Simplified}}$, we take into account the loss of signals due to post-selection, and only a portion of $\int_{\eta_T}^{1}p_{\eta_0,\sigma}(\eta)d\eta$ remains. This portion is always no larger than 1, and strictly decreases with $\eta_T$. On the other hand, $\langle \eta \rangle$ is always increasing with $\eta_T$, because we are post-selecting only the signals with higher transmittance. So, just like for the single photon case discussed in Section 3.2.1, we have an optimization problem, where the choice of $\eta_T$ is crucial to the rate we acquire. Using optimal threshold and applying post-selection, as we will later show in the numerical results in the next sections, can dramatically increase the rate over using no post-selection at all.

Therefore, using the simplified model, we can effectively treat the static channel QKD protocol as a ``black-box". We enjoy the same rate formula and security analysis as a static channel, while the only difference is that we use a higher $\langle \eta \rangle$ after post-selection as the input, and multiply a reduced portion $\int_{\eta_T}^{1}p_{\eta_0,\sigma}(\eta)d\eta$ to the output.\\

Now, let us compare the performance of the two models. From an information theory perspective, the rate-wise integration model makes use of all possible information on fluctuating transmittance (i.e. the whole PDTC), while the simplified model discards all distribution information and only acknowledges ``pass or fail", and keeps only the single mean transmittance after post-selection. Therefore, we expect that the rate-wise integration model, which makes use of the most information, would have a higher rate than the simplified model. We can write the relation as:

\begin{equation}
R^{\text{Rate-wise}} \geq R^{\text{Simplified}}
\end{equation}

This relation suggests that the Rate-wise integration model is an upper bound for the Simplified model key rate. This result can be shown rigorously by Jensen's Inequality (we include the detailed proof in Appendix B.2), under the condition that the rate function $R(\eta)$ is convex. Numerically, we show that (in the next section) the rate for single-photon BB84 and decoy-state BB84 are both convex. Therefore, the relation Eq. 3.7 always holds true.

The next question is, naturally, what is the optimal threshold to choose, such that $R^{\text{Simplified}}$ approaches the upper bound $R^{\text{Rate-wise}}$? Moreover, how closely can it approach the upper bound? We will discuss this optimal threshold in the next section, and show that it is only dependent upon $R(\eta)$ and independent of the PDTC.

\subsection{Optimal Threshold and Near-Tightness of Upper Bound}

In this section, we propose the ``Pre-fixed threshold Real Time Selection" (P-RTS) method, and show that the optimal threshold is independent of the PDTC and can be pre-fixed based on experimental parameters only. We also show that with this pre-fixed threshold the simplified model can approach its upper bound very closely.

Here, to describe the key rate function, we have to bring it into the context of an actual protocol model. We will first discuss single-photon BB84, using the Shor-Preskill \cite{Preskill} rate:

\begin{equation}
R=1-2h_2[QBER]
\end{equation}

\noindent here to keep the consistency of notations with following discussions, we will use parameters from Table 3.2 (which is also used as the channel model for decoy-state discussion), where detector dark count/background count rate is $Y_0$, basis misalignment is $e_d$, and total system transmittance is $\eta_{sys}=\eta\eta_d$:

\begin{equation}
R_{S-P}=(Y_0+\eta_{sys})\{1-2h_2[e(\eta_{sys})]\}
\end{equation}

\noindent while the single-photon QBER is

\begin{equation}
e(\eta_{sys})={{{1\over 2}Y_0+e_d \eta_{sys}}\over{Y_0+\eta_{sys}}}
\end{equation}

\begin{figure}[t]
	\includegraphics[scale=0.4]{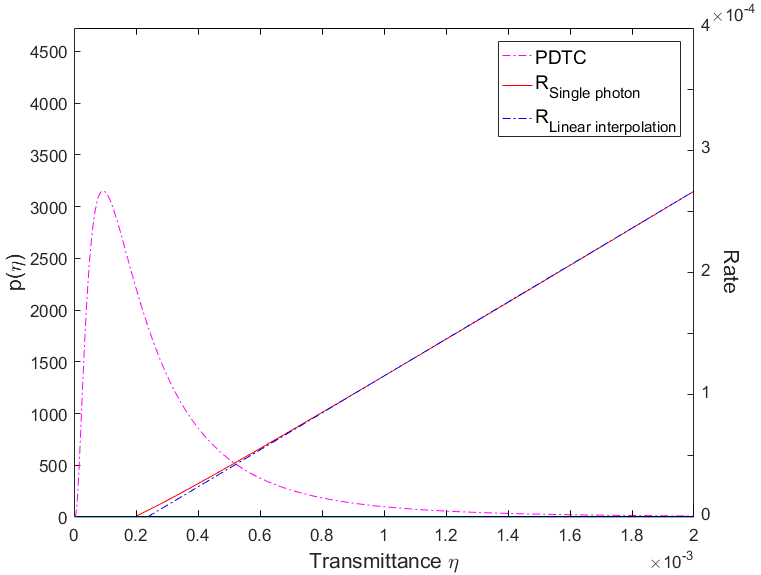}
	\caption{Single-photon rate and PDTC vs Transmittance $\eta$. As can be seen, there is an $\eta_{critical}$ such that $R_{S-P}(\eta)=0$ for all $\eta \leq \eta_{critical}$. For this example, we have plotted the single-photon rate R, using experimental parameters are listed in Table 3.2. We acquire an $\eta_{critical}=0.00020$ for single-photon case. Note that $\eta_{critical}$ is only determined by the experimental parameters of our devices (e.g. dark count rate, and misalignment, and the chosen intensities), and is independent of the actual PDTC. Linear interpolation of the asymptotic $\eta \gg Y_0$ case shows that the function is very close to linear. Here an instance of $p(\eta)$, the PDTC function, is also plotted for comparison. Reproduced from \cite{this_BB84} @2018 APS.}
	\label{fig:critical}
\end{figure}

A point worth noting is that $R_{S-P}(\eta)$ has the unique property of having an $\eta_{critical}$ such that $R_{S-P}(\eta)=0$ for all $\eta<\eta_{critical}$, and $R_{S-P}(\eta) \geq 0$ for $\eta\geq \eta_{critical}$. This critical position can be expressed as:

\begin{equation}
\eta_{critical}={Y_0 \over \eta_d}{{{1\over 2}-e_{critical}}\over{e_{critical}-e_d}}
\end{equation}

\noindent where $e_{critical}$ is the threshold QBER satisfying 
\begin{equation} 1-2h_2(e_{critical})=0 \end{equation}

\noindent that returns zero rate. For Shor-Preskill rate, this threshold is $e_{critical}=11\%$. More details can be seen in Appendix B.4.

As can be shown in Fig.\ref{fig:critical}, we plot out the single-photon rate $R_{S-P}(\eta)$, where a sharp turning point $\eta_{critical}$ exists. Moreover, within the $[\eta_{critical},1]$ region, numerical results show that $R(\eta)$ is slightly convex but very close to linear. (For larger $\eta \gg Y_0$, the rate is completely linear. Using this approximation we can make an interpolation of the ``linear" part of the rate. As shown in the plot, this linear interpolation is very close to the rate function itself.) 

This can lead to a very interesting result: We showed in Section 3.2.2 that $R^{\text{Rate-wise}}$ predicts the maximum possible performance of QKD with threshold post-selection in a turbulence channel. If we choose the threshold $\eta_T=\eta_{critical}$ for the simplified model, we can apply Jensen's Inequality for the truncated $p(\eta)$ distribution within region $[\eta_{critical},1]$, and acquire
\begin{equation}
R^{\text{Simplified}}(\eta_{critical}) \approx R^{\text{Rate-wise}}(\eta_{critical})
\end{equation}

\noindent given that $R_{S-P}(\eta)$ is very close to linear within the region (but still convex, so $R^{\text{Simplified}}$ is still slightly smaller), since Jensen's Inequality takes equal sign for a linear function. There is also no loss in $R^{\text{Rate-wise}}$ from truncating $[0,\eta_{critical})$, as $R(\eta)=0$ for all $\eta<\eta_{critical}$. 

\begin{equation}
R^{\text{Rate-wise}}(\eta_{critical}) = R^{\text{Rate-wise}}(0)
\end{equation}

\noindent Therefore, $R^{\text{Simplified}}$ can approximately reach the upper bound with $\eta_T=\eta_{critical}$, and the upper bound given by $R^{\text{Rate-wise}}$ is near-tight, due to the near-linearity of $R(\eta)$. A more rigorous proof showing that the optimal threshold for the simplified model is indeed $\eta_{critical}$ can be found in Appendix B.3.

Also, despite that there is no explicit analytical expression for $\eta_{critical}$, we can show that it depends more heavily on the background/dark count rate (approximately proportional to $Y_0$, if $\eta \ll 1$, and $Y_0 \ll \eta$). Details can be seen in Appendix B.4.

This result for the optimal threshold has two significant implications for using threshold post-selection and applying the simplified model:

\begin{itemize}
	\item  Since $R(\eta)$ is only a function of $\eta$, and not $(\eta_0, \sigma)$, this optimal threshold position $\eta_{critical}$ is only determined by the experimental parameters of the devices (e.g. detector efficiency, dark count rate, misalignment, and Alice's intensities - although here we make an assumption that the misalignment is independent of $\eta$), and thus $\eta_{critical}$ is \textit{\textbf{independent}} of the channel itself and its PDTC. This means that, regardless of the turbulence level, we can use the same threshold to get optimized performance - although the actual amount of performance improvement over not using post-selection \textit{will} be determined by the average loss and the amount of turbulence (i.e. the actual PDTC), as will also be shown in numerical results in the next section. 
	
	\item Given that we choose the optimal threshold and apply P-RTS, not only are we optimizing the rate for the simplified model, but we are also achieving the maximum possible performance for the turbulent channel, even if we make use of all information on transmittance fluctuations. This is because, at $\eta_{critical}$, the max value for $R^{\text{Simplified}}$ can almost reach the upper bound given by the rate-wise integration model - meaning that the upper bound is nearly tight. We will illustrate this point further with numerical results in the next section.
\end{itemize}

The significant implication is that, as long as we know the experimental parameters, we can determine the optimal threshold in advance, without the need to know any information about the channel (such as to measure the turbulence level), and perform post-selection in real time using the fixed threshold.

Therefore, we show that it is possible to perform post-selection on the channel transmittance with a pre-fixed threshold - which we will call ``Pre-fixed threshold Real Time Selection" (P-RTS). This is significantly more convenient than protocols that perform optimization of threshold after the experiment is done. It will substantially reduce the amount of data storage requirements in the experiment, since Bob doesn't need to store all data until after the experiment for optimization of the threshold, and will also save the computational resource since Bob no longer needs to perform optimization of the threshold.

\subsection{Numerical Results}

In this section we put the above models into a simulation program for single-photon BB84 in a turbulent channel. We use the experimental parameters from Ref. \cite{free2007}, as listed in Table 3.2. One note is that, though the dark count and stray light contribution is reported to be as high as 1700/s in the paper, because of the gated behaviour of the detector and the post-selection, only the counts within a 5.9ns time window (in 100ns period between two pulses, for the 10MHz source used) will affect the result. Therefore, here we take dark count rate as $Y_0=1\times 10^{-5}$ in the simulations.

\begin{table*}[t]
	\caption{Experimental parameters for free-space QKD over an 144km channel in Ref.\cite{free2007}}
	\begin{center}
		\begin{tabular}{ccccccc}            
			dark count rate $Y_0$ & pulse rate & detector efficiency $\eta_d$ & misalignment $e_d$ & error-correction efficiency $f$\\
			\hline
			$1\times 10^{-5}$ (per signal) & 10MHz & 25\% & 3\% & 1.22\\
		\end{tabular}
	\end{center}
\end{table*}

\begin{figure}[t]
	\includegraphics[scale=0.4]{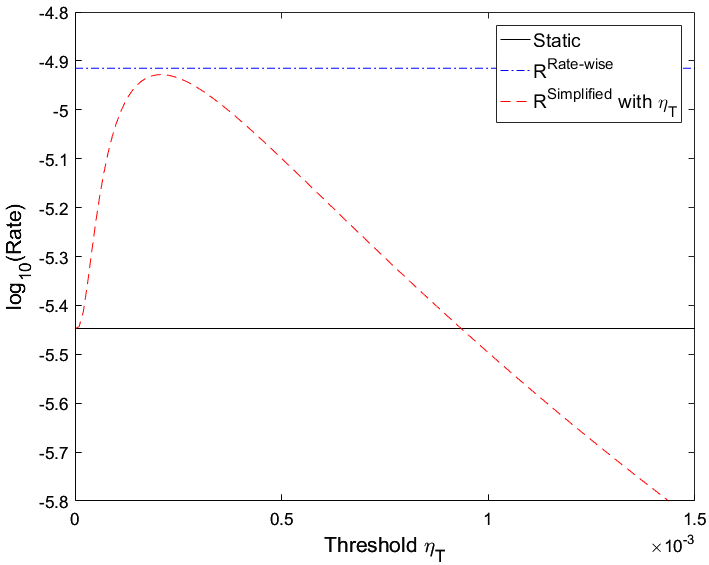}
	\caption{Comparison of the rate-wise integration model and simplified model vs no post-selection (static model), for single-photon case. Here we fix loss=36.5dB and $\sigma=0.9$, and scan through different threshold $\eta_T$. Experimental parameters are also from Table 3.2. As can be seen, choosing an optimal threshold, which is approximately $\eta_T=0.00020$ here, can allow the simplified model $R^{\text{Simplified}}$ to achieve a rate as much as $97\%$ of - though still lower than - the upper bound given by rate-wise integration model, $R^{\text{Rate-wise}}(0)$. Reproduced from \cite{this_BB84} @2018 APS.}
	\label{fig:threshold}
\end{figure}

Here, we first take a turbulence level of $\sigma=0.9$, and compare the performance of the two models plus the static model (which is a simplified model with no post-selection, i.e. $R_{S-P}(\eta_0)$) at a fixed loss of 37dB. We plot the results in Fig.\ref{fig:threshold}. As shown in the figure, $R^{\text{Simplified}}(\eta_T)$ first increases with the threshold $\eta_T$ (because of post-selecting high-transmittance signals) and then decreases when the threshold is further increased (because the decrease in rate due to loss of signals starts to dominate). 

Just as predicted in Section 3.2.3, the simplified model can achieve a very similar performance as the upper bound given by the rate-wise integration model, when the optimal threshold is chosen. For this case, at the optimal threshold $\eta_T=0.00020$, which, as we predicted, is the same as $\eta_{critical}=0.00020$ in Fig.\ref{fig:critical}, we get $R^{\text{Simplified}}=1.18 \times 10^{-5}$, very close to the upper bound $R^{\text{Rate-wise}}=1.22 \times 10^{-5}$ (only by $3\%$ difference - which is due to the rate above $\eta_{critical}$ not perfectly linear), and with dramatic increase in key rate compared with the default static model (using mean transmittance) $R^{\text{Static}}=3.5 \times 10^{-6}$, demonstrating the significant performance gain from using P-RTS in turbulence channel.

\begin{figure}[h]
	\includegraphics[scale=0.4]{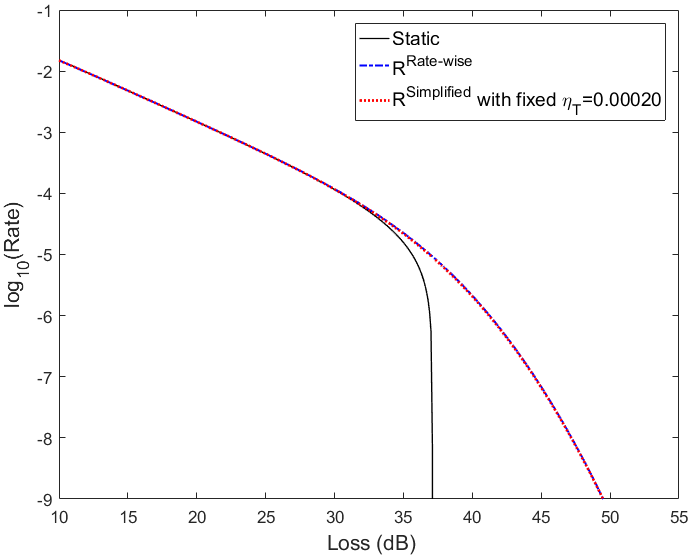}
	\caption{Comparison of the rate-wise integration model, simplified model with optimal threshold, and no post-selection (static model) under $\sigma=0.9$, for the single-photon case. Parameters are from Table 3.2. We can see that the simplified model, with optimized threshold, approaches the rate-wise integration model extremely closely, and both cases have significant improvement in key rate over static (no post-selection) model, especially in high-loss region. Reproduced from \cite{this_BB84} @2018 APS.}
	\label{fig:turbulence2}
\end{figure}

Furthermore, we compare the rate-wise integration model $R^{\text{Rate-wise}}$, the optimized $R^{\text{Simplified}}(\eta_T)$ with $\eta_T=\eta_{critical}$, and the non-post-selected model (whose rate is equivalent to static model, i.e. $R(\eta_0)$, as in Eq. 3.4) , by generating the rate vs loss relation for different average loss in the channel. Results can be seen in Fig. \ref{fig:turbulence2}. We see that indeed the rate for simplified model with fixed threshold is extremely close to its upper bound (as suggested in Eq. 3.13), the rate-wise integration model. Comparing with the static case, we see that the P-RTS method works best for high-loss regions, where post-selection can ``salvage" some rate where the static case would fail entirely, hence ``getting something out of practically nothing". Therefore, one of the major improvements we acquire from using P-RTS in free-space QKD is a dramatically increased maximum tolerant loss (which would mean longer maximum distance).

\section{Decoy-State BB84}

On the other hand, for decoy-state BB84 QKD, we follow decoy-state BB84 QKD theory from Ref. \cite{decoystate_Hwang,decoystate_LMC,decoystate_Wang}, and adopt the notations as in Lo, Ma, and Chen's Paper in 2005 \cite{decoystate_LMC}. Using the GLLP formula \cite{GLLP}, in the asymptotic limit of infinitely many data, we can calculate the secure key rate as:

\begin{equation}
\begin{aligned}
R_{GLLP} = q\{-f(E_\mu)Q_\mu h_2(E_\mu)+Q_1[1-h_2(e_1)]\}
\end{aligned}
\end{equation}

\noindent  where $h_2$ is the binary entropy function, $q={1\over 2}$ or $q\approx 1$ depending on whether efficient BB84 is used, and $f$ is the error-correction efficiency. $Q_\mu$ and $E_\mu$ are the observed Gain and QBER, while $Q_1$ and $e_1$ are the single-photon Gain and QBER contributions estimated using decoy-state. (For a more detailed recapitulation of decoy-state, see Appendix B.1.1. We have also discussed the channel model that we use for P-RTS in Appendix B.1.2).

\begin{figure}[h]
	\includegraphics[scale=0.4]{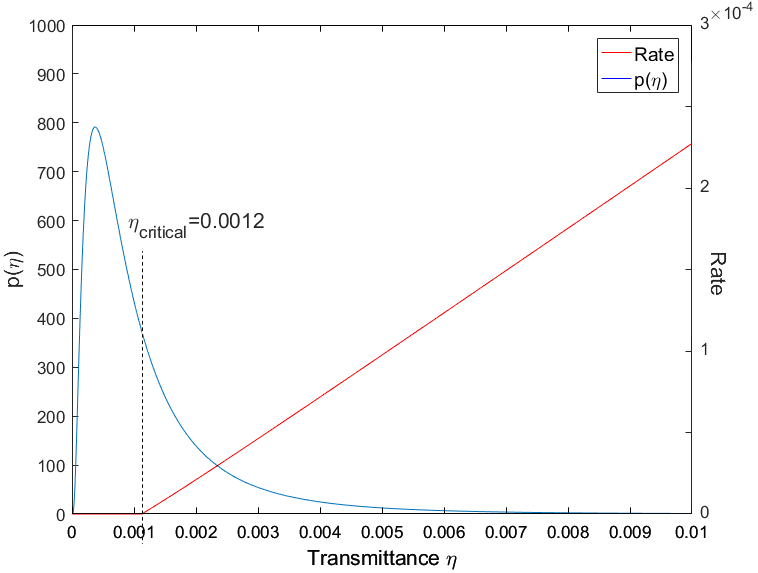}
	\caption{Rate and PDTC vs Transmittance $\eta$ for (asymptotic) decoy-state BB84 with infinite data size. Intensities are $\mu=0.3$, $\nu=0.05$, and experimental parameters are from Table 3.2. As can be seen, there is also an $\eta_{critical}=0.0012$ such that $R_{GLLP}(\eta)=0$ for all $\eta \leq \eta_{critical}$, just like for single photons. Here an instance of $p(\eta)$, the PDTC function, is also plotted for comparison. Reproduced from \cite{this_BB84} @2018 APS.}
	\label{fig:critical_decoy}
\end{figure}

Here for free-space decoy-state QKD. We fix the signal and decoy-state intensities as $\mu=0.3$, $\nu=0.05$, and the vacuum state $\omega=0$, and use the vacuum+weak method to estimate single-photon contribution, as in Ma et al.'s 2005 paper \cite{decoypractical} for practical decoy-state QKD. 

Like for the single-photon case, again we generate the rate vs $\eta$ function. As can be observed in Fig. \ref{fig:critical_decoy}, the decoy-state rate function $R_{GLLP}(\eta)$ behaves similarly as the single-photon rate $R_{S-P}(\eta)$, with a critical transmittance $\eta_{critical}$ ($\eta_{critical}=0.0012$ for this parameter set) such that all $\eta$ below it returns zero rate, and a nearly linear rate-transmittance relation for $\eta \geq \eta_{critical}$. Therefore, using the same proof from section 3.2, we can conclude that $\eta_{critical}$ is the optimal (and fixed) threshold for decoy-state BB84 with post-selection too.

\begin{figure}[h]
	\includegraphics[scale=0.4]{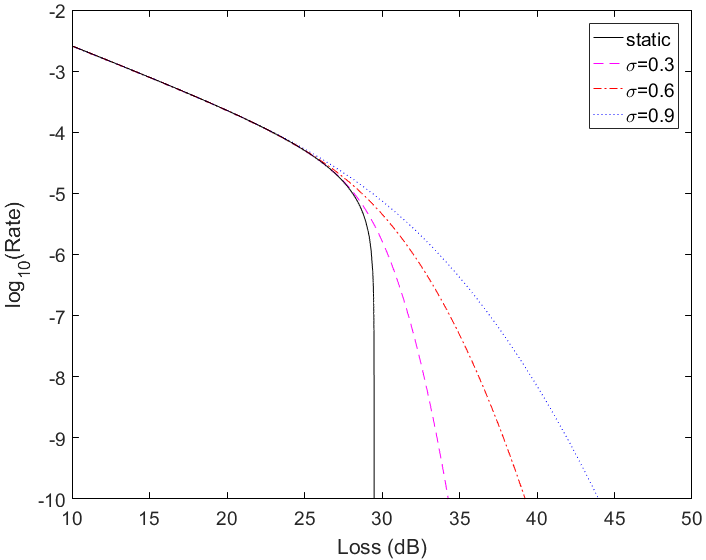}
	\caption{Comparison of the optimized Simplified model vs no post-selection (static model) under different levels of turbulence, for (asymptotic) decoy-state BB84 with infinite data size. Here we use $\sigma=0.3, 0.6, 0.9$ and $\eta_T=0.0012$. Intensities are $\mu=0.3$, $\nu=0.05$, and experimental parameters are from Table 3.2. We see that the improvement in rate from using P-RTS increases with the level of turbulence, and has a significant improvement over static model even under medium-level turbulence of $\sigma=0.6$. Reproduced from \cite{this_BB84} @2018 APS.}
	\label{fig:turbulence}
\end{figure}

Using the fixed threshold $\eta_T=\eta_{critical}$ to get the optimized rate $R^{\text{Simplified}}(\eta_{critical})$, we generate the rate vs loss relation for different levels of turbulence, as shown in Fig.\ref{fig:turbulence}. As can be seen, the P-RTS method works in the same way as decoy-sate BB84. We can also see that the higher the turbulence level is, the larger the performance gain from applying P-RTS will we be able to achieve. As described in Section 3.2.3, the optimal threshold is only determined by the parameters of the equipment, but the actual optimal \textit{performance} is determined by the amount of turbulence present in the channel that we can utilize. As can be seen in the plot, even for a medium-level turbulence of $\sigma=0.6$: for the same loss=29dB, $R^{\text{Simplified}}=8.453 \times 10^{-6}$, a 170\% increase over $R^{\text{Static}}=3.119 \times 10^{-6}$ at loss=29dB. Also, for a minimum rate of $R=10^{-7}$, the simplified model has a maximum tolerant loss of 34.4dB, versus 29.5dB for the static model, with a 5.1dB increase in tolerant loss. 

\begin{figure}[h]
	\includegraphics[scale=0.4]{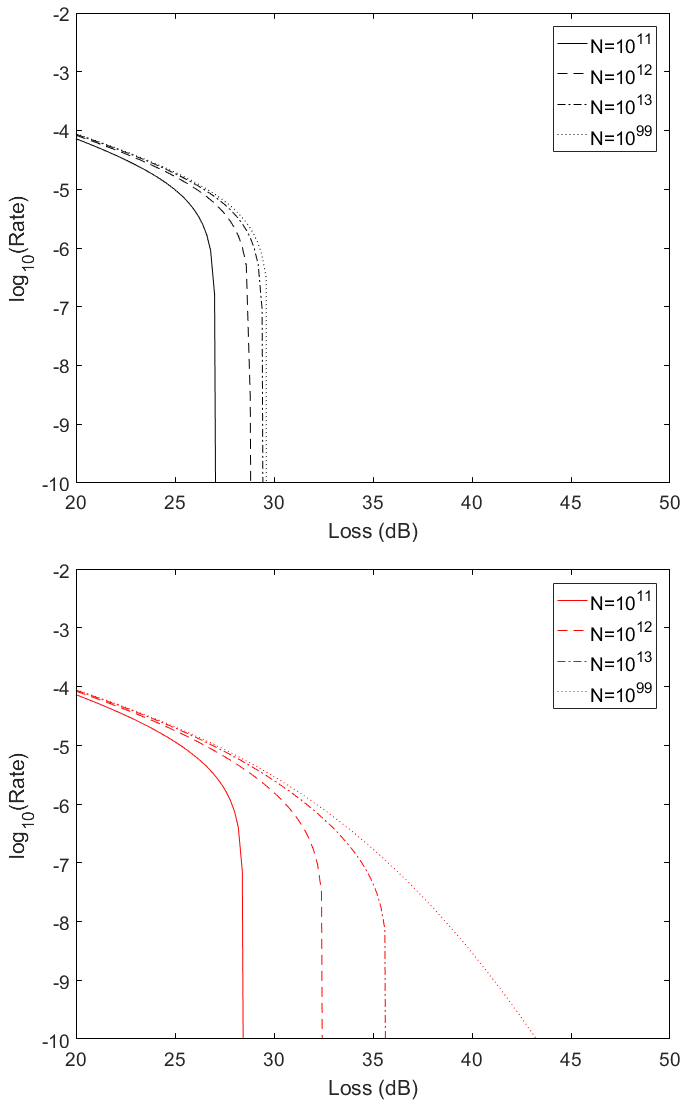}
	\caption{Comparison of the no post-selection static model (upper figure) vs optimized simplified model (lower figure), for decoy-state BB84 with finite-size effects. For each model, we test different data sizes $N=10^{11},10^{12},10^{13}$, and the near-asymptotic case $N=10^{99}$. Here we use a high turbulence of $\sigma=0.9$. The experimental parameters both follow Table 3.2, and intensities and probabilities used are $\mu=0.31$, $\nu=0.165$, $\omega=2\times 10^{-4}$, $p_{\mu}=0.5$, $p_{\nu}=0.36$, and the probability of sending X basis $q_x=0.75$. Lines in the upper figure shows the case where no post-selection is applied, while lines in the lower figure have post-selection applied with fixed $\eta_T=0.0012$. We can see from the figures that the improvement in rate from using P-RTS increases with the data size N, and at $R=10^{-7}$, maximum tolerant loss still increases by 3.5dB and 1.4dB respectively when $N=10^{12}$ and $10^{11}$. Reproduced from \cite{this_BB84} @2018 APS.}
	\label{fig:finite}
\end{figure}

\section{Decoy-State BB84 with Finite-size Effects}

We now turn to the case with finite data size, and apply the simplified model and P-RTS to decoy-state BB84 under finite-size effects. We also use simulations to numerically demonstrate the improvements in the key rate for the finite-size case. The protocol is based on C. Lim et al.'s finite-size decoy-state BB84 paper \cite{finitebb84}, and we have adopted the same channel model as in Ref. \cite{decoypractical}. Here we use the same experimental parameters (including dark count rate, detector efficiency and misalignment) as in Table 3.2, same as the ones used in our previous asymptotic-case simulations. Also, we fix the signal and decoy intensities to $\mu=0.31$, $\nu=0.165$ (in addition to the vacuum intensity $\omega=2\times 10^{-4}$), the probabilities of sending them $p_{\mu}=0.5$, $p_{\nu}=0.36$, and the probability of sending X basis $q_x=0.75$. Unlike in Ref. \cite{finitebb84}, however, we do not scan through the decoy-state intensities and probabilities to perform optimization. Instead, since we only concentrate on high-loss region, we use fixed parameters that are already very close to optimal (while changing them with distance does not provide much improvement in performance). Using intensities that do not change with channel loss also avoids changing the expression for $R_{GLLP}(\eta)$ (which depends on intensities $\mu$, $\nu$), and ensures that $\eta_{critical}$ is independent of the actual loss of the channel.

As described for the simplified model, we can use the same ``black box" idea, and simply substitute $R_{GLLP}$ for asymptotic BB84 with rate for finite-size BB84. However, one difference from the asymptotic case is that N, the number of signals sent by Alice, matters when calculating the rate, i.e. the rate becomes $R_{Finite-Size}(\eta,N)$ instead of $R_{GLLP}(\eta)$. Then, instead of using Eq. 3.5 for $R^{\text{Simplified}}$, we use 

\begin{equation}
\begin{aligned}
R^{\text{Simplified}}&=\int_{\eta_T}^{1}p_{\eta_0,\sigma}(\eta)d\eta \\
&\times R_{Finite-Size}(\langle \eta \rangle, N\times \int_{\eta_T}^{1}p_{\eta_0,\sigma}(\eta)d\eta)
\end{aligned}
\end{equation}

\noindent which means that the post-selection not only affects the overall rate due to the portion of lost signals, but also affects the rate for the \textit{selected} signals, since fewer signals than N are used to actually perform the protocol, and higher statistical fluctuations will be present among the selected signals. For rate-wise integration model, the finite-size effect is a bigger problem, because the data size will become smaller for each bin. Overall, this means that we need to be more prudent with post-selection when treating finite-size BB84.

The numerical results are shown in Fig.\ref{fig:finite}. As can be seen, P-RTS has a similar effect on finite-size BB84 as on the asymptotic case: we gain a significant advantage in the high-loss region, and have an improved maximum tolerant loss, when a minimum acceptable Rate is required. For instance, at $\sigma=0.9$ and for a minimum $R=10^{-7}$, the maximum loss increases by 1.4dB, 3.5dB, 5.2dB, and 6.2dB, respectively, for the cases with $N=10^{11}, 10^{12}, 10^{13}$, and near-asymptotic case, while not much improvement can be gained from P-RTS with N smaller than $10^{10}$. As shown, the improvement increases with the size of N (which is understandable, since the smaller N is, the more sensitive the rate will be to post-selection - because we are cutting off a portion from the already-insufficient number of signals and further aggravating the statistical fluctuations - while for the asymptotic case, for instance, the performance of selected signals does not depend upon how big is the selected portion of signals, and the only negative effect that post-selection has is the lost portion of signals). For a free-space QKD system with 100MHz repetition rate, $N=10^{11}$ would require about 17 minutes of communication.\\

\section{Conclusion and Discussions}

In this chapter we have proposed a post-selection method with pre-fixed threshold for QKD through a turbulent channel, and have also proposed a general framework for determining the optimal threshold beforehand and predicting the maximum possible performance. By choosing the threshold in advance, we can perform post-selection in real time regardless of the channel condition. This real-time post-selection also provides an additional benefit of reducing the amount of data that is required to be stored in the detector system on Bob's side. We also performed simulations to show the method's effectiveness in not only single-photon BB84, but also practical decoy-state QKD in both the asymptotic case and the case with finite-size effects.

This method is especially effective for regions of high turbulence and high loss, and can even ``salvage something out of nothing", when the secure key rate could have been zero without P-RTS method. In order to sample the real-time transmittance condition, the P-RTS method can use only an additional classical channel for each quantum channel, which would be easily implemented (or may even be already implemented as a beacon laser is often required for alignment in free-space QKD). Moreover, since our results only depend on post-selection of $\eta$, in essence our method is even possible without an additional classical channel, such as in Erven et al.'s SNRF setup \cite{SNRF} (which samples transmittance by observing quantum signal count rate). The thresholding, on the other hand, is purely implemented in post-processing, therefore does not require any additional resource, and could be readily deployed into existent infrastructure, and gain a ready increase in secure key rate performance over existing implementation for free-space QKD.

One requirement for the proposed method to work is that the transmittance for the quantum signal has good correlation with the transmittance of the classical signal. We provide some additional discussion on this point in Appendix B.6. 

Another requirement is that the turbulence time scale is not too much faster than what the classical sampling could observe. From the experiment of Ref. \cite{probetest}, it is mentioned that the repetition rate for the classical signal is 1kHz, which is faster than the time scale of e.g. beam wandering of 10-100ms, as mentioned in Ref. \cite{freespacethesis}, meaning that the post-selection method can effectively work. If one physically uses a classical probe signal alongside the quantum signal as in Ref. \cite{probetest} (rather than post-selecting the quantum signal itself as in Ref. \cite{SNRF}), in principle one can make the sampling rate of the classical laser even faster, as long as it is equal to or slower than the repetition rate of the quantum signal - which can be at the order of MHz or even GHz. In Ref. \cite{probetest} each classical signal corresponds to multiple quantum signals in a ``packet", but in principle each classical signal can even correspond to one signal timestamp of a quantum signal. Note that eventually the quantum signals are divided into just two bins: accepted or discarded (depending on whether the corresponding classical signal passed the threshold), and the accepted quantum signals are collected post-processed together as one data block. This means that increasing the sampling rate of the classical signal (and decreasing the number of quantum signals in each ``packet" being accepted or discarded) will not introduce more finite-size effect.

Lastly, throughout the text we have not discussed the possible cross-talk between the quantum channel and the classical channel (for which no information was provided in Ref. \cite{probetest} either). Taking the setup of Ref. \cite{probetest} as an example, the quantum signal comes from a WCP source at 850nm, while the classical signal is an 808nm 30mW laser. The channel is a 143km free-space channel between two observatories on La Palma and Tenerife Islands. In such a free space channel, the dominating effects on the signals are absorption, Rayleigh scattering, and turbulence-induced beam wandering and scintillation. Non-linear scattering such as Raman scattering is negligible in such a free-space scenario and for the energy level of the 30mW laser, so there is little cross-talk due to the frequency shift caused by Raman scattering. For the setup of e.g. Ref. \cite{SNRF} where the quantum signal count rate is used for post-selection, there is only a single channel and cross-talk is not present.

In the next chapter, we will apply a similar idea to the MDI-QKD protocol, which is able to eliminate all detector side channels. We will discuss how two independently fluctuating free-space channels, as well as a pre-fixed 2-dimensional threshold (instead of a single value for BB84) on the transmittances in the two channels, can affect the performance of free-space MDI-QKD in the presence of turbulence.
\chapter{MDI-QKD over Turbulent Free-Space Channels}

This chapter is largely reproduced from our draft paper Ref. \cite{this_MDI}, which is posted on the preprint server. The work is also presented at QCrypt 2019 as a poster.

In the previous chapter we have discussed the BB84 protocol through a turbulence channel. We showed that by applying a threshold on classical transmittance to perform selection on signals, we can greatly improve the performance of free-space BB84 in high-loss, high-turbulence scenarios. Moreover, we showed that the optimal value for such a threshold can be obtained before the experiment begins, independent of the channel condition. Such a prefixed threshold is convenient to implement, and also reduced data storage requirements for the receiver (as data not passing the threshold can be discarded on-the-fly).

In this chapter, we will apply a similar idea to the MDI-QKD protocol through free-space channels.

\section{Background}

As we introduced in Chapter 2, MDI-QKD eliminates all detector side channels and can prevent attacks on detectors. Here instead of Alice sending signals to Bob, they both send signals to a third-party Charles, who performs Bell-measurements on incoming signals and acts as an untrusted relay. 

Since its proposal, MDI-QKD has attracted much worldwide attention, and has seen many demonstrations in fibre systems and even fibre-based networks \cite{mdiexp1,mdiexp2,mdiexperiment,mdi404km,mdinetwork}. Meanwhile, another highly desirable (yet challenging) application of MDI-QKD would be its implementation over free-space channels, which could allow mobile platforms such as ships, planes, satellites to communicate without detector susceptibilities It would also allow these users over moving platforms to join a dynamic quantum network with untrusted relays based on MDI-QKD. However, up to now, an experimental demonstration of free-space MDI-QKD remains challenging. In addition to the high level of loss in free-space channels, the atmospheric turbulence - which causes fluctuations in channel transmittances - also plays a large part in affecting the secure key rate.

\begin{figure}[h]
	\includegraphics[scale=0.47]{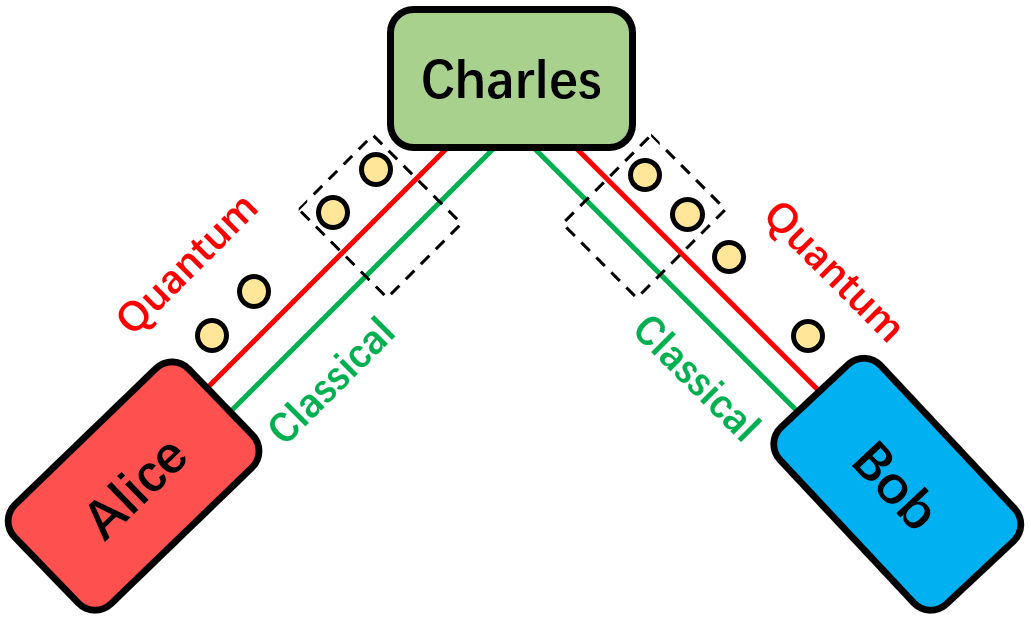}
	\caption{An illustration of MDI-QKD setup. Alice and Bob each send quantum signals to a third-party Charles, who acts as an untrusted relay. To perform real-time post-selection, Alice and Bob can each establish a classical channel alongside the quantum channel, to sample the channel transmittance in real time, and select only the sections where both channels have good transmittance. Note that, this classical channel could be either a strong laser at a slightly different wavelength, or observables such as the count rate of the quantum detectors, which could also serve as an indicator of channel transmittance.}
	\label{fig:setup}
\end{figure} 

In this chapter, we present two important results: Firstly, we show that in the presence of turbulence (scintillation of light, which causes transmittances to fluctuate in Alice's and Bob's channels), the key rate of MDI-QKD will decrease significantly due to turbulence-induced real-time asymmetry between the channels. This is contrary to many people's popular belief and different from the results we observe for BB84 (where fluctuation of transmittance does not affect the key rate - and post-selection can make use of turbulence to increase key rate). We show that, without post-selection, the key rate for MDI-QKD will drop significantly for turbulent channels. 

Secondly, we extend our P-RTS method to MDI-QKD, and show that by selecting a good threshold we can achieve a much higher key rate and an extended maximum tolerable channel loss. Moreover, our threshold does not depend on the channel condition and allows a semi-blind approach where ``bad" signals can be immediately discarded, which reduces storage and compute resource requirements. As atmospheric turbulence is very common in free-space channels, we believe that this work will be an important step towards the future experimental demonstration of MDI-QKD.

While Ref. \cite{this_MDI} is still under preparation, we notice a paper on a similar subject published at \cite{QIP_MDI} and made online in December 2018. While it also applies post-selection to free-space MDI-QKD, importantly, it only uses a rather naive model of the problem that does not consider the turbulence-induced asymmetry (and assumes the key rate does not change in the presence of turbulence), which we show is a rather inaccurate overestimation of the key rate. Moreover, it suggests a simple ``square" threshold for post-selection (which needs optimization and depends on channel condition), while we show that Charles in fact has a much larger parameter space for threshold choice, and we propose a threshold that can closely approach optimality, and can be pre-determined prior to the experiment without the need to know the channel condition.


\section{Theory}

In this section we define the models we use for the turbulent channel, for the post-selection, and also the models for a reliable estimation of the secure key rate. We point out an important point that, contrary to BB84 where fluctuation due to turbulence does not detrimentally affect the key rate, in MDI-QKD even without post-selection, the key rate will decrease due to turbulence-induced channel asymmetry in real-time. To address this, we then propose solutions to set a good threshold for the post-selection. We will show in the next section with numerical results the effectiveness of the post-selection with our proposed thresholds.

\subsection{Channel Model under Turbulence}


In a free-space channel subject to atmospheric turbulence, the transmittance fluctuates with time and follows a probability distribution, which is often denoted as a probability distribution of transmission coefficient (PDTC) \cite{PDTC}. There are multiple models for such a PDTC function, a commonly used model is the log-normal distribution \cite{laser}:

\begin{equation}
p_{\eta_0,\sigma}(\eta)={1 \over {\sqrt{2\pi}} \sigma\eta}e^{-{{[ln({\eta \over \eta_0})+{1\over 2}{\sigma}^2]^2}\over{2\sigma^2}}}
\end{equation}

\noindent where the channel is described by two parameters $(\eta_{0},\sigma)$ that respectively represent the mean transmittance and the variance of the channel. The log-normal distribution satisfies normalization condition when $\eta$ is small

\begin{equation}
\int_{0}^{1}p_{\eta_0,\sigma}(\eta)d\eta \approx 1
\end{equation}

\begin{figure*}[t]
	\includegraphics[scale=0.16]{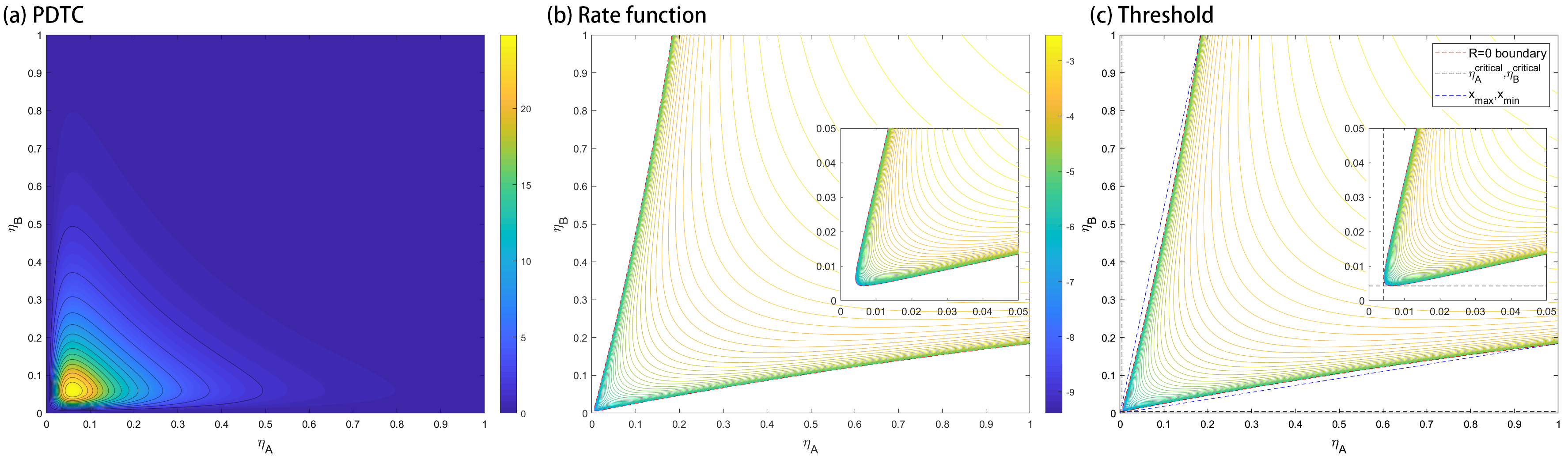}
	\caption{(a) A joint PDTC function for Alice's and Bob's real-time transmittances. (b) The contours of key rate function $R(\eta_A,\eta_B)$ for 4-intensity MDI-QKD protocol, for fixed intensities and infinite data size (plotted in $log_{10}$ scale). We can see that the $R=0$ contour follows a near-hyperbolic shape, whose asymptotic lines represent the maximum and minimum acceptable channel mismatch $x={\eta_A/\eta_B}$. There is also a ``gap" near the origin mainly determined by the noise (e.g. dark counts). (c) Choice of thresholds. We can choose the R=0 boundary as the threshold. Or for simplicity, we can also approximate it using the horizontal/vertical tangents $\eta_A^{critical},\eta_A^{critical}$ of the $R=0$ contour (which discard signals with low signal-to-noise ratio), combined with the asymptotic lines $x_{max},x_{min}$ (which discard signals with strong channel asymmetry). Importantly, all the information in this plot comes from the structure of $R(\eta_A,\eta_B)$ and are independent of the actual channel PDTC, i.e. they can be pre-determined before the experiment. In this plot $\eta_{A}^{critical}=\eta_{B}^{critical}=0.0042$, $x_{min}=1/x_{max}=0.184$.}
	\label{fig:freespace_MDI}
\end{figure*}

\noindent When $\eta$ is comparable to 1, there is a non-negligible portion of the probability that $\eta>1$, in this case we should calculate the integral from 0 to 1 to obtain a constant normalization factor (smaller than 1), and divide the PDTC with this factor, i.e. forming a truncated log-normal distribution. \footnote{Note that, although for simplicity of discussion we have used a truncated log-normal distribution here, it is an inaccurate representation of the PDTC for large $\eta_0$ close to 1. There have been proposals for more accurate PDTC models, e.g. in Ref. \cite{PDTC1}. Here we'd like to note that the threshold we choose (as we will show in Fig. \ref{fig:freespace_MDI} (c)) is independent of the channel model, hence a ``prefixed threshold", although the actual amount of performance gain will depend on the form of PDTC. Applying more accurate PDTC models will be a subject of future studies.}

By post-selecting $\eta$ with a threshold $\eta_T$, we can have a higher average transmittance among post-selected signals:

\begin{equation}
\langle \eta \rangle={{\int_{\eta_T}^{1}\eta p_{\eta_0,\sigma}(\eta)d\eta}\over{\int_{\eta_T}^{1}p_{\eta_0,\sigma}(\eta)d\eta}}
\end{equation}

\noindent Again, the post-selected signals follow a truncated log-normal distribution between $[\eta_T,1]$, hence a normalization factor (total probability within the post-selected region) is included.\\

Now, let us consider MDI-QKD, where Alice and Bob are each connected to Charles with a channel. Intuitively, here we can first assume that both channels are subject to atmospheric turbulence, and that their fluctuations are \textit{independent and non-correlated}. We can denote the channel transmittances as $\eta_A,\eta_B$, respectively. Then, the joint PDTC for the two channels can be written as

\begin{equation}
p_{AB}(\eta_A,\eta_B)=p_{\eta_{A_0},\sigma_A}(\eta_A)\times p_{\eta_{B_0},\sigma_B}(\eta_B)
\end{equation}

\noindent where the two channels are described by the $(\eta_{A_0},\sigma_A)$, $(\eta_{B_0},\sigma_B)$ channel condition parameters. The joint PDTC also follows the normalization condition:
\begin{equation}
\begin{aligned}
&\iint p_{AB}(\eta_A,\eta_B)d\eta_A d\eta_B\\
&= \int_{0}^{1}p_{\eta_{A_0},\sigma_A}(\eta_A)d\eta_A \times \int_{0}^{1}p_{\eta_{B_0},\sigma_B}(\eta_B)d\eta_B\\
&\approx 1
\end{aligned}
\end{equation}

The joint PDTC can be considered as a two-variable function on a plane defined by $(\eta_A,\eta_B)$, as shown in Fig. \ref{fig:freespace_MDI} (a). Now, we observe that for a post-selection on the signals received by Charles, he can actually observe both Alice's and Bob's channel transmittances $(\eta_A,\eta_B)$, and make a decision based on these two observables. Importantly, he does not have to independently set a threshold for each channel respectively (and select events where both transmittances pass the threshold), but rather, he is able to make a joint decision based on the two observables - for instance, selecting events based on a high level of symmetry between $(\eta_A,\eta_B)$ is present, instead of based on the respective signal strength of $\eta_A,\eta_B$ alone. Mathematically, Charles is selecting a domain $\Omega \subseteq \bm{R}^2$ in the 2D space defined by $(\eta_A,\eta_B)$.

In the selected domain, we can perform a 2D integral to obtain the expected values of the transmittances.

\begin{equation}
\begin{aligned}
\langle \eta_A \rangle&={{ \iint_{\Omega}\eta_A p_{AB}(\eta_A,\eta_B)d\eta_A d\eta_B}\over{ \iint_{\Omega}p_{AB}(\eta_A,\eta_B)d\eta_A d\eta_B }}\\
\langle \eta_B \rangle&={{ \iint_{\Omega}\eta_B p_{AB}(\eta_A,\eta_B)d\eta_A d\eta_B}\over{ \iint_{\Omega}p_{AB}(\eta_A,\eta_B)d\eta_A d\eta_B }} 
\end{aligned}
\end{equation}

In the simple case of a "square" threshold, i.e. 

\begin{equation}
\Omega^{square} = \{(\eta_A,\eta_B)\in \bm{R}^2: \eta_{A_T}\leq \eta_A \leq 1, \eta_{B_T}\leq \eta_B \leq 1\}
\end{equation}

\noindent the post-selection follows two independent thresholds $\eta_{A_T},\eta_{B_T}$. This is the simplest form of threshold Charles can implement, and the probability distribution can be decoupled between $\eta_A$ and $\eta_B$, hence one can simply use Eq. 4.3 to calculate the mean transmittances. However, note that there are more careful (and potentially better) ways to select such a threshold, to make use of Charles' joint knowledge of $(\eta_A,\eta_B)$, which we will discuss in later sections.

\subsection{Models for Key Rate}

In \cite{this_BB84}, for BB84 protocol with a single free-space channel, we have proposed two models:
\begin{equation}
\begin{aligned}
R^{\text{Simplified}}_{BB84}(\eta_T) &= \left(\int_{\eta_T}^{1}p_{\eta_0,\sigma}(\eta)d\eta\right) \times R(\langle \eta \rangle) \\
R^{\text{Integration}}_{BB84} &=\int_{0}^{1}R( \eta )p_{\eta_0,\sigma}(\eta)d\eta
\end{aligned}
\end{equation}

\noindent where $R(\eta)$ is the key rate function (where all other experimental parameters are fixed, e.g. dark count, detector efficiency, misalignment, etc.), and $\eta_T$ is the threshold used to post-select signals according to the real-time transmittance. The ``\textbf{simplified model}" $R^{\text{Simplified}}_{BB84}(\eta_T)$ finds the mean transmittance among the post-selected signals, and calculates the key rate with a static model with this new transmittance, i.e. it assumes that all the signals are transmitted with this mean transmittance. It is also multiplied by the proportion of selected signals (since the total number of signals decreases due to post-selection). On the other hand, the ``\textbf{rate-wise integration model}" $R^{\text{Integration}}_{BB84}$ (for simplicity in the following text we will just call it integration model in short) divides all signals into bins of $[\eta,\eta+\Delta \eta)$ and adds up the key rate in all bins. In the asymptotic (infinite-data) limit, the integration model can make use of the entire probability distribution's information, and always produces higher key rate than the simplified model. Effectively, it provides an upper-bound to the maximum key rate $R^{\text{Simplified}}_{BB84}(\eta_T)$ can achieve by adjusting the threshold $\eta_T$:

\begin{equation}
R^{\text{Simplified}}_{BB84}(\eta_T) \leq R^{\text{Integration}}_{BB84}
\end{equation}

For BB84, the near-linearity of the rate function $R(\eta)$ guarantees a fixed optimal threshold $\eta_{critical}$ exists, where $R=0$ for all $\eta \leq \eta_{critical}$. This optimal threshold position is calculated with $R(\eta)$ only, and is independent of the channel condition $(\eta_0, \sigma)$. Hence, we can find a prefixed threshold $\eta_{critical}$ that maximizes the performance of BB84 with post-selection, satisfying:

\begin{equation}
R^{\text{Simplified}}_{BB84}(\eta_{critical}) = R^{\text{Integration}}_{BB84}
\end{equation}\\

Now, for MDI-QKD, we can firstly extend the concepts in the BB84 case, and define the simplified model and the integration model as following:
\begin{equation}
\begin{aligned}
R^{\text{Simplified}}(\Omega) &= R(\langle \eta_A \rangle,\langle \eta_B \rangle) \\
&\times \left( \iint_{\Omega}p_{AB}(\eta_A,\eta_B)d\eta_A d\eta_B \right) \\
R^{\text{Integration}} &= \int_0^1 \int_0^1 R(\eta_A,\eta_B) p_{AB}(\eta_A,\eta_B)d\eta_A d\eta_B
\end{aligned}
\end{equation}

However, a crucial point here is that, for MDI-QKD, the simplified model \textit{does not} accurately represent the key rate. The reason is that MDI-QKD heavily depends on the \textit{symmetry} between channel transmittances (because it makes use of a two-photon interference in the X basis, and its quantum bit error rate (QBER) will depend on the interference visibility). Suppose the mean transmittances in the channels $\eta_{A_0},\eta_{B_0}$ are equal, and Alice and Bob choose the same intensities. Then, without post-selection, we can obtain $\langle \eta_A \rangle = \eta_{A_0}, \langle \eta_B \rangle=\eta_{B_0}$. This means that, when calculating the simplified model based on $R(\langle \eta_A \rangle,\langle \eta_B \rangle)$, we are assuming a perfectly symmetric setup (which will presumably result in low QBER in the X basis and a high estimated key rate). However, in reality, $\eta_A,\eta_B$ are independent variables, and they are very likely not equal in real-time for the majority of times. This means that, any deviation from $\eta_A=\eta_B$ will result in an increase in the QBER in the X basis. Overall, when one collects the observables (counts and error-counts), he/she will find a much larger-than-expected QBER in the X basis, preventing him/her from acquiring a good estimation of the phase-error rate and a good key rate. Therefore, the simplified model \textit{overestimates} the key rate.

In other words, we make the observation that turbulence-induced channel asymmetry in real-time will decrease the key rate of MDI-QKD. This is very different from what we observed for BB84, where key rate, gain, and error-gain are all near-linear functions, and any increase/decrease in error-gain due to fluctuations cancel out when computing the mean value. On the other hand, for a pair of channels with symmetric mean transmittances, fluctuation always decreases the visibility and increases the QBER.

Since simplified model is not an accurate model anymore for MDI-QKD, here we propose a better representation of the key rate in turbulence. Consider the process of obtaining key rate for MDI-QKD, for instance, for the 4-intensity MDI-QKD protocol \cite{mdifourintensity}:
\begin{equation}
\begin{aligned}
R=P_{s}^2 \{(s e^{-s})^2 Y_{11}^{X,L}[1-h_2(e_{11}^{X,U})]\\
-f_eQ_{ss}^Z h_2(E_{ss}^Z)\}
\end{aligned}
\end{equation}


\begin{figure*}[t]
	\includegraphics[scale=0.22]{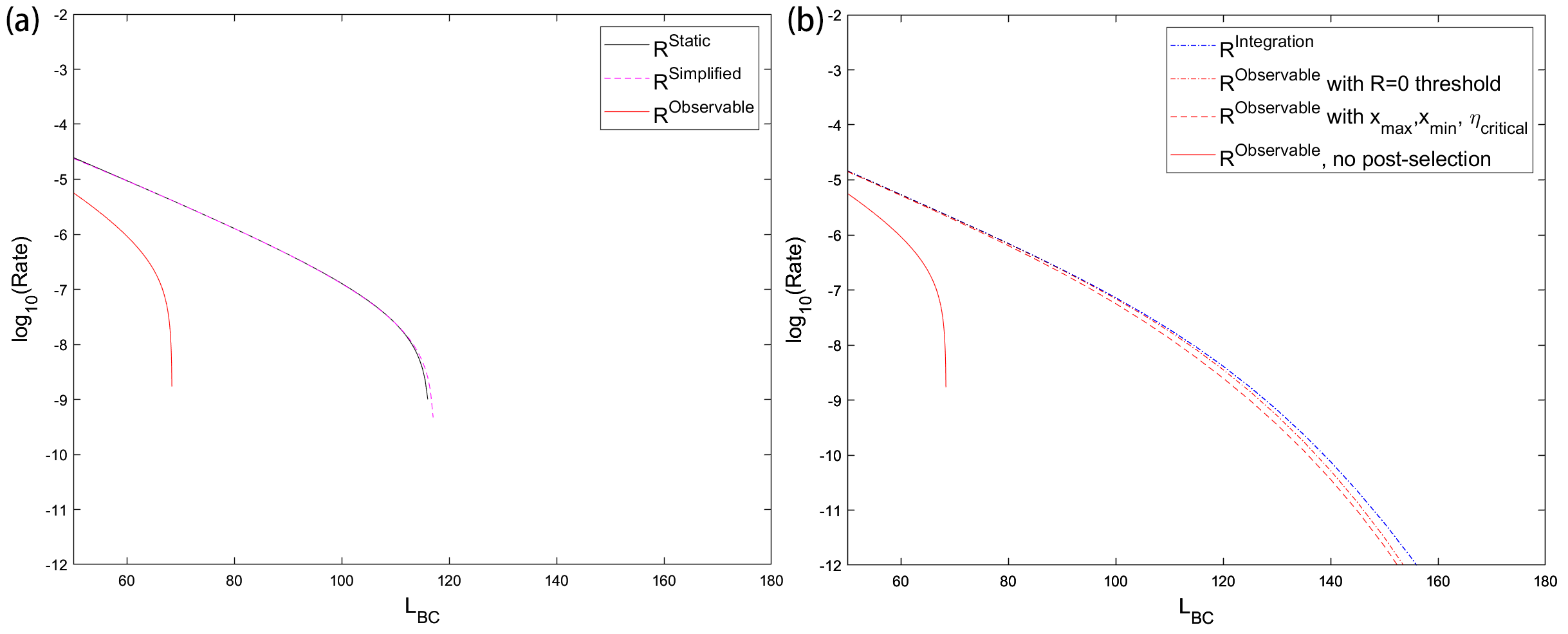}
	\caption{(a) Comparison of key rate models without post-selection. As can be seen, the simplified model incorrectly assumes the same rate as a static model, while from the observable model we can see that MDI-QKD key rate decreases significantly with turbulence. (b) Comparison of key rate obtained with different thresholds. As can be seen, both the R=0 boundary and the straight-line approximations ($x_{min},x_{max},\eta_A^{critical},\eta_B^{critical}$) greatly increase key rate and maximum distance/loss in the channel. Here for convenience we've plotted the key rate versus distance between Charles and Bob in standard optical fibre (0.2dB/km), but in terms of dB we can also see over 30dB of maximum increase in channel loss between Alice and Bob.}
	\label{fig:MDI_rate}
\end{figure*} 
\noindent This protocol uses one signal intensity $s$ for the Z basis to generate the key, and three decoy intensities each for Alice and Bob $\{\mu,\nu,\omega\}$ to perform decoy-state analysis in the X basis. Here the constants include the intensities and the probabilities of sending them, such as $s,P_{s}$, and the error-correction efficiency $f_e$. We can see that the ``variables" that change with $\eta_A,\eta_B$ are the observed gain and QBER in the Z basis $Q_{ss}^Z, E_{ss}^Z$, and the single-photon contributions $Y_{11}^{X,L},e_{11}^{X,U}$ estimated from the X basis observables $Q_{ij}^X, E_{ij}^X$ where $i,j \in \{ \mu,\nu,\omega \}$. Overall, the key rate can be considered as a function of the observables:

\begin{equation}
\begin{aligned}
R(\eta_A,\eta_B) = R[&Q_{ij}^X(\eta_A,\eta_B), T_{ij}^X(\eta_A,\eta_B),\\
&Q_{ss}^Z(\eta_A,\eta_B), T_{ss}^Z(\eta_A,\eta_B)]
\end{aligned}
\end{equation}

\noindent here $T_{ij}^X = Q_{ij}^X E_{ij}^X$ are the error-gains (which correspond to the actual observed error-counts) in the X basis. Similar goes for $T_{ss}^Z=Q_{ss}^Z E_{ss}^Z$. All the error-gains and gains are functions of $\eta_A,\eta_B$ too.

In an actual experiment, the users collect the corresponding counts and error-counts over the entire session, and divide them by the number of signals sent respectively for each intensity combination to acquire the average gain and error-gain. The X basis gain and error-gain are used in decoy-state analysis for privacy amplification, and the Z basis error-counts and counts follow error-correction and key generation. We can see that, since an experiment actually performs privacy amplification and error-correction on the \textit{observables} collected over the entire session and calculates their average values, we can define an \textbf{observable model} that accurately represents the expected key rate in an experiment:

\begin{equation}
\begin{aligned}
R^{\text{Observable}}(\Omega) = R[&\langle Q_{ij}^X \rangle, \langle T_{ij}^X \rangle, \langle Q_{ss}^Z \rangle, \langle T_{ss}^Z \rangle]
\end{aligned}
\end{equation}

\noindent This model represents the actual observables one would get in an experiment, and takes into consideration the effect turbulence-induced fluctuations can have on the average QBER (error-counts) in the X basis. We plot the observable model versus static and simplified model in Fig. \ref{fig:MDI_rate} (a) (without applying any post-selection). We can see that the simplified model fails to characterize the effect of turbulence and assumes the same key rate as a static channel, while the observable model shows that MDI-QKD key rate greatly decreases with turbulence if not addressed actively.

\subsection{Choice of Post-selection Thresholds}

In this subsection we discuss some good choices for the threshold $\Omega$ Charles uses when post-selecting signals.

Let us first plot out the key rate versus $\eta_A,\eta_B$ function in Fig. \ref{fig:freespace_MDI} (b). Note that this function is only determined by the experimental parameters (misalignment, dark count, detector efficiency) and the intensities Alice and Bob choose, and it \textit{does not depend on} the joint PDTC of the channels. 

From it we can see that the key rate follows a near-parabolic shape, with two asymptotic lines corresponding to the maximum and minimum channel asymmetry $x=\eta_A/\eta_B$ (which graphically correspond to the reciprocal of slope). This is reasonable because the QBERs (mainly $E_{ij}^X$, but also $E_{ss}^Z$ which is less sensitive but still affected) depend on the channel asymmetry, and the key rate becomes zero at two cut-off points $x_{max},x_{min}$. The existence of these two cut-off lines for asymmetry is analytically proven for the infinite-decoy case in Ref. \cite{this_asymMDI}, which shows that for $R=0$ there are two groups of solutions at $x_{max},x_{min}$, regardless of the actual amplitudes of $\eta_A,\eta_B$, while for the case of finite-decoys the result is numerically shown (although yet to be proven analytically because there is no simple analytical formula for $E_{ij}^X$).

In BB84, the optimal threshold we select was $\eta_{critical}$ such that all $\eta<\eta_{critical}$ satisfy $R(\eta)=0$. Similarly, for MDI-QKD, since we know all the information in the $R(\eta_A,\eta_B)$ plot, here we can propose to use a threshold $\Omega^{boundary}$ defined by where $ R(\eta_A,\eta_B) \geq 0$:

\begin{equation}
\Omega^{boundary} = \{(\eta_A,\eta_B)\in \bm{R}^2: R(\eta_A,\eta_B) \geq 0 \}
\end{equation}

To simplify the implementation, it's also possible to approximate this boundary (which takes a near-parabolic shape) with four straight lines, representing two characteristics:  $\eta_{A}^{critical},\eta_{B}^{critical}$ (which are mainly determined by the dark counts), and $x_{max},x_{min}$ (which are mainly affected by basis misalignment). We can then require the signals to jointly satisfy the conditions on signal-to-noise ratio and symmetry, i.e.

\begin{equation}
\begin{aligned}
\Omega^{joint} = \{(\eta_A,\eta_B)\in \bm{R}^2: &\eta_{A_T}\leq \eta_A \leq 1, \eta_{B_T}\leq \eta_B \leq 1, \\
&x_{min} \leq \eta_A/\eta_B \leq x_{max} \}
\end{aligned}
\end{equation}

\noindent The two thresholds are plotted in Fig. \ref{fig:freespace_MDI} (c). Importantly, the plot $R(\eta_A,\eta_B)$ is generated without any information of the PDTC, and all the above information including $R=0$ boundary, $\eta_{A}^{critical},\eta_{B}^{critical}$, and $x_{max},x_{min}$ are all acquired from the plot of $R(\eta_A,\eta_B)$ alone, which only depends on the intensities and the experimental parameters (misalignment, dark count rate, detector efficiency etc.), but are independent of the actual joint PDTC of the channel. This means that they can be very conveniently pre-determined before the experiment for real-time post-selection of signals (instead of having to optimize the threshold after the experiment).\\


\section{Numerical Results}

Here we present a numerical simulation for the post-selection method we proposed. For simplicity, here we consider a 4-intensity protocol with infinite-data size and fixed intensities.
First, we tested the key rate of MDI-QKD in the presence of turbulence without post-selection. As can be seen, the simplified model ``overestimates" the key rate since it assumes the same key rate as a static model. The observable model correctly captures the decrease in key rate due to turbulence-induced asymmetry. We can see that, without post-selection, the performance of MDI-QKD greatly decreases in the presence of turbulence.

In Fig. \ref{fig:MDI_rate}(b), we plot the observable model obtained from the two threshold methods versus no post-selection. As can be seen, the $R=0$ boundary $\Omega^{boundary}$ (and the approximated $\Omega^{joint}$ which has a key rate just slightly less than the former) captures the most information of the key rate function and results in a key rate very similar to the upper bound of integration model $R^{\text{Integration}}$, which asymptotically utilizes all information of the probability function and is expected to always produce higher key rate than models that utilize average values - i.e. an upper-bound for the observable model. Nonetheless, we can see that using the thresholds, the performance we obtain with observable model approaches this upper bound very closely, meaning that the thresholds we propose are near-optimal.\\

\section{Discussions}

In this chapter we make an important observation that turbulence-induced channel asymmetry decreases the key rate of MDI-QKD. This means that using post-selection is not only a potential means of improvement, but actually might be a necessity in acquiring a good rate, since simply not addressing the turbulence will result in a low rate. We then proposed the powerful solution of a prefixed-threshold post-selection method for MDI-QKD, which can greatly increase the maximum tolerable loss in MDI-QKD communications, paving the way towards future experimental implementations of MDI-QKD.

A few remaining questions include: (1) although we numerically show that certain threshold choices can result in a near-optimal key rate, it remains to be shown analytically why such a boundary of $R=0$ gives maximum key rate after post-selection. (2) In reality, the free-space channels between Alice and Bob are likely not of equal mean transmittances (e.g. due to different distances). The $R(\eta_A,\eta_B)$ map (and R=0 boundary) method in principle holds true for asymmetric cases where the two channels have different mean transmittances $\eta_{A_0} \neq \eta_{B_0}$ (which will be represented by a lopsided joint PDTC), and the users can also choose different intensities (which results in a lopsided $R(\eta_A,\eta_B)$ contour too). More testing remains to be shown for these cases. (3) finite-size effects are important considerations in practical QKD. For this case the intensities (and the probabilities of sending them) need to be highly optimized for a good key rate, and the optimal parameters change with distance - which changes the rate function too. Moreover, post-selection not only decreases the total amount of signals, but will also result in stronger statistical fluctuation among post-selected signals too, hence affecting the key rate. In this case a more careful discussion is needed. These questions will be the subject of our future studies.

\chapter{MDI-QKD over Asymmetric Channels}

This chapter is largely reproduced from our paper Ref. \cite{this_asymMDI}.

\section{Background}

As we introduced in Chapter 2, there is great interest in the community to apply QKD to a network setting, i.e. designing \textit{quantum networks} that can connect and provide service to numerous users, who may freely join or leave a network. There have been multiple field implementations of point-to-point QKD networks, e.g., Refs.\cite{quantumnetwork1,quantumnetwork2,QKDbackbonenetwork}, but they all rely on trusted relays, which could become susceptibilities in the security of the system. MDI-QKD addresses this problem by allowing untrusted relays. This makes MDI-QKD a great candidate for future quantum networks. A first three-user MDI-QKD network experiment has been reported in Ref. \cite{mdinetwork}.

\begin{figure}[t]
	\includegraphics[scale=0.5]{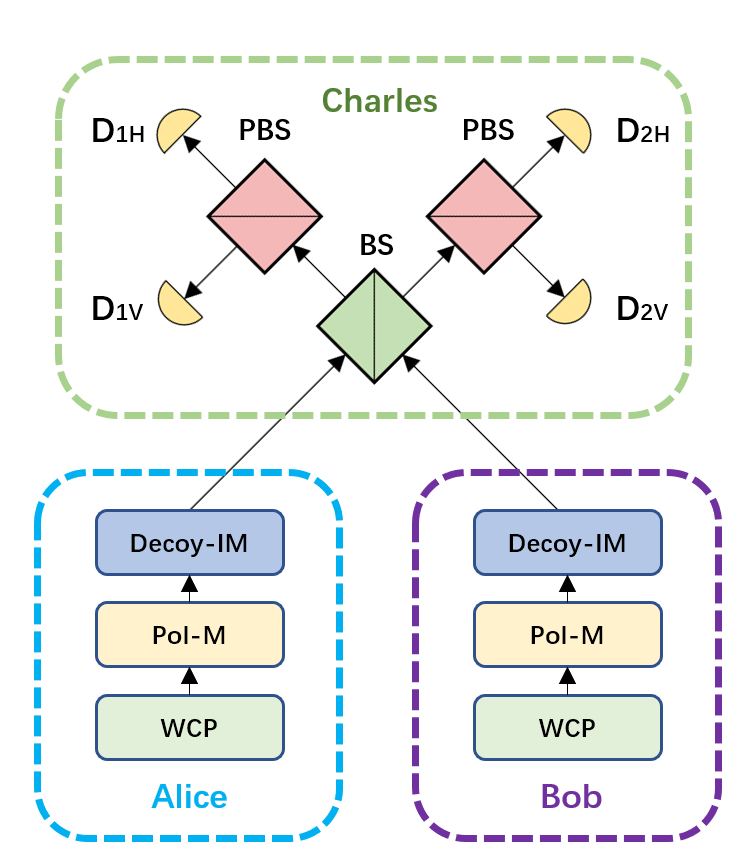}
	\caption{An example schematic setup of MDI-QKD, reproduced from Ref. \cite{mdiqkd}. Alice and Bob respectively send signals through \textit{two} channels, and Charles measures the signals with a Bell-state measurement (by observing the coincidence click events in detectors $D_{1H},D_{1V},D_{2H},D_{2V}$, behind the beam-splitter (BS) and polarizing beam-splitters (PBS)) and announces the results. Here weak coherent pulse (WCP) sources are used, in combination with decoy states created with intensity modulators (Decoy-IM). In this particular setup, polarization encoding is used (with polarization modulators (Pol-M)), but MDI-QKD can be performed with other degrees-of-freedom, such as time-bin phase encoding, too. Reproduced from \cite{this_asymMDI} @2019 APS.}
	\label{fig:mdiqkd}
\end{figure}

\begin{figure}[h]
	\includegraphics[scale=0.28]{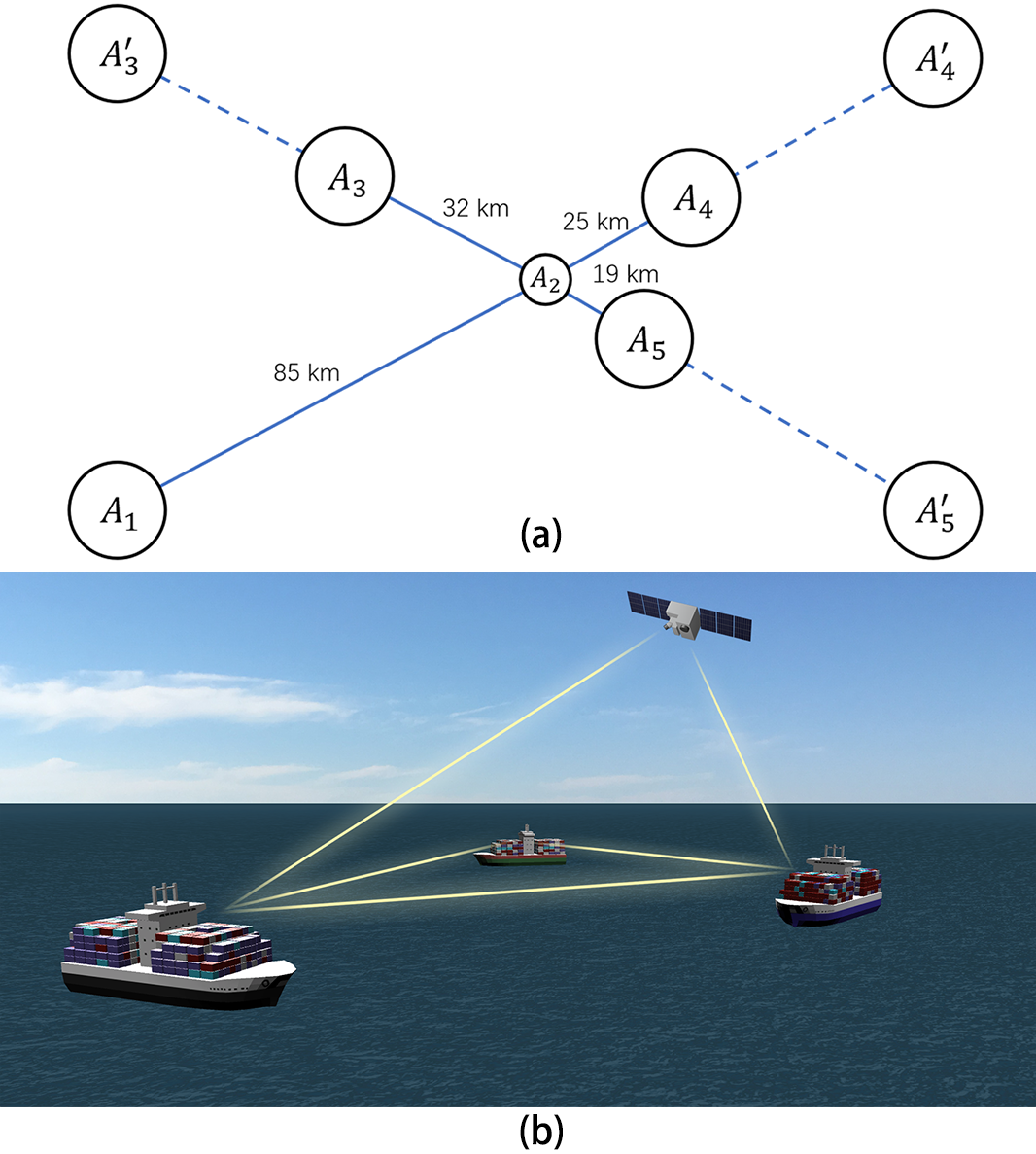}
	\caption{(a) Part of the QKD network setup from Ref.\cite{quantumnetwork1}. Here as an example, we focus on the five nodes with high asymmetry (Nodes $A_1,A_3,A_4,A_5$ connected with $A_2$, corresponding to nodes 1-5 in Ref.\cite{quantumnetwork1}), where $A_2$ can be set up as an untrusted relay. We keep the same topology and redraw it as a star-shaped MDI-QKD network with four users connected to a single untrusted relay. When performing MDI-QKD, all users need to accommodate for the longest channel (i.e. $A_1$) and add losses to their channels (e.g. extending to $A_3', A_4', A_5'$), if previous protocols are used. (b) Ship-to-ship communication and ground-satellite communication, where the participants' distances to the detector are constantly changing, and the channels will thus have quickly varying asymmetry. Reproduced from \cite{this_asymMDI} @2019 APS.}
	\label{fig:network}
\end{figure}

However, because MDI-QKD makes use of two-photon interference between the incoming signals of Alice and Bob, in order to obtain a good interference visibility, a major limitation is imposed on MDI-QKD: \textit{all users need to have near-identical (i.e. symmetric) distances to the untrusted relay} \cite{mdipractical,HOM}, and the key rate will quickly decrease (due to reduced interference visibility and increased QBER) with an increased level of asymmetry between channels.\footnote{Note that, there have also been proposals for continuous variable (CV) MDI-QKD~\cite{CVMDIQKD1,CVMDIQKD2}, which provides high key rate for short distances, but is typically limited to distances $<25km$ even when assuming a high detector efficiency of $98\%$. In this work we will focus on only discrete-variable MDI-QKD.} 

To circumvent this limitation, previous experiments of MDI-QKD either were performed in the laboratory over symmetric fibre spools~\cite{mdiexp1,mdiexp2,mdiexperiment,mdi200km,mdiexp3,mdi404km}, or had to resort to a makeshift solution of deliberately adding a tailored length of fibre to the shorter channel, which introduces additional loss in exchange for better symmetry~\cite{mdiPOP}. Such a strategy of adding additional fibres not only is cumbersome as it requires halting the system (and it is also not practical when there are many pairs of connections in a quantum network, or when channel loss is changing with time), but also results in suboptimal key rate when channels are asymmetric. An intuitive illustration of this can be found in Appendix C.1.

In reality, it is very likely for a quantum network to have asymmetric channels due to e.g. different geographical locations of sites. For instance, the channel losses in Ref.\cite{quantumnetwork1, quantumnetwork2} are largely different. Here we select 5 nodes from the Vienna QKD network~\cite{quantumnetwork1} and show them in Fig.~\ref{fig:network}(a), where the biggest difference between channels is as large as 66km. If we'd like to perform MDI-QKD over these locations, although one can add additional fibres to each channel to compensate for channel differences, users will have to accommodate for the lowest-transmittance channel -- just like in ``\textit{Liebig's barrel}" -- and have sub-optimal rate. Moreover, in a scalable network with large numbers of dynamically added/deleted users, it is not practical to add fibres and maintain symmetry between each pair of users all the time. Additionally, if one is to implement a MDI-QKD network over free-space between mobile platforms (e.g., satellite-based MDI-QKD~\cite{satelliteQKD} or maritime MDI-QKD between ships), the losses in the channels are constantly changing, and the channels will often be highly asymmetric, as shown in Fig.~\ref{fig:network}(b). In summary, in a practical quantum network, the requirement on symmetric channels will significantly limit the key rate of previous MDI-QKD protocols, and seriously hinder the widespread deployment of MDI-QKD.

The issue of MDI-QKD with asymmetric channel losses was first considered in Ref. [4], which provided a rule of thumb on the ratio of intensities between Alice's and Bob's signals. However, Ref. [4] assumes infinitely large data size, and was also restricted to protocols where the same set of intensities for the optical signals are used in the two bases, X and Z. In this paper, we make no such assumptions.


In this chapter, we present a new method to overcome this crucial limitation directly, and enable high-rate MDI-QKD with arbitrary user locations. First of all, our work provides an important conceptual insight: a common folklore in the field is that MDI-QKD relies on Hong-Ou-Mandel (HOM) dip and, therefore, it is important to use matched intensities at the beam-splitter of the receiver, Charles, in MDI-QKD. Here, we show that such a folklore is, in fact, a misconception. We show that there is an intrinsic asymmetry between the two bases of MDI-QKD: only the X-basis relies on the indistinguishability of photons from the two beams, while the Z basis does \textit{not}. We will later show that one can make use of such asymmetry to create protocols resilient against asymmetric channels. We also show that this is a general theoretical result applicable to many protocols, including various types of MDI-QKD protocols, and potentially other protocols such as MDI-quantum-digital-signature \cite{MDIQDS1,MDIQDS2}, and twin field QKD \cite{TFQKD} in asymmetric settings. 

Following this conceptual insight, we present a novel method in this chapter to combat channel asymmetry. We make use of the inherent asymmetry between bases in MDI-QKD, and propose a type of asymmetric MDI-QKD protocols where intensities are not only different for Alice and Bob, but also different in X and Z bases. In this way, by decoupling the bases and also allowing Alice and Bob to independently vary their intensities, the users can effectively compensate for channel asymmetry in one basis, and optimize the key generation rate in another basis, enabling a much higher key rate for our asymmetric protocols in the presence of channel asymmetry. Additionally, we present a technique that makes it possible to efficiently perform local search for high-speed parameter optimization over the extremely large parameter space for such asymmetric protocols (which would be otherwise impossible to optimize using previous algorithms such as in Refs. \cite{mdiparameter,mdipractical}).

The protocols we propose have important practical impacts. We show that, when channels are asymmetric, our protocols can provide a much higher key rate than previous protocols \cite{mdifourintensity}\footnote{A previous protocol of interest for MDI-QKD is the 4-intensity protocol proposed by Zhou et al. \cite{mdifourintensity}. In this protocol, Alice and Bob each use three intensities $\{\mu, \nu, \omega\}$ in the X basis to perform decoy-state analysis \cite{decoystate_LMC,decoystate_Hwang,decoystate_Wang}, and uses one signal intensity $\{s\}$ in the Z basis to generate the secret key. The 4-intensity protocol can greatly improve MDI-QKD performance under limited data size. However, it limits its discussions to the symmetric case only (Alice and Bob using identical parameters), which is suboptimal in an asymmetric setting. Although Ref. \cite{mdifourintensity} mentioned on passing the possibility of using different intensities of optical signals for Alice and Bob, little analysis on this important case was performed there. So, up till now, it has not been clear how exactly Alice and Bob could compensate for asymmetric channel losses with different signal intensities.} that were designed for symmetric channels (for instance, one to two orders of magnitudes higher rate at mid-to-close distances e.g. (60km, 10km) for (Alice's, Bob's) channels, with 50km difference in channel distances). Moreover, it enables a much larger region of possible combinations of channels: for instance, even at a small data size of $N=10^{11}$ ($N$ is defined as the total number of pulses sent by Alice and Bob), one can generate a high secret key rate of $R=10^{-7}$ per pulse even through an extremely asymmetric channel pair of (0km, 90km) for (Alice's, Bob's) channels, whereas with previous protocols no key could be generated at all. Using the type of protocol we proposed, one can completely remove the requirement of symmetric channels in MDI-QKD. This makes our proposal a powerful solution that enables high-rate MDI-QKD under arbitrary asymmetry, which paves the way for practical MDI-QKD networks where users can be placed at arbitrary locations.

The structure of this chapter is as follows: In section 5.2.1, we point out a theoretical insight that there is an inherent asymmetry between the two bases of MDI-QKD. In section 5.2.2 we make use of this insight and propose a type of asymmetric protocols that simultaneously have two kinds of asymmetries: the asymmetry between Alice and Bob, and the asymmetry between X and Z bases, which, together, enable the protocol to effectively compensate for different pairs of channels and maintain good key generation rate. We show the security of such a scheme in section 5.2.3. We then describe how to optimally choose the asymmetric parameters in Subsection II.D. While our proposal applies to a general type of MDI-QKD protocols, we also highlight a specific implementation, a ``7-intensity protocol", and show that it is a good trade-off between key rate and ease of implementation. Lastly, we present the simulation results to show the effectiveness of our protocol compared with prior protocols in section 5.3.

\section{Asymmetric Protocols}

In this section, we present a general theoretical framework for designing protocols that can effectively compensate for channel asymmetry and provide a good key rate.

Note that, our method proposed here is a general result that can be applied to any decoy-state MDI-QKD protocol with WCP source, for both asymptotic and finite-size cases, as long as (1) decoupled bases are used and (2) Alice and Bob have asymmetric intensities. We show in Appendices C.2 and C.3, that the scaling of key rate versus distance is determined by the signal states, so in principle any number of decoy states (e.g. two-decoy, three-decoy, and four-decoy) can be used so long as they can effectively estimate the single-photon contributions. In principle, such method can potentially even be applied to other types of protocols in asymmetric settings, such as MDI-quantum-digital-signature \cite{MDIQDS1,MDIQDS2}, and twin field QKD \cite{TFQKD}, which are also currently limited to symmetric intensities between Alice and Bob, and which also use two asymmetric bases X and Z.

\subsection{Asymmetry between Bases in MDI-QKD}

Here, we start by making a key theoretical observation on MDI-QKD:\\

\begin{figure}[!ht]
	\includegraphics[scale=0.18]{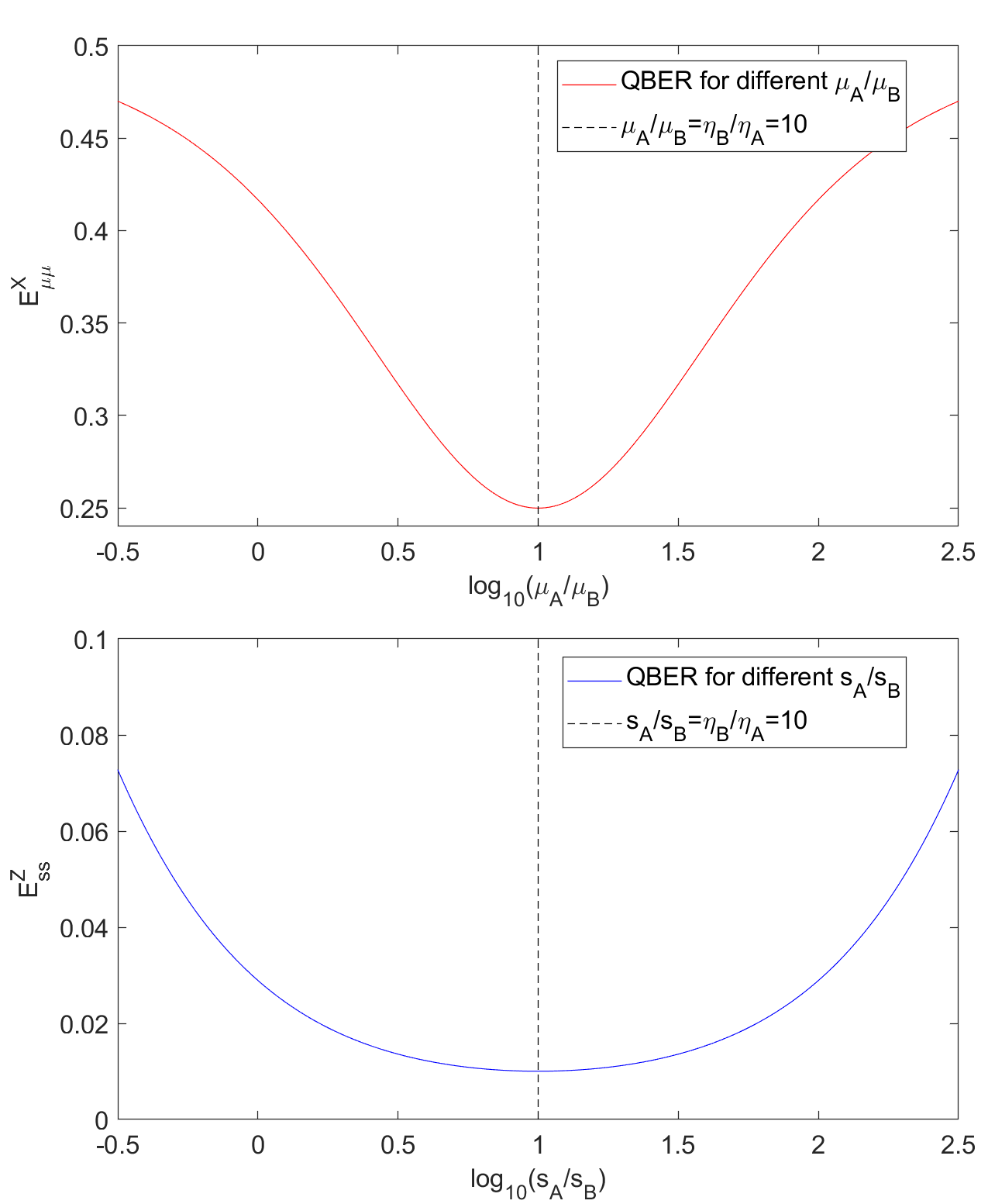}
	\caption{An example of the respective quantum bit error rate (QBER) in X basis and Z basis, $E^X_{\mu\mu}$ and $E^Z_{ss}$ (we consider a pair of decoy states with intensities $\mu_A,\mu_B$ in the X basis, and signal state intensities $s_A,s_B$ in the Z basis) versus ratio of intensities, for MDI-QKD using WCP sources. Parameters from Table 5.2 are used. Here we consider the case where the respective distances from Alice and Bob to Charles are $L_{\text{A}}=60km,L_{\text{B}}=10km$ (i.e. the ratio of transmittances in the two channels satisfies $\eta_{\text{B}}/\eta_{\text{A}}=10$). We fix $s_{\text{A}}\eta_{\text{A}}+s_{\text{B}}\eta_{\text{B}}=0.17$ (or $\mu_{\text{A}}\eta_{\text{A}}+\mu_{\text{B}}\eta_{\text{B}}=0.07$) and scan over different $s_{\text{A}}/s_{\text{B}}$ (or $\mu_{\text{A}}/\mu_{\text{B}}$). Specifically, we also mark out the position where $s_{\text{A}}\eta_{\text{A}}=s_{\text{B}}\eta_{\text{B}}$ ($\mu_{\text{A}}\eta_{\text{A}}=\mu_{\text{B}}\eta_{\text{B}}$). Because QBER in the X basis heavily depends on the visibility of two-photon interference, it is lowest when intensities arriving at Charles' beam-splitter are equal (Similar observation has been made in Ref.\cite{HOM}.) However, importantly, the Z basis does not require signal indistinguishability, and its QBER is determined mainly by misalignment. The misalignment makes the Z basis QBER also slightly dependent on the interference visibility, and lowest when arriving intensities are equal, but such QBER is much less sensitive to unbalanced intensities and is relatively low even if $s_{\text{A}}\eta_{\text{A}} \neq s_{\text{B}}\eta_{\text{B}}$. Therefore, by decoupling X and Z basis, we can maintain highly balanced decoy state intensities arriving at Charles in the X basis, while further optimizing signal intensities to obtain higher key rate. As a quantitative example of such difference in sensitivity, let us consider $L_{\text{A}}=60km$, $L_{\text{B}}=10km$ and $N=10^{11}$ (Table 5.4 line 1), and focus on two pairs of decoy states with intensities $\mu_A,\mu_B,\nu_A,\nu_B$. An optimal key rate of $R=3.1\times 10^{-5}$ can be achieved, where optimal decoy state intensities satisfy ${\mu_A / \mu_B}={\nu_A / \nu_B}=9\approx{\eta_B / \eta_A}$, and $E^X_{\mu\mu},E^X_{\nu\nu}$ are both close to 25\% (see Table 5.4 for the full list of intensities and probabilities). Deviating from ``balanced arriving intensities" results in low or zero rate (for instance when choosing symmetric ${\mu_A / \mu_B}={\nu_A / \nu_B}=1$, $E^X_{\mu\mu}$ and $E^X_{\nu\nu}$ are almost 42\%, and rate $R=0$). On the other hand, the optimal signal states satisfy $s_A / s_B=3.5$, which deviates from $\eta_B / \eta_A$, but $E^Z_{\mu\mu}$ is still a rather small $0.013$. In fact, here even if we choose $s_A=s_B=0.245$, we can still get $R=1.1\times 10^{-5}$ while $E^Z_{ss}=0.029$. Figure based on \cite{this_asymMDI} @2019 APS, but re-generated with fixed $s_{\text{A}}\eta_{\text{A}}+s_{\text{B}}\eta_{\text{B}}$ and $\mu_{\text{A}}\eta_{\text{A}}+\mu_{\text{B}}\eta_{\text{B}}$ to make the curves symmetric.}
	\label{fig:QBER}
\end{figure}

\noindent \textbf{Observation 1:} \textit{For MDI-QKD, there is an inherent asymmetry between the bases: only the diagonal (X) basis requires the indistinguishability of the signals from Alice and Bob, while the rectilinear (Z) basis does not.}\\

Such an observation is because, in MDI-QKD, Charles performs a Bell-state measurement with \textit{post-selection}, making the protocol different from a simple two-photon interference in standard Hong-Ou-Mandel (HOM) dip. Here let us follow the discussions in Ref. \cite{mdiqkd} (and consider the experimental setup from Fig. 1 in Ref. \cite{mdiqkd}). Note that while Alice and Bob randomly send signals in the X and Z bases, Charles always measures in the Z basis (as defined by his polarizing beam-splitter(PBS)) and post-selects detector click events that correspond to the two Bell states  $\ket{\psi^+}=1/\sqrt{2}(\ket{HV}+\ket{VH})$ and $\ket{\psi^-}=1/\sqrt{2}(\ket{HV}-\ket{VH})$. Such a post-selection results in an asymmetry between the two bases. In the Z basis, only events where Alice and Bob sent opposite states (e.g. $\ket{HV}$ or $\ket{VH}$) are accepted as bits. In these cases no photon interference takes place, and indistinguishability between the two input photon beams is not required, because each of the clicking detectors respectively receives only a signal from either Alice or Bob but never both. For WCP sources, in the ideal case with no misalignment or dark counts, the intensities of the pulses and even their spectrum and timing need not be matching at all. In the X basis, however, the events may correspond to identical states sent by Alice and Bob (e.g. $\ket{++}$ and $\ket{--}$ corresponding to $\ket{\psi^+}=1/\sqrt{2}(\ket{++}-\ket{--})$), which do interfere at the beam splitter.\footnote{Another case where Alice and Bob sent $\ket{+-}$ or $\ket{-+}$ corresponds to the other Bell state, $\ket{\psi^-}$. A two-photon interference happens not at the beam splitter but at the polarizing beam splitter (PBS) instead. This setup is slightly different from HOM interference but similar to that of Ref. \cite{Photon_Interference}, and also requires indistinguishability of e.g. spectrum, timing, and matching intensities. For simplicity, here we will use the term ``two-photon interference" for both cases.} To ensure that the correct events are triggered, a good visibility of such a two-photon interference is required. Note that for WCP sources, the interference visibility is at most 50\% (resulting in a 25\% observed QBER in the X basis even in the ideal case - for instance $E^X_{\mu\mu},E^X_{\nu\nu}$ when Alice and Bob use decoy states with intensities $\mu_A,\mu_B$ and $\nu_A,\nu_B$ - but we can perform decoy-state analysis to correctly estimate a low QBER among single photon components, $e_{11}^{X,U}$) and the visibility will quickly drop when intensities are mismatched, such as observed in \cite{HOM}. 

Therefore, a low QBER in the X basis heavily relies on the indistinguishability of the signals and the balance of incoming intensities at Charles, while such dependence is not present in the Z basis.\footnote{In the non-ideal case with basis misalignment, there may be a slight dependence in the Z basis too, as we see in Fig. \ref{fig:QBER}, because misalignment results in crosstalk between signals from the two bases, but it will be a much smaller dependence than that in the X basis.} Such a conclusion is rather general and also not dependent on the degree-of-freedom used for qubit encoding - such as polarization encoding or time-bin phase encoding (where $\ket{HV}$ and $\ket{VH}$ in the Z basis correspond to pairs of early and late pulses, which will similarly not interfere at the beam splitter since they have different timing).

\subsection{Using Decoupled Bases and Asymmetric Intensities}

Here, let us consider the key rate formula of MDI-QKD \cite{mdiqkd,mdifourintensity}:
\begin{equation}
\begin{aligned}
R=P_{s_{\text{A}}}P_{s_{\text{B}}} \{(s_{\text{A}} e^{-s_{\text{A}}})(s_{\text{B}} e^{-s_{\text{B}}}) Y_{11}^{X,L}[1-h_2(e_{11}^{X,U})]\\
-f_eQ_{ss}^Z h_2(E_{ss}^Z)\}
\end{aligned}
\end{equation}

\noindent where $s_A,s_B$ are the intensities of signal states, $Q_{ss}^Z, E_{ss}^Z$ are the gain and QBER in the Z (signal) basis, $Y_{11}^{X,L},e_{11}^{X,U}$ are the lower (upper) bounds of single-photon yield and QBER, estimated from the decoy state statistics in the X basis (i.e. the observed gain and QBER for decoy states $Q_{ij}^X,E_{ij}^X$, where $i,j$ are decoy intensities, such as in $\{\mu_{\text{A}},\nu_{\text{A}},\omega\}$ and $\{\mu_{\text{B}},\nu_{\text{B}},\omega\}$ if Alice and Bob each chooses three decoy states), $h_2$ is the binary entropy function, and $f_e$ is the error-correction efficiency. 

In the key rate formula, the first part corresponds to key generation (where the privacy amplification depends on the single-photon contributions estimated from decoy-state analysis), and the second part corresponds to error-correction for the signal states. We can make another key observation on the intensities used in the two bases:\\ 

\noindent \textbf{Observation 2:} \textit{In our protocol, the intensities of the signal states $\{s_{\text{A}}, s_{\text{B}}\}$ used in the Z basis are independent from those of the decoy states used in the X basis, which means that the privacy amplification process (to bound Eve's information on the final key, i.e. estimate the phase error rate) in the X basis is completely \textit{decoupled} from error-correction in the Z basis for key generation.}\\

This means that, it is possible for us to independently adjust the decoy states and the signal states in their respective bases, to compensate for channel asymmetry, or to optimize key rate.

For the decoy states, their role is to estimate the single-photon contributions as accurately as possible. As mentioned above, when channels are asymmetric, using the same intensities for Alice and Bob (hence different intensities arriving at Charles after the channels' attenuation) will result in poor interference visibility and high QBER in the X basis, and consequently poor estimation of $e^{X,U}_{11}$. For a good interference visibility, Alice and Bob should try to maintain similar intensities arriving at Charles, so the decoy intensities should be chosen to roughly satisfy
\begin{equation}
\mu_{\text{A}}\eta_{\text{A}}=\mu_{\text{B}}\eta_{\text{B}}
\end{equation}

\noindent where $\eta_{\text{A}}$ and $\eta_{\text{B}}$ are the channel transmittances in Alice's and Bob's channels. A similar equation holds true for $\nu_{\text{A}}$ and $\nu_{\text{B}}$.

In contrast, for the signal states, they are not involved in privacy amplification. On the other hand, they affect the signal state gain and QBER $Q^Z_{ss}, E^Z_{ss}$ (which determine the amount of error-correction), and the probability of sending single photons for key generation $s_{\text{A}} e^{-s_{\text{A}}} s_{\text{B}} e^{-s_{\text{B}}}$. The key point is, the QBER $E^Z_{ss}$ does \textit{not} require indistinguishability of the signals. If there is no misalignment or noise, $E^Z_{ss}$ would be zero regardless of incoming intensities. In practice, due to imperfections such as misalignment, the QBER $E_{ss}^Z$ (whose full expression can be found in Appendix C.3 Eq. C.7) still slightly depends on channel asymmetry and is also minimal if incoming intensities at Charles are balanced - but this is for a much different reason (due to misalignment) than that in the X basis (mostly due to two-photon interference). Furthermore, $E_{ss}^Z$ is much less sensitive to channel asymmetry than QBER in the X basis. We can observe this from Fig.\ref{fig:QBER}.

Note that, not only do signal intensities affect the signal state QBER, they also determine the probabilities of sending single photons, hence affecting key generation too. This means that, while having similar received signal intensities at Charles is surely one important criterion in achieving a good key rate, the optimal choice of signal state intensities requires a trade-off between the single photon probabilities and the error correction (and their optimal values can be found by numerical optimization). Generally speaking, the ratio of signal intensities ${s_{\text{A}} / s_{\text{B}}}$ does not satisfy a similar relation as Eq. 5.2, i.e. generally

\begin{equation}
s_{\text{A}}\eta_{\text{A}}\neq s_{\text{B}}\eta_{\text{B}}
\end{equation}

Therefore, the protocols we propose have two inherent asymmetries: an asymmetry between Alice and Bob (so that they can have different intensities, and establish good two-photon interference in the X basis), and an asymmetry between the X and Z bases (which allows decoy and signal states to be independently optimized). Such inherent asymmetries in the protocols allow us to have a novel choice of parameters and maintain a good key rate of MDI-QKD, even when Alice's and Bob's channels have very different levels of loss. A more detailed discussion on how such independent choices of decoy and signal states affect the key rate can be found in Appendix C.4.

We will discuss the security of such a scheme in section 5.2.3, and in section 5.2.4 we will discuss how to actually choose the optimal decoy and signal intensities. We introduce the main challenge in implementing such asymmetric protocols - performing efficient parameter optimization over a huge parameter space - and how we address this problem by proposing two important theoretical results for the key rate function of asymmetric MDI-QKD, and using them to design an efficient optimization algorithm.

\subsection{Security}

In this subsection, we show that the security of our protocol with decoupled bases and asymmetric intensities is not compromised compared to prior art protocols. Here we state two \textit{facts} that its security relies upon, both of which have been proven in established papers \cite{mdiqkd,decoystate_LMC}: 

(1) \textit{Given the same photon number i} in a pulse, Eve has no way of differentiating whether it came from the decoy states or the signal states in the same basis;

(2) the single photons pairs in X and Z bases cannot be distinguished from each other. 

The first fact is proven in Ref. \cite{decoystate_LMC}, which ensures the decoy-state analysis works even with asymmetric intensities, and the second fact is proven in Ref. \cite{mdiqkd}, which ensures that the decoupling of bases works.\\

For fact (1), note that the density matrices for the two states are the same, independent of whether they came from the decoy states or signal states. In quantum mechanics, whenever two density matrices are the same, they are indistinguishable. Similarly, for fact (2), note that the density matrices for the two cases are the same. Therefore, it is not possible to distinguish them even in principle.

In more detail, for fact (1), let us first consider a very similar process in traditional decoy-state BB84 \cite{decoystate_Hwang,decoystate_LMC,decoystate_Wang}. Consider Alice using a laser with intensities $\mu$ or $\nu$ to send weak coherent pulses to Bob. The photon number $i$ follows a Poissonian distribution, e.g. 

\begin{equation}
\begin{aligned}
p(i|\mu) &= e^{-\mu} {\mu^i \over i!}\\
p(i|\nu) &= e^{-\nu} {\nu^i \over i!}
\end{aligned}
\end{equation}

The crucial point is that, the conversion from such a probability distribution described by intensities $\mu$ or $\nu$ to a certain photon number $i$ is a \textbf{Markov process}, i.e. it is memoryless, and for any given photon number $i$ in the channel, it does not contain any information of the intensity it came from. Surely, Eve can guess with a conditional distribution, e.g. $p(\mu|i)$, the likelihood that it comes from a certain intensity, but whatever actions Eve chooses to perform on the signal (e.g. choosing different levels of yields $Y_i^{E1},Y_i^{E2},...$ with different probabilities for a given photon number $i$) will be completely independent of the intensities Alice chose when sending the signal. This means that, in the asymptotic case with infinite data\footnote{For the finite-size case, the yield and QBER in Eq. 5.5 are replaced with their \textit{expected values}, $\langle Y_i \rangle$ and $\langle e_i \rangle$, and the result still holds true, i.e. $\langle Y_i \rangle(\mu) = \langle Y_i \rangle(\nu)$, $\langle e_i \rangle(\mu) = \langle e_i \rangle(\nu)$. Similar applies to the yield and QBER in MDI-QKD. More details on finite-size analysis can be found in Appendix C.8.}, given the same photon number $i$ in a pulse, we will still always have yield $Y_i$ and QBER $e_i$ satisfying

\begin{equation}
\begin{aligned}
Y_i(\mu) &= Y_i(\nu) \\
e_i(\mu) &= e_i(\nu)
\end{aligned}    
\end{equation} 

\noindent when Alice uses intensities $\mu,\nu$. This is the exact observation as made in Ref. \cite{decoystate_LMC} (in Eqs. 4,5). 

Similarly, for MDI-QKD, for any given pair of i-photon and j-photon pulse, there is no information on which pair of intensity settings (e.g. $\mu_A^1,\mu_B^1$ or $\mu_A^2,\mu_B^2$) they came from. That is, the yield $Y_{i,j}$ and QBER $e_{i,j}$ will satisfy

\begin{equation}
\begin{aligned}
Y_{i,j}^X(\mu_A^1,\mu_B^1)&=Y_{i,j}^X(\mu_A^2,\mu_B^2)\\
e_{i,j}^X(\mu_A^1,\mu_B^1)&=e_{i,j}^X(\mu_A^2,\mu_B^2)\\
Y_{i,j}^Z(\mu_A^1,\mu_B^1)&=Y_{i,j}^Z(\mu_A^2,\mu_B^2)\\
e_{i,j}^Z(\mu_A^1,\mu_B^1)&=e_{i,j}^Z(\mu_A^2,\mu_B^2)
\end{aligned}
\end{equation}

\noindent for signals in each of the bases X and Z (the latter two equations are meaningful if one also uses multiple decoy intensities in the Z basis, although here we only use the first two equations as decoy-state analysis is only performed in X basis for our protocols). This, again, is a well-established result for decoy-state MDI-QKD as used in the original MDI-QKD paper \cite{mdiqkd}. Note that, this result (which simply comes from the fact that the sending of photon number i from an intensity $\mu$ is a Markov process) does not rely on the fact that Alice and Bob use the same intensities, and will remain unchanged for asymmetric intensities too, i.e. $\mu_A^1 \neq \mu_B^1$ and $\mu_A^2 \neq \mu_B^2$. Also, note that for successful decoy-state analysis we do not require the symmetry between the two bases, i.e. $Y^X_{i,j} (\mu_A^1,\mu_B^1 )=Y^Z_{i,j} (\mu_A^2,\mu_B^2)$ or $e^X_{i,j} (\mu_A^1,\mu_B^1 )=e^Z_{i,j} (\mu_A^2,\mu_B^2)$ for multi-photon pulses are not required.\\

Fact (2) stems from the fact that, for single-photon components, Alice and Bob send the same density matrices $\rho^X_{1,1}=\rho^Z_{1,1}$, that is, the single-photon pairs are basis-independent. This is an important result explicitly stated in the original MDI-QKD paper \cite{mdiqkd} (in the ``Security Analysis" section in the Supplemental Information). Using decoupled bases (i.e. a different set of intensities for X and Z bases) does not affect the single-photons themselves at all, but only affects the probability of sending these single-photon pairs. However, this process of sending single-photon pair is again a Markov process. That is, although the single-photon pairs might have different probabilities of coming from either X or Z bases (which Eve can fully be aware of, just like in ``efficient BB84" \cite{efficientBB84}, where basis choice probability is biased on purpose), for any given pair of single-photons that are sent, they are described by exactly the same density matrix and there is no information contained on which basis they came from, i.e. they are basis-independent. Therefore, we can safely conclude that $Y_{11}^X=Y_{11}^Z$, which is the reason we can perform decoy-state in the X basis only to estimate $Y_{11}^X$, and use $Y_{11}^Z=Y_{11}^X$ to obtain single-photon yield in the Z basis. 

The security of a scheme of decoupling the bases in MDI-QKD and using $Y_{11}^Z=Y_{11}^X$ has also been theoretically studied in Ref. \cite{mdifourintensity} and (in the Appendix of) Ref. \cite{mdiexp3} \footnote{The idea of decoupling the bases was first studied for BB84 in Ref. \cite{biasedQKD}. There the relation $Y_{1}^Z=Y_{1}^X$ was used for single photons instead of single photon pairs.}, and the scheme has also been experimentally demonstrated in Ref. \cite{mdi404km} and Ref. \cite{mdiexp3} - although all these works were focused on the scenario of symmetric channels only, and did not discuss the role of decoupled bases in compensating channel asymmetry, which is one of the main novelties of our work. However, physically, the only difference between the signals sent from Alice and Bob in our protocol and those in prior protocols will be the different intensities on the two arms (which we know, from fact (1) $Y^X_{i,j} (\mu_A^1,\mu_B^1 )=Y^X_{i,j} (\mu_A^2,\mu_B^2)$, will not affect the security of decoy-state analysis), and for decoupled bases we use the same result $Y_{11}^Z=Y_{11}^X$ for single-photon pairs, which is no less secure than prior works either.

\subsection{Parameter Optimization}

In this section we discuss how to perform efficient parameter optimization for such asymmetric protocols. Here we highlight an implementation that we denote as ``7-intensity protocol" (which is the case where three decoy intensities are used in the X basis). We show that it is a good trade-off between key rate and ease of implementation, and focus on this implementation when discussing parameter optimization (and the following numerical simulations). Nonetheless, we also show the generality of our method by including the results for other protocol cases (e.g. two-intensity and four-intensity) in Appendix C.5.

Note that, results in the previous subsection are general and not limited to the number of decoys Alice and Bob use in the X basis. For instance, while using signal states $\{s_A,s_B\}$ in the Z basis, in the X basis Alice and Bob can each use a different set of two decoy states $\{\mu,\nu\}$, three decoy states $\{\mu,\nu,\omega\}$, or even four decoy states $\{\mu,\nu,\nu_2,\omega\}$. The concept of asymmetric intensities between Alice and Bob can in principle also be applied to prior art protocols with non-decoupled bases, such as in Refs. \cite{mdipractical,decoyMDI_Wang} (where Alice and Bob use the same three decoy states $\{\mu,\nu,\omega\}$ for both bases, and the Z basis $\mu$ is used as the signal state for key generation) - it is just that such protocol will have lower key rate since $\mu$ cannot simultaneously satisfy asymmetry compensation and key rate optimization.

\begin{table*}[t]
	
	\caption{Example key rate comparison among MDI-QKD protocols where Alice and Bob use different numbers of decoy states in X basis (and each keeps one signal state in the Z basis). The protocol in Ref. \cite{mdipractical,decoyMDI_Wang} where bases are not decoupled is also included for comparison. We use parameters from Table 5.2, $L_{\text{A}}=60km$, $L_{\text{B}}=10km$, and $N=10^{11}$. We can see that, regardless of the protocol, using asymmetric intensities between Alice and Bob always provides a higher key rate when channels are asymmetric. The three-decoy protocol has a significantly higher key rate than either the prior art protocol (which also uses three decoy states but uses non-decoupled bases) or two-decoy case. While the asymmetric four-decoy case can provide the highest key rate, it provides a limited performance increase of 60\%, but comes at a cost of a more complex experimental implementation and more difficult data collection and analysis. Therefore, in the presence of channel asymmetry, the three-decoy case, whose asymmetric case corresponds to the ``7-intensity protocol" (marked in bold), provides a good trade-off between ease of implementation and performance.} 
	\begin{center}
		\begin{tabular}{ccccccc}        
			\hline \hline
			Parameters & prior art protocol in \cite{mdipractical,decoyMDI_Wang} & two-decoy & three-decoy & four-decoy \\
			\hline
			Symmetric & $6.834 \times 10^{-10}$& $0$ & $3.890\times 10^{-7}$ & $1.057 \times 10^{-5}$\\
			Asymmetric & $5.378\times 10^{-7}$ & $7.715 \times 10^{-6}$ & $\boldsymbol{3.106 \times 10^{-5}}$ & $4.932 \times 10^{-5}$\\
			\hline \hline
		\end{tabular}
	\end{center}
	
\end{table*}

As an example, in Table 5.1 we list a comparison between the key rate of using different numbers of decoy states (with and without asymmetric intensities between Alice and Bob) in the presence of asymmetric channels. We include the non-decoupled-bases case \cite{mdipractical,decoyMDI_Wang} too. We can see that, regardless of the protocol, using asymmetric intensities between Alice and Bob always provides a higher key rate when channels are asymmetric. Also, the three-decoy case provides significant performance improvement over either the two-decoy case or prior art protocol (which also has three decoy states, meaning that decoupled bases are crucial in the compensation for channel asymmetry). While the asymmetric four-decoy case can provide the highest key rate, it provides a limited performance increase (60\%) over three-decoy, but comes at a cost of more complex experimental implementation as well as more difficult data collection and analysis. See Appendix C.5 for a more detailed comparison between the protocols. Overall, we can see that the three-decoy case provides a good balance between ease of implementation and performance. 

Therefore, for practicality here, in the following text we will focus on the three-decoy case as a concrete example (whose symmetric case is the 4-intensity protocol \cite{mdifourintensity}), and generalize it to the asymmetric case by allowing Alice and Bob to have independent intensities and probabilities. This enables a ``7-intensity protocol" (with 3 independent $\{s,\mu,\nu\}$ for each of Alice and Bob, and the vacuum state $\omega_{\text{A}}=\omega_{\text{B}}=\omega=0$) in the asymmetric case.\\

For such a protocol, efficient and accurate parameter optimization is crucial for obtaining good key rate (especially when considering the finite-size effects). For the 7-intensity protocol we need to use a total of 12 parameters for a full finite-size parameter optimization:

\begin{equation} 
\vec{v}=[s_{\text{A}}, \mu_{\text{A}}, \nu_{\text{A}}, P_{s_{\text{A}}}, P_{\mu_{\text{A}}}, P_{\nu_{\text{A}}}, s_{\text{B}}, \mu_{\text{B}}, \nu_{\text{B}}, P_{s_{\text{B}}}, P_{\mu_{\text{B}}}, P_{\nu_{\text{B}}}]. \nonumber
\end{equation}

\noindent here we denote the parameters as a vector $\vec{v}$, and when all devices and channel parameters (e.g. channel loss, misalignment, dark count rate, detector efficiency, etc.) are fixed, the key rate is a function of the intensities and probabilities $R(\vec{v})$, and the question of intensity parameter optimization can be viewed as searching for:

\begin{equation}
\vec{v}_{\text{opt}}=\text{arg max}_{\vec{v} \in V}[R(\vec{v})]
\end{equation}

\noindent where $V$ is the search space for the parameters.

To provide a high key rate under finite-size effects, the optimal choice of parameters is very important in implementing the protocol. However, the 7-intensity protocol has an extremely large parameter space of 12 dimensions, for which a brute-force search is next to impossible. Therefore, to efficiently search over the parameters in a reasonable time, a local search algorithm must be applied. But, as we will show here, an important characteristic of asymmetric MDI-QKD is the discontinuity of first-order derivatives for the function $R(\vec{v})$ with respect to the intensity parameters in $\vec{v}$. This means that a straightforward local-search algorithm, such as previously proposed in~\cite{mdiparameter}, will inevitably fail to find the optimal point, since it requires continuous first-order derivatives of the searched function.

Here we will present two important theoretical results for the key rate versus parameter function, and propose a method to circumvent the problem of discontinuous derivatives and perform efficient and correct local search in parameter space. This method helps us overcome the biggest challenge in successfully implementing the 7-intensity protocol.

\begin{figure}[h]
	\includegraphics[scale=0.35]{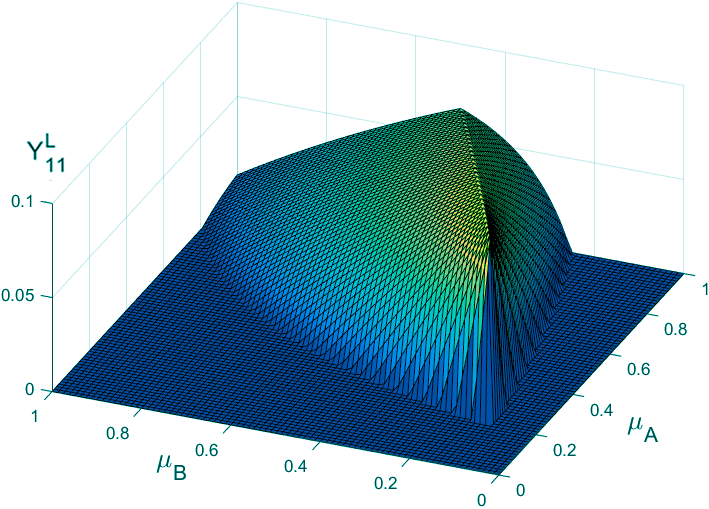}
	\caption{An example of the discontinuity of first-order derivatives of $Y_{11}^{L}$ vs $\mu_{\text{A}}, \mu_{\text{B}}$ function in decoy-state MDI-QKD, for fixed values of $\nu_{\text{A}}=0.2$, $\nu_{\text{B}}=0.1$. Note the ridge on the line ${\mu_{\text{A}} \over \mu_{\text{B}}}={\nu_{\text{A}} \over \nu_{\text{B}}}=2$. Reproduced from \cite{this_asymMDI} @2019 APS.}
	\label{fig:ridge}
\end{figure}

Firstly, we propose that there is an inherent symmetry constraint for the \textit{ratio} of optimal decoy intensities, that\\

\textbf{Theorem I}. \textit{for any arbitrary choice of device and channel parameters, the optimal decoy intensities $\mu_{\text{A}}^{\text{opt}}, \nu_{\text{A}}^{\text{opt}}, \mu_{\text{B}}^{\text{opt}}, \nu_{\text{B}}^{\text{opt}}$  that maximize the key rate always satisfy the constraint:}

\begin{equation}
{\mu_{\text{A}}^{\text{opt}} \over \mu_{\text{B}}^{\text{opt}} } = { \nu_{\text{A}}^{\text{opt}} \over \nu_{\text{B}}^{\text{opt}}}
\end{equation}

Secondly, we make an important observation that,\\

\textbf{Theorem II}. \textit{The key rate versus ($\mu_{\text{A}}, \mu_{\text{B}}$) function, for any given $\nu_{\text{A}}, \nu_{\text{B}}$, does not have continuous first-order derivatives.\\}

Both of these theorems result from the fact that the lower bound for single-photon yield, $Y_{11}^{L}$, in decoy-state analysis (whose expression can be found in Ref. \cite{mdipractical,decoyMDI_Wang}) is a piecewise function that depends on whether ${\mu_{\text{A}} \over \mu_{\text{B}}} \leq {\nu_{\text{A}} \over \nu_{\text{B}}}$, where a boundary line ${\mu_{\text{A}} \over \mu_{\text{B}}} = {\nu_{\text{A}} \over \nu_{\text{B}}}$ exists.

Theorem I states that, the optimal parameters that maximize the key rate must lie exactly on this boundary line, while Theorem II states that, the key rate does not have a continuous partial derivative with respect to $\mu_{\text{A}}$ or $\mu_{\text{B}}$ across this boundary line. This will cause the boundary line to behave like a sharp ``ridge", on which the gradient is not defined. An illustration for this ``ridge" can be seen in Fig. \ref{fig:ridge}. A rigorous proof for Theorems I and II in the asymptotic limit can be found in Appendix C.6.

\begin{figure}[h]
	\includegraphics[scale=0.195]{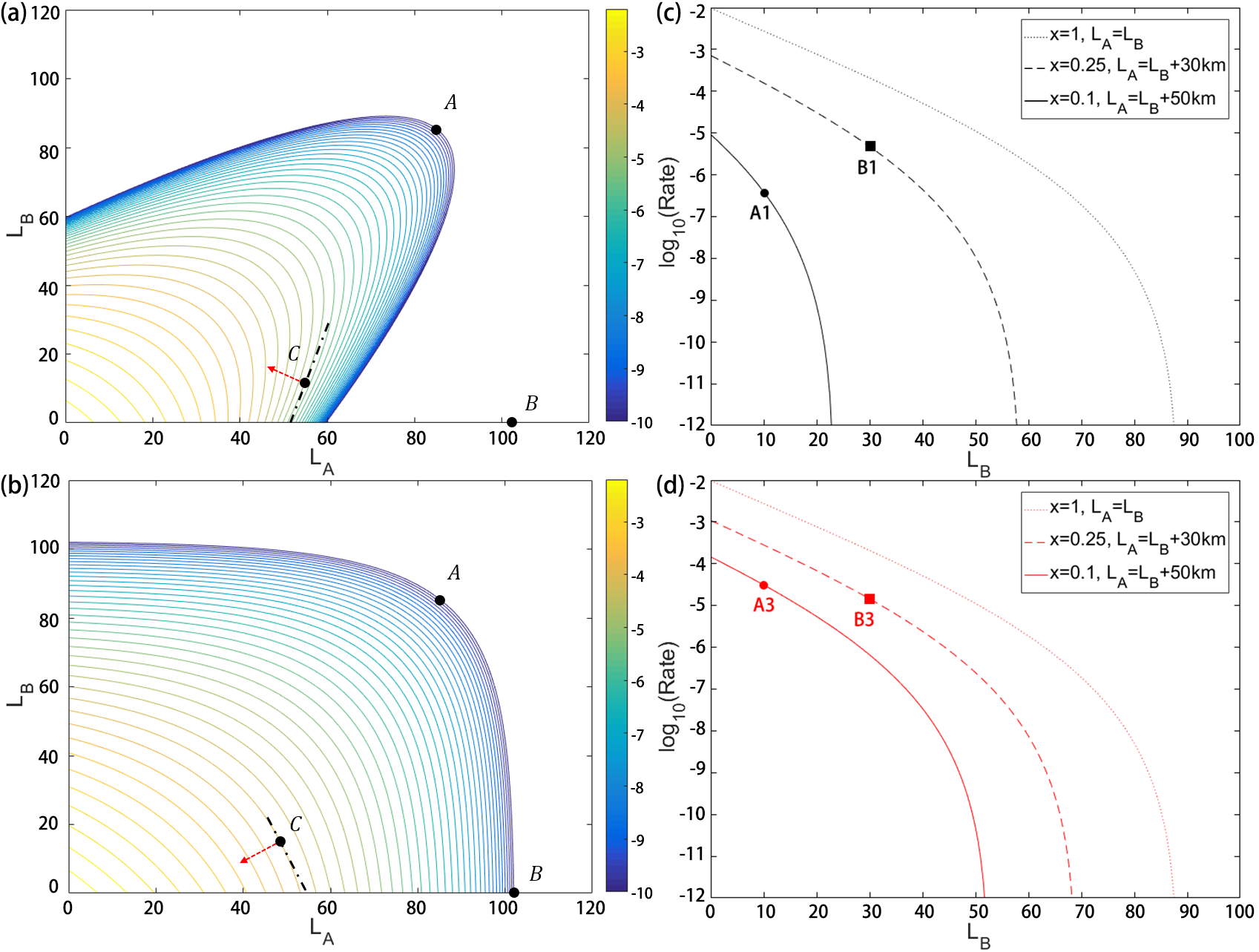}
	\caption{Left: Comparison of rate vs $(L_{\text{A}}, L_{\text{B}})$. The rates are plotted in contours in log-scale, from $10^{-2}$ to $10^{-10}$. We use the parameters from Table 5.2, and $N=10^{11}$. (a) using a previous 4-intensity protocol, (b) using our 7-intensity protocol. As can be seen, while 4-intensity MDI-QKD is limited to only high-symmetry regions, using 7-intensity can greatly increase the applicable region of MDI-QKD, even in extremely asymmetric regions such as $(L_{\text{A}},0)$ where one channel has zero distance (point B). Moreover, we see that with 7-intensity protocol both $L_{\text{A}}, L_{\text{B}}$ components of the gradient for key rate (red dotted arrow) are always negative, meaning that with 7-intensity protocol it is always optimal to only adjust the intensities, and never necessary to add any fibre, while for 4-intensity protocol, adding fibre (e.g. increasing $L_{\text{B}}$ at point C) will sometimes increase the rate. Right: Comparison of rate vs distance (Bob to Charles) for various fixed levels of mismatch $x={\eta_\text{A}\over\eta_\text{B}}$ where $\eta_\text{A},\eta_\text{B}$ are the channel transmittances, (c) using 4-intensity protocol (d) using 7-intensity protocol. As can be observed, the higher the mismatch, the more advantage 7-intensity protocol has (and only when the channels are symmetric will the two protocols perform identically). Data points from A1, A3, B1, B3 from Table 5.3 are also shown in the plots. Reproduced from \cite{this_asymMDI} @2019 APS.}
	\label{fig:2d_Results}
\end{figure}

{\color{black}
	Using Theorems I and II it is possible to transform the coordinates of the search variables, and eliminate the undefined gradient problem of the key rate function. More specifically, instead of expressing $(\mu_{\text{A}},\mu_{\text{B}}),(\nu_{\text{A}},\nu_{\text{B}})$ in Cartesian coordinates, we can express them in polar coordinates $(r_\mu^{\text{polar}}, \theta_\mu^{\text{polar}})$ and $(r_\nu^{\text{polar}}, \theta_\nu^{\text{polar}})$, where polar angles satisfy $\theta_\mu^{\text{polar}}=\theta_\nu^{\text{polar}}$ due to Theorem I. This means we can \textit{jointly} search for:
	\begin{equation}
	\theta_{\mu\nu}^{\text{polar}}=\theta_\mu^{\text{polar}}=\theta_\nu^{\text{polar}}
	\end{equation} 
	\noindent with respect to which the key rate is a smooth function (graphically, this is because we are now always searching along the ``ridge").} In Appendix C.7, we will describe in more detail how to perform efficiently an optimization of the parameters based on local-search to obtain a high secure key rate for our 7-intensity protocol. Our method allows extremely fast and highly accurate optimization for asymmetric MDI-QKD, and takes below 0.1s for each full local search (at any given distance) on a quad-core i7-4790k@4.0GHz PC. Such computing efficiency makes it possible for real-time optimization of intensities on-the-field, and also makes possible a dynamic MDI-QKD network that might add/delete new user nodes in real time. {\color{black}In addition, in Appendix C.7 we also discuss the effect of inaccuracies and fluctuations of the intensities and probabilities on the key rate, and show that our method is robust even in the presence of inaccuracies and fluctuations of the parameters.}

In summary, using our two Theorems and switching to polar coordinates as in Eq. 5.9 allows us to greatly simplify the optimization problem and allow standard coordinate descent method to be applied here.


\begin{table}[t]
	\caption{Parameters for numerical simulations, adopted from~\cite{mdi404km}, including detector dark count rate and efficiency $Y_0$, $\eta_d$, optical misalignment $e_d$, error-correction efficiency $f$, and failure probability $\epsilon$.}
	\begin{center}
		\begin{tabular}{ccccc}    
			\hline    \hline
			$Y_0$ & $\eta_d$ & $e_d$ & $f$ & $\epsilon$\\
			\hline
			$8 \times 10^{-7}$ & 65\% & 0.5\% & 1.16 & $10^{-7}$\\
			\hline \hline
		\end{tabular}
	\end{center}
\end{table}

\begin{table*}[t]
	\caption{Simulation results for asymmetric MDI-QKD in two scenarios: case A (10km, 60km) and case B (30km, 60km), using parameters from Table 5.2 and $N=10^{11}$. We define channel mismatch as $x={{\eta_{\text{A}}}\over{\eta_{\text{B}}}}$ where $\eta_{\text{A}},\eta_{\text{B}}$ are the channel transmittances. Note that in reality, Alice and Bob cannot modify the physical channels, and they can either add loss to the channels or keep them as-is, but cannot decrease channel loss. Three strategies are compared here: A1 and B1 represent using the old 4-intensity protocol directly. A2 and B2 (not in Fig. \ref{fig:2d_Results}) represent adding fibre to the shorter channel to match the longer channel, i.e. making the channels (60km, 60km). And A3, B3 represent using our new 7-intensity protocol without modifying the channels. As shown here, 7-intensity protocol always returns a higher rate than both strategies using 4-intensity protocol.}
	\begin{center}
		\begin{tabular}{ccccccc}        
			\hline \hline
			Protocol & Point &$x$ & $L_{\text{B}}$ & $L_{\text{A}}$ & Rate & Comparison with 4-intensity protocol\\
			\hline
			4-intensity & A1 & 0.1 & 10km & 60km & $3.891 \times 10^{-7}$ & -  \\
			4-intensity + fibre & A2 & 1 & 60km & 60km & $1.862 \times 10^{-6}$ & +379\% \\
			Our protocol  & A3 & 0.1 & 10km & 60km & $3.106 \times 10^{-5}$ & +7883\% \\
			
			4-intensity & B1 & 0.25 & 30km & 60km &  $4.746 \times 10^{-6}$ & - \\
			4-intensity + fibre & B2 & 1 & 60km & 60km & $1.862 \times 10^{-6}$ & -61\% \\
			Our protocol & B3 & 0.25 & 30km & 60km &  $1.445 \times 10^{-5}$ & +204\% \\
			\hline \hline
		\end{tabular}
	\end{center}
\end{table*}

\section{Simulation Results}

Now, we can proceed to study the performance of asymmetric MDI-QKD protocols with full parameter optimization. Again, we use the 7-intensity protocol as a concrete example as it provides a good trade-off between performance and practicality. We also include simulation results for protocols with alternative numbers of decoy states in Appendix C.5.

In the main text we focus on the practical case of having finite data size. The asymptotic case of infinitely many data size (and an analytical understanding of the ideal infinite-decoy case) is discussed in Appendices C.2 and C.3, and its simulation results can be found in Fig.\ref{fig:contours_compare}. 

Our finite-key analysis is described in more detail in Appendix C.8. For simplicity we consider a standard error analysis in numerical simulations, but it is important to note that our theory is fully compatible with composable security. See Appendix C.8 for discussions. {\color{black}In addition, compared to the ``joint-bound" analysis as proposed in Ref. \cite{mdifourintensity} (which jointly considers the statistical fluctuation of multiple observables. Such an analysis model increases the key rate, but introduces multiple maxima undesirable for local search), in the main text here we have chosen to use an ``independent-bound" analysis for our simulations, which considers each variable's statistical fluctuations independently, and is far more stable and faster in simulations. However, we specifically note here that all our methods are fully compatible with joint-bound analysis. We list some representative results generated with joint-bound analysis in Table 5.4 for comparison, and will discuss the different finite-size analysis models in more detail in Appendix C.8.}

Firstly, we consider the key rate for an arbitrary combination of $(L_\text{A}, L_\text{B})$, and perform a simulation of key rate over \textit{all} possible range of Alice and Bob's channels. This provides a bird's-eye view of how using 7-intensities can affect the performance in asymmetric channels. We show the results in Fig.~\ref{fig:2d_Results} (a)(b). From the plot we can make three important observations:

(1) Using 7-intensity protocol, we have a much wider applicable region for asymmetric MDI-QKD, and an acceptable key rate can be acquired even for highly asymmetric channels. In addition, 7-intensity protocol will always provide a higher key rate than 4-intensity protocol, except when channels are already symmetric.

(2) No matter what position one is at, there is never any necessity for adding loss when 7-intensity protocol is used, and optimizing on-the-spot always provides the highest rate. Details can be seen in Fig.~\ref{fig:2d_Results} caption.

(3) Using 7-intensity protocol, even extremely asymmetric scenarios, such as $(L, 0)$ where $L_{\text{B}}=0$, can be used to generate a good key rate. In fact, this provides an even higher rate than with symmetric channels such as $(L, L)$ (As the comparison between points A and B in Fig.~\ref{fig:2d_Results}).

Point 3 has an important practical implication: it can lead to a new type of ``single-arm" MDI-QKD setup. More details can be found in Appendix C.9 and Fig. \ref{fig:singlearm}.

{\color{black}
	Here for Points 1-3, we have a good physical understanding of why allowing different intensities for Alice and Bob can provide a larger region where the key rate is positive. As discussed in section 5.2.2, MDI-QKD requires highly balanced intensities arriving at Charles on the two arms in the X basis for good interference visibility, as well as roughly similar (but not necessarily balanced) levels of arriving intensities in the Z basis, which optimize a trade-off between error-correction and probability of sending single-photons. (The optimal choice of intensities is subject to numerical optimization as described in section 5.2.4). Prior methods with same intensities for Alice and Bob will suffer from high QBER both in X and Z basis, while our method decouples X and Z basis, and optimally chooses Alice and Bob's signal and decoy intensities respectively to compensate for channel asymmetry in both bases, ensuing low QBER and allowing for a much higher key rate under channel asymmetry. Such effect is present in both asymptotic and finite-key scenarios, and is the underlying reason that the 7-intensity protocol can allow high-rate MDI-QKD regardless of channel asymmetry.}

\begin{table*}[t]
	\caption{{\color{black}Examples of optimal parameters for the 7-intensity protocol, using simulation parameters from Table 5.2. The numerical values are rounded to the accuracy of 0.001 in the table here. As can be observed, Alice and Bob's intensities compensate for channel asymmetry, while their intensity probabilities are mostly identical - since the intensities have already compensated for the asymmetry - despite having some numerical noises (as the key rate is not sensitive to the probabilities near the maximum, the algorithm satisfies with them having close enough, rather than perfectly identical, values, so the optimal values found are still slightly different even when $x=1$). As shown in section 5.2, the optimal decoy state ratios are the same, i.e. ${\mu_{\text{A}}\over\mu_{\text{B}}}={\nu_{\text{A}}\over\nu_{\text{B}}}$. Moreover, we can observe that the ratio of decoy states more closely follows $1\over x$ than the ratio of signal intensities.}}
	\begin{center}
		{\color{black}
			\begin{tabular}{cccccccccccccccc}            
				\hline \hline
				$L_{\text{A}}$ & $L_{\text{B}}$ & x & $s_{\text{A}}$ & $\mu_{\text{A}}$ & $\nu_{\text{A}}$ & $P_{s_{\text{A}}}$ & $P_{\mu_{\text{A}}}$ & $P_{\nu_{\text{A}}}$ & $R$\\
				\hline
				60km & 10km & 0.1 & 0.662 & 0.522 & 0.100 & 0.600 & 0.033 & 0.255& $3.106 \times 10^{-5}$  \\
				60km & 30km & 0.25 & 0.593 & 0.457 & 0.089 & 0.581 & 0.036 & 0.266& $1.445 \times 10^{-5}$  \\
				60km & 60km & 1 & 0.402 & 0.305 & 0.063 & 0.478 & 0.047 & 0.330& $1.862 \times 10^{-6}$  \\
				\hline \hline
				$L_{\text{A}}$ & $L_{\text{B}}$ & x & $s_{\text{B}}$ & $\mu_{\text{B}}$ & $\nu_{\text{B}}$ & $P_{s_{\text{B}}}$ & $P_{\mu_{\text{B}}}$ & $P_{\nu_{\text{B}}}$ \\
				\hline
				60km & 10km & 0.1 & 0.202 & 0.058 & 0.011 & 0.600 & 0.031 & 0.256 \\
				60km & 30km & 0.25 & 0.294 & 0.125 & 0.024 & 0.580 & 0.034 & 0.269  \\
				60km & 60km & 1 & 0.402 & 0.305 & 0.063 & 0.480 & 0.047 & 0.329   \\
				\hline \hline
				
			\end{tabular}
		}
	\end{center}
\end{table*}

Additionally, we show that, when channels are highly asymmetric, the asymptotic key rate of the 7-intensity protocol scales quadratically with the \textit{lower} transmittance among the two channels - which means that, albeit always being able to provide higher key rate and being much more convenient than e.g. adding fibres when channels are asymmetric (which is a relation we rigorously prove in Appendix C.3.2 for the asymptotic case), the 7-intensity protocol will not change the asymptotic scaling properties of MDI-QKD key rate, which is still quadratically related to transmittance - {\color{black}physically, this is understandable, since although we effectively compensate for channel asymmetry with optimized intensities and allow good Hong-Ou-Mandel interference at Charles for decoy states, MDI-QKD still fundamentally depends on two single signal photons both passing through the channels, hence its key rate is quadratically related to transmittance, even in the asymmetric case with compensated intensities.} More detailed discussions and analytical proofs of the above observations can be found in Appendices C.2 and C.3.

Now, as a concrete example, let us consider two sets of channels at $(L_{\text{B}}=10km, L_{\text{A}}=60km)$ and $(L_{\text{B}}=30km, L_{\text{A}}=60km)$, through which Alice and Bob would like to perform MDI-QKD. We compare strategies of using the 4-intensity protocol, directly or with fibres added until channels are symmetric, with directly using our 7-intensity protocol. As can be seen in Table 5.3, in this specific example, using 7-intensity protocol can provide one or two magnitudes higher key rate, and its rate is also always higher than either strategy with 4-intensity protocol.

In fact, we can also show this by plotting key rate vs $L_{\text{B}}$ under a fixed mismatch $x={{\eta_{\text{A}}}\over{\eta_{\text{B}}}}$ where $\eta_\text{A},\eta_\text{B}$ are the channel transmittances (i.e. fixed difference between $L_{\text{A}}$ and $L_{\text{B}}$). This is also the scenario studied by Ref. \cite{mdipractical}. Results are shown in Fig.~\ref{fig:2d_Results} (c) and (d). The data points A1/A3 and B1/B3 in Table 5.3 are also plotted. As can be seen, the higher the asymmetry between channels, the more improvement we can gain from using 7-intensity protocol.

\begin{figure}[h]
	\includegraphics[scale=0.27]{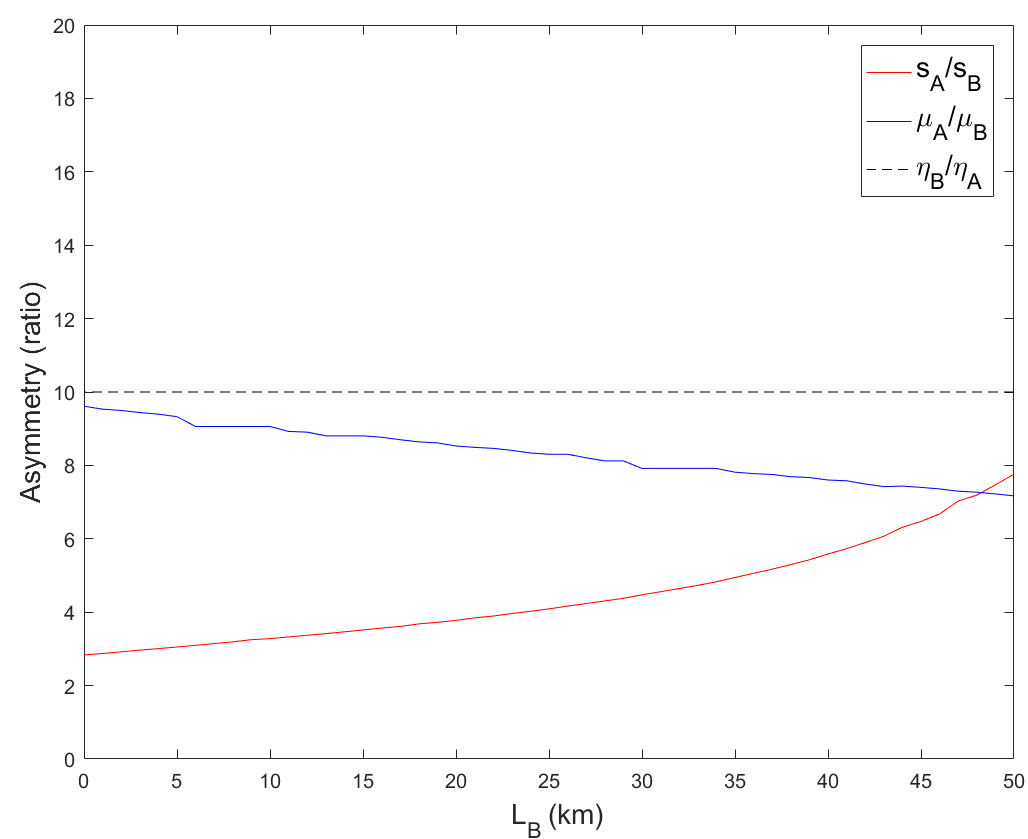}
	\caption{Here we plot the ratios of signal intensities and decoy intensities versus distance, when channel mismatch is fixed at $x=0.1$ (i.e. $L_A=L_B+50km$). The simulation parameters are from Table 5.2 (and this plot of intensities corresponds to the solid red key rate line in Fig. \ref{fig:2d_Results} (d)) We also include the line ${\eta_{\text{B}} \over \eta_{\text{A}}} = 10$ for comparison. We can observe that, the ratio of decoy states roughly follows ${\eta_{\text{B}} \over \eta_{\text{A}}}$ (to maintain good HOM interference visibility in X basis), while the optimal ratio of signal intensities varies greatly between $1$ (optimal for probability of sending single photon) and ${\eta_{\text{B}} \over \eta_{\text{A}}}$ (optimal for $E^Z_{ss}$). This is because signal states affect both key generation and error-correction, so having similar intensities arriving at Charles after channel attenuation is not the only criteria for good key rate, and optimal parameters do not necessarily satisfy $s_{\text{A}}\eta_{\text{A}} = s_{\text{B}}\eta_{\text{B}}$. In fact, since signal states in Z basis are decoupled from X basis, and $E^Z_{ss}$ is less sensitive to unbalanced arriving intensities, $s_{\text{A}}\over s_{\text{B}}$ can be much more freely optimized between $1$ and ${\eta_{\text{B}} \over \eta_{\text{A}}}$, allowing 7-intensity protocol to have higher key rate. Reproduced from \cite{this_asymMDI} @2019 APS.}
	\label{fig:ratio}
\end{figure}

Here, we also list some examples of optimal parameters found by the optimization algorithm, which are listed in Table 5.4. As we can observe from the table, Alice and Bob adjust their intensities to compensate for channel asymmetry. Physically, since MDI-QKD depends on Hong-Ou-Mandel interference of two WCP sources in the X basis, we expect the \textit{received} intensities for decoy states at Charles to be similar on the two arms to ensure good visibility (and consequently lower QBER) in the X basis, i.e. the ratio of decoy intensities ${\mu_{\text{A}} \over \mu_{\text{B}}}$ and ${\nu_{\text{A}} \over \nu_{\text{B}}}$ would roughly follow the rule-of-thumb of $\mu_{\text{A}}\eta_{\text{A}}=\mu_{\text{B}}\eta_{\text{B}}$, which is indeed what we can observe from Table 5.4 and Fig.\ref{fig:ratio}.

On the other hand, the ratio of signal intensities ${s_{\text{A}} \over s_{\text{B}}}$ deviates more from ${\eta_{\text{B}} \over \eta_{\text{A}}}$. This is because, as mentioned in section 5.2.1, signal intensities not only affect the Z basis QBER, but also need to optimize a trade-off between the single photon probabilities and error-correction. This makes it usually not follow $s_{\text{A}}\eta_{\text{A}}=s_{\text{B}}\eta_{\text{B}}$. An illustration of the ratios of decoy intensities and signal intensities can be seen in Fig. \ref{fig:ratio}.


Now, having demonstrated the new 7-intensity protocol, we proceed to introduce a powerful reality application for it: a scalable high-performance MDI-QKD network where any node can be dynamically added or deleted. We consider the channels from a real quantum network setup in Vienna, reported in Ref.\cite{quantumnetwork1}. We focus here on the high-asymmetry nodes, $A_1,A_2,A_3,A_4,A_5$, as shown in Fig. \ref{fig:network}(a). We found that our method leads to much higher key rates, and allows easy dynamic addition or deletion of nodes. Since intensities can be independently optimized for each pair of channels, the establishment of new connections does not affect any existing connections, hence providing good scalability for the network (compared to e.g. the case of using 4-intensity protocol with the strategy of adding fibres, where each channel needs to accommodate for the \textit{longest} link among all channels). See Appendix C.10 for numerical results.

\section{Conclusion}

In summary, in this chapter we have proposed a method of effectively compensating for channel asymmetry in MDI-QKD by adjusting the two users' intensities and decoupling the two bases. Such a method can drastically increase the scenarios MDI-QKD can be applied to while maintaining a good key rate. This study provides a powerful and robust software solution for a scalable and reconfigurable MDI-QKD network.

Our method is also a general result that can in principle be used for e.g. various numbers of decoys, or various finite-size analysis models (e.g. joint-bounds analysis, or composable security with Chernoff's bound). It is also potentially applicable to other types of quantum communication protocols, such as Twin-Field QKD \cite{TFQKD}, or MDI quantum digital signature \cite{MDIQDS1,MDIQDS2}, which both use WCP sources and decoy-state analysis. We hope that our proposal can inspire more future work on the study of asymmetric protocols. In fact, in the following Chapter 6, we will discuss the successful application of our method to TF-QKD in asymmetric settings.

After the completion of an earlier version of our manuscript, we have also experimentally implemented our new protocol in \cite{asymmetric_experiment} (see also \cite{QCrypt}), thus demonstrating clearly the practicality of our work.

\chapter{TF-QKD over Asymmetric Channels}

This chapter is largely reproduced from our paper Ref. \cite{this_asymTF}.

In this chapter, we consider the Twin-Field (TF) QKD protocol in an asymmetric setting. We apply our results from Chapter 5 (for asymmetric MDI-QKD) to TF-QKD, and show that our method can work effectively here too, enabling high key rate for TF-QKD even when channels are highly asymmetric. Just like for MDI-QKD, these results enable a potential quantum network with untrusted relays implemented with TF-QKD, which has the advantage of both high key rate and measurement-device-independence.

\section{Background}

As we have introduced in Chapter 2, the Twin-Field (TF) QKD protocol \cite{TFQKD} is able to beat the fundamental rate-distance trade-off bound (the linear bound). Moreover, it also provides security against attacks on measurement devices similar to MDI-QKD. Because of these advantages, TF-QKD has attracted much attention worldwide since its proposal. Since a rigorous security proof is not provided in the original proposal, several papers have improved the protocol and provided security proof \cite{TFQKD01,TFQKD02,TFQKD03,simpleTFQKD,TFQKD04}. Also, recently there have been multiple reports of TF-QKD demonstrated experimentally \cite{TFexperiment01,TFexperiment02,TFexperiment03,TFexperiment04}.

However, all the above security proofs and experimental demonstrations only consider the symmetric case where Alice's and Bob's channels have the same amount of loss. In reality, though, in a network setting, due to e.g. geographical locations, or Alice and Bob being situated on moving free-space platforms (such as ships or satellites), it is very likely that Alice's and Bob's channels are not symmetric. In the future, if a quantum network is built around the protocol - e.g. a star-shaped network where numerous users (senders) are connected to one central node with measurement devices, asymmetry will be an even more severe problem since it is difficult to maintain the same channel loss for all users (and users might join/leave a network at arbitrary locations). If channels are asymmetric, for prior art protocols, users would either have to suffer from much higher quantum-bit-error-rate (QBER) and hence lower key rate, or would have to deliberately add fibre to the shorter channel to compensate for channel asymmetry, which is inconvenient (since it requires physically modifying the channels) and also provides sub-optimal key rate.

A similar limitation to symmetric channels has been observed in MDI-QKD. In Ref. \cite{this_asymMDI}, we have proposed a method to overcome this limitation, by allowing Alice and Bob to adjust their intensities (and use different optimization strategies for two decoupled bases) to compensate for channel loss, without having to physically adjust the channels. The method has also been successfully experimentally verified for asymmetric MDI-QKD in Ref. \cite{asymmetric_experiment}. 

In this chapter, we will apply our method to TF-QKD and show that it is possible to obtain good key rate through asymmetric channels by adjusting Alice's and Bob's intensities - in fact, we will show that, Alice and Bob only need to adjust their signal intensities to obtain optimal performance. We show that the security of the protocol is not affected, and that an order of magnitude higher (than a symmetric protocol) or 2-3 times higher (than adding fibre) key rate can be achieved with the new method. Furthermore, we show with numerical simulation results that our method works well for both finite-decoy and finite-data cases with practical parameters, making it a convenient and powerful method to improve the performance of TF-QKD through asymmetric channels in reality.

While we use the same main idea of allowing Alice and Bob to use asymmetric intensities to compensate for asymmetric channel losses as in Ref. \cite{this_asymMDI}, there are some key differences that need to be addressed for asymmetric TF-QKD. Firstly, the security proof needs to be discussed, to show that the introduction of asymmetric channels and intensities do not affect security. Secondly, as we will later show in Section 6.4, there is an interesting distinction between the MDI-QKD protocol in \cite{this_asymMDI} and the TF-QKD protocol in \cite{simpleTFQKD} in how the two bases X and Z respectively react to asymmetric incoming intensities, which makes the optimal strategies for asymmetric-intensity MDI-QKD and TF-QKD different. We will discuss this in more detail in Section 6.4.

The idea of TF-QKD protocol through asymmetric channels has also been discussed in a recent paper \cite{tf01}. Our work Ref. \cite{this_asymTF} (which this chapter is based on) is different from Ref. \cite{tf01} in several aspects. First, Ref. \cite{tf01} starts with a different protocol---``sending or not sending protocol". Second, Ref. \cite{tf01} is mostly numerical. In contrast, we start with the ``simple TF-QKD" protocol in \cite{simpleTFQKD}  and consider its asymmetric-intensity version. We include both analytical and numerical reasoning. We also provide a detailed discussion about the physics behind the security of our asymmetric-intensity protocol.

The layout of the chapter is as follows: In Section 6.2, we define the setup and the protocol we use. In Section 6.3 we will extend the security proof in Ref. \cite{simpleTFQKD} to the case with asymmetric intensities and channels. In Section 6.4, we discuss how the performance of TF-QKD is affected by channel asymmetry and asymmetric intensities (and for the latter, what are the best strategies for choosing the intensities). We show the effectiveness of our method with simulation results in Section 6.5.

\section{Protocol}

Here we consider a similar TF-QKD setup as in Ref. \cite{simpleTFQKD} ``Protocol 3". Alice and Bob choose two bases X and Z randomly. When X basis is chosen, Alice (Bob) sends states $\ket{\alpha}_a$ ($\ket{\alpha}_b$) for bit $b_A=0$ ($b_B=0$) or states $\ket{-\alpha}_a$ ($\ket{-\alpha}_b$) for bit $b_A=1$ ($b_B=1$). When Z basis is chosen, Alice and Bob send phase-randomized coherent states $\rho_{a,\beta_A}$ ($\rho_{b,\beta_B}$), where the decoy state intensities are $\{\beta_A,\beta_B\}$. Note that here Alice and Bob have a common phase reference for X basis signals. After the signals are sent to Charles, the detector events are denoted as $k_c,k_d$ (0 denotes no click, and 1 denotes a click).

\begin{figure}[h]
	\centering
	\includegraphics[scale=0.25]{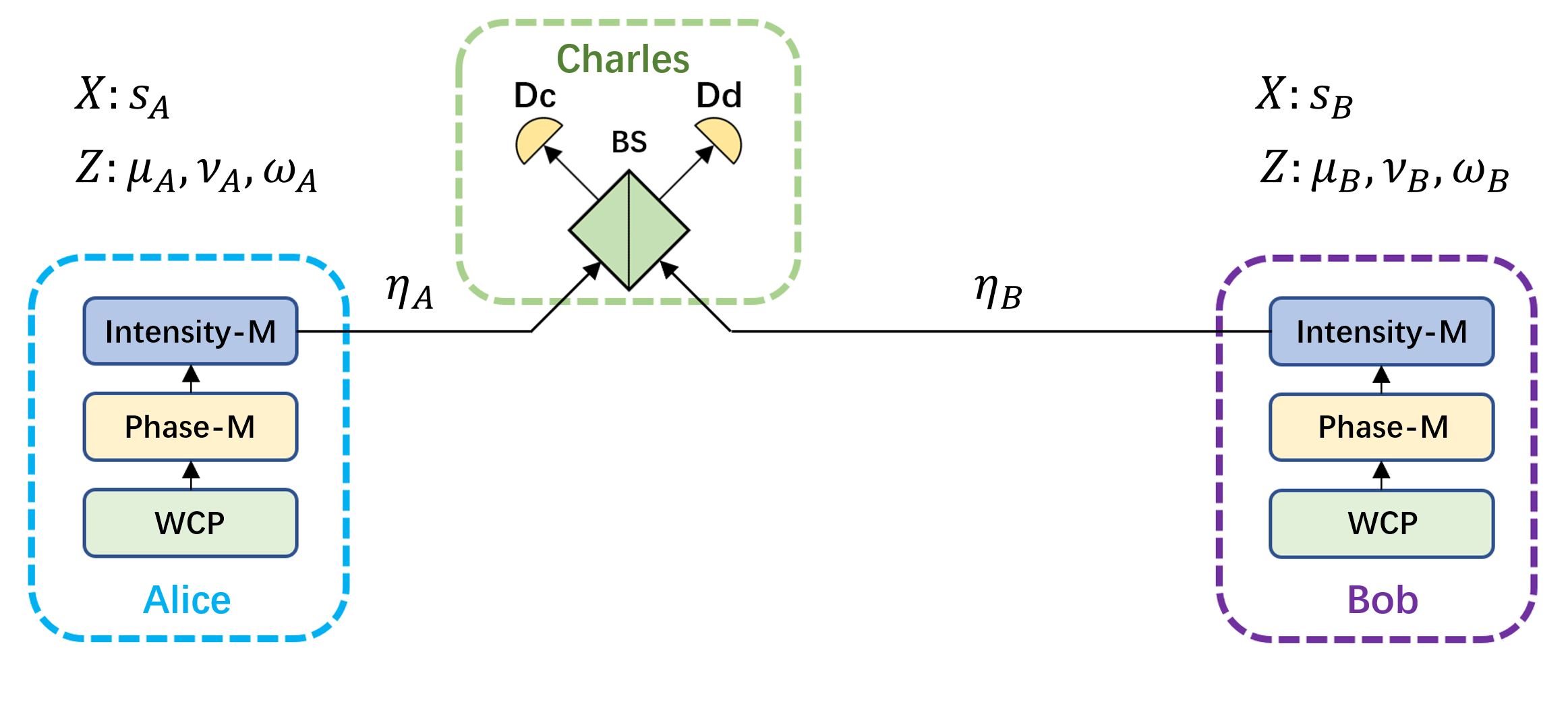}
	\caption{An example setup for a Twin-Field QKD system. Alice and Bob send signals in X and Z bases randomly. In X basis, Alice and Bob send coherent states with amplitudes $\alpha_A,\alpha_B$ (with intensities $s_A = \alpha_A^2$, $s_B = \alpha_B^2$ - in the asymmetric case, we allow $s_A$ to be different from $s_B$), phase-modulated by $\{0,\pi\}$ depending on the encoded bit. In Z basis, Alice and Bob send signals in phase-randomized coherent states, with intensities chosen from $\{\mu_A,\nu_A,\omega_A\}$ and $\{\mu_A,\nu_A,\omega_A\}$, respectively. Charles performs a swap test on the incoming signals and reports the click events in his two detectors $D_c,D_d$ (denoted by $k_c,k_d$). By choosing different intensities between Alice and Bob (and using the decoupled X and Z bases), the protocol can have high key rate even if Alice and Bob's channels have different transmittances, $\eta_A,\eta_B$. Reproduced from \cite{this_asymTF} @2019 NJP (CC-BY 3.0 license).}
	\label{fig:setup}
\end{figure}

The papers \cite{TFQKD,simpleTFQKD} consider only the case where the channels between Alice (Bob) and Charles have equal transmittances. In reality, it is possible that the channels might have different levels of loss, due to e.g. geographical locations or moving platforms. Here we are interested in three questions for TF-QKD with asymmetric channels:

\noindent 1. Does channel asymmetry affect security?

\noindent 2. How does channel asymmetry affect the quantum bit error rate (QBER) and hence key rate?\footnote{In the Supplementary Materials of Ref. \cite{asymmetric_experiment}, we and our collaborators presented a preliminary study on this point, and showed that asymmetry decreases single-photon interference visibility - which will in turn increase observable QBER for TF-QKD.}

\noindent 3. Can we improve the performance of the protocol under channel asymmetry?\\

We will use our method from Ref. \cite{this_asymMDI} and apply it to Protocol 3 in Ref. \cite{simpleTFQKD}, to make an ``\textit{asymmetric-intensity}" TF-QKD protocol that works well even when channels are highly asymmetric. Similar to MDI-QKD, the protocol in \cite{simpleTFQKD} has decoupled X and Z bases. Here we allow Alice and Bob to have different intensities in the X and the Z bases respectively, such that in X basis Alice (Bob) now send states $\ket{\alpha_A}_a$ ($\ket{\alpha_B}_b$) for bit $b_A=0$ ($b_B=0$) or states $\ket{-\alpha_A}_a$ ($\ket{-\alpha_B}_b$) for bit $b_A=1$ ($b_B=1$). We can denote the signal intensities as $s_A = \alpha_A^2$, $s_B = \alpha_B^2$. In the Z basis, the amplitudes for the phase-randomized coherent states, $\{\beta_A,\beta_B\}$, can be different for Alice and Bob too (we can denote the intensities as $\{\beta_A^2,\beta_B^2\}$, and for the three-decoy case, the sets of intensities can be specifically written as $\{\mu_A,\nu_A,\omega_A\}$ and $\{\mu_A,\nu_A,\omega_A\}$). An example setup can be found in Figure \ref{fig:setup}.

We will answer the above three questions by showing in the following text three main pieces of results:\\

(1) Neither asymmetric channels nor asymmetric intensities between Alice and Bob affect security. 

(2) The X basis (signal state) QBER will increase with channel asymmetry, and greatly reduce the key rate of TF-QKD if no compensation is performed - on the other hand, the Z basis gain (as well as the upper bound to the yield and phase-error rate derived from the observable data in the Z basis) is little affected by channel asymmetry.

(3) We can use different intensities between Alice and Bob to compensate for channel asymmetry and get a good key rate - in fact, using only different signal states between Alice and Bob (and keeping all decoy states and probabilities identical for Alice and Bob) can already effectively compensate for channel asymmetry and allow good key rate for asymmetric TF-QKD.

\section{Security}

In this section we will show that neither asymmetric channels nor asymmetric intensities between Alice and Bob affect security. Following the discussion in \cite{simpleTFQKD}, the key is generated from events in the X basis, and the secure key rate is bounded using the bit-error rate and the phase-error rate. The X basis bit-error rate is directly obtained as an observable, hence the key part of the security proof lies in the estimation of X basis phase-error rate (equivalent to the Z basis bit-error rate) based on the Z basis observables - which, since Z basis signals are phase-randomized, is not directly obtainable.

In the security proof in Ref. \cite{simpleTFQKD}, the phase-error rate is obtained by upper-bounding the phase-error rate using the estimated \textit{yields} of given photon numbers $\{m,n\}$ (which can be upper-bounded by using decoy-state analysis, based on observed count rates, i.e. the gains, in the Z basis). 

The key message we'd like to point out is that, this entire estimation process of the phase error rate \textit{does not} rely on the fact that Alice and Bob use the same amplitude $\alpha$ for their signal states, or that the channels have the same transmittance. Therefore, here we will follow the proof in Ref. \cite{simpleTFQKD} step-by-step, but with asymmetric intensities and channel transmittances, to show that the security proof can be easily extended to the asymmetric case.

A small note is that, the formulation of the original security proof in \cite{simpleTFQKD} appears to assume a single Kraus operator for the channel (i.e. a pure state after the measurement, which could involve measurements from Charles and/or Eve), but the proof can, in fact, be extended to cover the general case where the state could be a mixed state after passing through the channel and potentially being disturbed by Eve.\footnote{The explicit discussion about using a single Kraus operator for each announcement outcome was also previously made in Ref. \cite{TFQKD04} in a different security proof. So, this is a known result.} We have discussed with the original authors, and we thank Koji Azuma for pointing out this fact \cite{Koji} and the details of incorporating multiple Kraus operators to represent a mixed state after the measurement (and applying Cauchy-Schwarz inequality to the mixed state to obtain the bounds on the phase error rate). In the following text we will use the formulation of multiple Kraus operators and density matrices as in Koji Azuma's clarification of the original proof.

We can start by imagining a virtual scenario where Alice (Bob) prepares entangled states between a local qubit A (B) and a signal a (b) to be sent to Charles. After Charles performs a measurement on the signals, the X basis phase-error rate (or Z basis bit-error rate) can be obtained by Alice (Bob) measuring their local qubits in the Z basis. The initial states can be written as:

\begin{equation}
\begin{aligned}
\ket{\psi_X^A}_{Aa} &= {1\over \sqrt{2}}(\ket{+}_A\ket{\alpha_A}_a + \ket{-}_A\ket{-\alpha_A}_a)\\
\ket{\psi_X^B}_{Bb} &= {1\over \sqrt{2}}(\ket{+}_B\ket{\alpha_B}_b + \ket{-}_B\ket{-\alpha_B}_b)
\end{aligned}
\end{equation}

\noindent here $\ket{\pm}={1\over \sqrt{2}}(\ket{0}\pm\ket{1})$ are the X basis states. Here in the asymmetric-intensity case, we allow $s_A = \alpha_A^2$ to be different from $s_B = \alpha_B^2$. For convenience, here let us write $\ket{\psi_X}_{AaBb} = \ket{\psi_X^A}_{Aa} \ket{\psi_X^B}_{Bb}$. We can then write the density matrix of the initial state as

\begin{equation}
\rho_{AaBb} = \ket{\psi_X}_{AaBb} \prescript{}{AaBb}{\bra{\psi_X}}
\end{equation}

Now, the process of signals a and b going through their respective channels and Charles making a measurement can be represented by a set of Kraus operators $\{\hat{M}^{ab}_{k_ck_d,e}\}$, where $k_c,k_d$ are Charles' detector events, and $e$ is the (implicit) measurement results of a potential eavesdropper Eve. The superscript $ab$ represents that this operator only acts on the systems $a,b$ (pulses sent to Charles) and not on $A,B$ (local qubits in Alice's and Bob's labs). From the perspective of Alice and Bob, as $e$ is not announced, the state they obtain is equivalent to Eve having discarded all measurement results $e$. After signals pass through the channels and Charles announces the measurement result $k_c,k_d$, the conditional state becomes:

\begin{equation}
\begin{aligned}
\rho'_{AaBb}={{\sum_e \hat{M}^{ab}_{k_ck_d,e} \rho_{AaBb} \hat{M}^{ab\dagger}_{k_ck_d,e}}\over{p_{XX}(k_c,k_d)}}
\end{aligned}
\end{equation}

\noindent here $p_{XX}(k_c,k_d)$ is the X basis Gain for detection events $k_c,k_d$ (which can be $0,1$ or $1,0$ for a detection event to be considered successful). Note that, this set of operator $\{\hat{M}^{ab}_{k_ck_d,i}\}$ includes all information of the channels, detectors (and the eavesdropper) and is a general representation of their joint effects, and, importantly, it \textit{does not} require that the channels are symmetric at all.

By measuring their local qubits in the Z basis, Alice and Bob can obtain the Z basis bit-error rate $e_{ZZ,k_ck_d}$ (i.e. the X basis phase-error rate):

\begin{equation}
\begin{aligned}
e_{ZZ,k_ck_d}=\sum_{j=0,1}\prescript{}{AB}{\bra{jj}} \rho'_{AB} \ket{jj}_{AB}
\end{aligned}
\end{equation}

\noindent where $\rho'_{AB}$, the state of the local qubits $A,B$, can be obtained by performing a partial trace over the systems $a,b$

\begin{equation}
\begin{aligned}
\rho'_{AB} = tr_{ab}(\rho'_{AaBb})
\end{aligned}
\end{equation}

Now, the key observation in the proof of \cite{simpleTFQKD} is that, Alice and Bob making a measurement on the local qubits A and B after sending signals a and b and Charles making a measurement should be equivalent to the time-reversed scenario where Alice and Bob first make local Z basis measurements on the initial pure states $\ket{\psi_X^A}_{Aa},\ket{\psi_X^B}_{Bb}$, and then send the signal systems a and b to Charles. After Alice and Bob make the local measurements, the states become 

\begin{equation}
\begin{aligned}
{}_A\bra{0}\ket{\psi_X}_{Aa}&=\ket{C_0^A}_a\\
{}_A\bra{1}\ket{\psi_X}_{Aa}&=\ket{C_1^A}_a\\
{}_B\bra{0}\ket{\psi_X}_{Bb}&=\ket{C_0^B}_b\\
{}_B\bra{1}\ket{\psi_X}_{Bb}&=\ket{C_1^B}_b\\
\end{aligned}
\end{equation}

\noindent which are cat states:

\begin{equation}
\begin{aligned}
\ket{C_0^A}_a &= e^{-{{\alpha^2_A}\over{2}}} \sum_{n=0}^{\infty} {{\alpha^{2n}_A}\over{\sqrt{2n}}}\ket{2n}_a= \sum_{n=0}^{\infty}c_n^{A,(0)}\ket{n}_a\\
\ket{C_1^A}_a &= e^{-{{\alpha^2_A}\over{2}}} \sum_{n=0}^{\infty} {{\alpha^{2n+1}_A}\over{\sqrt{2n+1}}}\ket{2n+1}_a= \sum_{n=0}^{\infty}c_n^{A,(1)}\ket{n}_a\\
\ket{C_0^B}_b &= e^{-{{\alpha^2_B}\over{2}}} \sum_{n=0}^{\infty} {{\alpha^{2n}_B}\over{\sqrt{2n}}}\ket{2n}_b= \sum_{n=0}^{\infty}c_n^{B,(0)}\ket{n}_b\\
\ket{C_1^B}_b &= e^{-{{\alpha^2_B}\over{2}}} \sum_{n=0}^{\infty} {{\alpha^{2n+1}_B}\over{\sqrt{2n+1}}}\ket{2n+1}_b= \sum_{n=0}^{\infty}c_n^{B,(1)}\ket{n}_b\\
\end{aligned}
\end{equation}

\noindent here the even (odd) cat states only contain nonzero amplitudes for even (odd) photon numbers. Nonetheless we can still write the amplitudes as $c_n^{A,(0)}, c_n^{B,(0)}$ ($c_n^{A,(1)}, c_n^{B,(1)}$) for all photon number states, where the coefficients are zero for odd (even) photon number states in an even (odd) cat state. 

Note that here in the asymmetric-intensity case, Alice and Bob's cat states are not the same, because they use different signal intensities (hence different amplitudes $\alpha_A,\alpha_B$), but as we will show below, the derivation of the upper bound for the phase error rate does not depend on the fact that Alice and Bob have the same cat states. Therefore, the security is not compromised by using asymmetric signal intensities. \footnote{The performance, however, does depend on signal intensities, as we will show in Section 6.4. The protocol favours smaller $\alpha_A,\alpha_B$ for lower phase error rate, which becomes one of the factors - but not the only factor - that affect the optimal choice of signal intensities)}.

For Alice and Bob's local Z basis measurement results $i,j \in \{0,1\}$ and for detection events $k_c,k_d$:

\begin{equation}
\begin{aligned}
\prescript{}{AB}{\bra{ij}} \rho'_{AB} \ket{ij}_{AB} &= \prescript{}{AB}{\bra{ij}} tr_{ab}(\rho'_{AaBb}) \ket{ij}_{AB}\\
&= tr_{ab}(\prescript{}{AB}{\bra{ij}} \rho'_{AaBb} \ket{ij}_{AB})\\
&= {1\over{p_{XX}(k_c,k_d)}}tr_{ab}(\prescript{}{AB}{\bra{ij}} \sum_e \hat{M}^{ab}_{k_ck_d,e} \rho_{AaBb} \hat{M}^{ab\dagger}_{k_ck_d,e} \ket{ij}_{AB}) \\
&= {1\over{p_{XX}(k_c,k_d)}}\sum_e tr_{ab}(\hat{M}^{ab}_{k_ck_d,e} \prescript{}{AB}{\bra{ij}}  \rho_{AaBb}  \ket{ij}_{AB}\hat{M}^{ab\dagger}_{k_ck_d,e}) \\
&= {1\over{p_{XX}(k_c,k_d)}}\sum_e tr_{ab}(\hat{M}^{ab}_{k_ck_d,e} \ket{C_i^A}_a \ket{C_j^B}_b \prescript{}{a}{\bra{C_i^A}} \prescript{}{b}{\bra{C_j^B}} \hat{M}^{ab\dagger}_{k_ck_d,e}) \\
&= {1\over{p_{XX}(k_c,k_d)}}\sum_e \prescript{}{a}{\bra{C_i^A}} \prescript{}{b}{\bra{C_j^B}} \hat{M}^{ab\dagger}_{k_ck_d,e} \hat{M}^{ab}_{k_ck_d,e} \ket{C_i^A}_a \ket{C_j^B}_b  \\
&= {1\over{p_{XX}(k_c,k_d)}} \prescript{}{a}{\bra{C_i^A}} \prescript{}{b}{\bra{C_j^B}} \sum_e\hat{M}^{ab\dagger}_{k_ck_d,e} \hat{M}^{ab}_{k_ck_d,e} \ket{C_i^A}_a \ket{C_j^B}_b  \\
\end{aligned}
\end{equation}

\noindent which means that, the probabilities for local Z basis measurement results $i,j \in \{0,1\}$ (which determine the phase-error rate) can be acquired by observing the gain if Alice and Bob sent cat states. However, Alice and Bob are not really sending cat states - when Z basis is chosen, they are sending phase-randomized coherent states. Using decoy-state analysis, what Alice and Bob acquire are the yields for phase-randomized photon number states, $p_{ZZ}(k_c,k_d|n_A,n_B)=\prescript{}{a}{\bra{n_A}} \prescript{}{b}{\bra{n_B}} \sum_e\hat{M}^{ab\dagger}_{k_ck_d,e} \hat{M}^{ab}_{k_ck_d,e} \ket{n_A}_a \ket{n_B}_b $. The yields for photon number states are linked to Equation 6.8 using the Cauchy-Schwarz inequality that upper-bounds the gains for cat states (and subsequently the phase-error rate):

\begin{equation}
\begin{aligned}
&\prescript{}{a}{\bra{C_i^A}} \prescript{}{b}{\bra{C_j^B}} \sum_e\hat{M}^{ab\dagger}_{k_ck_d,e} \hat{M}^{ab}_{k_ck_d,e} \ket{C_i^A}_a \ket{C_j^B}_b \\
=&\sum_{m_A,m_B,n_A,n_B=0}^{\infty}c_{m_A}^{A,(i)}c_{m_B}^{B,(j)}c_{n_A}^{A,(i)}c_{n_B}^{B,(j)} \times\sum_e\prescript{}{a}{\bra{m_A}}\prescript{}{b}{\bra{m_B}}\hat{M}^{ab\dagger}_{k_ck_d,e} \hat{M}^{ab}_{k_ck_d,e}\ket{n_A}_{a}\ket{n_B}_{b}\\
\leq
&\sum_{m_A,m_B,n_A,n_B=0}^{\infty}c_{m_A}^{A,(i)}c_{m_B}^{B,(j)}c_{n_A}^{A,(i)}c_{n_B}^{B,(j)}\\
&\times\sum_e \sqrt{\prescript{}{a}{\bra{n_A}} \prescript{}{b}{\bra{n_B}} \hat{M}^{ab\dagger}_{k_ck_d,e} \hat{M}^{ab}_{k_ck_d,e} \ket{n_A}_a \ket{n_B}_b} \sqrt{\prescript{}{a}{\bra{m_A}} \prescript{}{b}{\bra{m_B}} \hat{M}^{ab\dagger}_{k_ck_d,e} \hat{M}^{ab}_{k_ck_d,e} \ket{m_A}_a \ket{m_B}_b}\\
\end{aligned}
\end{equation}

\noindent where we have applied Cauchy-Schwarz inequality to the two vectors: $\hat{M}^{ab}_{k_ck_d,e} \ket{n_A}_a \ket{n_B}_b$ and 

\noindent $\hat{M}^{ab}_{k_ck_d,e} \ket{m_A}_a \ket{m_B}_b$. We can then write:

\begin{equation}
\begin{aligned}
&\sum_{m_A,m_B,n_A,n_B=0}^{\infty}c_{m_A}^{A,(i)}c_{m_B}^{B,(j)}c_{n_A}^{A,(i)}c_{n_B}^{B,(j)}\\
&\times\sum_e \sqrt{\prescript{}{a}{\bra{n_A}} \prescript{}{b}{\bra{n_B}} \hat{M}^{ab\dagger}_{k_ck_d,e} \hat{M}^{ab}_{k_ck_d,e} \ket{n_A}_a \ket{n_B}_b} \sqrt{\prescript{}{a}{\bra{m_A}} \prescript{}{b}{\bra{m_B}} \hat{M}^{ab\dagger}_{k_ck_d,e} \hat{M}^{ab}_{k_ck_d,e} \ket{m_A}_a \ket{m_B}_b}\\
&\leq\sum_{m_A,m_B,n_A,n_B=0}^{\infty}c_{m_A}^{A,(i)}c_{m_B}^{B,(j)}c_{n_A}^{A,(i)}c_{n_B}^{B,(j)}\\
&\times \sqrt{\prescript{}{a}{\bra{n_A}} \prescript{}{b}{\bra{n_B}} \sum_e\hat{M}^{ab\dagger}_{k_ck_d,e} \hat{M}^{ab}_{k_ck_d,e} \ket{n_A}_a \ket{n_B}_b} \sqrt{\prescript{}{a}{\bra{m_A}} \prescript{}{b}{\bra{m_B}} \sum_e\hat{M}^{ab\dagger}_{k_ck_d,e} \hat{M}^{ab}_{k_ck_d,e} \ket{m_A}_a \ket{m_B}_b}\\
=&\left[\sum_{n_A,n_B=0}^{\infty}c_{n_A}^{A,(i)}c_{n_B}^{B,(j)} \sqrt{p_{ZZ}(k_c,k_d|n_A,n_B)}\right]^2\\
\end{aligned}
\end{equation}

\noindent where we again use Cauchy-Schwarz inequality by considering two vectors $\vec{u},\vec{v}$ whose $e$-th components are 

\begin{equation}
\begin{aligned}
u_e &= \sqrt{\prescript{}{a}{\bra{n_A}} \prescript{}{b}{\bra{n_B}} \hat{M}^{ab\dagger}_{k_ck_d,e} \hat{M}^{ab}_{k_ck_d,e} \ket{n_A}_a \ket{n_B}_b}\\
v_e &= \sqrt{\prescript{}{a}{\bra{m_A}} \prescript{}{b}{\bra{m_B}} \hat{M}^{ab\dagger}_{k_ck_d,e} \hat{M}^{ab}_{k_ck_d,e} \ket{m_A}_a \ket{m_B}_b}
\end{aligned}
\end{equation}

\noindent respectively, and apply $\vec{u} \cdot \vec{v} \leq \sqrt{\vec{u} \cdot \vec{u}}\sqrt{\vec{v} \cdot \vec{v}}$.

This means that, the phase-error rate can be upper-bounded by the yields for photon number states $p_{ZZ}(k_c,k_d|n_A,n_B)$:

\begin{equation}
\begin{aligned}
&p_{XX}(k_c,k_d)e_{ZZ}(k_c,k_d)\\
=&p_{XX}(k_c,k_d)\sum_{j=0,1}\prescript{}{AB}{\bra{jj}} \rho'_{AB} \ket{jj}_{AB}\\
\leq& \sum_{j=0,1}\left[\sum_{n_A,n_B=0}^{\infty}c_{n_A}^{A,(j)}c_{n_B}^{B,(j)} \sqrt{p_{ZZ}(k_c,k_d|n_A,n_B)}\right]^2\\
\end{aligned}
\end{equation}

\noindent this phase-error rate, combined with the bit error rate in the X basis, can be used to perform privacy amplification on the error-corrected raw keys and obtain the secure key.

The key point is that, the above proof that upper bounds the phase error rate does not require the fact that $\alpha_A=\alpha_B$ at all. The different signal intensities will cause Alice and Bob to have different cat states, but these states are independently used to obtain the inner product with $\prescript{}{A}{\bra{i}}$ and $\prescript{}{B}{\bra{j}}$ respectively. With the Cauchy-Schwarz inequality, the joint cat states are reduced to a mixture of photon number states, and there are no cross-terms between the two cat states. 

This means that, using asymmetric intensities between Alice and Bob will not affect the estimation of the phase-error rate. Moreover, as we described in Equation 6.3, $\{\hat{M}^{ab}_{k_ck_d,e}\}$ is a general representation of the channels and detection, and does not require that $\eta_A=\eta_B$ either, i.e. asymmetric channels do not affect the security proof either.\\

Additionally, the decoy intensities $\{\beta_A^2,\beta_B^2\}$ might be different for Alice and Bob too, but these states are only used to estimate the yields of photon number states $p_{ZZ}(k_c,k_d|n_A,n_B)$ using decoy-state analysis, which is exactly the same process as in MDI-QKD. As long as Eve cannot distinguish pulses from different intensity settings, this decoy-state analysis is secure, even in the asymmetric setting - since the sending of a given photon number $n$ given the Poisson distribution $P(n|\mu)=e^{-\mu}{{\mu^n}\over{n!}}$ is a Markov process, i.e. memoryless process, Eve has no way of telling which intensity setting the photon number state came from, therefore using asymmetric intensities does not affect the estimation of yields for photon number states $p_{ZZ}(k_c,k_d|n_A,n_B)$.

Therefore, overall, we conclude that neither asymmetric channel losses, nor asymmetric intensities Alice and Bob use (for signal states or decoy states), will affect the security of the protocol. Asymmetry will only affect the performance of the protocol (which will be the subject of discussion in the next section) - asymmetric channels will result in higher QBER and subsequently lower key rate, and asymmetric intensities can compensate for channel asymmetry and enable high key rate for the protocol even when channels are highly asymmetric.

\section{Performance}

In this section we will discuss how channel asymmetry, and asymmetric intensities, can affect the performance of TF-QKD.

\subsection{Channel Model}

We will first discuss the channel model in the asymmetric case. Again, we extend the expressions in the Appendix of Ref. \cite{simpleTFQKD}, and consider asymmetric intensities and channel transmittances.

To obtain the secure key rate, three sets of observables are needed: the X basis gain $p_{XX}(k_c,k_d)$, the X basis bit-error rate $e_{XX}(k_c,k_d)$, and the Z basis gain $p_{ZZ}(k_c,k_d|\beta_A,\beta_B)$ (for all combinations of $\{\beta_A,\beta_B\}$). 

Now, let us suppose Alice and Bob send signals with intensities $s_A,s_B$, and channels between Alice/Bob and Charles have transmittances $\eta_A,\eta_B$. For simplicity we can write:

\begin{equation}
\begin{aligned}
\gamma_A &= s_A \eta_A\\
\gamma_B &= s_B \eta_B\\
\end{aligned}
\end{equation}

\noindent for signal states, and

\begin{equation}
\begin{aligned}
\gamma'_A &= \mu^i_A \eta_A\\
\gamma'_B &= \mu^j_B \eta_B\\
\end{aligned}
\end{equation}

\noindent for decoy states, where $\mu_A^i$ and $\mu_B^j$ are selected from the set of decoy intensities.

The other imperfections in the channel include the dark count rate $p_d$, the polarization misalignment between Alice and Bob $\theta$, and the phase mismatch $\phi$ between Alice and Bob. If we first do not consider dark counts and phase mismatch, the intensities arriving at the detectors C and D at Charles can be written as (similar to the discussions in Ref. \cite{mdipractical}):

\begin{equation}
\begin{aligned}
D_c &= {{1}\over{2}}\left( \gamma_A + \gamma_B - 2\sqrt{\gamma_A\gamma_B} cos\theta \right)\\
D_d &= {{1}\over{2}}\left( \gamma_A + \gamma_B + 2\sqrt{\gamma_A\gamma_B} cos\theta \right)
\end{aligned}
\end{equation}

\noindent the probability that one detector clicks and the other doesn't (e.g. C clicks and D doesn't) can be written as

\begin{equation}
\begin{aligned}
&(1-e^{-D_c})e^{-D_d}\\
=&e^{-D_d} - e^{-(D_c+D_d)}\\
=&e^{-{{1}\over{2}}\left[ \gamma_A + \gamma_B + 2\sqrt{\gamma_A\gamma_B} cos\theta \right]} - e^{-(\gamma_A+\gamma_B)}\\
\end{aligned}
\end{equation}

Including the phase mismatch and dark counts, we can write the X basis gain and QBER in a similar form as Ref. \cite{simpleTFQKD}:

\begin{equation}
\begin{aligned}
p_{XX}(k_c,k_d)=&{{1}\over{2}}(1-p_d) \left( e^{-\sqrt{\gamma_A\gamma_B} cos\phi cos\theta} + e^{\sqrt{\gamma_A\gamma_B} cos\phi cos\theta} \right)\\
\times& e^{-{{1}\over{2}}(\gamma_A+\gamma_B)} -(1-p_d)^2 e^{-(\gamma_A + \gamma_B)}\\
\end{aligned}
\end{equation}

\begin{equation}
\begin{aligned}
e_{XX}(k_c,k_d)=&{{ e^{-\sqrt{\gamma_A\gamma_B} cos\phi cos\theta}-(1-p_d)e^{-{{1}\over{2}}(\gamma_A+\gamma_B)}}  \over  {e^{-\sqrt{\gamma_A\gamma_B} cos\phi cos\theta} + e^{\sqrt{\gamma_A\gamma_B} cos\phi cos\theta}-2(1-p_d)e^{-{{1}\over{2}}(\gamma_A+\gamma_B)}}}\\
\end{aligned}
\end{equation}

\noindent and the Z basis gain is the integral over all possible (random) relative phases:

\begin{equation}
\begin{aligned}
&p_{ZZ}(k_c,k_d|\beta_A,\beta_B)\\
=&(1-p_d)\left[e^{-{{1}\over{2}}(\gamma'_A+\gamma'_B)}I_0(\sqrt{\gamma'_A\gamma'_B}cos\theta) - e^{-(\gamma'_A+\gamma'_B)}\right]+p_d(1-p_d)e^{-(\gamma'_A+\gamma'_B)}\\
\end{aligned}
\end{equation}

\noindent where $I_0(x)$ is a modified Bessel function of the first kind.

The Z basis gain can be used in decoy-state analysis to obtain $m,n$ photon yields $p_{ZZ}(k_c,k_d|n_A,n_B)$. Here for simplicity we first consider the infinite-decoy case, where $p_{ZZ}(k_c,k_d|n_A,n_B)$ can be assumed to be perfectly known (similar to Supplementary Information Equations 6.18, 6.19 in Ref. \cite{simpleTFQKD} but with asymmetric channel transmittances):

\begin{equation}
\begin{aligned}
p_{ZZ}(k_c,k_d|n_A,n_B)= & (1-p_d)q_{ZZ}(k_c,k_d|n_A,n_B) + (1-p_d)p_d(1-\eta_A)^{n_A}(1-\eta_B)^{n_B}\\
\end{aligned}
\end{equation}

\noindent where

\begin{equation}
\begin{aligned}
q_{ZZ}(k_c,k_d|n_A,n_B)=& \sum_{k=0}^{n_A} {n_A\choose k} \sum_{l=0}^{n_B} {n_B\choose l} {{\eta_A^k \eta_B^l (1-\eta_A)^{n_A-k}(1-\eta_B)^{n_B-l}}\over{2^{k+l}k!l!}} \\
&\sum_{m=0}^{k} {k \choose m}  \sum_{p=0}^{l} {l \choose p}\sum_{q=max(0,m+p-l)}^{min(k,m+p)} {k \choose q}{l \choose m+p-q}\\
&(m+p)!(k+l-m-p)!cos^{m+q}(\theta_A)cos^{m+p-q}(\theta_B)\\
&sin^{2k-m-q}(\theta_A)sin^{2l-m-2p+q}(\theta_B) - (1-\eta_A)^{n_A} (1-\eta_B)^{n_B} \\
\end{aligned}
\end{equation}

In the case with finite decoys (e.g. 3 decoy states for each of Alice and Bob), we can use linear programming to upper-bound the yields, which is described in more detail in Appendix D.1.

Afterwards, the phase-error rate can be upper-bounded using these yields:

\begin{equation}
\begin{aligned}
p_{XX}(k_c,k_d)e_{ZZ}(k_c,k_d) \leq& \sum_{i=0,1}\left[\sum_{n_A,n_B=0}^{\infty}c_{n_A}^{A,(i)}c_{n_B}^{B,(i)} \sqrt{p_{ZZ}(k_c,k_d|n_A,n_B)}\right]^2\\
\end{aligned}
\end{equation}

With the the X basis gain $p_{XX}(k_c,k_d)$, the X basis bit-error rate $e_{XX}(k_c,k_d)$, and the phase-error rate $e_{ZZ}(k_c,k_d)$, we can obtain the final secure key rate:

\begin{equation}
\begin{aligned}
R_{k_c k_d}&=p_{XX}(k_c,k_d) \times[1-h_2(e_{XX}(k_c,k_d))-h_2(e_{ZZ}(k_c,k_d))]\\
\end{aligned}
\end{equation}

\noindent where $h_2(x)=-xlog_2(x) - (1-x)log_2(1-x)$ is the binary entropy function.

\subsection{Effect of Channel and Intensity Asymmetry on Gain and QBER}

In the estimation of key rate, only three sets of observables are used: the X basis gain $p_{XX}(k_c,k_d)$, the X basis bit-error rate $e_{XX}(k_c,k_d)$, and the set of Z basis gain for each combination of decoy intensities $p_{ZZ}(k_c,k_d|\beta_A,\beta_B)$. Here we note that, the X basis gain and Z basis gain do not explicitly depend on the symmetry of incoming signal strengths $\gamma_A/\gamma_B$, and only the X basis QBER is affected by $\gamma_A/\gamma_B$.

For simplicity, here let us consider the second-order approximation for the Bessel function and exponential function, and for now ignore the phase mismatch and dark count rate: 

\begin{equation}
\begin{aligned}
I_0(x) &= 1 + {1\over 4}x^2 + O(x^4)\\
e^x &= 1 + x + {1\over 2}x^2 + O(x^3)\\
\end{aligned}
\end{equation}
\begin{figure*}[t]
	\begin{minipage}{\textwidth}
		\includegraphics[scale=0.21]{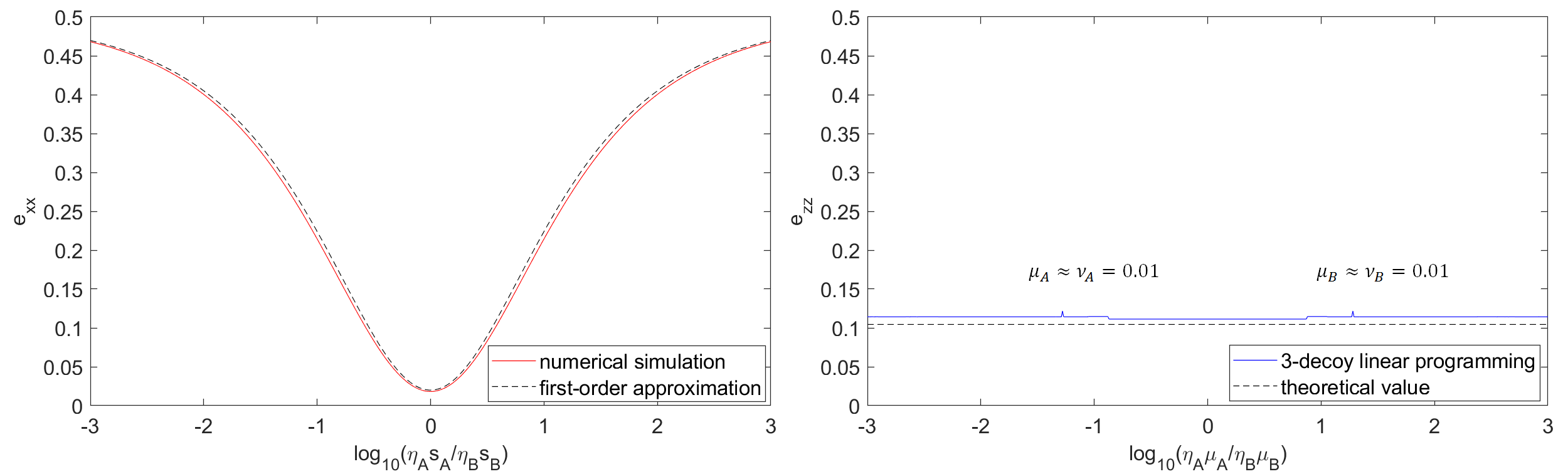}
		\caption{QBER versus asymmetry in arriving intensities at Charles, in the X and Z bases. Here we consider $k_c,k_d=0,1$, while the other case of $k_c,k_d=1,0$ has exactly the same values. Both plots use parameters $\eta_A=\eta_B=1$ (as well as detector efficiency $100\%$) and assume misalignment of $2\%$ from Alice and Bob each, and no dark counts or phase mismatch. (a) We plot the X basis QBER (acquired from full expression in Equation 6.14) as well as its first-order approximation (acquired from Equation 6.24). Here we vary $s_A/s_B$ while keeping $s_A+s_B=0.2$ to test different levels of asymmetry. As can be seen, the X basis QBER heavily depends on the symmetry between arriving intensities $s_A\eta_A$ and $s_B\eta_B$. Physically, this is because key generation depends on single-photon interference and therefore requires indistinguishability of incoming signals. When $\eta_B/\eta_A\neq 1$, X basis QBER will increase drastically if intensities are symmetric, while one can adjust $s_A/s_B$ such that $s_A/s_B=\eta_B/\eta_A$ to obtain minimal X basis QBER. (b) We plot the Z basis bit-error rate (i.e. X basis phase-error rate) obtained from linear programming using data from 3 decoy states. Here we set $\nu_A=\nu_B=0.01$, $\omega_A=\omega_B=0$, and signal states $s_A=s_B=0.1$. We fix $\mu_A+\mu_B=0.2$ and vary $\mu_A/\mu_B$ to test asymmetry. As can be seen, the upper-bounded phase error rate depends very little on the asymmetry between $\mu_A\eta_A$ and $\mu_B\eta_B$, and the linear program can effectively bound the error rate - in fact rather close to the theoretical value obtained with infinite-decoys - as long as $\mu_A,\mu_B$ are of reasonable values {\color{red}$^a$}. The physical intuition is clear too: the yields are estimated by linear programming, which usually has redundant information (in the 9 cross terms $Q_{\mu_i\mu_j}$ where Alice and Bob each uses one of their three decoys) and is insensitive to asymmetry, and the phase-error rate (Equation 6.17) is a linear combination of the yields, which makes it insensitive to asymmetry just like the yields. Reproduced from \cite{this_asymTF} @2019 NJP (CC-BY 3.0 license), re-generated with $s_A+s_B$ and $\mu_A+\mu_B$ fixed (rather than $s_B$ and $\mu_B$ fixed) to maintain symmetry of the curves.}
		\label{fig:QBERTF}

		
		\footnotetext[1]{Note that the yields obtained from linear programming are numerical solutions, and there is higher noise near $\mu_A/\mu_B=10^{1.28}$ and $\mu_A/\mu_B=10^{-1.28}$ which are the cases where $\mu_A \approx \nu_A=0.01$, or $\mu_B \approx \nu_B=0.01$ - if the decoy intensities are too close, the constraints from observable data containing $\mu_A$ (or $\mu_B$) provides little useful information, and the linear program essentially has to use the data from one less decoy state, which is why the numerical solution is less stable near these two points.}
	\end{minipage}
\end{figure*}

We can then rewrite the X basis gain as:

\begin{equation}
\begin{aligned}
p_{XX}(k_c,k_d)=&{{1}\over{2}}\left( e^{-\sqrt{\gamma_A\gamma_B} cos\theta} + e^{\sqrt{\gamma_A\gamma_B} cos\phi cos\theta} \right) \times e^{-{{1}\over{2}}(\gamma_A+\gamma_B)} - e^{-(\gamma_A + \gamma_B)}\\
\approx& {1\over 2}(\gamma_A+\gamma_B) - {1 \over 8}\left[3 \gamma_A^2 + 3\gamma_B^2 + (2+4e_d)\gamma_A \gamma_B \right]\\
\end{aligned}
\end{equation}

\noindent and the Z basis gain as:

\begin{equation}
\begin{aligned}
p_{ZZ}(k_c,k_d|\beta_A,\beta_B)=&e^{-{{1}\over{2}}(\gamma'_A+\gamma'_B)}I_0(\sqrt{\gamma'_A\gamma'_B}cos\theta) - e^{-(\gamma'_A+\gamma'_B)}\\
\approx& {1\over 2}(\gamma'_A+\gamma'_B) - {1 \over 8}\left[3 \gamma_A^{'2} + 3\gamma_B^{'2} + (4+2e_d)\gamma'_A \gamma'_B \right]\\
\end{aligned}
\end{equation}

\noindent where the terms higher than second order are omitted, and $\theta$ is the total polarization misalignment angle between Alice and Bob satisfying $\theta = 2sin^{-1}(\sqrt{e_d}))$ (suppose Alice-Charles and Bob-Charles each has misalignment error $e_d$, but with misalignment angles in different directions). We can see that, the gain in both X and Z basis is dominated by the term ${1\over 2}(\gamma_A + \gamma_B)={1\over 2}(s_A\eta_A + s_B\eta_B)$ or ${1\over 2}(\gamma'_A + \gamma'_B)={1\over 2}(\mu_A^i\eta_A + \mu_B^j\eta_B)$, i.e. taking first-order approximation:

\begin{equation}
\begin{aligned}
&p_{XX}(k_c,k_d)\approx {1\over 2}(\gamma_A+\gamma_B)\\
&p_{ZZ}(k_c,k_d|\beta_A,\beta_B)\approx {1\over 2}(\gamma'_A+\gamma'_B)\\
\end{aligned}
\end{equation}

\noindent which means that the gain scales with the \textit{average} of arriving intensities through Alice's and Bob's channels - this is different from MDI-QKD, where the gain only contains the second-order terms $\gamma_A^2,\gamma_B^2,\gamma_A\gamma_B$. We can also see that the gain does not depend on the asymmetry of arriving intensities, e.g. $\gamma_A/\gamma_B$.

On the other hand, the QBER in X basis depends on the balance of arriving intensities:

\begin{equation}
\begin{aligned}
e_{XX}(k_c,k_d) =&{{ e^{-\sqrt{\gamma_A\gamma_B} cos\phi cos\theta}-(1-p_d)e^{-{{1}\over{2}}(\gamma_A+\gamma_B)}}  \over  {e^{-\sqrt{\gamma_A\gamma_B} cos\phi cos\theta} + e^{\sqrt{\gamma_A\gamma_B} cos\phi cos\theta}-2(1-p_d)e^{-{{1}\over{2}}(\gamma_A+\gamma_B)}}}\\
\approx& {{{1 \over 2}(\gamma_A+\gamma_B) - \sqrt{\gamma_A \gamma_B}cos\theta + {1\over 2}\gamma_A\gamma_Bcos^2\theta - {1\over 8}(\gamma_A + \gamma_B)^2}\over {(\gamma_A + \gamma_B) + \gamma_A\gamma_Bcos^2\theta - {1\over 4} (\gamma_A + \gamma_B)^2}}
\end{aligned}
\end{equation}

\noindent which, in the first-order approximation\footnote{The first order approximation for $e_{XX}(k_d,k_d)$ assumes that $\gamma_A,\gamma_B$ are much smaller than 1 - which is reasonable, since to get a good phase-error rate estimation, usually $s_A,s_B$ are smaller or equal to $0.1$, and for positions of interest where TF-QKD beats PLOB bound, the loss in each channel is usually larger than 10dB, which means that $\eta_A$ and $\eta_B$ are much smaller than 1 too - for instance 10dB channel loss corresponds to $0.1$ transmittance.}, can be simplified as:

\begin{equation}
\begin{aligned}
e_{XX}(k_c,k_d)\approx& {{{1 \over 2}(\gamma_A+\gamma_B) - \sqrt{\gamma_A \gamma_B}cos\theta}\over {\gamma_A + \gamma_B}}\\
=& {{{1 \over 2}({{\gamma_A \over \gamma_B}}+1) - \sqrt{{\gamma_A \over \gamma_B}}cos\theta}\over {{\gamma_A \over \gamma_B} + 1}}
\end{aligned}
\end{equation}

We can see that here the X basis QBER does depend on asymmetry - more precisely, it depends on how much the arriving intensities at Charles, $\gamma_A=\eta_As_A$ and  $\gamma_B=\eta_Bs_B$ are balanced. This is understandable physically, since the X basis key generation depends on single-photon interference and relies on the indistinguishability of incoming signals. This means that, in the case that channels are not symmetric, compensating for the channel asymmetry with different signal intensities for Alice and Bob and aiming for $\eta_As_A=\eta_Bs_B$  can help minimize the X basis QBER. 

On the other hand, in the Z basis, the bit-error rate (i.e. the X basis phase-error rate) cannot be directly measured, but is instead upper-bounded using the observable gain data from the decoy states. As we mentioned above, the Z basis gain (in the first-order approximation) scales with ${1\over 2}(\gamma'_A + \gamma'_B)={1\over 2}(\mu_A^i\eta_A + \mu_B^j\eta_B)$ and does not depend on the symmetry between incoming intensities. Moreover, the yields $p_{ZZ}(k_c,k_d|n_A,n_B)$ are estimated using linear programming. For instance, for three decoys where Alice and Bob respectively use $\{\mu_A,\nu_A,\omega_A\}$, $\{\mu_B,\nu_B,\omega_B\}$ as their decoy states, there are nine sets of observable gains, $\{Q_{\mu\mu},Q_{\mu\nu},Q_{\mu\omega},Q_{\nu\mu},Q_{\nu\nu},Q_{\nu\omega},Q_{\omega\mu},Q_{\omega\nu},Q_{\omega\omega}\}$, each of which constitutes a constraint for the linear program that helps bound the yields $p_{ZZ}(k_c,k_d|n_A,n_B)$. Such a structure makes the linear program relatively robust against asymmetry in the decoy states, and the linear program can fairly accurately upper-bound the yields as long as the intensities are of reasonable values (i.e. $\mu_A\neq \nu_A$, $\mu_B\neq \nu_B$, and none of the intensities are too large e.g. $>1$). 

{\color{black}
	The phase error rate, as shown in Equation 6.17, is based on a linear combination of the square root of the yields. It is therefore also very little affected by asymmetry, and almost always reaches a good value (at least in the infinite-data case) so long as the intensities are within a reasonable range, regardless of the asymmetries in channel transmittances or decoy intensities.
}

We plot the QBER in the X and the Z bases versus asymmetry in arriving intensities (e.g. $s_A\eta_A/s_B\eta_B$ or $\mu_A\eta_A/\mu_B\eta_B$) in Figure \ref{fig:QBERTF}. As can be seen, the X basis QBER depends heavily on asymmetry and is minimal when $s_A\eta_A/s_B\eta_B=1$, while the upper-bounded Z basis QBER (i.e. phase-error rate) is hardly affected by asymmetry.

Therefore, a viable strategy for TF-QKD in asymmetric channels is to compensate for the channel asymmetry with signal intensities $\{s_A,s_B\}$ only, while the decoy intensities $\{\mu_A,\nu_A,\omega_A\}$, $\{\mu_B,\nu_B,\omega_B\}$ can be still kept symmetric. However, note that the signal intensities not only determines (1) X basis QBER, it also affects (2) X basis gain (which determines the raw key generation rate, and favours large $s_A,s_B$), as well as (3) upper-bound of phase error rate (since the cat states are determined by signal intensities, and the estimation favours small $s_A,s_B$ - typically $<0.1$ - for a tighter upper bound on phase error rate). Criteria (1-3) cannot be simultaneously satisfied, therefore an optimization for $\{s_A,s_B\}$ is required for the highest key rate.

Interestingly, we can compare this with the case of MDI-QKD. As described in Ref. \cite{this_asymMDI}, the 4-intensity protocol (and 7-intensity protocol in the extended asymmetric case) has decoupled X and Z bases, where Z basis is used for key generation and X basis uses decoy states to estimation phase-error rate. In MDI-QKD, the X basis data depends on two-photon interference and requires balanced arriving intensities (or else the X basis QBER will increase dramatically), while the Z basis does not require indistinguishability of the signals, and is therefore insensitive to channel asymmetry. In MDI-QKD, all the X basis decoy states should satisfy e.g. $\mu_A\eta_A = \mu_B\eta_B$, while the signal states $s_A,s_B$ can be chosen to simply optimize key generation rate. (Due to misalignment, there is a slight dependence of Z basis QBER to asymmetry too, hence optimal $s_A,s_B$ are still not equal, but this is a much weaker dependence on symmetry than in the X basis, and optimal $s_A/s_B$ is much closer to 1 than $\eta_B/\eta_A$ in MDI-QKD.) 

While our approach works both for MDI-QKD and TF-QKD, a key difference is that states that compensate for channel asymmetry are the signal states in TF-QKD (while this responsibility lies on decoy states in MDI-QKD), which are also involved in key generation and phase error estimation. This means that in TF-QKD, it is more difficult to simultaneously keep a low X basis QBER and a good key generation rate \& low phase error rate. Perhaps due to this reason, the advantage of asymmetric-intensity protocols is somewhat less pronounced in TF-QKD - nonetheless, it still provides about an order of magnitude higher key rate than completely symmetric protocols and still 2-3 times higher key rate than adding fibre - which means that it still is the strategy that provides highest key rate when channels are asymmetric.

\section{Numerical Results}
\begin{figure}[h]
	\centering
	\includegraphics[scale=0.23]{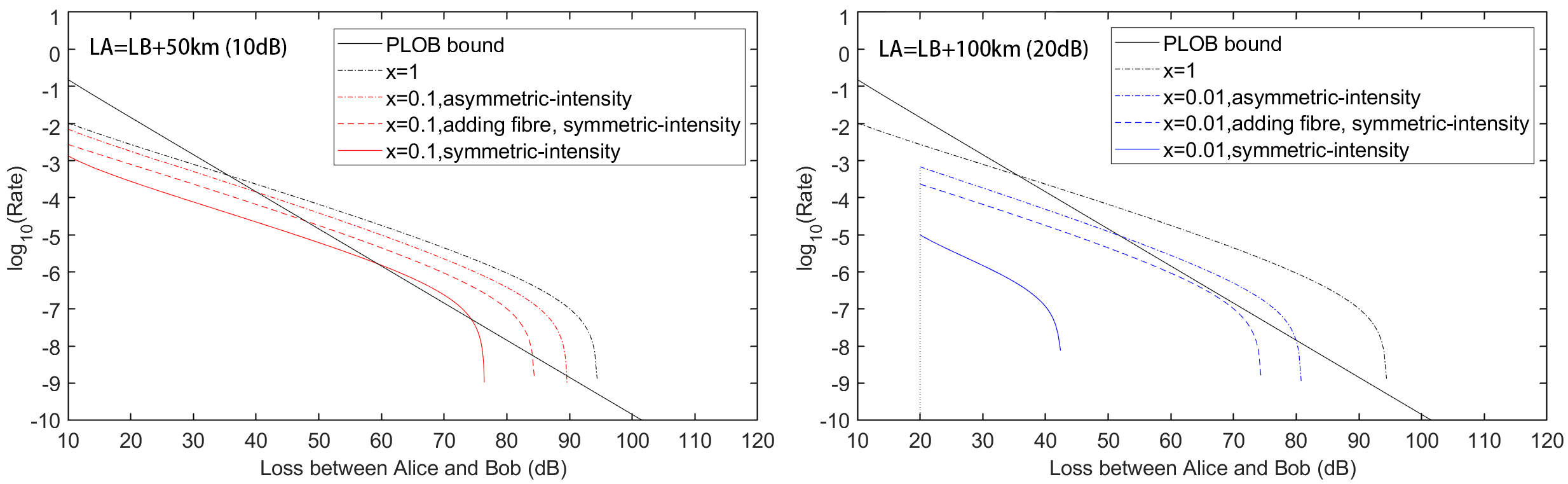}
	\caption{Key rate versus loss between Alice and Bob, for protocols with symmetric intensities ($s_A=s_B$), symmetric intensities with fibre added until channels are equal, and asymmetric intensities ($s_A,s_B$ fully optimized). The PLOB \cite{PLOB} linear bound is included for comparison. The channel mismatch $x$ is fixed at $x=\eta_A/\eta_B=0.1$ (left) and $\eta_A/\eta_B=0.01$ (right), i.e. Alice-Charles always has 10dB (20dB) higher loss than Bob-Charles. Note that the right plot starts from 20dB of total distance (which is the case where Bob-Charles loss is 0dB and Alice-Charles loss is 20dB). The dark count rate is set to $10^{-8}$, misalignment is $2\%$ for Alice and Bob each, and detector efficiency is incorporated into channel loss. As can be seen, allowing the use of asymmetric intensities greatly improves key rate when channels are asymmetric, and compared with a symmetric protocol, it can consistently provide approximately one order of magnitude higher key rate when there is a 10dB channel mismatch, and two orders of magnitude when there is a 20dB mismatch, for most distances. Interestingly, adding fibre can improve the key rate considerably too - but it still has a lower key rate than the asymmetric-intensity protocol (the latter has about 2-3 times higher key rate), and has the additional inconvenience of having to modify the physical channel. Reproduced from \cite{this_asymTF} @2019 NJP (CC-BY 3.0 license).}
	\label{fig:asym_rate}
\end{figure}

In this section we use the technique described above - to compensate for channel asymmetry simply with different $s_A,s_B$ for Alice and Bob. We first compare our method with prior art techniques and study the numerically optimized intensities for the asymptotic (infinite-decoy, infinite-data) case. Then, we also show that our method works with finite decoys and also finite data size.

We plot the simulation results for asymptotic TF-QKD in Figure \ref{fig:asym_rate}. As can be seen, for the two cases where channel mismatch $x=\eta_A/\eta_B=0.1$ and $\eta_A/\eta_B=0.01$, our method consistently have much higher key rate than TF-QKD with symmetric intensities. Interestingly, we show that adding fibre can help users obtain a higher key rate, but it comes with the additional inconvenience of having to physically modify the channel, and also it still has a lower key rate than our method of simply adjusting signal intensities.

\begin{figure}[h]
	\centering
	\includegraphics[scale=0.25]{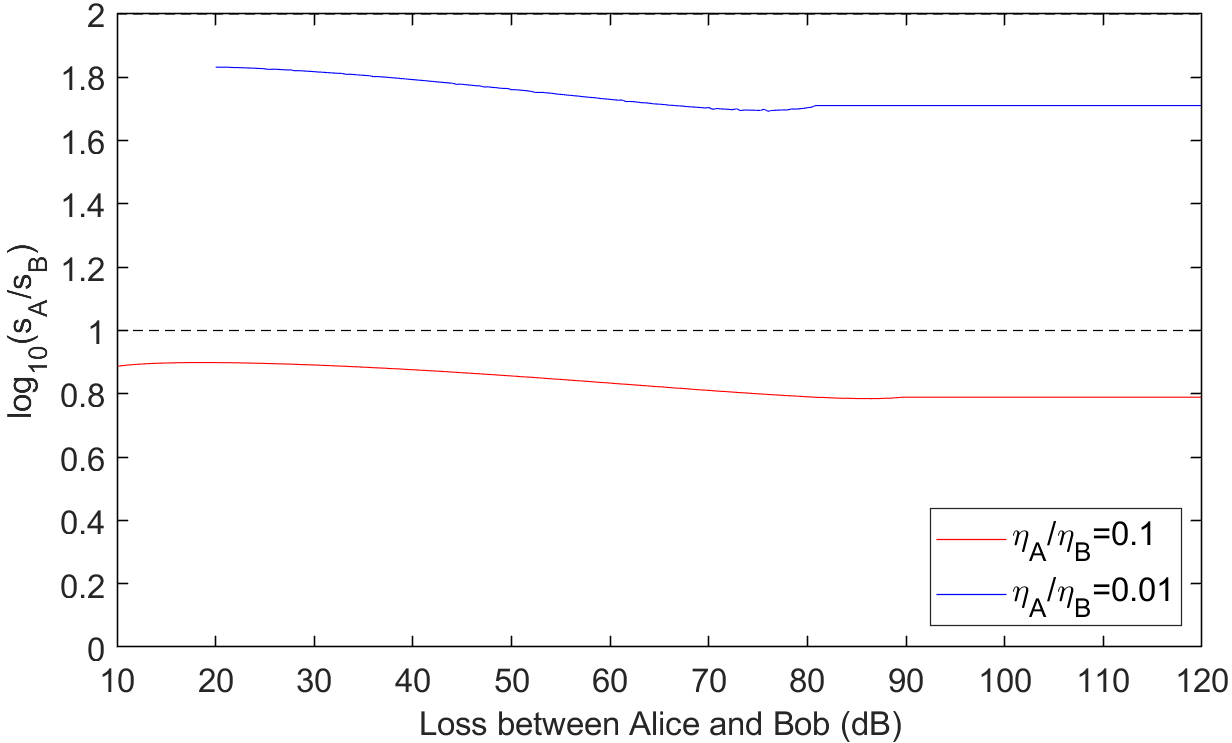}
	\caption{Ratio of optimal intensities $log_{10}(s_A/s_B)$ over loss between Alice and Bob. Here we test two cases where the channel mismatch $x=\eta_A/\eta_B=0.1$ and $\eta_A/\eta_B=0.01$. As can be seen, $s_A/s_B$ is rather close to $\eta_B/\eta_A$ (here respectively $10^1$ and $10^2$). However, due to signal states being involved in key generation and phase-error rate estimation too, it slightly deviates from the value that minimizes X basis QBER (and instead takes the value that maximizes key rate). Reproduced from \cite{this_asymTF} @2019 NJP (CC-BY 3.0 license).}
	\label{fig:compare_sA_sB}
\end{figure}

We also plot the ratio of optimal signal intensities in Figure \ref{fig:compare_sA_sB}. As we have predicted, the optimal signal intensities are rather close to the relation of $s_A\eta_A=s_B\eta_B$, in order to maintain a lower X basis QBER. However, as we discussed, since signal states are also involved in key generation and phase error rate estimation (based on the imaginary cat states), they prevent the signal states from taking the values that minimize QBER (but rather, makes it choose the value that maximizes the overall key rate).

\begin{figure}[h]
	\centering
	\includegraphics[scale=0.25]{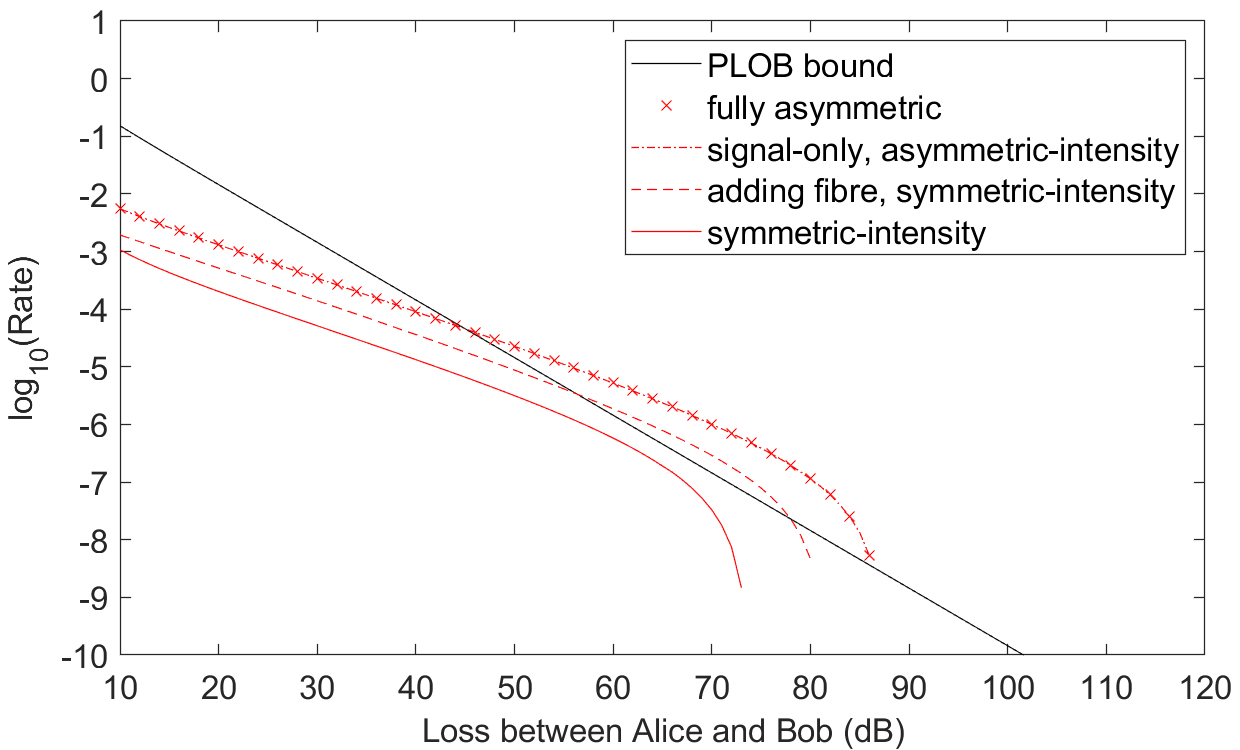}
	\caption{Key rate versus loss between Alice and Bob, for protocols with symmetric intensities and probabilities, with symmetric intensities and fibre added until channels are equal, with only asymmetric signal intensities (while all other parameters are symmetric between Alice and Bob), and with fully optimized parameters (all intensities and probabilities are freely optimized). The PLOB \cite{PLOB} linear bound is included for comparison. Here we consider $10^{-8}$ dark count rate, a $2\%$ misalignment for Alice and Bob each, $N=10^{12}$ total pulses sent, and channel mismatch of $x=\eta_A/\eta_B=0.1$. Again, detector efficiency is incorporated into channel loss. As we can see, similar to the asymptotic case, using asymmetric intensities can greatly improve the key rate. Perhaps more interestingly, we can see that allowing asymmetry in signal intensities alone is sufficient in obtaining a good key rate through asymmetric channels (its key rate almost entirely overlaps with the fully optimized case). Reproduced from \cite{this_asymTF} @2019 NJP (CC-BY 3.0 license).}
	\label{fig:rate_finite}
\end{figure}

Additionally, we also plot our results for the practical case with a finite number of decoys (here we use three decoys each for Alice and Bob: $\{\mu_A,\nu_A,\omega_A\}$, $\{\mu_B,\nu_B,\omega_B\}$) and finite data size. {\color{black} The upper-bounding of photon number yields using linear programming, as well as the finite-key analysis, are both described in more detail in Appendix D.1.} We can see that similar result holds - our method has an advantage over either using symmetric intensities directly or adding fibre. More interestingly, we include both the case where we only allow $s_A,s_B$ to be asymmetric, versus the case where all intensities and probabilities can be optimized, and as shown in the plot, we see that using asymmetric signal intensities alone is sufficient to compensate for channel asymmetry.

\section{Conclusion}

In this chapter we present a simple method to obtain good performance for TF-QKD even if channels are asymmetric. We present a theoretical understanding of why signal states (and not decoy states) should be adjusted to compensate for asymmetry, and we also show that the method is still compatible with existing security proofs. With our method, there is no need to add additional fibre, and Alice and Bob can implement the method in software-only. This provides great convenience for TF-QKD in practice - where realistic channels might likely be asymmetric - and can also be used in fibre-based quantum networks (where adding fibre for each pair of users is impractical) where a central service-provider can easily optimize the intensities for each pair of users. The results in this chapter make possible a high-rate quantum network with untrusted relays using TF-QKD.

We have very recently performed an experimental demonstration for TF-QKD protocol over asymmetric channels, using the method we proposed in this chapter (in Ref. \cite{this_asymTF}). The work is posted on arXiv as a draft paper \cite{this_asymTF_experiment} and also selected as a talk at CLEO2020. We also plan to implement a TF-QKD network with multiple users (and channels of different levels of loss) in the future.\\

We thank Koji Azuma for providing the corrected security proof \cite{Koji} that incorporates mixed states after the measurement.

During the preparation of the manuscript for Ref. \cite{this_asymTF}, it came into our knowledge that another work on asymmetric TF-QKD is under preparation \cite{tf02}, which is independently completed from Ref. \cite{this_asymTF}. The two works are posted simultaneously on the preprint server \cite{this_asymTF_arxiv,tf02}.

\chapter{Machine Learning in QKD}

This chapter is largely reproduced from our paper Ref. \cite{this_ML}.

In the previous Chapters 3-6, we have discussed the topics of free-space QKD and fibre-based QKD network, and how to apply techniques to address some important practical challenges, i.e. turbulence in free-space, and channel asymmetry in quantum networks. In this chapter, we will discuss an additional challenge that is present in both free-space QKD and quantum network - Namely, efficient parameter optimization (for e.g. decoy-state intensities, and probabilities of sending each intensity/basis). Parameter optimization is generally a computationally intensive task, that might be a limiting factor for low-power free-space systems, or quantum networks with many users. In this chapter we show that, by applying neural networks (a type of machine learning algorithm), we can greatly accelerate parameter optimization, by up to 2-4 orders of magnitude and retaining very high accuracy compared with traditional algorithms. Such a method enables real-time parameter optimization for future free-space systems or quantum networks, which can significantly reduce latency, and improve key rate compared to using fixed parameters.

\section{Background}

\subsection{Parameter Optimization in QKD}

As we have discussed in Chapter 2, in reality, a QKD experiment always has a finite transmission time, which means that the total number of signals is finite. Therefore, when estimating the single-photon contributions with decoy-state analysis, one needs to take into consideration the statistical fluctuations of the observables: the Gain and Quantum Bit Error Rate (QBER), which might deviate from their respective expected values (and the amount of deviation is what one needs to upper-bound). This is called the finite-key analysis of QKD. In such a finite-key scenario, the choice of intensities and probabilities of sending these intensities is crucial to getting the optimal rate. Therefore, we would need to perform an optimization on the parameters to search for values that maximize the key rate.

Note that in this chapter, by ``parameter optimization", we mainly discuss the optimization of the intensities of laser signals and the probabilities of choosing each intensity setting, specifically in the finite-size scenario (similar to the model outlined in Ref. \cite{mdiparameter} for MDI-QKD). There is also previous literature \cite{QKD1,QKD2,QKD3} discussing e.g. the optimization of the \textit{number} of decoy states, but the subject of study is different here. Also, some of the above literature \cite{QKD1} discusses optimization of intensities in the asymptotic (infinite-data) limit, but here in the finite-size case that we study, the parameter space is much larger, making the problem much more computationally challenging. Additionally, in this chapter we discuss a broader picture where our method can be applied to various kinds of QKD protocols (and potentially other optimization problems outside QKD or even in classical systems).

There have been various studies on BB84 \cite{finiteQKD2,finitebb84}, MDI-QKD \cite{finiteQKD1,mdiparameter,mdifourintensity}, and TF-QKD \cite{TFfinite1} under finite-size effects. Here in the chapter, we will employ Ref.\cite{finitebb84}'s method for BB84 under finite-size effects, and use a standard error analysis for MDI-QKD \cite{mdifourintensity,this_asymMDI} and asymmetric TF-QKD \cite{this_asymTF}. Note that, our method is not really dependent on the security analysis model - in fact it is not really dependent on any specific protocol at all - and in principle Chernoff bound can be applied too, but for simplicity in this chapter we only use a simple Gaussian assumption for the probability distribution of observables.

Traditionally, parameter optimization is implemented as either a brute-force global search for smaller numbers of parameters, or a local search for larger numbers of parameters. For instance, in several papers studying MDI-QKD protocols with symmetric \cite{mdiparameter} and asymmetric channels \cite{this_asymMDI}, a local search method called coordinate descent algorithm is used to find the optimal set of intensities and probabilities.

\begin{figure}[t]
	\includegraphics[scale=0.4]{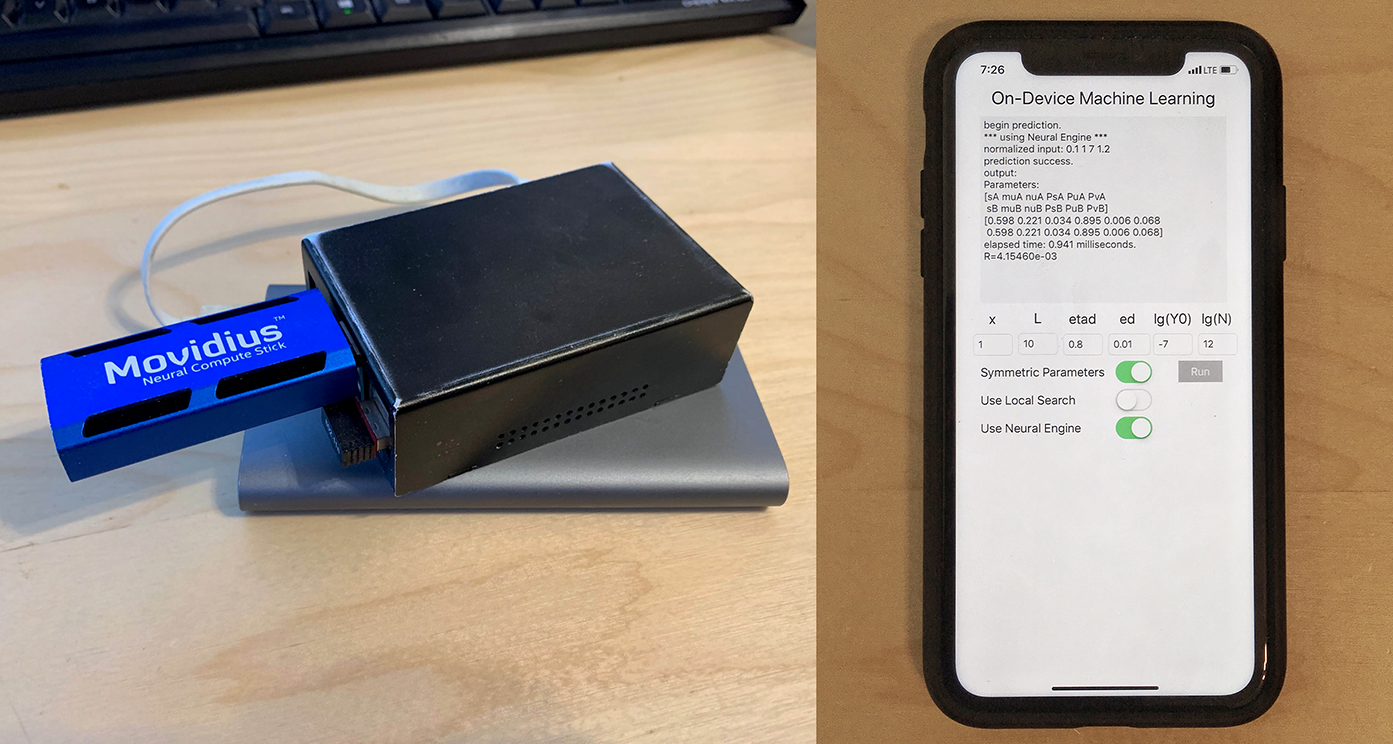}
	\caption{Left: Raspberry Pi 3 single-board computer equipped with an Intel Movidius neural compute stick. Right: A smartphone (iPhone XR) running parameter prediction for a QKD protocol with an on-device neural network. In the same app one can also choose to run local search on the device (and compare its running time to that of neural networks). Reproduced from \cite{this_ML} @2019 APS.}
	\label{fig:mobile}
\end{figure} 

\begin{table*}[t]
	\begin{minipage}{\textwidth}
		\caption[Caption for LOF]{Time benchmarking between a previous local search algorithm and our new algorithm using neural network (NN) for parameter optimization on various devices. Here as an example we consider two protocols: symmetric MDI-QKD \cite{mdifourintensity} and asymmetric TF-QKD \cite{this_asymTF}. Devices include a desktop PC with an Intel i7-4790k quad-core CPU equipped with an Nvidia Titan Xp GPU, a modern mobile phone Apple iPhone XR with an on-board neural engine, and a low-power single-board computer Raspberry Pi 3 with quad-core CPU, equipped with an Intel Movidius neural compute stick {\color{red}$^a$}. As can be seen, the neural network generally can provide over 2-4 orders of magnitude higher speed than local search depending on the protocol, enabling millisecond-level parameter optimization. Moreover, note that the smartphone and single-board computer provide similar performance with only less than $1/70$ the power consumption of a PC, making them ideal for free-space QKD or a quantum internet-of-things with portable devices. More details on the benchmarking are provided in Section 7.4.}
		\begin{center}
			\begin{tabular}{cccccc}            
				Protocol & Device & NN accelerator & Local search & NN & Power Consump.\\
				\hline
				MDI-QKD & Desktop PC & Titan Xp GPU & 0.1s & 0.5-1.0ms & $\sim$350w\\
				TF-QKD & Desktop PC & Titan Xp GPU & 2s & 0.5-1.0ms & $\sim$350w\\
				MDI-QKD & iPhone XR & on-board neural engine & 0.2s & $\sim$1ms & $<$5w\\
				TF-QKD & iPhone XR & on-board neural engine & N/A & $\sim$1ms & $<$5w\\
				MDI-QKD & Raspberry Pi 3 & neural compute stick & 3-5s & 2-3ms & $<$5w\\
				TF-QKD & Raspberry Pi 3 & neural compute stick & 15-16s & 3ms & $<$5w\\
			\end{tabular}
		\end{center}
		
		\footnotetext[1]{The CPU on an iPhone XR has dual big cores + four small cores, but here for simplicity we use a single-threaded program for local search, since OpenMP multithreading library is not supported on Apple devices. OpenMP is supported on the PC and on Raspberry Pi 3, so multithreading is used for local search on these devices. Also, TF-QKD requires a linear solver, but all commercial linear solvers cannot be used on iPhone (while open-source solvers in principle can be compiled for iPhone, in practice the porting is highly non-trivial and very difficult). Therefore local search for TF-QKD cannot be performed on an iPhone - but note that the neural network can still predict the optimal parameters regardless of software library limitations, which is in fact an additional advantage of using a neural network. For TF-QKD, the PC and Raspberry Pi respectively use the commercial solver Gurobi \cite{gurobi} and the open-source solver Coin-OR \cite{CoinOR} (as commercial solvers are not available for Raspberry Pi either).}
	\end{minipage}
	
\end{table*}

However, optimization of parameters often requires a significant amount of computational power. This means that, a QKD system either has to wait for an optimization off-line (and suffer from delay), or use sub-optimal or even unoptimized parameters in real-time. Due to the amount of computing resource required, so far parameter optimization is usually only performed on relatively powerful devices such as a desktop PC.

As we have discussed in Chapter 2, a promising future direction of QKD is to implement it over free-space on mobile platforms, such as drones \cite{freespace_drone}, handheld systems \cite{freespace_handheld}, and even satellites \cite{freespace_satellite1}. Such devices (e.g. single-board computers and mobile system-on-chips) are usually limited in computing power. At the same time, such free-space applications often require low-latency. However, fast and accurate parameter optimization based on a changing environment in real time is a difficult task on such low-power platforms.

Another highly attractive future direction of QKD is a quantum ``internet of things" that connects multiple devices, each of which can be portable and mobile, and numerous connections can be present at the same time. An increased number of connections will present a great computational challenge for the controller of a quantum network, which can be quickly overloaded with many pairs of users (where real-time optimization might simply be infeasible for even a moderate number of connections).

In recent years, machine learning technologies based on neural networks have seen much development in both hardware and software, and they have attracted a huge amount of attention from both the academia and the industry. There is also an increasing number of new low-power devices implementing on-board acceleration chips for neural networks. Here in this chapter we present a new method of using neural networks to help predict optimal parameters efficiently on low-power devices. We test our machine learning algorithm on real-life devices such as a single-board computer and a smart phone (see Fig. \ref{fig:mobile}), and find that with our method they can easily perform parameter optimization in milliseconds, within a power consumption of less than 5 watts. We list some time benchmarking results in Table 7.1. Such a method makes it possible to support \textit{real-time} parameter optimization for both free-space QKD systems and large-scale QKD networks with potentially thousands of connections.

\subsection{Neural Network}

\begin{figure}[h]
	\includegraphics[scale=0.33]{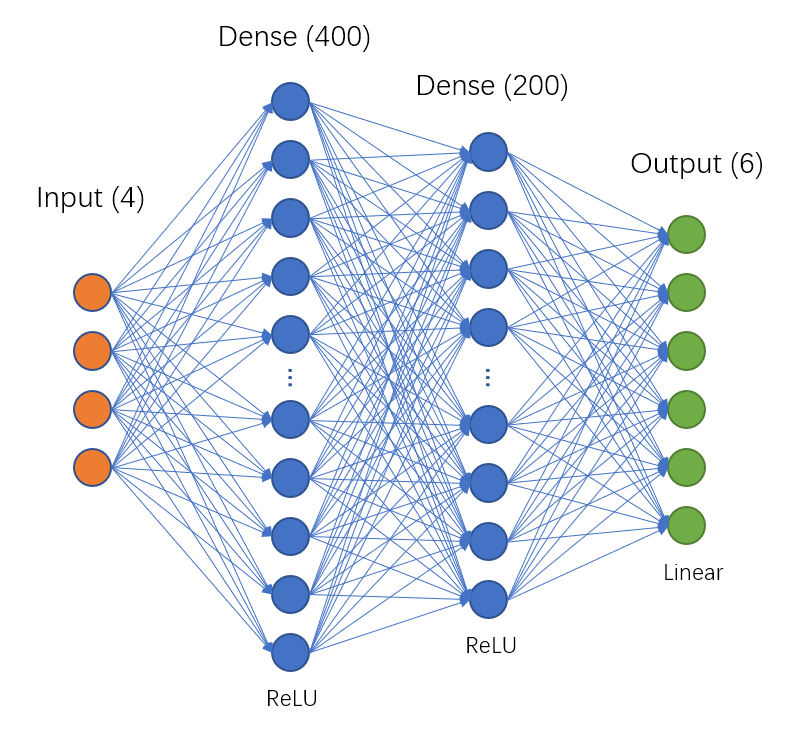}
	\caption{An example of a neural network (in fact, here it is an illustration of the neural network used in our work). It has an input layer and an output layer of 4 and 6 neurons, respectively, and has two fully-connected ``hidden" layers with 400 and 200 neurons with rectified linear unit (ReLU) function as activation. The cost function (not shown here) is mean squared error. Reproduced from \cite{this_ML} @2019 APS.}
	\label{fig:NN}
\end{figure} 

In this subsection we present a very brief introduction to machine learning with neural networks.

Neural networks are multiple-layered structures built from ``neurons", which simulate the behaviour of biological neurons in brains. Each neuron takes a linear combination of inputs $x_i$, with weight $w_i$ and offset $b$, and calculates the activation. For instance:

\begin{equation}
\sigma(\sum w_i x_i+b)={1\over{1+e^{-(\sum w_i x_i+b)}}}
\end{equation}

\noindent where the example activation function is a commonly used sigmoid function $\sigma(x)={1\over{1+e^{-x}}}$, but it can have other forms, such as a rectified linear unit (ReLU) \cite{RELU} function $max(0,\sum w_i x_i+b)$, a step function, or even a linear function $y=x$.

Each layer of the neural network consists of many neurons, and after accepting input from the previous layer and calculating the activation, it outputs the signals to the next layer. Overall, the effect of the neural network is to compute an output $\vec{y}=N(\vec{x})$ from the vector $\vec{x}$. A ``cost function" (e.g. mean squared error) is defined on the output layer by comparing the network's calculated output $\{\vec{y}^i\}=\{N(\vec{x_0^i})\}$ on a set of input data $\{\vec{x_0^i}\}$, versus the set of desired output $\{\vec{y_0^i}\}$. It uses an algorithm called ``backpropagation"\cite{BP} to quickly solve the partial derivatives of the cost function to the internal weights in the network, and adjusts the weights accordingly via an optimizer algorithm such as stochastic gradient descent (SGD) to minimize the cost function and let $\{\vec{y^i}\}$ approach $\{\vec{y_0^i}\}$ as much as possible. Over many iterations, the neural network will be able to learn the behaviour of $\{\vec{x_0^i}\}\rightarrow{\{\vec{y_0^i}\}}$, so that people can use it to accept a new incoming data $\vec{x}$, and predict the corresponding $\vec{y}$. The \textit{universal approximation theorem} of neural network \cite{Universal} states that it is possible to infinitely approximate any given bounded, continuous function on a given defined domain with a neural network with even just a single hidden layer, which suggests that neural networks are highly flexible and robust structures that can be used in a wide range of scenarios where such mappings between two finite input/output vectors exist.

There is increasing interest in the field in applying machine learning to improve the performance of quantum communication. For instance, Ref. \cite{CVQKD1,CVQKD2} respectively apply machine learning to continuous-variable (CV) QKD to improve the noise-filtering and to improve the prediction/compensation of intensity evolution of light over time. 

In this work, we apply machine learning to predict the optimal intensity and probability parameters for QKD (based on given experimental parameters, such as channel loss, detector efficiency, misalignment, dark count rate, and data size), and show that with a simple fully-connected neural network with two layers, we can very accurately and efficiently predict parameters (that can achieve e.g. 95-99\%, or even 99.9\% the key rate depending on the protocol).

Our work demonstrates the feasibility of deploying neural networks on actual low-power devices, to make them perform fast QKD parameter optimization in real time, with up to 2-4 orders of magnitudes higher speed. This enables potential new applications in free-space or portable QKD devices, such as on a satellite\cite{freespace_satellite1}, drone \cite{freespace_drone}, or handheld \cite{freespace_handheld} QKD system, where the power consumption of devices is a crucial factor and computational power is severely limited, and traditional CPU-intensive optimization approaches based on local or global search are infeasible.

Additionally, we point out that with the higher optimization speed, we can also enable applications in a large-scale quantum internet-of-things (IoT) where many small devices can be interconnected (thus generating a large number of connections), and now with neural networks, even low-power devices such as a mobile phone will be able to optimize the parameters for hundreds of users in real-time in a quantum network.

This chapter is organized as follows: In Section 7.2 we will describe how we can formulate parameter optimization as a function that can be approximated by a neural network. We then describe the structure of the neural network we use, and how we train it such that it learns to predict optimal parameters. In Section 7.3 we test our neural network approach with four example protocols, and show that neural networks can accurately predict parameters, which can be used to obtain a near-optimal secure key rate for the protocols. In Section 7.4 we describe two important use cases for our method and benchmark them: enabling real-time parameter optimization on low-power and low-latency portable devices, and paving the road for large-scale quantum networks. We conclude our results in Section 7.5.

\section{Methods}

In this section we describe the process of training and validating a neural network for parameter optimization. As mentioned in Sec. I, the \textit{universal approximation theorem} implies that the approach is not limited to any specific protocol. Here for simplicity, in this section when describing the methods we will first use a simple symmetric ``4-intensity MDI-QKD protocol" \cite{mdifourintensity} as an example protocol. Later in the next section when presenting the numerical results, we also include three other protocols, the asymmetric ``7-intensity" MDI-QKD protocol\cite{this_asymMDI}, the BB84 protocol (under finite-size effects) \cite{finitebb84}, and the asymmetric TF-QKD protocol \cite{this_asymTF} to show that the method applies to them effectively too. 

\subsection{Optimal Parameters as a Function}

Let us consider the symmetric-channel case for MDI-QKD. Alice and Bob have the same distance to Charles; hence they can choose the same parameters. When taking finite-size effects into consideration, the variables to be optimized will be a set of 6 parameters, $[s, \mu, \nu, P_s, P_\mu, P_\nu]$, where $s, \mu, \nu$ are the signal and decoy intensities, and $P_s, P_\mu, P_\nu$ are the probabilities of sending them. Here there are also two other parameters, the vacuum state $\omega$ and the vacuum state probability $P_\omega$, but we assume the former is a constant, and the latter satisfies $P_\omega=1-P_s-P_\mu-P_\nu$, so neither are included as variables. Note that, since only the signal intensity $s$ in the Z basis is used for key generation, and $\mu, \nu$ in X basis are used for parameter estimation, $P_s$ is also the basis choice probability.  We will unite these 6 parameters into one parameter vector $\vec{p}$.

The calculation of the key rate depends not only on the intensities and the probabilities, but also on the experimental parameters, namely the distance $L$ between Alice and Bob (or equivalently $L_{BC}$, the distance between the relay Charles and Bob), the detector efficiency $\eta_d$, the dark count probability $Y_0$, the basis misalignment $e_d$, the error-correction efficiency $f_e$, and the number of signals $N$ sent by Alice. We will unite these parameters into one vector $\vec{e}$, which we call the ``experimental parameters".

Therefore, we see that the QKD key rate can be expressed as 

\begin{equation}
Rate=R(\vec{e},\vec{p})
\end{equation}

\noindent which is a function of the experimental parameters $\vec{e}$, which cannot be controlled by the users, and the ``user parameters" $\vec{p}$ (or just ``parameters" for short, in the rest of the chapter if not specifically mentioned), which can be adjusted by the users.

However, this only calculates the rate for a given fixed set of parameters and experimental parameters. To calculate the optimal rate, we need to calculate

\begin{equation}
R_{max}(\vec{e})=max_{\vec{p} \in P} R(\vec{e},\vec{p})
\end{equation}

\noindent which is the optimal rate for a given $\vec{e}$. By maximizing R, we also end up with a set of optimal parameters $\vec{p_{opt}}$. Note that $\vec{p_{opt}}$ is a function of $\vec{e}$ only, and the key objective in QKD optimization is to find the optimal set of $\vec{p_{opt}}$ based on the given $\vec{e}$:

\begin{equation}
\vec{p_{opt}}(\vec{e}) = argmax_{\vec{p} \in P} R(\vec{e},\vec{p})
\end{equation} 

Up so far, the optimal parameters are usually found by performing local or global searches \cite{mdiparameter,this_asymMDI}, which evaluate the function $R(\vec{e},\vec{p})$ many times with different parameters to find the maximum. However, we make the key observation that the functions $R_{max}(\vec{e})$ and $\vec{p_{opt}}(\vec{e})$ are still \textit{single-valued, deterministic functions} (despite that their mathematical forms are defined by $max$ and $argmax$ and not analytically attainable).\\

As mentioned in Section 7.1, the universal approximation theorem of neural network states that it is possible to infinitely approximate any given bounded, continuous function on a given defined domain with a neural network (with a few or even a single hidden layer). Therefore, this suggests that it might be possible to use a neural network to fully described the behaviour of the aforementioned optimal parameters function $\vec{p_{opt}}(\vec{e})$. Once such a neural network is trained, it can be used to directly find the optimal parameters and the key rate based on any input $\vec{e}$ by evaluating $\vec{p_{opt}}(\vec{e})$ and $R(e,\vec{p_{opt}})$ once each (rather than the traditional approach of evaluating the function $R(\vec{e},\vec{p})$ many times), hence greatly accelerating the parameter optimization process. As we will later show, this method works for several common types of protocol as long as we can formulate a good analytical form of the key rate function. Nonetheless, this method also relies on the fact that the given protocol has a convex key rate versus parameters function (and has a bounded domain - which in practice is mostly just a ``square" domain with acceptable constant upper/lower bound values for each dimension of $\vec{p}$), such that the optimization problem is a convex optimization and a local search works (or that the function is not too highly non-convex such that some simple global search techniques can address the non-convexity).

\subsection{Design and Training of Network}

\begin{figure}[h]
	\includegraphics[scale=0.3]{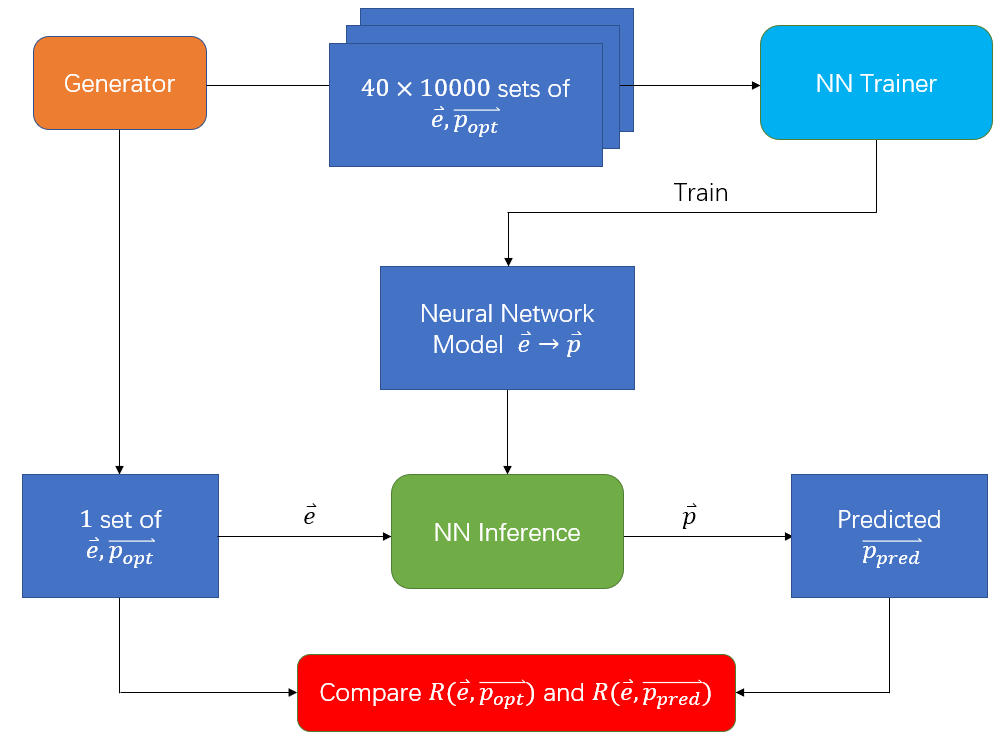}
	\caption{Data flow of the training and testing of the neural network (NN). The rounded-corner boxes represent programs, and rectangular boxes represent data. The generator program generates many random sets of experimental parameters $\vec{e}$ and calculates the corresponding optimal parameters $\vec{p_{opt}}$. These data are used to train the neural network. After the training is complete, the network can be used to predict optimal parameters based on arbitrary new sets of random experimental data and generate $\vec{p_{pred}}$ (for instance, to plot the results of Fig. 4 and Fig. 5, for each protocol a single random set of data is used as input). Finally, another ``validation" program calculates the key rate based on the actual optimal parameters $\vec{p_{opt}}$ found by local search and the predicted $\vec{p_{pred}}$ respectively, and compares their performances. Reproduced from \cite{this_ML} @2019 APS.}
	\label{fig:dataflow}
\end{figure} 

Here we proceed to train a neural network to predict the optimal parameters. We first write a program that randomly samples the input data space to pick a random combination of $\vec{e}$ experimental parameters, and use local search algorithm \cite{mdiparameter} to calculate their corresponding optimal rate and parameters. The experimental parameter - optimal parameter data sets (for which we generate 10000 sets of data for 40 points from $L_{BC}=$0-200km, over the course of 6 hours) are then fed into the neural network trainer, to let it learn the characteristics of the function  $\vec{p_{opt}}(\vec{e})$. The neural network structure is shown in Fig.\ref{fig:NN}. With 4 input and 6 output elements, and two hidden layers with 200 and 400 ReLU neurons each. We use a mean squared error cost function. 

For input parameters, since $\eta_d$ is physically no different from the transmittance (e.g. having half the $\eta_d$ is equivalent to having 3dB more loss in the channel - note that here we will assume that loss from detector efficiency can be controlled by Eve and therefore can be merged into channel loss, and that all detectors have equal $\eta_d$), here as an example we fix it to $80\%$ to simplify the network structure (so the input dimension is 4 instead of 5) - when using the network for inference, a different $\eta_d$ can be simply multiplied onto the channel loss while keeping $\eta_d=80\%$. We also normalize parameters by setting 

\begin{equation}
\begin{aligned}
e_1&=L_{BC}/100\\
e_2&=-log_{10}(Y_0)\\
e_3&=e_d\times 100\\
e_4&=log_{10}(N)\\
\end{aligned}
\end{equation}

\noindent to keep them at a similar order of magnitude of 1 (which the neural network is most comfortable with) - what we're doing is a simple scaling of inputs, and this pre-processing doesn't modify the actual data. The output parameters (intensities and probabilities) are within $(0,1)$ to begin with (we don't consider intensities larger than 1 since these values usually provide poor or zero performance) so they don't need pre-processing. When generating random sets of experimental parameters for training, here as an example we use a range of common values for $e_1\in[0,2]$, $e_2\in[5,7]$, $e_3\in[1,3]$, $e_4\in[11,14]$ which correspond to $L_{BC}\in[0,200]$, $Y_0\in[10^{-7},10^{-5}]$, $e_d\in[0.01,0.03]$, $N\in[10^{11},10^{14}]$. The normalized parameters are sampled uniformly from the range (i.e. some parameters $Y_0,N$ are uniformly sampled in log scale). Also, note that, the range we set here (with commonly encountered values in experiment) is a testing example, but in practice one can train with a wider range for the input values to encompass more possible scenarios for experimental parameters (and reasonably with more sample training data).

We can also easily modify the setup to accommodate for other protocols by adjusting the number of input and output parameters. For the asymmetric MDI-QKD scenario, one can add an additional input parameter, the channel mismatch $x=\eta_A/\eta_B$, where $\eta_A,\eta_B$ are the transmittances in Alice's and Bob's channels. We can normalize the mismatch too and make it an additional input variable:

\begin{equation}
\begin{aligned}
e_5&=-log_{10}(x)\\
\end{aligned}
\end{equation}

For the random training data, we sample $e_5\in[0,2]$, i.e. channel mismatch $x\in[0.01,1]$. In this case the output parameter vector $\vec{p}$ would be $[s_A, \mu_A, \nu_A, P_{s_A}, P_{\mu_A}, P_{\nu_A}, s_B, \mu_B, \nu_B, P_{s_B}, P_{\mu_B}, P_{\nu_B}]$.\\

For BB84 under finite-size effects, the input vector is the same as in symmetric MDI-QKD, while the output parameter vector $\vec{p}$ would be $[\mu, \nu, P_\mu, P_\nu, P_X]$, where vacuum+weak decoy states are used (i.e. intensities are $[\mu,\nu,\omega]$, which correspond respectively to the signal, weak decoy, and vacuum states) and only one basis - for instance the X basis - is used for encoding. Here $P_X$ is the probability of choosing the X basis. Since the parameter space of BB84 is slightly non-convex, when generating the training set, we have modified the local search to start from multiple random starting points (and choose the highest local maximum). This is a simple form of global search, and can mostly overcome the small non-convexity for the key rate versus parameters function for BB84.

For asymmetric TF-QKD, the input vector is the same as in asymmetric MDI-QKD, while the output parameter vector $\vec{p}$ is $[s_A, s_B, \mu, \nu, P_{s}, P_{\mu}, P_{\nu}]$, where, as shown in Ref. \cite{this_asymTF}, to compensate for channel asymmetry, we employ asymmetric signal states $s_A,s_B$ while using the identical vacuum+weak decoy states (and probabilities) for Alice and Bob, while leads to $2+5$ output parameters. For the detector efficiency $\eta_d$, we assume it is part of the channel loss and merge it as part of $L_{BC}$, i.e. in the program $\eta_d$ is set to $100\%$ (and like for MDI-QKD, we assume the two detectors have equal detector efficiency). An additional note for TF-QKD is that it uses a linear program to calculate the key rate $R(\vec{e},\vec{p})$, which makes it over an order of magnitude slower than e.g. MDI-QKD, which uses analytical functions to solve for the key rate. Due to the large amount of computation involved, we performed the data generation on the Niagara supercomputer (using about 4 nodes $\times$ 8 hours, or 1280 core hours as each node has 40 Intel Xeon cores, after which we chose 4500 sets of random data collected from multiple runs), but note that, these computations can be considered offline. That is, once one takes the time to generate the training set and obtain a neural network, end users can simply deploy the neural network to compute all future sets of data online in milliseconds.

We train the neural network using Adam \cite{ADAM} as the optimizer algorithm for 120 epochs (iterations), which takes roughly 40 minutes on an Nvidia Titan Xp GPU.

Note that, here to prevent overfitting, we have crudely employed early-stopping when training the model, by checking the validation set (20\% of the data) and stopping the training when the validation set loss no longer decreases (despite that the training set still shows increasingly smaller error), which helps with preventing overfitting. We test in increments of 60 epochs at a time, and choose 120 epochs as the stopping point. We have also tested with adding e.g. Dropout layers, but the results change very little. Likely, since the data size itself (41 samples $\times$ 10000 data sets) is larger than the number of weights in the network (at the order of $200\times 400$), the overfitting problem is not severe here.

Another point worth noting is that, the universal approximation theorem suggests that the neural network is able to approximate a function over a given domain accurately. However, we did notice from numerical testing that at the boundaries of the domain (e.g. $L_{BC}=0$ for $L_{BC}\in[0,200]$) there is some level of deviation for the neural network output from the ``ground truth" optimal value. One reason might be that the approximation might be less accurate at the boundaries (which could potentially be alleviated by selecting a wider domain for the training values - also, the boundary values such as zero channel loss $L_{BC}=0$ are not feasible in practice anyway)\footnote{Another more technical potential cause for this behaviour might be that, when we generate the large set of training data, the local search algorithm is set to have a relatively low maximum iteration count limit to reduce computation time - meaning that at extreme values such as $L_{BC}=0$ or very long distance where the rate is close to zero, which require more iterations, the generated training data might not be the optimal value because the local search has not run enough iterations, and this difference is reflected in the neural network trained by these data, too.}, while for most values except the boundaries, the approximation is extremely accurate, as we will show with numerical results in the next section. 

In terms of the security of the neural network, we do not yet know if e.g. adversarial attacks (specific input values designed such that the neural network outputs an incorrect result) might exist for our neural network, which has continuous output values (and not discrete ones like in a classification task). Nonetheless, even if the neural network outputs inaccurate or incorrect values, it will not affect the security of the QKD protocol - since the decoy-state analysis is valid as long as Alice (and/or Bob) knows accurately what intensities she sent, even if the intensity values are poorly chosen. In the case a ``bad" set of intensities and probabilities are suggested by the neural network, it can at most be considered a denial-of-service attack, and Alice and Bob can also find out the issue before using the values, as they can easily test the expected key rate by evaluating $R(\vec{e},\vec{p})$ with the neural network output first.

\section{Numerical Results}

\begin{figure}[t]
	\includegraphics[scale=0.15]{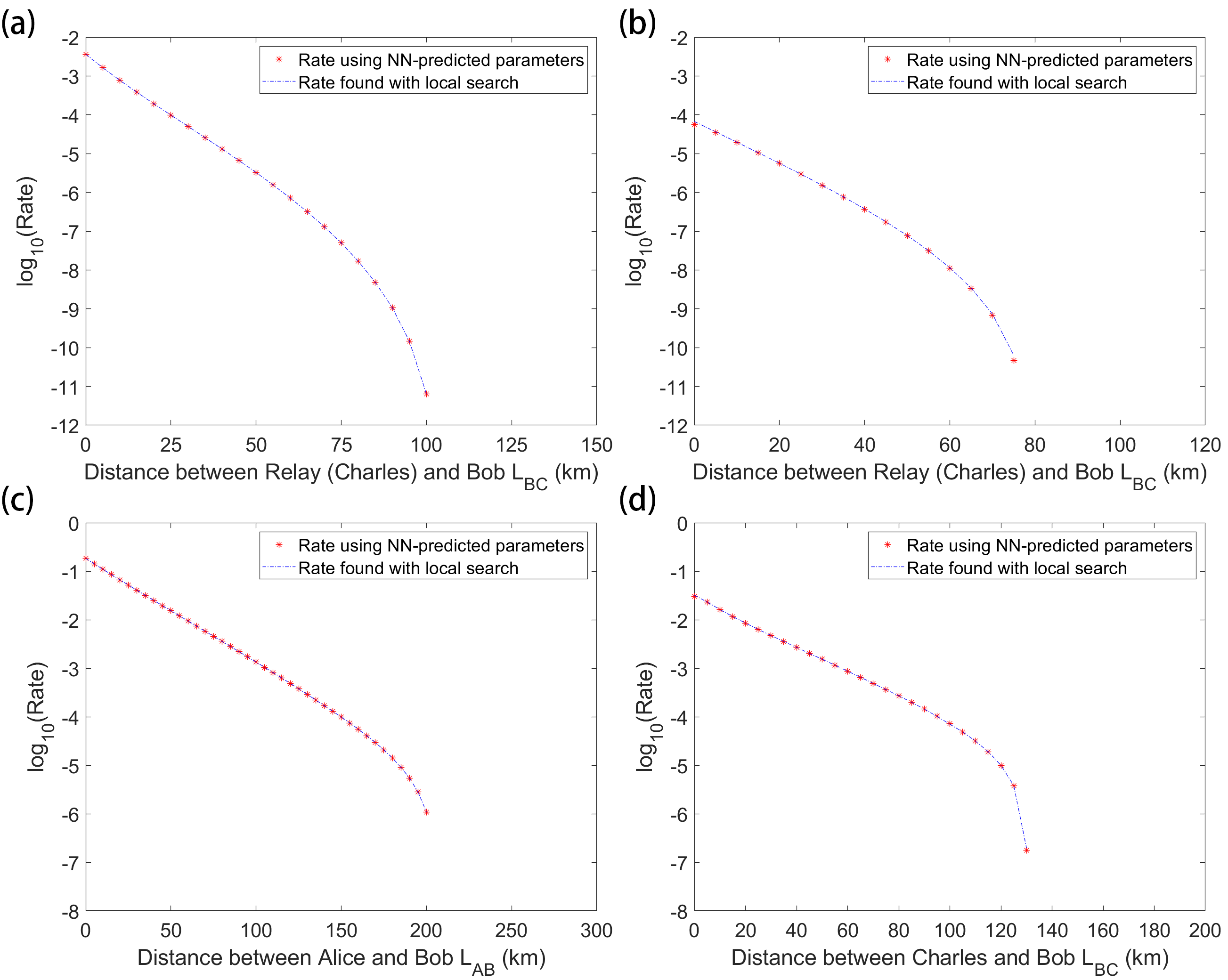}
	\caption{Comparison of expected key rate using neural network (NN) predicted parameters vs using optimal parameters found by local search for various protocols, using experimental parameters from Table 7.3, at different distances between Alice (or Charles) and Bob. We compare the key rate generated with either sets of parameters (dots with NN-predicted parameters, and lines with local search generated parameters). We tested four protocols: (a) symmetric MDI-QKD (4-intensity protocol) \cite{mdifourintensity}, (b) asymmetric MDI-QKD (7-intensity protocol) \cite{this_asymMDI}, (c) BB84 protocol \cite{finitebb84}, and (d) asymmetric TF-QKD protocol \cite{this_asymTF}. As can be seen, the key rate obtained using predicted parameters is very close to that obtained from using optimal parameters found with local search. Reproduced from \cite{this_ML} @2019 APS.}
	\label{fig:rate}
\end{figure} 

\begin{figure}[t]
	\includegraphics[scale=0.15]{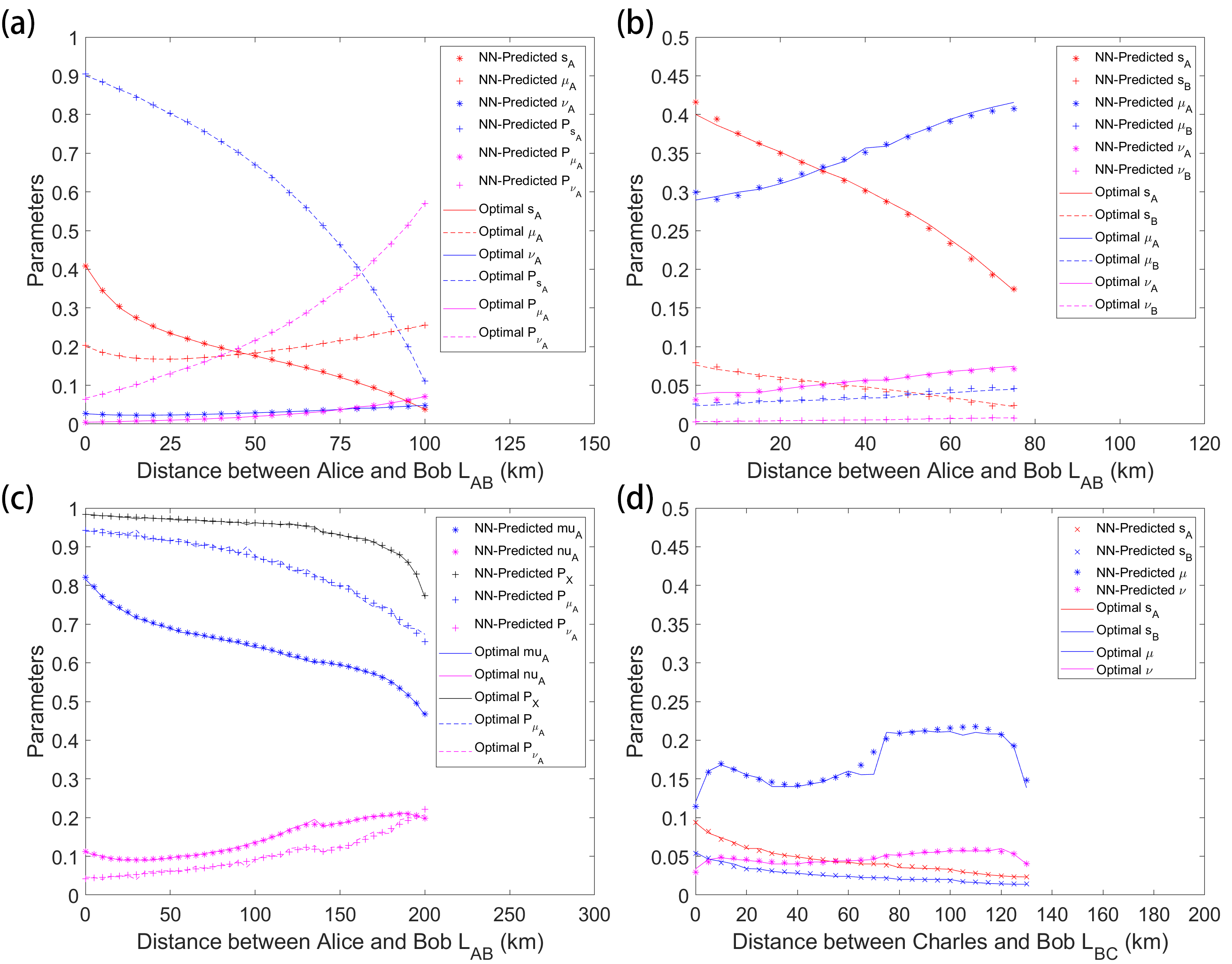}
	\caption{Comparison of neural network (NN) predicted parameters vs optimal parameters found by local search for various protocols, using experimental parameters from Table 7.3, at different distances between Alice (or Charles) and Bob. We compare neural network (NN) predicted parameters (dots) versus optimal parameters found by local search (lines), for four protocols: (a) symmetric MDI-QKD (4-intensity protocol), (b) asymmetric MDI-QKD (7-intensity protocol), (c) BB84 protocol, and (d) asymmetric TF-QKD protocol. Similar to Fig. \ref{fig:rate}, as can be seen, the NN-predicted parameters are very close to optimal values found with local search. Note that there is some noise present for the BB84 protocol. This is because the key rate versus parameters function shows some levels of non-convexity, and we combined local search with a randomized approach (similar to global search) that chooses results from multiple random starting points. Therefore there is some level of noise for the probability parameters (which are insensitive to small perturbations), while the neural network is shown to learn the overall shape of the global maximum of the parameters and return a smooth function. Similar applies for TF-QKD, which has small levels of non-convexity due to linear solvers being inherently non-convex, although due to the limitation in training time, we did not apply global search for TF-QKD. Reproduced from \cite{this_ML} @2019 APS.}
	\label{fig:param}
\end{figure}

\begin{table*}[t]
	\caption{Optimal parameters found by local search vs neural network (NN) predicted parameters for symmetric MDI-QKD (4-intensity protocol) using three different random sets of experimental parameters (set b we include here is the same one used for Fig. \ref{fig:rate}(a) and Fig. \ref{fig:param}(a), as listed in Table 7.3), at the same distance $L_{BC}$ of 50km between Charles and Bob. $Y_0$ is the dark count probability, $e_d$ is the basis misalignment, and $N$ is the number of signals sent by Alice. Here for simplicity, the detector efficiency is fixed at $\eta_d=80\%$ (since it is equivalent to channel loss). Fibre loss per km is assumed to be $\alpha=0.2 dB/km$, the error-correction efficiency is $f_e=1.16$, and finite-size security failure probability is $\epsilon=10^{-7}$. As can be seen, the predicted parameters from our neural network are very close to the optimal parameters found by local search, within a $1\%$ error. Moreover, the key rate is even closer, where the rate calculated with predicted parameters can achieve up to 99.99\% (sometimes even higher than) the key rate found by local search for this protocol. Due to limited width of the page, this table (reproduced from Table 7.2 in Ref. \cite{this_ML}) has been split into two parts, showing experimental input parameters (top) and optimal intensity parameters (bottom) separately.}
	\begin{center}

		\begin{tabular}{ccc|cccc}            
			Set & Method & R  &$L_{BC}$ &$Y_0$ & $e_d$ & N \\
			\hline\hline
			a & Local search & $5.5390\times 10^{-5}$& 50km & $7.50 \times 10^{-6}$ & 0.0115 & $3.14 \times 10^{13}$\\
			a & NN & $5.5385\times 10^{-5}$ & 50km & $7.50 \times 10^{-6}$ & 0.0115 & $3.14 \times 10^{13}$\\
			
			b & Local search & $3.2559\times 10^{-6}$ & 50km & $1.75 \times 10^{-7}$ & 0.0287 & $2.99 \times 10^{12}$\\
			b & NN & $3.2559\times 10^{-6}$ & 50km & $1.75 \times 10^{-7}$ & 0.0287 & $2.99 \times 10^{12}$\\
			
			c & Local search & $2.7738\times 10^{-6}$ & 50km & $4.29 \times 10^{-7}$ & 0.0196 & $2.11 \times 10^{11}$\\
			c & NN & $2.7739\times 10^{-6}$  & 50km & $4.29 \times 10^{-7}$ & 0.0196 & $2.11 \times 10^{11}$\\
			\vspace{5pt}
		\end{tabular}
		
		\begin{tabular}{ccc|cccccc}            
			Set & Method & R & $s$ & $\mu$ & $\nu$ & $P_s$ & $P_{\mu}$ & $P_{\nu}$ \\
			\hline\hline
			a & Local search & $5.5390\times 10^{-5}$&  0.431& 0.170&0.0256 & 0.862& 0.00736& 0.0906\\
			a & NN & $5.5385\times 10^{-5}$ &  0.431& 0.169&0.0257 &0.861& 0.00729& 0.0901\\
			
			b & Local search & $3.2559\times 10^{-6}$ & 0.176 &0.183& 0.0290& 0.670 & 0.0200 & 0.216\\
			b & NN & $3.2559\times 10^{-6}$ & 0.177& 0.184 & 0.0289& 0.670&0.0193 & 0.216\\
			
			c & Local search & $2.7738\times 10^{-6}$ & 0.209& 0.241 & 0.0442 & 0.540 & 0.0339 & 0.298\\
			c & NN & $2.7739\times 10^{-6}$  & 0.210 & 0.242 & 0.0441 & 0.538 & 0.0338 & 0.298\\
			
		\end{tabular}
	\end{center}
\end{table*}

\begin{table}[t]
	\caption[]{Random experimental parameter sets we use for simulation of Fig. \ref{fig:rate} and Fig. \ref{fig:param} for the four protocols (symmetric/asymmetric MDI-QKD, BB84, TF-QKD). $Y_0$ is the dark count probability, $e_d$ is the basis misalignment, and $N$ is the number of signals sent by Alice (and Bob, in MDI-QKD and TF-QKD). Here for simplicity, the detector efficiency is fixed at $\eta_d=80\%$ for MDI-QKD and BB84, and $100\%$ for TF-QKD (detector loss is included in the channel loss). The channel mismatch x for MDI-QKD and TF-QKD is the ratio of transmittances between Alice's and Bob's channels, $\eta_A/\eta_B$.}
	\begin{center}
		\begin{tabular}{c|ccccc}            
			Protocol & x & $e_d$ & $Y_0$  & $N$ & $\eta_d$\\
			\hline
			4-intensity MDI & 1 & 0.029 & $1.7\times 10^{-7}$ & $3.0\times 10^{12}$ & 80\% \\
			7-intensity MDI & 0.10 & 0.026 & $2.7\times 10^{-6}$ & $2.6\times 10^{13}$ & 80\% \\
			BB84 & - & 0.011 & $3.6\times 10^{-6}$ & $3.2\times 10^{12}$ & 80\% \\
			TF-QKD & 0.54 & 0.024 & $1.8\times 10^{-6}$ & $1.9\times 10^{13}$ & 100\% \\
		\end{tabular}
	\end{center}
\end{table}

After the training is complete, we use the trained network for 4-intensity MDI-QKD protocol to take in three sets of random data, and record the results in Table 7.2. As can be seen, the predicted parameters and the corresponding key rate are very close to the actual optimal values obtained by local search, with the NN-predicted parameters achieving up to 99.99\% the optimal key rate. 

Here we also fix one random set of experimental parameters as seen in Table 7.3, and scan the neural network over $L_{BC}=$0-200km. The results are shown in Fig.\ref{fig:rate}(a) and \ref{fig:param}(a). As we can see, again the neural network works extremely well at predicting the optimal values for the parameters, and achieves very similar levels of key rate compared to using the traditional local search method.

We also use a similar approach to select a random set of input parameters and compare the predicted key rate versus optimal key rate for each of 7-intensity (asymmetric MDI-QKD), BB84, and TF-QKD protocol. The results are included in Fig.\ref{fig:rate}(b-d) and \ref{fig:param}(b-d). As can be seen, the accuracy of the neural network is very high in these cases too, with up to 95-99\% the key rate for 7-intensity protocol, up to 99.99\% for BB84, and $\sim$80-95\% for asymmetric TF-QKD (the accuracy for TF-QKD is smaller, likely because of either the smaller training set, or the linear solvers used in TF-QKD bringing in inherent non-convexities, which make the data noisier and more difficult to fit).

\section{Applications and Benchmarking}

In the previous section we have demonstrated that a neural network (NN) can be trained to very accurately simulate the optimal parameter function $\vec{p_{opt}}(\vec{e})$ and be used to effectively predict the optimal parameters for QKD. The question is, since we already have an efficient coordinate descent (CD) algorithm, what is the potential use for such a NN-prediction approach? Here in this section, we will discuss two important use cases for the neural network.\\

\textbf{1. Real-time optimization on low-power devices}. While it takes considerable computing power to ``train" a neural network (e.g. on a dedicated GPU), using it to predict (commonly called ``inference") is computationally much cheaper, and will be much faster than performing a local search, even if the neural network is run on the same CPU. Moreover, in recent years, with the fast development and wide deployment of neural networks, many manufacturers have opted to develop dedicated chips that accelerate NN-inference on mobile low-power systems. Such chips can further improve inference speed with very little required power, and can also offload the computing tasks from the CPU (which is often reserved for more crucial tasks, such as camera signal processing or motor control on drones, or system operations and background applications on mobile phones). 

Therefore, it would be more power-efficient (and much faster) to use a neural network running on inference chips, rather than using the computationally intensive local search algorithm with CPU on low-power devices. This can be especially important for free-space QKD scenarios such as drone-based, handheld, or satellite-ground QKD, which not only have a very limited power budget, but also require low latency in real-time.

Note that, some protocols we use for demonstration here (BB84, and MDI-QKD with ``independent bounds" finite key analysis as in \cite{this_asymMDI}) are pretty fast to optimize to begin with, on the order of seconds even on a single-board computer. However, there are cases where optimization itself is much slower, for instance when global search and/or linear solver are needed. For instance, the asymmetric TF-QKD protocol \cite{this_asymTF} or the ``9-intensity" MDI-QKD protocol (an asymmetric MDI-QKD protocol where Alice and Bob each use four instead of three decoy states, as described in Ref. \cite{this_asymMDI}, which requires using both linear solver and global search) would respectively take 2 seconds and 11 seconds to generate just one point even on a fast desktop PC. On single-board computers it is generally 10-30 times slower, meaning that an optimization would likely take tens of seconds to even minutes, which is quite long since many free-space sessions might only have a window of minutes (e.g. satellite-ground or handheld QKD). Also, some alternative finite-key analysis, such as the ``joint bounds" analysis (as opposed to using ``independent bounds") proposed in \cite{mdifourintensity}, will introduce similar problems, as it involves linear solver and non-convexities in the key rate versus parameters function (which often necessitate global search).

Moreover, there are some practical reasons where software/hardware limitations might favour using a neural network over performing local search on CPU on low-power platforms. For instance, as mentioned above, performing local search uses up all CPU resource, which would be non-ideal for drones and handheld systems that need the CPU for the control system, while a neural network running on a separate accelerator chip offloads the computational requirement. Also, software-wise, computing the key rate and performing optimization on CPU mean requiring the entire software stack to be set up on the mobile device - however, many software libraries, e.g. all commercial linear solver libraries, don’t even work on mobile architecture such as ARM CPUs – but a neural network pre-trained with data generated by linear solvers can still be run on these platforms. This can be shown in Table 7.1 where iPhones don't support linear solvers required for calculating the key rate, but a neural network can be used to directly output the parameters (without needing to calculate the key rate first).

Lastly, atmospheric turbulence causes the channel transmittance to quickly fluctuate at a time scale of 10-100 ms order \cite{freespacethesis} (and QKD over moving platforms can have average channel losses constantly changing with time e.g. due to changing distances). With a neural network it would be potentially feasible to quickly tune laser intensities based on sampled channel loss in real time. Nonetheless, a set of changing laser intensities would require modified finite-size analysis, so we only propose it as a possibility here and will leave the details for future discussions.

To benchmark the performance of neural networks on low-power devices, as examples, we choose the 4-intensity MDI-QKD protocol and the TF-QKD protocol, and test our neural network models on two popular mobile low-power platforms: a single-board computer, and an ordinary mobile phone, as shown in Fig. \ref{fig:mobile}. We implement both CPU-based local search algorithm and neural network prediction on the devices, and list the running time in Table 7.1, where we compare neural networks to local search on the portable devices and on a powerful desktop PC. As shown in Table 7.1, using neural networks, we can find the optimal parameters in milliseconds regardless of the protocol \footnote{Note that, for neural networks it generally takes some time to load the model into the device when first used (about 0.2-0.3s on Titan Xp GPU and neural engine on the iPhone, and 3s on Raspberry Pi with the neural compute stick), but this only needs to be done once at boot time, and can be considered part of the startup time of the device - once the network is running, the predictions can be performed on many sets of data taking only milliseconds for each operation.} (which is 2-4 orders of magnitude faster than local search on CPU), in a power footprint less than $1/70$ that of a desktop PC.

In Table 7.1 we used the 4-intensity MDI-QKD protocol and the TF-QKD protocol as two examples, although note that for other protocols, e.g. 7-intensity MDI-QKD or BB84 protocol, the advantage of NN still holds, since the NN prediction time is little affected by the input/output size (for instance, in Fig. \ref{fig:NN}, there are $400\times 200$ connections between the two middle hidden layers, and only $4\times 400$ and $6\times 200$ connections involving output or input neurons. This means that the numbers of input/output nodes have little impact on the overall complexity of the network), while local search time increases almost linearly with the number of output (searched) parameters. For instance, running 7-intensity MDI-QKD protocol, which has 12 output parameters, takes about 0.4s using local search on an iPhone XR - which is double the time of the 4-intensity MDI-QKD protocol, which has 6 output parameters - but with a NN it still takes about 1ms (making the advantage of using NN even greater in this case).

Additionally, note that even without neural network acceleration chips, many devices can still choose to (1) run the neural network on CPU (at the expense of some CPU resource), and this option is still much faster than local search (for instance, running the neural network on iPhone XR with CPU takes between $1.3-2.0$ms, which is not that much slower than the dedicated neural accelerator chip). 
(2) generate a static ``lookup table" for all possible inputs down to a given resolution. This is ideal for systems with extremely limited computing power or with software/hardware restrictions, such that neural networks cannot be run in real time. The lookup table can be generated using a GPU on a desktop computer first and stored on a mobile system to check when needed. This is slower than directly running a neural network, but it is still considerably faster than performing a local search. More details are discussed in Appendix E.1. \\

\textbf{2. Quantum networks}. In addition to free-space QKD applications that require low-power, low-latency devices, the neural network can also be very useful in a network setting, such as a quantum internet-of-things (IoT) where numerous small devices might be interconnected in networks as users or relays. For an untrusted relay network, MDI-QKD or TF-QKD protocol are desirable. However, the number of pairs of connections between users will increase quadratically with the number of users, which might quickly overload the computing resources of the relay/users.

With the neural network, any low-power device such as a single-board computer or a mobile phone can easily serve as a relay that connects to numerous users and optimizes e.g. $\sim$5000 pairs of connections (100 users) in under 5 seconds for MDI-QKD or TF-QKD. This is a task previously unimaginable even for a powerful desktop PC, for which even supporting a network with 20 users would take from 20s (MDI-QKD) to 6 minutes (TF-QKD) to even 30 minutes (``9-intensity MDI-QKD"), if the protocol chosen is difficult to optimize. Therefore, our new method can greatly reduce the required compute power of devices and the latency of the systems when building a quantum Internet of Things.

\section{Conclusion and Discussions}

In this chapter we have presented a simple way to train a neural network that accurately and efficiently predicts the optimal parameters for a given QKD protocol, based on the characterization of devices and channels. We show that the approach is general and not limited to any specific form of protocol, and demonstrate its effectiveness for four examples: symmetric/asymmetric MDI-QKD, BB84, and TF-QKD.

We show that an important use case for such an approach is to enable efficient parameter optimization on low-power devices. We can achieve 2-4 orders of magnitude faster optimization speed compared to local search, with a fraction of the power consumption. Our method can be implemented on either the increasingly popular neural network acceleration chips, or on common CPUs that have relatively weak performance. This can be highly useful not only for free-space QKD applications that require low latency and but have a limited power budget, but also for a quantum internet-of-things (IoT) where even a small portable device connected to numerous users in a quantum network can easily optimize the parameters for all connections in real-time.

Here we have demonstrated that the technique of machine learning can indeed be used to optimize the performance of QKD protocols. The effectiveness of this simple demonstration suggests that it may be possible to apply similar methods to other optimization tasks, which are common in the designing and control of practical QKD systems, such as determining the optimal threshold for post-selection in free-space QKD, tuning the polarization controller motors for misalignment control, etc.. Such a method might even be applicable for optimization and control tasks in classical systems, to use a pre-trained neural network in accelerating well-defined but computationally intensive tasks. We hope that our work can further inspire future works in investigating how machine learning could help us in building better performing, more robust QKD systems.\\

All training data are generated from simulations based on models in Refs. \cite{mdifourintensity,this_asymMDI,finitebb84,this_asymTF}, using local/global search algorithm. The generated datasets, and the neural networks trained from them, are available upon reasonable request by email (see Ref. \cite{this_ML} for author contact information).\\

\textit{Note added:} After our posting of a first draft of Ref. \cite{this_ML} on the preprint server \cite{arxiv}, another work on a similar subject was subsequently posted on the preprint server \cite{arxiv_JOSAB_AI} and later published at \cite{JOSAB_AI}. While both our work and the other work \cite{this_ML,JOSAB_AI} have similar approaches in parameter optimization with neural networks, and observe the huge speedup neural network has over CPU local search, a few important differences remain. Firstly, we show that the neural network method is a general approach not limited to any specific protocol (and show its versatile applications with four examples), while Ref. \cite{arxiv_JOSAB_AI,JOSAB_AI} is limited to discussing asymmetric MDI-QKD only. Secondly, we point out that a key use case of this approach would be performing parameter optimization on low-power devices with neural networks. This was only briefly mentioned in passing in Ref. \cite{arxiv_JOSAB_AI,JOSAB_AI}. In contrast, we perform testing and benchmarking on real hardware devices. Our work not only will allow more types of smaller portable devices to join a network setting, but also can be important for free-space QKD applications where low power consumption and low latency are crucial.

\chapter{Conclusion}

\section{Conclusions of this Thesis}

\begin{table}[t]
	
	\caption{ An recapitulation of the topics of chapters in this thesis, and their corresponding applications (for free-space QKD or fibre-based networks). Some terms (MDI-QKD, TF-QKD, turbulence, asymmetry) are abbreviated due to limited space here.} 
	\begin{center}
		\begin{tabular}{cc|ccc|ccc|cc}		
			\hline \hline
			\multirow{2}{*}{Chapter} & \multirow{2}{*}{Ref.} & \multicolumn{3}{c|}{Protocol} & \multicolumn{3}{c|}{Topic} & \multicolumn{2}{c}{Application}\\
			& & BB84 & MDI & TF & Turb. & Asym. & Machine Learning & Free-space & Network\\
			\hline
			3 & \cite{this_BB84} & $\checkmark$ & - & - & $\checkmark$ & - & - & $\checkmark$ & - \\
			4 & \cite{this_MDI} & - & $\checkmark$ & - & $\checkmark$ & - & - & $\checkmark$ & - \\
			\hline
			5 & \cite{this_asymMDI} & - & $\checkmark$ & - & - & $\checkmark$ & - & - & $\checkmark$ \\
			6 & \cite{this_asymTF} & - & - & $\checkmark$ & - & $\checkmark$ & - & - & $\checkmark$ \\
			\hline
			7 & \cite{this_ML} & $\checkmark$ & $\checkmark$ & $\checkmark$ & - & - & $\checkmark$ & $\checkmark$ & $\checkmark$ \\
			\hline \hline
		\end{tabular}
		
	\end{center}
	
\end{table}

In this thesis, we have discussed two promising future directions of QKD: free-space QKD and fibre-based QKD networks. We have focused on three important challenges in practical QKD:

\begin{enumerate}
	\item atmospheric turbulence (in free-space QKD)
	\item asymmetric channel losses (in QKD networks with untrusted relays)
	\item efficient parameter optimization (for all QKD systems)
\end{enumerate}

\noindent and we have proposed adaptive techniques that greatly improve the practical performance of QKD when the above challenges are present. For (1), we have proposed a real-time selection method that can conveniently use a pre-fixed threshold and greatly increase the maximum tolerable channel loss. For (2), we have proposed a method that allows two users to use different intensity settings to compensate for channel loss. For (3), we have proposed the use of machine learning techniques to greatly accelerate parameter optimization, enabling low-power devices for free-space QKD and fast optimization for large-scale networks. A list of the topics of which each chapter corresponds to and their applications is shown in Table 8.1. We believe that such techniques, aimed at solving some of the key challenges in some important scenarios for QKD, will help pave the way towards future implementations of free-space QKD and fibre-based QKD networks. 

Here we believe it would also be beneficial to introduce some of the methodologies behind our work. In this thesis, we have taken an approach that combines both engineering/numerical techniques (such as post-selection, local/global optimization, and machine learning) as well as a strong physical/mathematical intuition behind our choice. For instance, we have proposed some key theoretical observations behind our methods: 

\begin{itemize}
	\item Chapter 3: For free-space BB84, optimal data post-selection threshold does not depend on the condition of turbulent channel, allowing us to perform real-time selection with a pre-fixed threshold, in a ``semi-blind" approach regardless of channel.
	\item Chapter 4: Free-space MDI-QKD will have turbulence-induced channel asymmetry in real-time that worsens its performance, which necessitates post-selection on both transmittance and asymmetry.
	\item Chapter 5: MDI-QKD has an inherent asymmetry between its two bases (and differs from a simple Hong-Ou-Mandel interference), allowing us to follow different strategies in two decoupled bases when compensating for channel asymmetry.
	\item Chapter 6: Similar to MDI-QKD, for TF-QKD single-photon interference visibility also depends on channel asymmetry (and can be improved by adjusting signal intensity).
	\item Chapter 7: The overall optimization process of a function can be viewed as a ``super-function". Such a super-function can be simulated efficiently with neural networks.
\end{itemize}

Notably, the results for Chapters 5 and 6 (asymmetric MDI-QKD and TF-QKD) might be especially of interest to a broader Physics community. Our physical insight\footnote{The following discussion here is based on helpful suggestions from HK Lo and from Ref. \cite{insight}.} is that QKD is a playground for demonstration of effects in quantum mechanics. For instance, in TF-QKD, we are interested in the effect of single-photon interference whereas MDI-QKD relies on two-photon interference.

In more detail, in TF-QKD, to achieve such an interference, it is important for the two laser pulses to be indistinguishable (same spectral property and timing info). [Incidentally, effects on misalignment of basis have been studied in the literature, such as Ref. \cite{mdipractical} for MDI-QKD, and Ref. \cite{simpleTFQKD} for TF-QKD.]

For MDI-QKD, we are interested in two-photon interference. Our work shows that, in one of the bases (X-basis), the two laser pulses need to be indistinguishable to achieve low QBER. This is directly analogous to the standard HOM dip. On the other hand, in the other basis (Z-basis), the QBER will remain low even when the two laser pulses are distinguishable from each other.

Asymmetry plays a key role in our work. The fact that Charlie is performing a measurement in a particular basis (Z-basis) in MDI-QKD means that Charlie's measurement is \textit{not} symmetric with respect to the basis choice by Alice and Bob. Such an asymmetry allows us to decouple the intensities of the two bases completely.

We also allow asymmetry between the signal states and the decoy states. Perhaps, we could name it as a principle of ``Efficiency from Asymmetry". Such a principle is also effectively applied in Chapter 6 to TF-QKD, where again one of the basis (X-basis) requires indistinguishability, but this time for single-photon interference, and the other basis (Z-basis) uses phase-randomized pulses and does not consider QBER. The choice of intensities are again decoupled completely for the two bases.\\

Another interesting physical observation we have made is the prediction of turbulence-induced channel asymmetry for free-space MDI-QKD, which will greatly decrease interference visibility if two channels are independently fluctuating. In this case, the channel asymmetry is changing quickly with time and we cannot effectively compensate for it with choice of intensities. We instead proposed that we can post-select the signals and keep only ones obtained under high channel symmetry. This is an effect - to our knowledge - not previously studied for free-space MDI-QKD, and we believe it will be helpful for implementations of MDI-QKD or simply HOM interference experiments through free-space channels. In fact, in Ref. \cite{freespaceHOM}, people have reported a HOM interference experiment between one free-space and one fibre channel, and they have observed such turbulence-induced decrease in visibility, for which they also implemented a post-selection technique to improve visibility.\\

Last but not least, many topics of this thesis are cross-disciplinary projects that bridge Computer Science with Physics, where efficient and novel software design, combined with a good understanding of the physics behind the problems, plays a key role in the solution for many of the problems in this thesis. We believe it is an approach with increasing importance in designing practical QKD systems (and perhaps in other physical fields too), and hope to apply such an approach of combining Physics with computing techniques (optimization, parallel computing, machine learning) to more future projects to come.

\section{Outlook and Future Works}

In this section, we discuss some remaining issues to be solved, and some potential future directions of research.

Some near-term topics: 

\begin{itemize}
	\item \textbf{Model for PDTC of a turbulent channel}: In Chapters 3 and 4 we have focused on a simple channel model (PDTC) of a truncated log-normal distribution. This model is applicable to the scintillation of signals, but does not apply well to beam-wandering, which is also an important affecting factor for QKD under turbulence. There have been studies on better PDTC models, such as Ref. \cite{PDTC1}. In principle, as we stated in the main text, the optimal threshold value itself does not depend on the form of PDTC, but the actual performance does depend on the channel. It would be interesting to see a rigorous discussion and estimation of the performance of the methods in Chapters 3 and 4 with a more comprehensive turbulence model.
	
	\item \textbf{Finite-size effects and asymmetry in free-space MDI-QKD}: In Chapter 4 we have only studied the infinite-data scenario and have not finished the study on MDI-QKD with finite-size effects (and a full optimization of parameters, including both intensities and probabilities) yet. Part of the reason is the limitation in computational power (since, as we stated in Chapter 4, the current channel model involves 2D integrations every time the key rate function is calculated), which would require more efficient algorithm design and more powerful computing devices, such as a computer cluster. A comprehensive study of free-space MDI-QKD with fully optimized parameters and finite-size analysis would be an interesting topic (with practical importance) to be studied in the future. 
	
	Also, up so far we have discussed \textit{symmetric} expected values for the channel losses and the same intensities for Alice and Bob. It would be interesting to discuss the general case of asymmetric channels and intensities, by combining the methods in Chapters 4 and 5.
	
	\item \textbf{Experimental implementation of asymmetric TF-QKD and TF-QKD network}: Together with our collaborators we have successfully implemented MDI-QKD over highly asymmetric channels in experiment, demonstrating its usefulness for a future network setting. With the theory for asymmetric TF-QKD completed (as shown in Chapter 6), we are interested in experimentally demonstrating TF-QKD over asymmetric channels, and exploring a TF-QKD network setting (which, compared to MDI-QKD, can increase the key rate and extend maximum distance between users). We have already completed a preliminary demonstration of asymmetric TF-QKD \cite{this_asymTF_experiment}, which is also accepted as a talk at the CLEO 2019 conference.
	
	\item \textbf{Machine learning accelerated optimization in more fields} In Chapter 7 we have demonstrated that neural networks can be used to accelerate parameter optimization for QKD. As we discussed in the chapter, the key idea of our method is in fact rather general, and in principle applies to any convex optimization problem (or even non-optimization problem - as long as there is a single-valued function that is computationally intensive to evaluate, but is needed in real-time applications), not necessarily only for QKD, but potentially for other fields of Physics and even outside Physics in other fields that require optimization. For instance, we note that very recently, an article with a similar approach has been posted on arXiv \cite{threebody} that makes use of a neural network, trained from pre-generated input-solution pairs, to accelerate the calculation of numerical solutions to the three-body problem. We hope to see such an approach of machine learning acceleration finding uses in more fields in and outside quantum information.

\end{itemize}

Some long-term topics: 

\begin{itemize}
		\item \textbf{Model for Hong-Ou-Mandel interference through two free-space channels}: In Chapter 5, we have focused on a model where the two channels are independently fluctuating (each following a PDTC based on a single channel). This is based on the assumption that the signals from Alice and Bob respectively reach two receiving telescopes coupled to two single-mode fibres. The signals in the fibres then interfere at an in-fibre beam-splitter. In this way, the interference is physically no different from a fully fibre-based system, while the turbulent channels can be modelled as two independent PDTCs, plus the coupling loss to the single-mode fibres. This setup has been physically implemented in a current experiment (presented as a poster at the QCrypt 2019 conference \cite{freespaceHOM}), albeit it performs MDI-QKD through one fibre channel and one free-space channel (where the free-space signal couples back to a single-mode fibre), meaning that the model is in fact feasible. 
		
		However, an alternative setup is to either couple to multi-mode fibre, or directly interfering signals coming from free-space channels at a free-space beam splitter. In such cases, the channels can no longer be modelled as just independent fluctuations in the channel losses, but rather, the turbulent modes and the distorted wavefronts will directly affect the Hong-Ou-Mandel interference visibility. Such a study will involve more background knowledge and further theoretical studies on atmospheric physics and quantum light through turbulence, and will also require more experimental verifications.
		
		Additionally, one more approach to addressing turbulence in free-space MDI-QKD could be using adaptive optics (e.g. deformable mirrors) that directly compensate for the distorted wavefronts and reduce the effect of turbulence in hardware. This is an alternative to the approach of receiving non-compensated signals, and improving signal-to-noise ratio in software by post-processing as discussed in this thesis.
		
		We thank the helpful discussions from Prof. Thomas Jennewein and Mr. Shuang-Lin Li on some of the above topics and ideas in free-space HOM interference.
		
		\item \textbf{Machine learning in finding new physics}: In Chapter 7 we presented a very simple demonstration that novel techniques introduced from other fields such as machine learning can be highly useful for practical QKD systems. Nonetheless, it is based on given analytical and numerical models of QKD systems, and simply improving upon existing knowledge, by accelerating the computation of a function we already know. In fact, there is an increasing interest in bridging the fields of machine learning and Physics. For instance, there have been proposals to use reinforcement learning (another type of machine learning algorithm, without requiring pre-generated training data) to find new optical experimental setups that create a desired quantum state \cite{ML_RL}, or to increase the stability of quantum gates on physical superconducting qubits \cite{ML_Quantum}. We believe that a deeper understanding of machine learning techniques and the introduction of it to the field of Physics (especially quantum communication and computation) can help us find more new physics and improve the practical QKD/Quantum Computing systems in fundamental and new ways.
\end{itemize}

\newpage{\pagestyle{empty}\cleardoublepage}

\appendix

\chapter{Supplemental Information for Background on QKD}

This Appendix contains supplemental information for Chapter 2.

\section{Variants of TF-QKD}

Since the proposal of the original TF-QKD protocol, there have been multiple papers that propose variants of the TF-QKD protocol that aim at providing a rigorous security analysis. As mentioned in the main text, an important underlying idea of these proofs is the division of signals into an ``encoding part" that generates the key, and a ``testing part" that bounds the phase error rate. The variants of TF-QKD type protocol differ in their approach of choosing ``encoding" and ``testing" phases as well as the respective usage/absence of phase randomization in the two bases. In this section we will briefly introduce some of the representative protocols.

For instance, in the ``TF-QKD* protocol" \cite{TFQKD01}, the authors divide signals into Code and Test phases that both contain signals in X and Z bases. The former phase announces the global phase publicly and generates the key, while the latter phase randomizes the phase, but only asks Alice and Bob to sample the gains (without requiring QBER) of signals. The gains can be used to perform a decoy-state analysis and estimate the yields of each photon number state, which can be used to bound the phase error rate with an ``unbalanced quantum coin", whose bias represents the amount of disturbance in the channel and is considered invariant between Test and Code phases. This coin bias can then be used to bound the phase error rate for the Code phase.

Another proposal, \cite{simpleTFQKD} (``simple TF-QKD"), assigns the X and Z bases respectively to encoding and testing phases. The X basis uses a fixed (zero) global phase without randomization and encodes the key in $0, \pi$ phase modulations, while the Z basis uses phase-randomized pulses to perform decoy-state analysis. In the non-phase-randomized X basis, it is proven that the phase-error rate can be estimated from the gain of ``cat states" (superpositions of either odd-numbered or even-numbered photon number states). It is then shown that the gain of cat states can be upper bounded using the yields of phase-randomized photon number states (which is invariant across the bases), where these yields can be estimated from the decoy-state analysis in the Z basis.

The Phase-Matching (PM) QKD \cite{TFQKD02} protocol doesn't physically divide signals into encoding and testing phases, but rather employs a ``virtual" protocol, where Alice and Bob did not announce the global phase, and performed decoy-state analysis. The virtual protocol has the same experimental observables as the real protocol. This means that Alice and Bob can use the real observables to estimate the yields of ``photon number states" (which only exist in the virtual protocol) and derive a phase error rate in the virtual protocol. The phase error rate is assumed to be invariant across the virtual and real protocols; therefore it can be used to estimate the key rate for the real protocol.

The Sending-or-not-Sending (SNS) QKD \cite{TFQKD03} uses Z and X bases for encoding and testing. However, a difference here is that SNS-QKD uses randomized coherent pulses in the Z basis (and instead of encoding information in the phase, Alice and Bob uses a small probability $\epsilon$ of sending a pulse to encode bit 1, and not sending $1-\epsilon$ to encode bit 0). The X basis uses phase-randomized pulses, but announces the phase. In the X basis, SNS-QKD employs a similar argument as above mentioned for PM-QKD, that suppose there is a virtual protocol where the phases were not announced, the observables as well as the phase error rate in the virtual protocol should be invariant, i.e. the same as the real protocol. Like PM-QKD, this allows SNS-QKD to estimate the single-photon yield and phase error rate based on X basis data and estimate the key rate.

There are also other protocols, such as MDI-TF-QKD, and PM-MDI-QKD, which we haven't discussed in detail here. In the review paper Ref. \cite{reviewQKD2019}, the authors classify the TF-QKD-like protocols into two types, the ones that use ``BB84-type two-basis analysis" (which fundamentally use single-photons as information carriers), such as TF-QKD* \cite{TFQKD01}, SNS-QKD \cite{TFQKD03}, and the ones that directly use coherent states as information carriers, such as PM-QKD \cite{TFQKD02}, simple TF-QKD \cite{simpleTFQKD}, MDI TF-QKD \cite{TFQKD05}, and PM-MDI QKD \cite{TFQKD04}.

\chapter{Supplemental Information for Free-Space BB84}

This Appendix contains supplemental information for Chapter 3.

\section{Decoy-State BB84 Rate Function}

\subsection{Standard Channel Model}

Here we present a brief recapitulation of the decoy-state BB84 model we used. We follow the notations as in Lo, Ma, and Chen's Paper in 2005 \cite{decoystate_LMC}.

Alice uses a WCP source at intensity $\mu$, which sends pulses with a Poissonian photon number distribution: $P_i={\mu^i \over i!}e^{-\mu}$. We will first consider using the standard channel model (as in the original paper Ref.\cite{decoystate_LMC}), where for each i-photon pulse $\ket{i}$, the transmittance, yield $Y_i$, gain $Q_i$, and QBER $e_i$ are:
\begin{equation}
\begin{aligned}
\eta_i&=1-(1-\eta)^i\\
Y_i&\approx Y_0+\eta_i=Y_0+1-(1-\eta)^i\\
Q_i&=Y_i{\mu^i \over i!}e^{-\mu}\\
e_i&={{e_0Y_0+e_d \eta _i}\over Y_i}\\
\end{aligned}
\end{equation}

\noindent where $Y_0$, $e_d$ are the dark count rate and misalignment, respectively, and $e_0={1\over 2}$. The overall Gain $Q_{\mu}$ and QBER ${E_{\mu}}$ for this intensity $\mu$ are:

\begin{equation}
\begin{aligned}
Q_{\mu}&=\sum_{i=0}^{\infty}Y_i{\mu^i \over i!} e^{-\mu}=\sum_{i=0}^{\infty}Y_iP_i \\
E_\mu &= {1 \over Q_\mu} \sum_{i=0}^{\infty}e_i Y_i{\mu^i \over i!} e^{-\mu}={1 \over Q_\mu} \sum_{i=0}^{\infty}e_iY_iP_i
\end{aligned}
\end{equation}

\noindent where $Q_{\mu}$ and $E_{\mu}$ are simulated here for rate estimation using known channel transmittance $\eta$, while in experiment they will be measured observables. 

For this standard channel model, we assume that the photon number distribution after passing through the channel would still be Gaussian. Using decoy-state technique to combine $Q_{\mu}$ and $E_{\mu}$ for different intensities, we can estimate the single-photon contributions $Q_1$ and $e_1$. The achievable secure key rate is at least

\begin{equation}
\begin{aligned}
R_{GLLP} = q\{-f(E_\mu)Q_\mu h_2(E_\mu)+Q_1[1-h_2(e_1)]\}
\end{aligned}
\end{equation}

\noindent as given by the GLLP formula\cite{GLLP}, where $h_2$ is the binary entropy function, $q={1\over 2}$ or $q\approx 1$ depending on whether efficient BB84 is used, and $f$ is the error-correction efficiency.

\subsection{Channel Model after Post-Selection}

However, one thing worth noting is that although photon number distribution is Gaussian after the signals pass through the standard channel model, it is no longer necessarily so if we perform post-selection, in which case the photon number distribution might change, and thus the decoy-state key rate form in Eq. B.3 (which depends on a Gaussian distribution model) might no longer be adequate.

To show that this will not be a concern for us, we will explicitly discuss how the post-selection from P-RTS will affect the yield for each photon number. From Eq. B.1, before post-selection, the yield for pulses with a given photon number $i$ is

\begin{equation}
\begin{aligned}
Y_i(\eta)=Y_0+1-(1-\eta)^i
\end{aligned}
\end{equation}

For simplified model, among the post-selected signals, we have replaced $\eta$ in Eq. B.4 with 
\begin{equation}
\begin{aligned}    
\langle \eta \rangle={{\int_{\eta_T}^{1}\eta p_{\eta_0,\sigma}(\eta)d\eta}\over{\int_{\eta_T}^{1}p_{\eta_0,\sigma}(\eta)d\eta}}
\end{aligned}
\end{equation}

\noindent thus the yield for i-photon pulse is assumed to be:

\begin{equation}
\begin{aligned}
Y_i^{Simplified}(\eta_T)=Y_0+1-(1-\langle \eta \rangle)^i
\end{aligned}
\end{equation}

\noindent in which case, we are simply replacing the $\eta$ with a higher expected value $\langle \eta \rangle$, but the expression is in the same form as Eq. B.4, and the received photon number distribution is still Gaussian. (Hence decoy-state analysis and key rate expression still hold).

However, if we consider the more realistic case, post-selection might have a different effect on pulses with different photon number $i$. Therefore, to estimate the yield for each photon number, and analyze the photon number distribution after the channel and the post-selection, we should group up pulses with the same given photon number, and calculate the expected value of the yield for each given $i$. We can call this the ``pulse-wise integration" model.

\begin{equation}
\begin{aligned}
Y_i^{Pulse-wise}(\eta_T)={{\int_{\eta_T}^{1}Y_i(\eta) p_{\eta_0,\sigma}(\eta)d\eta}\over{\int_{\eta_T}^{1}p_{\eta_0,\sigma}(\eta)d\eta}}=\langle Y_i(\eta) \rangle
\end{aligned}
\end{equation}

\noindent and the Gain $Q_{\mu}$ and QBER ${E_{\mu}}$ would become:

\begin{equation}
\begin{aligned}
Q_{\mu}&=\sum_{i=0}^{\infty}\langle Y_i(\eta) \rangle{\mu^i \over i!} e^{-\mu}=\sum_{i=0}^{\infty}\langle Y_i(\eta) \rangle P_i \\
E_\mu &= {1 \over Q_\mu} \sum_{i=0}^{\infty}e_i \langle Y_i(\eta) \rangle{\mu^i \over i!} e^{-\mu}={1 \over Q_\mu} \sum_{i=0}^{\infty}e_i \langle Y_i(\eta) \rangle P_i
\end{aligned}
\end{equation}

In the case of this ``pulse-wise integration" model, $Q_\mu$ and $E_\mu$ can no longer be considered as from a Gaussian distribution with intensity $\eta\mu$, which is seemingly warning us that the decoy-state analysis might not hold true anymore. However, here we make the observation that for $i=0$, trivially, 
\begin{equation}
\begin{aligned}
Y_0^{Simplified}=Y_0=Y_0^{Pulse-wise}
\end{aligned}
\end{equation}

\noindent and for $i=1$, the yield is a linear function of $\eta$, hence

\begin{equation}
\begin{aligned}
Y_1^{Simplified}=Y_0+\langle \eta \rangle = \langle(Y_0+\eta)\rangle=Y_1^{Pulse-wise}
\end{aligned}
\end{equation}

While for all multi-photon cases where $i\geq 2$, the function 

\begin{equation}
\begin{aligned}
Y_i(\eta)=Y_0+1-(1-\eta)^i
\end{aligned}
\end{equation}

is a strictly concave function on the domain $[0,1]$. Therefore, from Jensen's Inequality, the expected value of a concave function is strictly smaller than the function ($Y_i$) of the expected value, i.e.

\begin{equation}
\begin{aligned}
Y_i^{Pulse-wise}=\langle Y_i(\eta)\rangle < Y_i(\langle \eta \rangle) = Y_i^{Simplified}, i\geq 2
\end{aligned}
\end{equation}

This means that, with the simplified model, with the Gaussian photon number distribution assumption and the standard decoy-state key rate analysis, we are correctly estimating the vacuum and single-photon contributions, but always \textit{over-estimating} the multi-photon contributions. This will in fact result in an \textit{under-estimated} key rate for the simplified model than the realistic case (yield-wise integration model). Therefore, we make the ``validity argument" here that, despite post-selection will result in a non-Gaussian photon number distribution, by using the simplified model and the same decoy-state analysis, we will never incorrectly over-estimate the key rate, and can be confident in the improvement in performance from using P-RTS.

\section{Proof of Rate-Wise Integration Model as Upper Bound}

To better compare the models, let us first simply the notations, and define $\langle f(\eta)\rangle$ operator as taking the expected value of $f(\eta)$ over $p_{\eta_0,\sigma}(\eta)$ (in the case of using post-selection, the distribution is truncated, and will be normalized by dividing by $\int_{\eta_T}^{1}p_{\eta_0,\sigma}(\eta)d\eta$). The expected value $\langle f(\eta) \rangle$ can be expressed as:

\begin{equation}
\langle f(\eta) \rangle={{\int_{\eta_T}^{1}f(\eta) p_{\eta_0,\sigma}(\eta)d\eta}\over{\int_{\eta_T}^{1}p_{\eta_0,\sigma}(\eta)d\eta}}
\end{equation}

Then, we can easily see that, mathematically, the two models we proposed so far, the rate-wise integration model and the simplified model, are only different in that they apply the ``expected value" operator at different levels of the function. We can simply write $R^{\text{Rate-wise}}$ and $R^{\text{Simplified}}$ as:

\begin{equation}
\begin{aligned}
R^{\text{Rate-wise}} (\eta_T)&=\langle R(\eta) \rangle\\
R^{\text{Simplified}} (\eta_T)&=R(\langle \eta \rangle)
\end{aligned}
\end{equation}

Now, we introduce the \textit{Jensen's Inequality}: \\

\textit {For a random variable X following a probability distribution p(X), and for any given convex function f(x), we always have}
\begin{equation}
\langle f(X) \rangle \geq f(\langle X \rangle)
\end{equation}

\noindent the equal sign is taken when the function $f(x)$ is linear.

For decoy-state BB84, $R_{GLLP}(\eta)$ is a convex (and increasing) function of $\eta$, therefore we have $\langle R(\eta) \rangle \geq R(\langle \eta \rangle)$, i.e. 

\begin{equation}
R^{\text{Rate-wise}}(0) \geq R^{\text{Simplified}}(0)
\end{equation}

\noindent This holds true even after a threshold is applied, too, since we can simply replace the distribution $p(\eta)$ with the truncated distribution on domain $[\eta_T,1]$, and normalize it by dividing by the constant $\int_{\eta_T}^{1}p_{\eta_0,\sigma}(\eta)d\eta$. Since $R(\eta)$ is non-concave on all sections of $[0,1]$, the Jensen's Inequality always holds true, regardless of the threshold. i.e.

\begin{equation}
R^{\text{Rate-wise}}(0) \geq R^{\text{Rate-wise}}(\eta_T) \geq R^{\text{Simplified}}(\eta_T)
\end{equation}

\noindent here we also include Eq. 3.3's result that $R^{\text{Rate-wise}}(\eta_T)$ is non-increasing with $\eta_T$.

Therefore, we see that $R^{\text{Rate-wise}}$ serves as an upper bound for the possible rate in a turbulent channel, as it is the maximum achievable rate when we know all transmittance information and make use of the entire PDTC. The simplified model always has no higher rate than this upper bound. This means that, when we use $R^{\text{Simplified}}$ to calculate the rate, we \textit{never overestimate} the performance of the protocol. When we demonstrate the improvements we gain by using P-RTS in decoy-state BB84, the actual possible rate will be even higher, thus the validity argument for the usage of the simplified model in estimating the rate. 

\section{Proof of Optimality of Critical Transmittance as Threshold for Simplified Model}

Following the argument in Section 3.2.3, here we give a rigorous proof that $\eta_T=\eta_{critical}$ is indeed the optimal threshold for the simplified model, given that $R_{S-P}(\eta)$ (and similarly for $R_{GLLP}(\eta)$) is nearly linear. For the simplified model, we showed that

\begin{equation}
R^{\text{Simplified}}(\eta_T)=\int_{\eta_T}^{1}p_{\eta_0,\sigma}(\eta)d\eta \times R_{S-P}(\langle \eta \rangle)
\end{equation}

\noindent where $\langle \eta \rangle$ satisfies:

\begin{equation}
\langle \eta \rangle={{\int_{\eta_T}^{1}\eta p_{\eta_0,\sigma}(\eta)d\eta}\over{\int_{\eta_T}^{1}p_{\eta_0,\sigma}(\eta)d\eta}}
\end{equation}

\noindent Then, using the Leibniz Integration Rule, and taking derivative with respect to $\eta_T$ (here we omit the subscript of $\eta_0,\sigma$ for the PDTC, and $S-P$ (or $GLLP$) for the rate), we have

\begin{equation}
\begin{aligned}
{d \over d\eta_T}{\langle \eta \rangle}={{p(\eta_T)}\over{\int_{\eta_T}^{1}p(\eta)d\eta}}({\langle \eta \rangle} - \eta_T)
\end{aligned}
\end{equation}

\noindent using the chain rule, 

\begin{equation}
\begin{aligned}
{d \over d\eta_T}R(\langle \eta \rangle) &= {dR(\eta) \over {d\langle \eta \rangle}} {{d\langle \eta \rangle}\over d\eta_T}\\
&= R'(\langle \eta \rangle) {{p(\eta_T)}\over{\int_{\eta_T}^{1}p(\eta)d\eta}}({\langle \eta \rangle} - \eta_T)\\
\end{aligned}
\end{equation}

\noindent Maximizing $R^{\text{Simplified}}$ requires that 
\begin{equation}
{d \over d\eta_T}R^{\text{Simplified}}(\eta_T)=0
\end{equation}

\noindent expanding the derivative using Eq. B.18 gives us
\begin{equation}
\begin{aligned}
&{d \over d\eta_T}R^{\text{Simplified}}(\eta_T)\\
&= \left( \int_{\eta_T}^{1}p(\eta)d\eta \right) \times {d \over d\eta_T}R(\langle \eta \rangle) - p(\eta_T)R(\langle \eta \rangle) \\
&= R'(\langle \eta \rangle) {p(\eta_T)}({\langle \eta \rangle} - \eta_T) - p(\eta_T)R(\langle \eta \rangle) \\
&= p(\eta_T)[({\langle \eta \rangle} - \eta_T) R'(\langle \eta \rangle) - R(\langle \eta \rangle)]
\end{aligned}
\end{equation}

\noindent Therefore, the optimal threshold requires that 

\begin{equation}
({\langle \eta \rangle} - \eta_T) R'(\langle \eta \rangle) = R(\langle \eta \rangle)
\end{equation}\\

When $R(\eta)$ is a linear function on the domain $[\eta_{critical}, 1]$ and $R(\eta_{critical})=0$, there is

\begin{equation}
R'(\langle \eta \rangle) = {{R(\langle \eta \rangle) - R(\eta_T)}\over{({\langle \eta \rangle} - \eta_T)}}
\end{equation}

\noindent combined with Eq. B.24, we have

\begin{equation}
R(\eta_T)=0
\end{equation}

\noindent for $\eta \in [\eta_{critical}, 1]$, there is one and only one point satisfying $R(\eta_T)=0$, that is 
\begin{equation}
\eta_T=\eta_{critical}
\end{equation}

\noindent For $\eta \in [0, \eta_{critical})$, on the other hand,

\begin{equation}
R(\langle \eta \rangle) - R(\eta_T) < R'(\langle \eta \rangle) ({\langle \eta \rangle} - \eta_T) = R(\langle \eta \rangle)
\end{equation}

\noindent which becomes

\begin{equation}
R(\eta_T) > 0
\end{equation}

\noindent but $R(\eta)=0$ for all $\eta \leq \eta_{critical}$, so no $\eta_T\in [0, \eta_{critical})$ satisfies the zero derivative requirement. Which means that, when $R_{GLLP}(\eta)$ is near linear on $[\eta_{critical}, 1]$, we have

\begin{equation}
\eta_T=\eta_{critical}
\end{equation}

\noindent as the one and only optimal threshold for $R^{\text{Simplified}}$.\\

Additionally, if we do not ignore the convexity of $R(\eta)$, consider the tangent line for $R(\eta)$ at $\langle \eta \rangle$, since $R(\eta)$ is a convex function of $\eta$,

\begin{equation}
({\langle \eta \rangle} - \eta_T) R'(\langle \eta \rangle) > R(\langle \eta \rangle) - R(\eta_T)
\end{equation}

\noindent optimal threshold requires that 

\begin{equation}
R(\langle \eta \rangle) > R(\langle \eta \rangle) - R(\eta_T)
\end{equation}

\noindent i.e. $R(\eta_T) > 0$, which means that the optimal threshold position will be shifted rightward from $\eta_{critical}$, the actual amount of shift depends on how much $R$ deviates from linearity (in numerical simulations, we see that since $R(\eta)$ is very close to linear, this shift is very small). Also, although $R^{\text{Rate-wise}}$ is not affected for a threshold no larger than $\eta_{critical}$, using a threshold larger than $\eta_{critical}$ will cause $R^{\text{Rate-wise}}$ to decrease, since ``bins" with positive rate are discarded. Therefore, the maximum point for $R^{\text{Simplified}}$ is no longer the maximum $R^{\text{Rate-wise}}$, but slightly smaller than it. This also explains why in the numerical results, the optimal $R^{\text{Simplified}}$ is always slightly lower than upper bound, due to non-linearity of $R(\eta)$.

Also, a small note is that, the Jensen's Inequality asks the function to be differentiable at every point, while the turning point of $R$ at $\eta_{critical}$ is a sharp point. To address this, we can construct another $R_2$ with an infinitesimally small yet smooth ``turn" at $\eta_{critical}$ to replace the sharp point, but as the ``turn" is infinitely small, integrating $R$ and $R_2$ over any region will yield infinitely close results. Therefore the turning point's structure does not affect the above results.

\section{Analytical Expression for Optimal Threshold}

\subsection{Single-Photon Case}

Let $\eta_{sys}=\eta\times \eta_d$. The single photon Shor-Preskill rate is

\begin{equation}
R_{S-P}=(Y_0+\eta_{sys})\{1-2h_2[e(\eta_{sys})]\}
\end{equation}

where the single-photon QBER is

\begin{equation}
e(\eta_{sys})={{{1\over 2}Y_0+e_d \eta_{sys}}\over{Y_0+\eta_{sys}}}
\end{equation}

For the rate to be zero, we require:

\begin{equation}
R_{S-P}=0
\end{equation}

hence

\begin{equation}
1-2h_2[e(\eta_{sys})]=0
\end{equation}

or, $h_2[e(\eta_{sys})]={1\over 2}$. This numerically corresponds to $e(\eta_{sys})=11\%=e_{critical}$ (which is the QBER threshold for Shor-Preskill rate). Therefore, substituting into Eq. B.34, we have

\begin{equation}
\eta_{sys}={{{1\over 2}-e_{critical}}\over{e_{critical}-e_d}}Y_0
\end{equation}

expressing it in channel transmittance $\eta$

\begin{equation}
\eta_{critical}={Y_0 \over \eta_d}{{{1\over 2}-e_{critical}}\over{e_{critical}-e_d}}
\end{equation}

\noindent or, if we substitute $\eta_{critical}=11\%$ into the equation, we have

\begin{equation}
\eta_{critical}={Y_0 \over \eta_d}{0.39\over{0.11-e_d}}
\end{equation}

This is the analytical expression for the critical transmittance for the single-photon case. Also, we can see that the critical transmittance is proportional to the background count (i.e. noise) in the system. i.e.

\begin{equation}
\eta_{critical} \propto{Y_0 \over \eta_d}
\end{equation}

\subsection{Decoy-State BB84}

Consider the asymptotic case of decoy-state BB84, with infinite number of decoys (i.e. the only significant intensity is the signal intensity $\mu$). Using the GLLP rate,

\begin{equation}
\begin{aligned}
R_{GLLP} = q\{-fQ_\mu h_2(E_\mu)+Q_1[1-h_2(e_1)]\}
\end{aligned}
\end{equation}

\noindent we would like to find $\eta_{critical}$ such that 
\begin{equation}
R(\eta_{critical})=0
\end{equation}

\noindent hence
\begin{equation}
fQ_\mu h_2(E_\mu)=Q_1[1-h_2(e_1)]
\end{equation}

\noindent or

\begin{equation}
h_2(e_1)+f{Q_\mu \over Q_{1}} h_2(E_\mu)=1
\end{equation}

Let $\eta_{sys}=\eta\times \eta_d$, the observables and single-photon contributions can be written as:

\begin{equation}
\begin{aligned}
Q_\mu &= Y_0+1-exp(-\mu\eta_{sys})\\
E_\mu &= {{{1\over 2}Y_0+e_d(1-exp(-\mu\eta_{sys}))}\over{Y_0+1-exp(-\mu\eta_{sys})}}\\
Q_1 &=\mu exp(-\mu)(Y_0+\eta_{sys})\\
e_1 &={{{1\over 2}Y_0+e_d\eta_{sys}}\over{Y_0+\eta_{sys}}}
\end{aligned}
\end{equation}

Now, if $\eta \ll 1$, we can use the approximation $1-exp(-\mu\eta_{sys})=\mu\eta_{sys}$. If the dark/background count rate $Y_0$ also satisfies $Y_0 \ll \eta_{sys}$ (which is a reasonable approximation, since with parameters in Table 3.2, $Y_0$ is at the order of $10^{-5}$, while $\eta_d \eta_{critical}$ is at the order of $10^{-3}$), we can write

\begin{equation}
\begin{aligned}
{Q_\mu \over Q_1 }&\approx {{Y_0+\mu\eta_{sys}}\over{\mu exp(-\mu)(Y_0+\eta_{sys})}} \approx exp(\mu)\\
e_1 &\approx {1\over 2}{Y_0 \over \eta_{sys}}+e_d \\
E_\mu &\approx {1\over {2\mu}}{Y_0 \over \eta_{sys}}+e_d
\end{aligned}
\end{equation}

\noindent substituting back into Eq. B.44, and defining

\begin{equation}
x={Y_0 \over \eta_{sys}}={Y_0\over{\eta_d\eta}}
\end{equation}

\noindent we can have

\begin{equation}
h_2({1\over 2}x+e_d)+fe^\mu h_2({1\over {2\mu}}x+e_d)=1
\end{equation}

\noindent which is a function that is only determined by $e_d$ and $\mu$. We can write its solution for x as

\begin{equation}
x_{critical}=\mathcal{F}(e_d,\mu)
\end{equation}

Then the critical transmittance (i.e. optimal threshold position) can be written as

\begin{equation}
\eta_{critical}={{Y_0}\over{\eta_d}} [{1\over\mathcal{F}(e_d,\mu)}]
\end{equation}

\noindent where $\mathcal{F}(e_d,\mu)$ does not have an explicit analytical expression, because $h_2$ function cannot be analytically expanded. (One can, however, numerically use linear fit to expand $h_2$, if given the approximate range of the experimental parameters $e_d$ and $\mu$). The important observation here, however, is that for the decoy-state case, we can still have:

\begin{equation}
\eta_{critical} \propto{Y_0 \over \eta_d}
\end{equation}

\noindent which points out that the critical threshold is directly proportional to the dark (or background) count rate of the experimental devices, and inversely proportional to the detector efficiency.

\section{PDTC parameters}

In our simulations, we have fixed several typical values for $\sigma$ for free-space QKD, corresponding to the case of weak-to-medium level turbulence, and have considered the PDTC to be a fixed distribution for a given $\sigma$ regardless of the channel loss. In reality, though, $\sigma$ is distance-dependent, too. A commonly used estimation for $\sigma$ is the ``Rytov Approximation"\cite{rytov}

\begin{equation}
\sigma^2 = 1.23 C_n^2 k^{7/6} L^{11/6}
\end{equation}

\noindent which relates $\sigma$ both to the distance $L$ and the refractive index structure constant $C_n^2$ (which is determined by atmospheric conditions). 

Also, with simulation software such as MODTRAN\cite{modtran}, it is possible to simulate the relationship between $\eta_0$ and $L$ for a given free-space channel. Therefore, one necessary next step would also be to estimate performance for cases with realistic values for $\eta_0$ and $\sigma$, both from literature and from simulations, as well as to study the possible correlation $\sigma$ and $\eta_0$ (both related to $L$) in simulations.

\section{Note about Correlations between Quantum and Classical Signals}

As we have described in Chapter 2, the real-time selection method we discuss does not depend on the specific means with which we obtain the channel transmittance. For some specific setups, such as one based on a classical signal (e.g. the setup in Ref. \cite{probetest}), the effectiveness of the real-time selection depends on a good correlation of transmittances for the classical signal and the quantum signal. In Ref. \cite{probetest} Fig. 2, the authors of that paper has provided an illustrative comparison between the classical signal strength and the quantum count rate, and concluded that the correlation was good (despite that a quantitative analysis on the temporal correlation of data was not provided - except a brief discussion that the fluctuation $\sigma$ was of different values for the classical and quantum signals).

If the correlation between classical and quantum signals are not perfect, we would like to point out that this can be considered as an ``imperfect threshold". Suppose we calculate an optimal pre-fixed threshold of $\eta_T=\eta_{critical}=0.0012$ based on our device parameters. In principle, we are supposed to select all signals with $\eta>\eta_{T}=0.0012$. However, what we're actually sampling and selecting upon is the classical detector voltage $V>V_T$. For a given $V_T$, the actual threshold $\eta_T$ we apply on the quantum transmittance might be e.g. a value slightly different from $\eta_{critical}=0.0012$ (due to a bias in the classical reading), or a value fluctuating around $\eta_{critical}=0.0012$ (due to fluctuation in the classical reading, or the correlation changing in real time). This means that, imperfect correlation can be thought of as an imperfect threshold $\eta_T$ which could be shifted from the optimal position, or even could be following a probability distribution. Such an imperfect threshold will decrease the performance, but from Fig. \ref{fig:threshold} and Appendix B.3, we can see that $\eta_{critical}$ is the position where ${d \over d\eta_T}R^{\text{Simplified}}(\eta_T)=0$, this means that a slight perturbation on $\eta_T$ near this optimal point will have little effect on the key rate, i.e. our method is robust against small inaccuracies on the threshold.

We can perform a quick estimation based on Rayleigh scattering (which depends quadratically on the inverse of wavelength, i.e. $I\propto {1\over \lambda^4}$). Given e.g. classical channel of 808nm and quantum channel of 850nm, there will be a $22.5\%$ difference in the scattering intensity, i.e. quantum channel has about 0.9dB smaller loss than the classical channel. If the entire channel loss comes from Rayleigh scattering (while in practice it actually also depends on absorption, Mie scattering, and the loss due to scope geometry and turbulence), this means that e.g. setting $\eta_{classical}=0.0012$ would result in an actual threshold of $\eta_{quantum}=0.00147$. Using parameters in Table 3.2 and $\mu=0.3,\nu=0.05, \sigma=0.9$, at 35dB using $\eta_T=0.00147$ instead of $0.0012$ results in only a $26.5\%$ decrease in key rate ($6.2\times 10^{-7}$ versus $4.9\times 10^{-7}$, while at this location $R=0$ without post-selection), and only $0.6dB$ smaller maximum tolerable loss (44.4dB versus 43.8dB, while the case without post-selection only reaches 31dB). This means that indeed an inaccurate threshold has a limited impact on the effectiveness of our method. For a more comprehensive atmospheric transmittance model, we also tested simulating a channel of 30km using MODTRAN (relative humidity $RH=95\%$, ``Ocean View" profile), from which we obtain $\eta=0.0201$ at 808nm, and $\eta=0.0253$ at 850nm, which means about $26\%$ higher transmittance for the quantum signal. Again, this level of inaccuracy for the threshold would not affect the key rate too much, which we have observed is very robust against perturbations on the threshold near the optimal position. \footnote{The level of wavelength dependency of transmittance is affected by e.g. humidity, which determines the amount of water absorption - which makes the dependency deviate from $I\propto {1\over \lambda^4}$ from only Rayleigh scattering. In some more extreme cases, such as 143km with relative humidity $RH=10\%$, the wavelength dependency is stronger, and the transmittances for classical/quantum channels are $0.0008$ and $0.0016$, meaning that the quantum transmittance is twice as high. Even so, for the same BB84 test case in the previous paragraph, using \textit{double} the value of optimal threshold $\eta_T=0.0024$ only decreases maximum loss to 42.4dB instead of 44.4dB (from using optimal threshold), which is still much better than the 31dB maximum loss of not using post-selection; and e.g. at the same 35dB position, even when using double the value of optimal threshold, key rate is $2.1 \times 10^{-7}$, about $1/3$ that of $R=6.2\times 10^{-7}$ with optimal threshold, but it is still rather effective, as using no post-selection cannot generate key at this position.}

\begin{figure}[h]
    \includegraphics[scale=0.55]{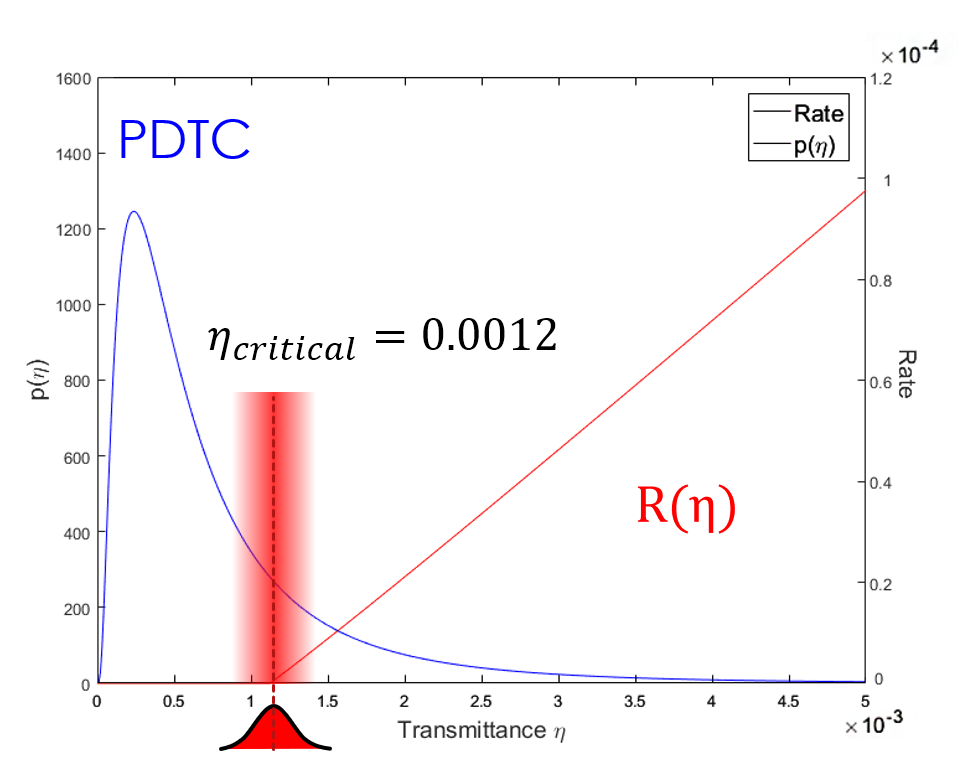}
    \caption{Illustration of a case where the correlation between classical and quantum signal is time-dependent and the deviation is a random variable. In this case we can consider the threshold as a ``fuzzy threshold" that follows a probability distribution. In this case the proportion of post-selected signals and the expected value of transmittances among these signals need to be integrated over this probability distribution of the threshold value. Nonetheless, if the amount fluctuation is small, such imperfections will likely have little impact on the key rate, which is robust against small perturbations of threshold from the optimal value.}
    \label{fig:fuzzy_threshold}
\end{figure}

If the correlation is time-dependent and the deviations are randomized, we can think of $\eta_T$ as a ``fuzzy" threshold which follows a probability distribution of its own, see Fig. \ref{fig:fuzzy_threshold}. For instance a simple $p_{\eta_{T0},\sigma_T}(\eta_T)$, which can be e.g. a Gaussian distribution determined by expected value of threshold $\eta_{T0}$, and standard deviation $\sigma_T$ of the distribution. If we write the cumulative PDTC as $CDF(\eta_T)=\int_{\eta_T}^{1}p_{\eta_0,\sigma}(\eta)d\eta$, and expected value after post-selection $E(\eta_T)=\int_{\eta_T}^{1} \eta p_{\eta_0,\sigma}(\eta)d\eta$, we can obtain portion of selected signals and expected value of transmittance among these signals:

\begin{equation}
\begin{aligned}
    P &= \int_{0}^{1} p_{\eta_{T0},\sigma_T}(\eta_T) CDF(\eta_T) d\eta_T\\
    \langle \eta \rangle &= {1 \over P} \int_{0}^{1} p_{\eta_{T0},\sigma_T}(\eta_T) E(\eta_T) d\eta_T\\
\end{aligned}
\end{equation}

\noindent Of course, the actual form of $p_{\eta_{T0},\sigma_T}(\eta_T)$ needs experimental data to validate, and the optimality of the threshold might need more careful discussions. (While Ref. \cite{probetest} mentioned that the probability distributions for quantum and classical signals have different $\sigma$ - i.e. they do have somewhat different levels of fluctuations - their temporal correlations are not quantified.) Nonetheless, the key point here is that having inaccurate or ``fuzzy" threshold values will likely have little impact on the key rate (due to the robustness of the key rate against small perturbations near the optimal point), and it will not affect the security either (as the channel transmittance is public information - Even can even directly manipulate the classical signal and control Bob's post-selection, but this is at most a denial-of-service attack that reduces the key rate but not the security of the protocol).

\begin{figure}[h]
    \includegraphics[scale=0.55]{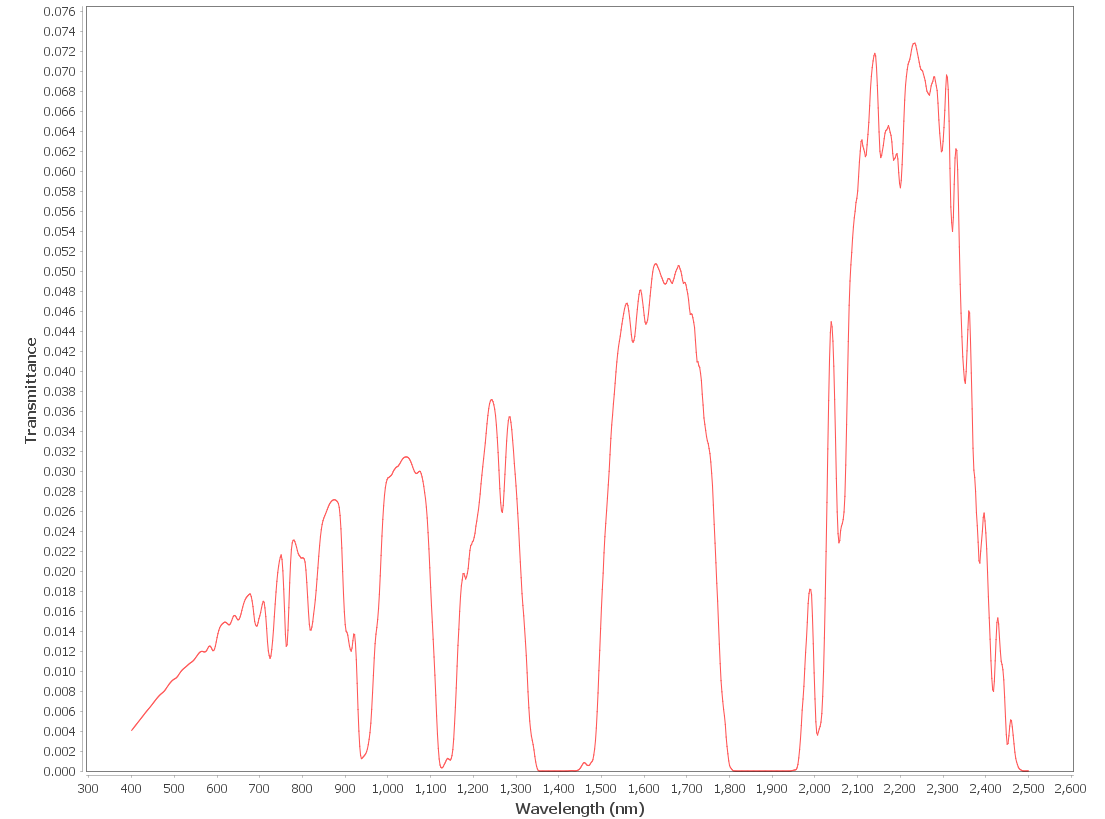}
    \caption{An example transmittance versus wavelength plot for a 30km horizontal free-space channel, simulated with MODTRAN.}
    \label{fig:BB84_MODTRAN}
\end{figure}

\chapter{Supplemental Information on Asymmetric MDI-QKD}

This Appendix contains supplemental information for Chapter 5.

\section{Note about Adding Fibre}

\begin{figure}[h]
	\includegraphics[scale=0.35]{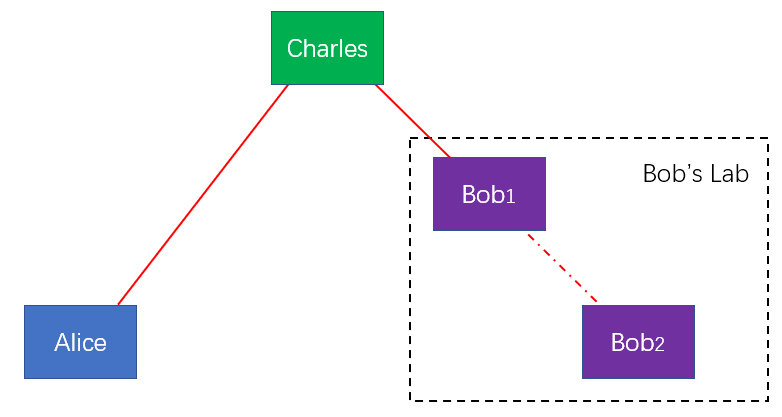}
	\caption{Setup for asymmetric MDI-QKD. When channels are highly asymmetric (e.g. Alice and Bob1), to increase the symmetry in the channel, sometimes one adds additional loss to the system in Bob's lab\cite{mdiPOP}, in exchange for better symmetry. When estimating the key rate, Bob assumes that both Charles-Bob1 and Bob1-Bob2 channels are controlled by Eve. This is therefore a pessimistic estimation of the key rate, and is not necessarily the optimal strategy. Reproduced from \cite{this_asymMDI} @2019 APS.}
	\label{fig:setup}
\end{figure}

In this section we provide an intuitive description of why adding additional loss is suboptimal, and how our method works better with asymmetric channels.

Previously, when Alice and Bob have asymmetric channels, a common solution is to add fibre (thus adding loss) to the shorter channel in exchange for better symmetry, such as in Ref. \cite{mdiPOP}. Afterwards one selects symmetric intensities for Alice and Bob and acquires a higher rate. However, the added fibre lies in Bob's lab, and is in fact securely under the control of Bob. But by assuming a symmetric setup, we are effectively relinquishing its control to Eve, and pessimistically estimating the key rate. Therefore, intuitively, this is not necessarily the optimal strategy. We will show with our new protocol that, when the channels are asymmetric, Alice and Bob can independently choose their optimal intensities, and that optimizing intensities and probabilities alone is sufficient to compensate for the different channel losses.

\section{Scaling of Key Rate with Transmittance}

{\color{black}In this section we discuss the scaling properties of key rate versus transmittance, for prior protocols with the same parameters for Alice and Bob, and our new protocol that uses different intensities for Alice and Bob. We will show in Appendices B and C that the scaling of the key rate versus distance is mainly determined by the signal states (so long as we have good single photon estimation from decoy states). This also means that, the advantage of our method is really not dependent on the number of decoy states used or the finite-size analysis model used (or lack thereof, in the asymptotic case), and our results are in principle applicable to any protocol that decouples the signal and decoy states in the Z and X bases and allows different intensities for Alice and Bob.}

The transmittance of the two channels are $(\eta_{\text{A}}, \eta_{\text{B}})$, and the asymmetry (mismatch) $x$ is defined as 

\begin{equation}
x={{\eta_{\text{A}}}\over{\eta_{\text{B}}}}
\end{equation}

\subsection{Single-Photon Source}

Now, let us consider a single-photon case first. That is, suppose Alice and Bob both send perfect single photons only, and the key is generated from two-photon interference. If we ignore the dark counts, the asymptotic key rate can be written as \cite{Preskill}:

\begin{equation}
R_{SP}=\eta_{\text{A}}\times \eta_{\text{B}} \times [1-2h_2(e_{11})]
\end{equation}

\noindent where $h_2$ is the binary entropy function and $e_{11}$ is the QBER (which is a quantity that, when dark count rate is ignored, is independent of the transmittance). This means that in the perfect single-photon case, the key rate is proportional to $\eta_{\text{A}} \eta_{\text{B}}$, and the mismatch $x$ does not explicitly appear in its expression:

\begin{equation}
R_{SP} \propto \eta_{\text{A}}\eta_{\text{B}}
\end{equation}

In fact, for a given total distance $L_{\text{A}}+L_{\text{B}}=L$, any positioning of the untrusted relay Charles (e.g. at the midpoint, in Alice's lab, or in Bob's lab) would not affect the key rate, since $\eta_{\text{A}} \eta_{\text{B}}$ only depends on $L$.

\subsection{Weak Coherent Pulse Source}

The previous discussion for single-photon MDI-QKD suggests that, by nature, there is not really any limitation on symmetry for MDI-QKD, at least for the ideal single photon case. Then, where does this dependence of key rate on channel symmetry which we observed come from? In this section, we will show that the scaling of the key rate depends on the signal states' trade-off between error-correction and probabilities of sending single-photons, when using WCP sources, rather than privacy amplification (which depends on the estimation of single-photon contributions).

More concretely, (as we will prove in the next section) for protocols with symmetric intensities, there are two sharp cut-off values for the mismatch, $x^{max}$ and $x^{min}$, that prevent the protocol from acquiring any key rate when $x>x^{max}$ or $x<x^{min}$ (and optimizing identical intensities $s_{\text{A}}=s_{\text{B}}$ cannot circumvent this problem). This is why protocols such as 4-intensity protocol are limited to near-symmetric positions. 

On the other hand, when a protocol allows independent intensities for Alice and Bob (such as our new 7-intensity protocol described in the main text), we show that the mismatch can always be compensated by optimizing intensities $s_{\text{A}}$ and $s_{\text{B}}$ (hence lifting the limitations $x^{max}$ and $x^{min}$). In fact, we show that for positions with high asymmetry, key rate no longer depends on mismatch $x={{\eta_{\text{A}}}\over{\eta_{\text{B}}}}$ at all, and the optimal key rate only scales with the \textit{smaller} of the two channel transmittances. That is, 

\begin{equation}
R_{optimal} \propto min(\eta_{\text{A}}^2, \eta_{\text{B}}^2)
\end{equation}

\noindent which means that, the biggest advantage of protocols with independent intensities for Alice and Bob (e.g. 7-intensity protocol) is to completely lift the limitation on channel asymmetry. When compared with adding fibre to maintain asymmetry, we see that its scaling property is still the same, i.e. quadratically related to the (smaller of) channel transmittances, although our method will always perform better (by a constant coefficient) than adding fibre. Moreover, it provides the convenience of not needing additional fibre, which may not be feasible in free-space channels, or when channel mismatch is changing.

Proofs for the above scaling properties can be found in the next section.

\section{Proof of Scaling Properties of Key Rate with Transmittance}

In this section we outline the analytical proofs for the observations on the scaling properties of asymptotic MDI-QKD key rate versus transmittance in the presence of asymmetry, described in Appendix C.2. We also discuss how the finite-decoy and finite-size effects can be considered as imperfections in the infinite-decoy, infinite-data case, and that the scaling properties are still approximately the same - which are only determined by the signal states' trade-off between error correction and probabilities of sending single photons, and not affected by decoy states.

To simplify the discussion, it is convenient to first use a few crucial approximations as described in Ref.\cite{mdipractical}:\\ 

1. We consider the asymptotic case with infinite data size.

2. We assume an infinite number of decoy states, i.e. Alice and Bob can perfectly estimate the single photon gain $Y_{11}$ and QBER $e_{11}$. In this case, Alice and Bob only need to choose appropriate signal intensities $s_{\text{A}}$, $s_{\text{B}}$.

3. We ignore the dark count rate $Y_0$, when studying the scaling properties with distance (as background noise only affects the maximum transmission distance where transmittance is at the same order as the dark count rate, but does not affect the overall scaling properties of key rate versus distance).

4. When describing the channel model to estimate the observable gain and QBER $Q_{ss}^Z$ and $E_{ss}^Z$ (which affect the error-correction), we make second-order approximations to two functions:

\begin{equation}
\begin{aligned}
I_0(x)&\approx 1 + {x^2\over 4} + O(x^4) \\
e^x &\approx 1+x+{x^2\over 2} + O(x^3)
\end{aligned}
\end{equation}

\noindent where $I_0$ is the modified bessel function of the first kind. This approximation is relatively accurate when $s_{\text{A}}\eta_{\text{A}}\eta_d$ and $s_{\text{B}}\eta_{\text{B}}\eta_d$ are both small, where $\eta_d$ is the detector efficiency.\\

With the above approximations, one can write the key rate conveniently as (excerpting Eq. C.1 and C.2 from Ref.\cite{mdipractical}):

\begin{equation}
R = {{\eta_{\text{B}}^2 \eta_d^2}\over 2} G(x,s_{\text{A}},s_{\text{B}})
\end{equation}

\noindent where $G(x,s_{\text{A}},s_{\text{B}})$ is a function determined by $(s_{\text{A}}, s_{\text{B}})$ and the asymmetry $x$ only:

\begin{equation}
\begin{aligned}
&G(x,s_{\text{A}},s_{\text{B}}) = x s_{\text{A}} s_{\text{B}} e^{-(s_{\text{A}}+s_{\text{B}})}[1-h_2(e_d-{e_d^2\over 2})] \\
&- {{2xs_{\text{A}}s_{\text{B}}+(s_{\text{B}}^2+x^2s_{\text{A}}^2)(2e_d-e_d^2)} \over 2} \times f_eh_2(E_{ss}^Z(x,s_{\text{A}},s_{\text{B}})) \\
&E_{ss}^Z(x,s_{\text{A}},s_{\text{B}}) = {{(s_{\text{B}}+xs_{\text{A}})^2(2e_d-e_d^2)}\over{2[2xs_{\text{A}}s_{\text{B}}+(s_{\text{B}}^2+x^2s_{\text{A}}^2)(2e_d-e_d^2)]}}
\end{aligned}
\end{equation}

\noindent where $h_2$ is the binary entropy function.

Now, having described the key rate function, we are interested in how it scales with the transmittances $\eta_{\text{A}}$, $\eta_{\text{B}}$, using different optimization strategies for the intensities. We will discuss two cases:\\

1. $R_{symmetric}$, where Alice and Bob use the same intensity $s=s_{\text{A}}=s_{\text{B}}$, and optimize $s$.

2. $R_{optimal}$, where Alice and Bob fully optimize a pair of intensities $s_{\text{A}}, s_{\text{B}}$, which can take different values.\\

\subsection{Symmetrically Optimized Intensities}

Let us consider the case where Alice and Bob use the same intensity $s=s_{\text{A}}=s_{\text{B}}$, and optimize $s$. This is the case discussed by previous protocols (such as the 4-intensity protocol, although here to simplify the proof we focus on the infinite-decoy case and only consider signal intensities).

In this case, the function $G$ is optimized over $s$ (and is a function of $x$ only). The rate satisfies

\begin{equation}
R_{symmetric} = \max\limits_{s}R \propto \eta_{\text{B}}^2 \max\limits_{s}G(x,s,s)
\end{equation}

\noindent therefore, $R_{symmetric}$ is proportional to $\eta_{\text{B}}^2$ when channel mismatch $\eta_{\text{A}} \over \eta_{\text{B}}$ is fixed. 

Moreover, since $R_{symmetric}$ is also proportional to $G(x)$, we will have $R_{symmetric}=0$ if $G(x)=0$. Note that, we can rewrite the signal state QBER $E_{ss}^Z$ as:

\begin{equation}
\begin{aligned}
&E_{ss}^Z(x) = {{(1+x)^2(2e_d-e_d^2)}\over{2[2x+(1+x^2)(2e_d-e_d^2)]}}
\end{aligned}
\end{equation}

\noindent since the equal intensities are canceled out, i.e. $E_{ss}^Z$ is only a function of x. In fact, $E_{ss}^Z$ is a function that minimizes at $x=1$ and reaches $50\%$ (where $R_{symmetric}$ is naturally zero) when $x \rightarrow 0$ or $x\rightarrow \infty$. Therefore, if $G(x)\neq 0$ at $x=1$, there must exist some critical values of $x^{max}$ and $x^{min}$ which result in a sufficiently large QBER such that $G(x)=0$ (and $R_{symmetric}=0$). 

This means that, $R_{symmetric}$ is quadratically related to $\eta_{\text{B}}$ (or $\eta_{\text{A}}$) when mismatch $\eta_{\text{A}} \over \eta_{\text{B}}$ is fixed, but also has two cut-off positions for critical levels of mismatch, beyond which no key can be generated. These two critical mismatch positions are what limit previous MDI-QKD protocols to near-symmetric positions. Also, as we have previously mentioned, we see that this critical dependence on mismatch actually comes from the error-correction part (which involves $E_{ss}^Z$).

\subsection{Fully Optimized Intensities}

\begin{figure}[h]
	\includegraphics[scale=0.4]{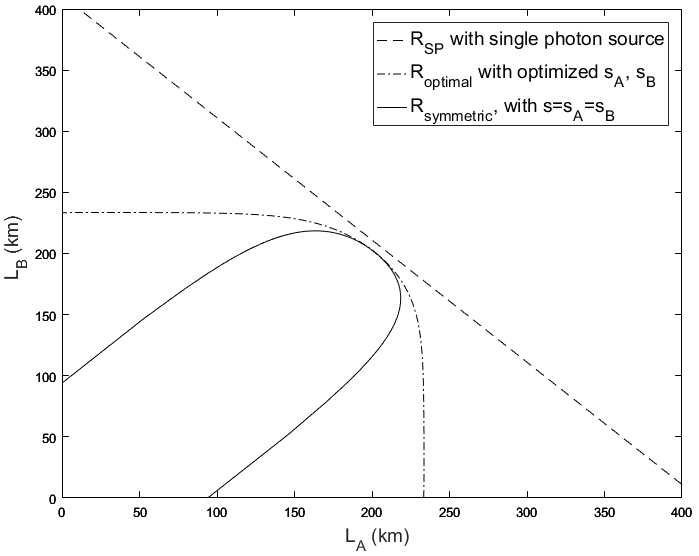}
	\caption{Rate vs distance contours for single photon MDI-QKD $R_{SP}$, decoy-state MDI-QKD with symmetric intensities $R_{symmetric}$, and with fully optimized intensities $R_{optimal}$. We plot the contour line of $R=10^{-9.5}$. Here for a better comparison with WCP sources, we arbitrarily set a probability $P_{11}=s_{\text{A}}s_{\text{B}}\times e^{-(s_{\text{A}}+s_{\text{B}})}$ (where $s_{\text{A}}=s_{\text{B}}=0.6533$) of single photon pairs being sent when calculating $R_{SP}$. For the decoy-state case, as described in Appendix C.3, we assume infinite decoys, infinite data size, ignore dark count rate, and take second-order approximation when calculating gain and QBER (so that we only focus on the ideal scaling properties of key rate with distance). As can be seen, $R_{SP}$ is not limited by asymmetry, and takes constant value for any fixed $L_{\text{A}}+L_{\text{B}}$ (meaning that the dependence of key rate on asymmetry does not come from single photon contributions in the privacy amplification part when using WCP sources). For decoy-state MDI-QKD, we can clearly see $R_{symmetric}$ being limited by the two cut-off lines where $|L_{\text{A}}-L_{\text{B}}|$ takes maximum value (which corresponds to critical values of channel mismatch $x^{max}$ and $x^{min}$). On the other hand, $R_{optimal}$ is not limited by asymmetry, and has contours nearly perpendicular to the axes when asymmetry is high (meaning that, when one channel is significantly longer than the other, $R_{optimal}$ is only dependent on the longer channel). Reproduced from \cite{this_asymMDI} @2019 APS.}
	\label{fig:contours}
\end{figure}

Now, let us consider the case where Alice and Bob are allowed to fully optimize their intensities $s_{\text{A}},s_{\text{B}}$ (such as in the 7-intensity protocol, although again, here we only focus on the signal states in the infinite-decoy case).

In this case, the function $G$ is optimized over $s_{\text{A}}, s_{\text{B}}$. The rate satisfies

\begin{equation}
R_{optimal} = \max\limits_{s_{\text{A}},s_{\text{B}}}R \propto \eta_{\text{B}}^2 \max\limits_{s_{\text{A}},s_{\text{B}}}G(x,s_{\text{A}},s_{\text{B}})
\end{equation}

Now, let us focus on the properties of $G(x,s_{\text{A}},s_{\text{B}})$. Looking at its expression Eq. C.7 in the previous section, we make the important observation that, except for the term $e^{-(s_{\text{A}}+s_{\text{B}})}$ in the single photon probabilities, every other term is only a function of $s_{\text{B}}$ and $xs_{\text{A}}$ (rather than $x$ and $s_{\text{A}}$ separately). We can re-write $G(x,s_{\text{A}},s_{\text{B}})$ as 

\begin{equation}
\begin{aligned}
&G'(x,s_{\text{A}}',s_{\text{B}}) = s_{\text{A}}' s_{\text{B}} e^{-s_{\text{A}}'\over x}e^{-s_{\text{B}}}[1-h_2(e_d-{e_d^2\over 2})] \\
&- {{2s_{\text{A}}'s_{\text{B}}+(s_{\text{B}}^2+s_{\text{A}}'^2)(2e_d-e_d^2)} \over 2} \times f_eh_2(E_{ss}^Z(s_{\text{A}}',s_{\text{B}})) \\
&E_{ss}^Z(s_{\text{A}}',s_{\text{B}}) = {{(s_{\text{B}}+s_{\text{A}}')^2(2e_d-e_d^2)}\over{2[2s_{\text{A}}'s_{\text{B}}+(s_{\text{B}}^2+s_{\text{A}}'^2)(2e_d-e_d^2)]}}
\end{aligned}
\end{equation}

\noindent where we define \textit{equivalent intensity} $s_{\text{A}}'$ as 

\begin{equation}
s_{\text{A}}'={s_{\text{A}} \times x}
\end{equation}

Moreover, if $\eta_{\text{A}} \gg \eta_{\text{B}}$ (i.e. mismatch $x \gg 1$), we can approximately assume that 

\begin{equation}
e^{-s_{\text{A}}'\over x} \approx 1
\end{equation}

\noindent which means that we can rewrite $\max\limits_{s_{\text{A}},s_{\text{B}}}G(x,s_{\text{A}},s_{\text{B}})$ as

\begin{equation}
\begin{aligned}
G^{max}=\max\limits_{s_{\text{A}}',s_{\text{B}}} &G'(s_{\text{A}}',s_{\text{B}})\\
\end{aligned}
\end{equation}

\noindent which, importantly, is a constant value \textit{not} dependent on the value of $x$, when $x \gg 1$. The actual value of $s_{\text{A}}$ equals 

\begin{equation}
s_{\text{A}}={s_{\text{A}}' \over x}
\end{equation}

\noindent Physically, this means that, when there is asymmetry between Alice and Bob's channels, we can compensate for this asymmetry by adjusting the intensities, to keep the same "equivalent intensity" received by Charles and keep $E_{ss}^Z$ at a low value. In this case, $E_{ss}^Z$ is no longer limited by the mismatch $x$, and we can perform MDI-QKD at arbitrary values of asymmetry. 

Also, the key rate is now given by:

\begin{equation}
R_{optimal} \propto \eta_{\text{B}}^2 G^{max}
\end{equation}

\noindent This means that, when $\eta_{\text{A}} \gg \eta_{\text{B}}$ (e.g. the ``single-arm" case previously mentioned where $L_{\text{A}}$ is much shorter than $L_{\text{B}}$), the key rate of asymmetric MDI-QKD is only related to $\eta_{\text{B}}$ and still quadratically scales with $\eta_{\text{B}}$. When $\eta_{\text{B}} \gg \eta_{\text{A}}$, though, we can rewrite $x'={\eta_{\text{B}}\over \eta_{\text{A}}}$, and rewrite

\begin{equation}
R_{optimal} \propto \eta_{\text{A}}^2 \max\limits_{s_{\text{B}}',s_{\text{A}}} G'(s_{\text{B}}',s_{\text{A}})
\end{equation}

\noindent Therefore, overall, 

\begin{equation}
R_{optimal} \propto min(\eta_{\text{A}}^2, \eta_{\text{B}}^2)
\end{equation}

Now, we plot the two cases (symmetric intensities and fully optimized intensities) in a contour plot. As we can observe in Fig.\ref{fig:contours}, the key rate $R_{symmetric}$ has two cut-off mismatch positions beyond which the key rate is zero. This limitation is removed when full optimization of intensities is implemented. Moreover, for $R_{optimal}$, we see that the contours are perpendicular to the axes in high asymmetry regions, which means that the key rate only scales with the longer of the two channels.\\

{\color{black}
	Also, note that, from Eqs. (C4), (C6), we can also make the observation that there is never any need to add fibre to the shorter channel when fully optimizing the intensities, and our new method always provides higher key rate than prior art technique of adding fibre till channels are symmetric, while using same intensities for Alice and Bob. 
	
	To show this, consider the system having a fixed longer channel $L_{\text{B}}$ (i.e. suppose $\eta_{\text{B}}$ is fixed and $\eta_{\text{A}} > \eta_{\text{B}}$, $x ={\eta_{\text{A}} \over \eta_{\text{B}}} > 1$). Adding loss to $\eta_{\text{A}}$ is equivalent to decreasing $x$.
	
	With symmetric intensities (and adding loss till $\eta_{\text{A}}=\eta_{\text{B}}$), the key rate can be written as:
	
	\begin{equation}
	R_{symmetric} = {{\eta_d^2  \eta_{\text{B}}^2 } \over 2} \max\limits_{s}G(1,s,s)
	\end{equation}
	
	Suppose we fully optimize the intensities for this case with added fibre, we will obtain the same key rate (since for $x=1$, i.e. symmetric setup, the optimal choice of intensities satisfies $s_{\text{A}}=s_{\text{B}}$):
	
	\begin{equation}
	\max\limits_{s}G(1,s,s) = \max\limits_{s_{\text{A}},s_{\text{B}}}G(1,s_{\text{A}},s_{\text{B}})
	\end{equation}
	
	However, let us compare it with the case of using fully optimized intensities and no additional loss:
	
	\begin{equation}
	R_{optimal} = {{\eta_d^2  \eta_{\text{B}}^2 } \over 2} \max\limits_{s_{\text{A}},s_{\text{B}}}G(x,s_{\text{A}},s_{\text{B}})
	\end{equation}
	
	As described in Eq. C.11, we can re-write $G(x,s_{\text{A}},s_{\text{B}})$ as $G'(x,s_{\text{A}}',s_{\text{B}})$ (recall that the equivalent intensity $s_{\text{A}}'$ is defined as $xs_{\text{A}}$). We make the observation that $G'(x,s_{\text{A}}',s_{\text{B}})$ strictly increases with $x$. That is, for any two given values of $s_{\text{A}}',s_{\text{B}}$ and $x>1$,
	
	\begin{equation}
	G'(x,s_{\text{A}}',s_{\text{B}}) > G'(1,s_{\text{A}}',s_{\text{B}})
	\end{equation}
	
	\noindent hence after optimization we also have 
	
	\begin{equation}
	\max\limits_{s_{\text{A}}',s_{\text{B}}}G'(x,s_{\text{A}}',s_{\text{B}}) > \max\limits_{s_{\text{A}}',s_{\text{B}}}G'(1,s_{\text{A}}',s_{\text{B}})
	\end{equation}
	
	\noindent which means that, when fully optimizing Alice and Bob's intensities (which already compensate for the mismatch between channels), it is always optimal not to add any additional loss to the channels. Moreover, combining Eqs. (C15), (C16), (C17), (C19), we can see that
	
	\begin{equation}
	\begin{aligned}
	R_{optimal} &= {{\eta_d^2  \eta_{\text{B}}^2 } \over 2} \max\limits_{s_{\text{A}},s_{\text{B}}}G(x,s_{\text{A}},s_{\text{B}})\\ 
	&> {{\eta_d^2  \eta_{\text{B}}^2 } \over 2} \max\limits_{s}G(1,s,s) = R_{symmetric}
	\end{aligned}
	\end{equation}

	That is, compared to the case where one adds loss to $\eta_{\text{A}}$ until $\eta_{\text{A}}=\eta_{\text{B}}$, our new protocol always provides higher key rate as long as the channels are asymmetric. Intuitively, this is because adding fibre while using same intensities for Alice and Bob is in fact a suboptimal subset of the overall set of strategies Alice and Bob can take (which includes adjusting Alice and Bob’s intensities independently, as well as adding any length of fibres to change $x$). Even when considering adding fibre as one of the valid variables, we have shown that the optimal point is always located where no fibre is added. Therefore, our method is a better-optimized strategy than adding fibre because it considers a larger parameter space.
	
	Note that, fully optimizing Alice and Bob's intensities does not change the fundamental scaling property - the key rate is still \textit{quadratically} related to transmittance in the longer arm - However, it always provides better key rate than prior art techniques, and also offers the great convenience of not having to physically add loss to the channels and being able to implement everything in software.
}
\subsection{Practical Imperfections}

\begin{figure}[h]
	\includegraphics[scale=0.32]{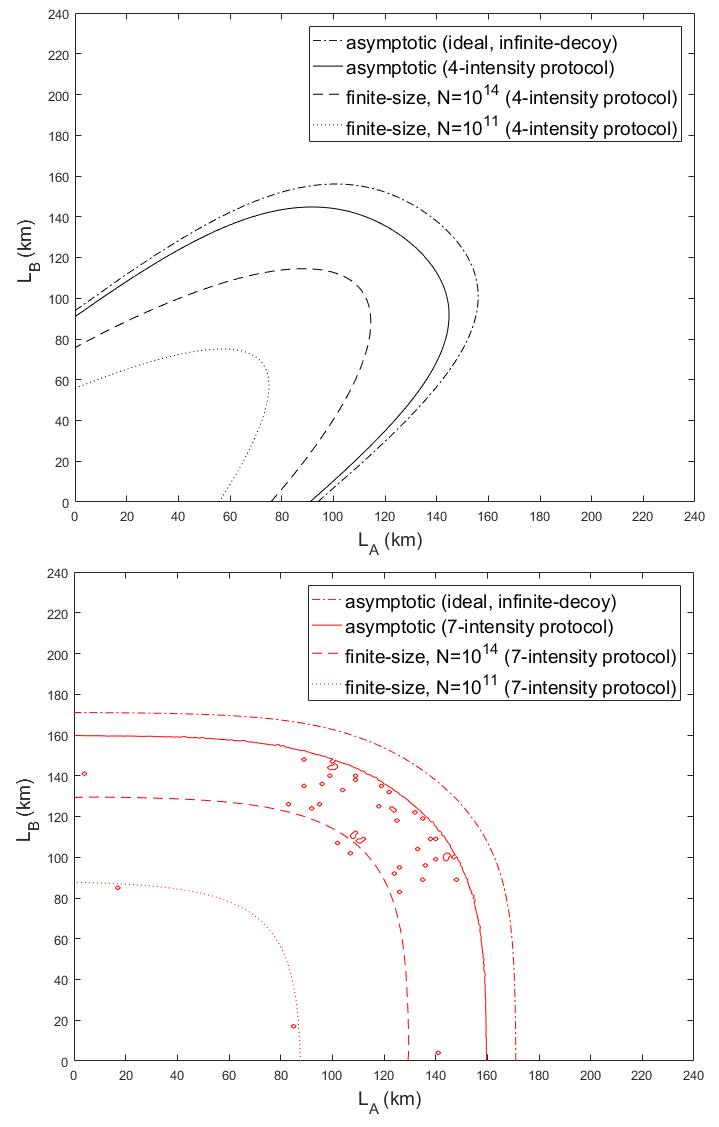}
	\caption{Contours of rate vs distance for decoy-state MDI-QKD, under different assumptions for practical imperfections, for the key rates for the asymptotic case with infinite decoys (and ideal assumption of zero dark count rate and 100\% detector efficiency), the asymptotic case with 4-intensity/7-intensity protocol (with practical device parameters), and the finite-size case with 4-intensity/7-intensity protocol. Top: protocols with identical intensities for Alice and Bob, Bottom: protocols with fully optimized intensities. (Note that in the bottom plot there are some noises in the asymptotic 7-intensity protocol key rate. This is because the optimal $\nu$ can take a very small value in the ideal case where data size is infinitely large. This results in some numerical noises in computer simulations). We plot the contour lines of $R=10^{-7}$. As can be observed here, the finite number of decoys, the non-ideal experimental parameters, and the finite-size effects are all imperfections that reduce the key rate. However, the overall shapes of the contours still remain largely the same, and follow the upper bounds given by the ideal infinite-decoy case. (Except for 4-intensity protocol under finite-size effect, which no longer has two clear cut-off mismatch positions, but is still severely limited by channel asymmetry, while 7-intensity protocol lifts this constraint completely). Reproduced from \cite{this_asymMDI} @2019 APS.}
	\label{fig:contours_compare}
\end{figure}

Up to here we have analytically shown how choosing to fully optimize the intensities can affect the key rate, for the asymptotic, infinite-decoy case. The behavior of contours as shown in Fig.\ref{fig:contours} is a result of $s_{\text{A}},s_{\text{B}}$ compensating for the difference in channel loss. However, we have so far assumed perfect knowledge of single-photon contributions, and have not yet discussed the decoy-state intensities. Moreover, non-ideal experimental parameters (including dark count rate and detector efficiency), and finite-size effects will both affect the key rate. Here in this subsection, we compare the key rate under more practical assumptions, and show that the above factors can be considered as \textit{imperfections} that reduce the key rate, but maintain similar contour shapes and scaling properties for the key rate - that is, we will still observe a high dependence on asymmetry for protocols with identical intensities for Alice and Bob, and fully optimizing intensities can completely lift this limitation.

In practice, with a finite number of decoys (for instance, for 4-intensity and 7-intensity protocols, where Alice and Bob choose respectively three decoy intensities, $\mu,\nu,\omega$), the estimation of $Y_{11}$ and $e_{11}$ is not perfect; therefore the key rate will be slightly lower than the aforementioned infinite-decoy case. Moreover, to accurately estimate $Y_{11}$ and $e_{11}$, the decoy intensities need to be optimized to compensate for channel loss, too. As described in section 5.2.1 in the main text, the decoy states should maintain balanced arriving intensities at Charles (e.g. $\mu_{\text{A}}\eta_{\text{A}}=\mu_{\text{B}}\eta_{\text{B}}$), to ensure good HOM visibility and low QBER in the X basis. Note that, the optimization of decoy intensities has a very different purpose from that of the signal intensities $s_{\text{A}},s_{\text{B}}$ - the signal intensities are optimized so as to reduce $E_{ss}^Z$ (while keeping single photon probability $s_{\text{A}}s_{\text{B}}e^{-(s_{\text{A}}+s_{\text{B}})}$ high) and maximize the key rate, while the decoy intensities are optimized to estimate $Y_{11}^{L}$ and $e_{11}^{U}$ as accurately as possible, whose ideal values $Y_{11}$ and $e_{11}$ (used in the infinite-decoy case above) provide an upper bound for the practical key rate with finite number of decoys. As we see in Fig.\ref{fig:contours_compare}, the asymptotic key rate with a finite number of decoys follows a similar shape as its upper bound, the infinite-decoy case. 

Additionally, the detector efficiency (which is equivalent to channel loss) contributes to a uniformly shifted key rate in both $L_{\text{A}}$ and $L_{\text{B}}$ directions, while dark counts reduce the key rate more significantly in the higher loss region (both of which we have ignored in the ideal case as described at the beginning of this section). However, as observed in Fig.\ref{fig:contours_compare} (the solid lines consider both finite-decoys and practical parameters), these factors do not change the overall shape of the contours either.

Lastly, finite-size effect will reduce the key rate significantly. As observed in Fig.\ref{fig:contours_compare} bottom plot, while the key rate is reduced, the contour shapes remain largely unchanged (meaning that even under finite-size effect, the 7-intensity protocol can still effectively compensate for channel asymmetry effectively). In Fig.\ref{fig:contours_compare} top plot, we can find similar observations, that finite-size effect reduces the overall key rate. However, note that, under finite-size effect, the shapes of key rate contours for the 4-intensity protocol are somewhat different, and no longer follow the two cut-off positions $x^{upper}$, $x^{lower}$ for channel mismatch (which appear as straight lines in e.g. Fig.\ref{fig:contours}). This is because, though the key rate is still limited by $E_{ss}^Z$ (which causes the cut-off mismatch positions), it is also limited by the estimation of $Y_{11}^{L}$ and $e_{11}^{U}$ using the decoy states. Compared to the asymptotic case, here under finite-size effect, the increased $e_{11}^U$ is likely a more severe limiting factor than $E_{ss}^Z$, and not being able to choose independent intensities for Alice and Bob prevents an accurate estimation of $Y_{11}^{L}$ and $e_{11}^{U}$ (due to poor HOM visibility in X basis caused by unbalanced intensities). Therefore, here the dependence of key rate on channel asymmetry is present in both privacy amplification and error-correction terms, and the shapes of contours are a result of both effects. (The difference in contour shape from the infinite-decoy case is more prominent for the finite-size case, likely because the key rate is more sensitive to $e_{11}^{U}$ here). Importantly, under finite-size effects, the key rate for 4-intensity protocol is still highly limited by channel asymmetry, while 7-intensity protocol completely removes such a constraint and allows two channels with arbitrary asymmetry between them.

\section{Note about Decoupling Signal and Decoy Intensities}

In this section we provide a simple intuitive explanation for why our protocol provides a better choice of decoy and signal intensities. 

Let us recall again the key rate formula of MDI-QKD \cite{mdiqkd,mdifourintensity}:
\begin{equation}
\begin{aligned}
R=P_{s_{\text{A}}}P_{s_{\text{B}}} \{(s_{\text{A}} e^{-s_{\text{A}}})(s_{\text{B}} e^{-s_{\text{B}}}) Y_{11}^{X,L}[1-h_2(e_{11}^{X,U})]\\
-f_eQ_{ss}^Z h_2(E_{ss}^Z)\}
\end{aligned}
\end{equation}

Here there are three criteria that determine whether a MDI-QKD protocol generates good key rate in the presence of channel asymmetry:\\

(a) Similar arriving intensities at Charles in the X basis, in order to have good HOM interference and keep QBER low in the X basis (which is important for a good estimation of $e_{11}^{X,U}$).

(b) Similar arriving intensities at Charles in the Z basis, in order to keep QBER $E_{ss}^Z$ low in the Z basis (which is due to misalignment), although this term is much less sensitive to the difference in intensities than (a).

(c) A high enough probability of sending single-photons, $s_{\text{A}} e^{-s_{\text{A}}}s_{\text{B}} e^{-s_{\text{B}}}$. Note that both criteria (b) and (c) involve the signal states $s_{\text{A}}$, $s_{\text{B}}$, so there is a trade-off between (b) and (c).\\

Prior protocols require Alice and Bob to use the same set of intensities. This overly constrains the solution space (because Alice and Bob try to use the same set of intensities to satisfy (a), (b) and (c) simultaneously), and leaves high QBER in both the X and Z bases, and thus resulting in low key rate when channels are asymmetric. 

By relaxing this constraint (allowing Alice and Bob to have different intensities), and decoupling criteria (a) from criteria (b) and (c) by allowing independent decoy and signal intensities, we can satisfy (a) nicely, while simultaneously achieving a good trade-off between (b) and (c), hence ensuring a high key rate.

\textit{\textbf{Remark}}: for more detail on the trade-off between (b) and (c), here (b) is optimal when arriving intensities are matched, i.e. $s_{\text{A}}/s_{\text{B}}=\eta_{\text{B}}/\eta_{\text{A}}$, and (c) is independent of asymmetry and is optimal when signal intensities are both 1. In fact, since $E^{Z}_{ss}$ is much less sensitive to $s_{\text{A}}/ s_{\text{B}}$, such a trade-off between two terms favors (c) more than (b), thus the optimal $s_{\text{A}}/s_{\text{B}}$ is often closer to 1 than $\eta_{\text{B}}/\eta_{\text{A}}$. The actual optimal signal intensities can be found by numerical optimization, as described in section 5.2.4. An example of $\mu_{\text{A}}/\mu_{\text{B}}$ and $s_{\text{A}}/s_{\text{B}}$ can also be seen in Fig.\ref{fig:ratio}, where we observe that  $\mu_{\text{A}}/\mu_{\text{B}}$ follows $\eta_{\text{B}}/\eta_{\text{A}}$ rather closely, while $s_{\text{A}}/ s_{\text{B}}$ has relatively much more freedom in its optimization (between $1$ and $\eta_{\text{B}}/\eta_{\text{A}}$).\\

\section{Generality of Our Method: MDI-QKD Protocols other than Three Decoy States}

In the main text we have focused on the 7-intensity protocol, where Alice and Bob each uses one signal intensity $s_A$ ($s_B$), and three decoy intensities $\mu_A,\nu_A, \omega$ ($\mu_B,\nu_B, \omega$). However, the core idea of our protocol lies in two key points: (1) X and Z bases are decoupled, where decoy-states in the X basis bound Eve's information and signal state in the Z basis encodes the key, and (2) Alice and Bob use different intensities to compensate for channel asymmetry. This means that our protocol is not limited to the 7-intensity protocol, but can easily be applied to other protocols too, as long as points (1) and (2) are satisfied. 

In this section, we demonstrate the generality of our method by actually applying it to other kinds of MDI-QKD protocols where Alice and Bob use a different number of decoy intensities in the X basis, and show that similar advantages as with the 7-intensity protocol can be observed when using asymmetric intensities. We will also show with numerical results that, although these alternative protocols will certainly work, the 7-intensity protocol provides a good balance between performance and ease of experimental implementation.\\

Here we compare three cases:

1. Alice and Bob each uses two decoy intensities $\mu_A,\nu_A$ ($\mu_B,\nu_B$) in the X basis. We denote this case as a \textbf{6-intensity protocol} (including the two signal intensities), where the parameter choices are: 

\begin{equation}
\begin{aligned}
&[s_A,\mu_A,\nu_A,P_{s_A},P_{\mu_A},P_{\nu_A},\\
&s_B,\mu_B,\nu_B,P_{s_B},P_{\mu_B},P_{\nu_B}]
\end{aligned}
\end{equation}

\noindent Here $P_{\nu_A}=1-P_{s_A}-P_{\mu_A}$ and $P_{\nu_B}=1-P_{s_B}-P_{\mu_B}$. This is similar to the ``one-decoy" setup that was discussed in Ref. \cite{mdiparameter}. Note that here it's not a ``5-intensity" protocol, because using $\mu_A,\mu_B,\omega$ alone is not sufficient to satisfactorily bound the single-photon contributions and will result in low or zero key rate. Therefore, in this setup, the vacuum state is not used, and Alice and Bob each use two non-zero decoy states.\\

2. Alice and Bob each use three decoy intensities $\mu_A,\nu_A, \omega$ ($\mu_B,\nu_B, \omega$) in the X basis. This case is the \textbf{7-intensity protocol} we discussed in the main text, where the parameter choices are 

\begin{equation}
\begin{aligned}
&[s_A,\mu_A,\nu_A,\omega,P_{s_A},P_{\mu_A},P_{\nu_A},P_{\omega_A},\\
&s_B,\mu_B,\nu_B,\omega,P_{s_B},P_{\mu_B},P_{\nu_B},P_{\omega_B}]
\end{aligned}
\end{equation}

\noindent Here $P_{\omega_A}=1-P_{s_A}-P_{\mu_A}-P_{\nu_A},P_{\omega_B}=1-P_{s_B}-P_{\mu_B}-P_{\nu_B}$, and $\omega$ is the vacuum state (for simplicity we can assume $\omega=0$).\\

3. Alice and Bob each use four decoy intensities $\mu_A,\nu_A, \nu_{2A}, \omega$ ($\mu_B,\nu_B, \nu_{2B}, \omega$) in the X basis. We denote this case as a \textbf{9-intensity protocol}, where the parameter choices are 

\begin{equation}
\begin{aligned}
&[s_A,\mu_A,\nu_A,\nu_{2A}, \omega, P_{s_A},P_{\mu_A},P_{\nu_A}, P_{\nu_{2A}},P_{\omega_A},\\
&s_B,\mu_B,\nu_B,\nu_{2B}, \omega, P_{s_B},P_{\mu_B},P_{\nu_B}, P_{\nu_{2B}},P_{\omega_B}]
\end{aligned}
\end{equation}

\noindent Here $P_{\omega_A}=1-P_{s_A}-P_{\mu_A}-P_{\nu_A}-P_{\nu_{2A}},P_{\omega_B}=1-P_{s_B}-P_{\mu_B}-P_{\nu_B}-P_{\nu_{2B}}$, and $\omega$ is the vacuum state.

Note that in all of these three protocols, the key rate formula stays the same as Eq. 5.1:

\begin{equation}
\begin{aligned}
R=P_{s_{\text{A}}}P_{s_{\text{B}}} \{(s_{\text{A}} e^{-s_{\text{A}}})(s_{\text{B}} e^{-s_{\text{B}}}) Y_{11}^{X,L}[1-h_2(e_{11}^{X,U})]\\
-f_eQ_{ss}^Z h_2(E_{ss}^Z)\}
\end{aligned}
\end{equation}

\noindent What is changing here is only the estimation of the single-photon contributions, namely the yield $Y_{11}^{X,L}$ and QBER $e_{11}^{X,U}$. While we have analytical bounds for decoy-state analysis \cite{mdipractical} for the 7-intensity protocol, we use a linear programming approach to numerically estimate $Y_{11}^{X,L}$ and QBER $e_{11}^{X,U}$ in the 6-intensity and 9-intensity cases. Such an approach has been widely discussed in literature as in Refs.\cite{MDIAnalytical,mdiparameter,mdiChernoff}.

Now, we perform numerical simulations with the 6-intensity, 7-intensity and 9-intensity protocols, and show that they all have much higher performance than their symmetric-intensity counterparts when channel asymmetry is present. This demonstrates the generality of our method as using asymmetric intensities can always improve the performance of MDI-QKD with asymmetric channels.

We also compare the performances of the three protocols with each other, and show that using more decoy intensities can always guarantee higher or equal performance than using fewer decoy intensities, regardless of data size and asymmetry. The 7-intensity always provides no smaller key rate than the 6-intensity protocol, and although 9-intensity protocol can potentially provide an even higher key rate, the advantage is small, and the 7-intensity protocol we used in the main text is a good balance between key rate performance and ease of experimental implementation.

\begin{figure}[h]
	
	\includegraphics[scale=0.15]{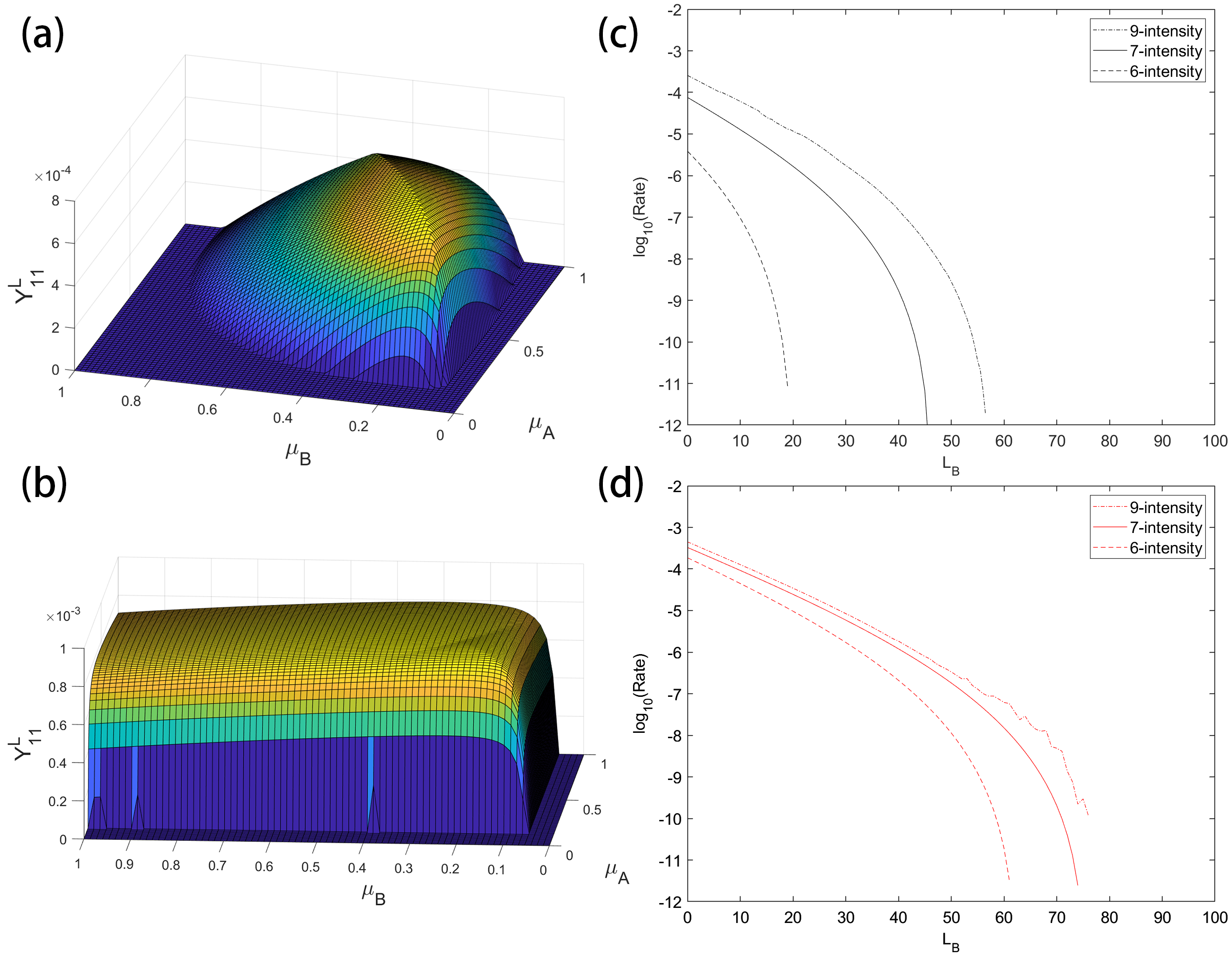}
	\caption{Left: Examples of $Y_{11}^{X,L}$ versus $\mu_A$ and $\mu_B$ where other parameters are all fixed for (a) 6-intensity protocol and (b) 9-intensity protocol. Just like for 7-intensity protocol, we can see a ridge along ${\mu_A \over \mu_B} = {\nu_A \over \nu_B}$, or ${\mu_A \over \mu_B} = {\nu_A \over \nu_B} = {\nu_{2_A} \over \nu_{2_B}}$. Note that the ridge is a lot less obvious for 9-intensity protocol (and sometimes is not visible) likely because two proportional pairs of decoy-states can estimate single-photon contribution reasonably well, so the third pair ($\mu_A,\mu_B$) here has more freedom in the choice of intensities. Right: Comparison of rate vs $L_{\text{B}}$ for 6-intensity, 7-intensity and 9-intensity protocols, where mismatch $x=0.1$, i.e. $L_{A} = L_{B} + 50km$ (assuming fibre loss $0.2dB/km$). The rates are plotted in log-scale. We use the parameters from Table 5.2, and $N=10^{12}$. (c) using symmetric intensities for Alice and Bob, (d) using fully optimized asymmetric parameters for Alice and Bob. As can be seen, using asymmetric intensities can greatly improve the key rate for all three protocols, when channel asymmetry is present. Note that there is a higher amount of noise present for the 9-intensity case due to the numerical instability brought by linear program solvers (similar to that of joint-bound finite-size analysis, which will be discussed in Appendix C.8), but the key points here are that the 9-intensity protocol also benefits considerably from using asymmetric intensities, and that the 9-intensity protocol does not have a significant advantage over the 7-intensity protocol despite being more complex to implement. Reproduced from \cite{this_asymMDI} @2019 APS.}
	\label{fig:generality_rate}
	
\end{figure}

Interestingly, as observed in Fig. \ref{fig:generality_rate} (a)(b), for the 6-intensity and 9-intensity protocols, although the yield $Y_{11}^{X,L}$ and QBER $e_{11}^{X,U}$ are estimated numerically using linear programming, there is still a ``ridge" (discontinuity in first-order derivatives) along ${\mu_A \over \mu_B} = {\nu_A \over \nu_B}$, and  ${\nu_A \over \nu_B} = {\nu_{2_A} \over \nu_{2_B}}$ as we saw for 7-intensity protocol in the main text. For 6-intensity protocol, the ridge is very clearly shown. For 9-intensity protocol, the ridge exists but is less prominent, and sometimes not visible (likely because, e.g. if two pairs of proportional decoy states ${\nu_A \over \nu_B} = {\nu_{2_A} \over \nu_{2_B}}$ already provide good estimation of $Y_{11}^{X,L},e_{11}^{X,U}$, the third pair $\mu_A,\mu_B$ has more freedom, and wouldn't affect the decoy-state analysis or the key rate too much even if it doesn't provide good HOM visibility and results in high $E_{\mu\mu}^X$).

We plot the simulated key rate for the protocols in Fig. \ref{fig:generality_rate} (c)(d). We first consider a similar scenario as Fig. \ref{fig:2d_Results} (c)(d), using parameters from Table 5.2, a channel mismatch of ${\eta_A \over \eta_B} = x = 0.1$, and data size of $N=10^{12}$ (here we use a larger data size than in the main text, since for $N=10^{11}$, 6-intensity protocol with symmetric intensities cannot generate key rate even at $L_B=0km$ so a comparison is not immediately clear in the plot). As shown in Fig. \ref{fig:generality_rate}, for each protocol, allowing asymmetric intensities provides a much higher key rate than using symmetric intensities only, demonstrating the general effectiveness of our method for different protocols under channel asymmetry.

We also make an important observation here: The more decoy intensities one uses, the higher the key rate one can obtain after parameter optimization, even with finite-size effects considered (e.g. the 9-intensity protocol always has a higher key rate than 7-intensity, and 7-intensity also always has a higher rate than 6-intensity). This is because, for instance, the 6-intensity protocol can in fact be considered as a special case of the 7-intensity protocol, just with $P_{\omega_A}$ and $P_{\omega_B}$ infinitely close to zero, and with 9 instead of 4 constraints for e.g. the gains $Q_{ij}^X$ when estimating $Y_{11}^{X,L}$. With close to zero data, the 5 new constraints are obviously very loose (with very large finite-size fluctuation) and will not provide any useful information, but the key point is, in a linear program these loose constraints \textit{will not decrease} the key rate. Therefore, any optimal set of parameters for the 6-intensity protocol can also be considered as a valid set of parameters for the 7-intensity protocol, i.e. the parameter space of 6-intensity protocol is a subset of that of the 7-intensity protocol, and the latter always provides \textit{no smaller} key rate than the former (and often the 7-intensity protocol can find a better parameter set in the larger parameter space, resulting in higher key rate).

The same goes for the 9-intensity protocol, but as we have seen in Fig. \ref{fig:generality_rate}, the advantage it provides over the 7-intensity protocol is rather small (compared to e.g. 6-intensity versus 7-intensity), while requiring more complex control of the intensity modulators in the experimental setup, and more complicated data collection and analysis: the users need to collect 16 sets of gains and error-gains, and the parameter optimization is also a lot slower and more unstable (evaluating the linear program is on average slower than analytical expression by about 50 times, and linear programs also introduce numerical instabilities). Similar observations have been made for the symmetric case in Ref. \cite{mdiparameter} (although in this paper the signal states are not decoupled from decoy states so the protocols are slightly different) that using decoy states $\{\mu,\nu,\omega\}$ provides higher key rate than $\{\mu,\nu\}$, but adding one more decoy-state $\nu_2$ provides little further advantage. 

Therefore, our conclusion is that, while our method of asymmetric intensities and decoupled bases surely works well with other protocols such as 6-intensity and 9-intensity protocols, the 7-intensity protocol we introduced in the main text strikes a good balance between key rate performance and the ease of both experimental implementation and data analysis.

\section{Decoy State Intensities}

In this section we will describe Theorems I and II in more detail, and show their theoretical proofs in the asymptotic limit of infinite data size (Moreover, numerically, we found that Theorems I and II in fact hold true even under finite-size effects).

\subsection{Symmetry of Optimal Decoy Intensities}

\begin{figure}[h]
	\includegraphics[scale=0.31]{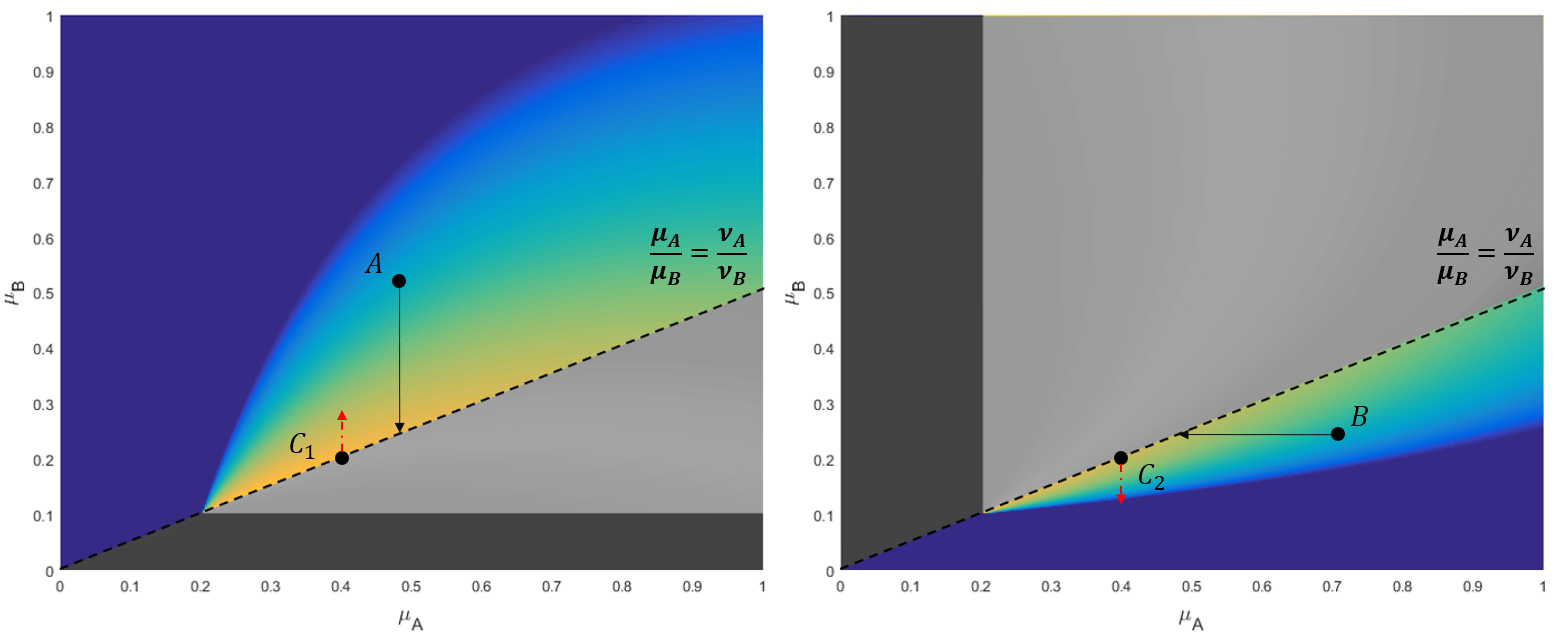}
	\caption{An example of the two difference cases of $Y_{11}^{X,L}$ vs $\mu_{\text{A}}, \mu_{\text{B}}$ function, for fixed values of $\nu_{\text{A}}=0.2$, $\nu_{\text{B}}=0.1$. Left: $Y_{11}^a$, where ${\mu_{\text{A}} \over \mu_{\text{B}}} \leq {\nu_{\text{A}} \over \nu_{\text{B}}}$ (``case 1"); Right: $Y_{11}^b$, where ${\mu_{\text{A}} \over \mu_{\text{B}}} > {\nu_{\text{A}} \over \nu_{\text{B}}}$ (``case 2"). Allowed regions are marked in color for either cases. In case 1, we show that ${{\partial Y_{11}^a} \over {\partial \mu_{\text{B}}}} <0$, so any given point A can descend along $\mu_{\text{B}}$ axis (the solid black arrow) to get higher rate, until it reaches boundary line ${\mu_{\text{A}} \over \mu_{\text{B}}} = {\nu_{\text{A}} \over \nu_{\text{B}}}$ where $\mu_{\text{B}}$ is highest. Similarly, in case 2, ${{\partial Y_{11}^b} \over {\partial \mu_{\text{A}}}} <0$, so any given point B can descend along $\mu_{\text{A}}$ axis until ${\mu_{\text{A}} \over \mu_{\text{B}}} = {\nu_{\text{A}} \over \nu_{\text{B}}}$ to get highest rate. Therefore, the optimal $(\mu_{\text{A}}^{opt}, \mu_{\text{B}}^{opt})$ that maximize the piecewise function $Y_{11}^{X,L}$ always occur on the boundary line. Moreover, for any given point $C(\mu_{\text{A}},\mu_{\text{B}})$ on the boundary line, the function values of $Y_{11}^a$, $Y_{11}^b$ are the same. However, we show that ${{\partial Y_{11}^a} \over {\partial \mu_{\text{A}}}}$ at $C_1$ is not equal to ${{\partial Y_{11}^b} \over {\partial \mu_{\text{A}}}}$ at $C_2$ (along the dot-dash red lines). Therefore, the piecewise function $Y_{11}^{X,L}$ is not smooth. Reproduced from \cite{this_asymMDI} @2019 APS.}
	\label{fig:Y11}
\end{figure}

To prove Theorem I, here we will actually propose an even stronger assumption for $\mu_{\text{A}}, \mu_{\text{B}}$:\\

\textit{\textbf{Theorem III}.
	for any arbitrary choice of device and channel parameters, and any two given values of $\nu_{\text{A}}, \nu_{\text{B}}$, the optimal decoy intensities $\mu_{\text{A}}^{opt}, \mu_{\text{B}}^{opt}$ that maximize $R$ always satisfy the constraint:}

\begin{equation}
{\mu_{\text{A}}^{opt} \over \mu_{\text{B}}^{opt} } = { \nu_{\text{A}} \over \nu_{\text{B}}}
\end{equation}

\textit{\textbf{Remark:}} as will be shown below, Theorem I is simply a corollary from Theorem III.\\

\textit{\textbf{Proof for Theorem III:}} Here for convenience, we first limit the discussion to the asymptotic case (i.e. infinite data size), and we assume that the vacuum intensity is indeed $\omega=0$. Throughout the rest of the text, we will use $Q_{ij}^k$ and $E_{ij}^k$ to denote the observed gain and QBER, where, if not specified, the first subscript is Alice's intensity, and the second is Bob's intensity, which can be chosen from $\{s_{\text{A}},\mu_{\text{A}},\nu_{\text{A}},\omega\}$ and $\{s_{\text{B}},\mu_{\text{B}},\nu_{\text{B}},\omega\}$ for Alice and Bob, respectively. The superscript $k$ signifies the basis X or Z (although, here we only explicitly write the basis for illustration purposes, since the basis is already implied by the choice of intensities).

First, looking at the key rate expression \cite{mdiqkd,mdifourintensity}:

\begin{equation}
\begin{aligned}
R=P_{s_{\text{A}}}P_{s_{\text{B}}} \{(s_{\text{A}} e^{-s_{\text{A}}})(s_{\text{B}} e^{-s_{\text{B}}}) Y_{11}^{X,L}[1-h_2(e_{11}^{X,U})]\\
-f_eQ_{ss}^Z h_2(E_{ss}^Z)\}
\end{aligned}
\end{equation}

\noindent we can see that only the term $Y_{11}^{X,L}[1-h_2(e_{11}^{X,U})]$, i.e. the decoy-state analysis and privacy amplification, is determined by the decoy intensities (and probabilities, if finite-size effect is considered) only, and not affected by the signal intensities $s_{\text{A}}, s_{\text{B}}$. This is an important and very convenient characteristic of the 4-intensity/7-intensity protocol, that the signal state is only concerned with key generation, while the decoy states are only responsible for privacy amplification. That is, the optimization of decoy-state intensities is decoupled from the key generation. Now, we can make an observation that, under given device parameters and channel loss, the optimization of the decoy intensities is independent of $s_{\text{A}}, s_{\text{B}}$, and its only goal is to maximize $Y_{11}^{X,L}[1-h_2(e_{11}^{X,U})]$.

Furthermore, to perform the decoy state analysis, we note that the upper bound for single-photon QBER satisfies the form of:

\begin{equation}
\begin{aligned}
e_{11}^{X,U} = f(Y_{11}^{X,L}, \nu_{\text{A}}, \nu_{\text{B}})
\end{aligned}
\end{equation}

\noindent where $e_{11}^{X,U}$ is only determined by $Y_{11}^{X,L}$, $\nu_{\text{A}}$ and $\nu_{\text{B}}$. The full expression, as in \cite{mdipractical}, is listed below:

\begin{equation}
\begin{aligned}
e_{11}^{X,U} = {1\over {\nu_{\text{A}}\nu_{\text{B}} Y_{11}^{X,L}}}(e^{\nu_{\text{A}}+\nu_{\text{B}}}Q_{\nu\nu}^XE_{\nu\nu}^X-e^{\nu_{\text{A}}}Q_{\nu\omega}^XE_{\nu\omega}^X\\
-e^{\nu_{\text{B}}}Q_{\omega\nu}^XE_{\omega\nu}^X+Q_{\omega\omega}^XE_{\omega\omega}^X)
\end{aligned}
\end{equation}.

Now, suppose we first fix two arbitrary values of $\nu_{\text{A}},\nu_{\text{B}}$, and try to maximize $Y_{11}^X[1-h_2(e_{11}^X)]$ by optimizing $\mu_{\text{A}}, \mu_{\text{B}}$, we can see that maximizing $Y_{11}^{X,L}$ will suffice, since it will simultaneously minimize $e_{11}^{X,U}$, whose only component dependent on $\mu_{\text{A}}, \mu_{\text{B}}$ is $Y_{11}^{X,L}$. The question now becomes simply finding:

\begin{equation}
({\mu_{\text{A}}^{opt}, \mu_{\text{B}}^{opt} }) = argmax(Y_{11}^{X,L}(\mu_{\text{A}},\mu_{\text{B}}))
\end{equation}

A very important characteristic of $Y_{11}^{X,L}$ is that, its expression is dependent upon whether ${\mu_{\text{A}} \over \mu_{\text{B}}} \leq {\nu_{\text{A}} \over \nu_{\text{B}}}$, i.e. it is a piecewise function, as described in Ref.\cite{mdipractical}:\\

\textbf{Case 1}: If ${\mu_{\text{A}} \over \mu_{\text{B}}} \leq {\nu_{\text{A}} \over \nu_{\text{B}}}$:

\begin{equation}
Y_{11}^{X,L} = Y_{11}^a = {1 \over (\mu_{\text{A}}-\nu_{\text{A}})} [{{\mu_{\text{A}}}\over{\nu_{\text{A}}\nu_{\text{B}}}}Q_{\nu\nu}^{M1}-{{\nu_{\text{A}}}\over{\mu_{\text{A}}\mu_{\text{B}}}}Q_{\mu\mu}^{M2}]
\end{equation}

\textbf{Case 2}: otherwise, if ${\mu_{\text{A}} \over \mu_{\text{B}}} > {\nu_{\text{A}} \over \nu_{\text{B}}}$:

\begin{equation}
Y_{11}^{X,L} = Y_{11}^b = {1 \over (\mu_{\text{B}}-\nu_{\text{B}})} [{{\mu_{\text{B}}}\over{\nu_{\text{A}}\nu_{\text{B}}}}Q_{\nu\nu}^{M1}-{{\nu_{\text{B}}}\over{\mu_{\text{A}}\mu_{\text{B}}}}Q_{\mu\mu}^{M2}]
\end{equation}

\noindent where we denote the two expressions of $Y_{11}^{X,L}$ in the two cases as $Y_{11}^a$ and $Y_{11}^b$, and the two terms $Q_{\nu\nu}^{M1}$ and $Q_{\mu\mu}^{M2}$ are linear combinations of the observable Gain, and are functions of $(\nu_{\text{A}},\nu_{\text{B}})$ and $(\mu_{\text{A}},\mu_{\text{B}})$ only, respectively. Their full expressions can be found in Appendix C.6.3. Also, note that if ${\mu_{\text{A}} \over \mu_{\text{B}}} = {\nu_{\text{A}} \over \nu_{\text{B}}}$, the two cases $Y_{11}^a=Y_{11}^b$.

Now, we can make a key observation, that in case 1, for any given $\mu_{\text{A}}$, the partial derivative ${\partial Y_{11}^a} \over {\partial \mu_{\text{B}}}$ always satisfies

\begin{equation}
{{\partial Y_{11}^a} \over {\partial \mu_{\text{B}}}} < 0
\end{equation}

\noindent (The actual expression of the partial derivative and proof of its positivity are shown in Appendix C.6.3). However, in case 1, $\mu_{\text{B}}$ is bounded by $\mu_{\text{B}} \geq {{\mu_{\text{A}} \nu_{\text{B}}} \over \nu_{\text{A}}}$, so the only optimal case is to take the boundary condition

\begin{equation}
\mu_{\text{B}}^{opt} = {{\mu_{\text{A}} \nu_{\text{B}}} \over \nu_{\text{A}}}
\end{equation}

This means that, in the region of ${\mu_{\text{A}} \over \mu_{\text{B}}} \leq {\nu_{\text{A}} \over \nu_{\text{B}}}$, any two optimal value pair $(\mu_{\text{A}}^{opt},\mu_{\text{B}}^{opt})$ must satisfy ${\mu_{\text{A}}^{opt} \over \mu_{\text{B}}^{opt} } = { \nu_{\text{A}} \over \nu_{\text{B}}}$, or else we can always decrease $\mu_{\text{B}}$ to get a higher rate, meaning that the previous point is not the actual maximum. We illustrate this behavior in Fig.\ref{fig:Y11}.

Similarly, for case 2, the partial derivative with respect to $\mu_{\text{A}}$ satisfies

\begin{equation}
{{\partial Y_{11}^b} \over {\partial \mu_{\text{A}}}} < 0
\end{equation}

\noindent and $\mu_{\text{A}}$ is bounded by $\mu_{\text{A}} > {{\mu_{\text{B}} \nu_{\text{A}}} \over \nu_{\text{B}}}$. In the same way, in case 2 for any given $\mu_{\text{B}}$, we can acquire:

\begin{equation}
\mu_{\text{A}}^{opt} = {{\mu_{\text{B}} \nu_{\text{A}}} \over \nu_{\text{B}}}
\end{equation}

Up to here, we have proven that Theorem III is indeed correct.  \qedsymbol \\

\textit{\textbf{Proof for Theorem I:}} Now, following the same idea, any four optimal value pair $(\mu_{\text{A}}^{opt},\mu_{\text{B}}^{opt},\nu_{\text{A}}^{opt},\nu_{\text{B}}^{opt})$ must satisfy ${\mu_{\text{A}}^{opt} \over \mu_{\text{B}}^{opt} } = { \nu_{\text{A}}^{opt} \over \nu_{\text{B}}^{opt}}$, or else we can always vary $(\mu_{\text{A}}, \mu_{\text{B}})$ while keeping $(\nu_{\text{A}}, \nu_{\text{B}})$ fixed, and let ${\mu_{\text{A}} \over \mu_{\text{B}}} = {\nu_{\text{A}} \over \nu_{\text{B}}}$ to get a higher rate, meaning that the previous point is not the actual maximum. Therefore, we have shown that Theorem I is indeed correct, that the optimal decoy intensities always satisfy

\begin{equation}
{\mu_{\text{A}}^{opt} \over \mu_{\text{B}}^{opt} } = { \nu_{\text{A}}^{opt} \over \nu_{\text{B}}^{opt}}
\end{equation}

\qedsymbol\\

Note that the same conclusion doesn't hold true for traditional 3-intensity MDI-QKD (i.e. using $\{\mu, \nu, \omega\}$ for both X and Z basis and using $\mu$ in Z basis to generate the key), that is because the key rate for 3-intensity depends on $\mu$ for both key generation and error-correction, such that $Q_{\mu\mu}^Z, E_{\mu\mu}^Z$ terms and the single-photon probability $\mu e^{-\mu}$ both depend on $\mu$, hence optimizing only $Y_{11}^L$ is no longer sufficient. Therefore, this independence of $s$ from $\mu, \nu$ is an additional advantage that the 4-intensity/7-intensity protocol can provide, under asymmetric conditions.

Also, one thing to note is that, although the above theorem provides us with a way to constrain ${\mu_{\text{A}} \over \mu_{\text{B}}}, {\nu_{\text{A}} \over \nu_{\text{B}}}$, the actual values of these ratios still need to be found by optimization. In Ref. \cite{mdipractical}, the authors have proposed a rule-of-thumb formula for finding optimal intensities:

\begin{equation}
\mu_{\text{A}} \eta_{\text{A}} = \mu_{\text{B}} \eta_{\text{B}}
\end{equation}

\noindent for which we now have a good understanding of the reason - such a relation keeps the arriving intensities balanced at Charles, in order to maintain good HOM visibility in the X basis and low QBER.

However, this is still only a rough approximation, and is an exact relation only when the dark count rate $Y_0$ is ignored, data size is infinite, and infinite number of decoys are used (Ref. \cite{mdipractical} considered the case where $\mu$ is both the signal and decoy intensity, and only proved Eq. C.42 to be exact in the ideal infinite-decoy case with no noise). For a general case, $\mu_{\text{A}}/\mu_{\text{B}}$ is not always exactly equal to $\eta_{\text{B}}/\eta_{\text{A}}$ (and does not only depend on the mismatch $x=\eta_{\text{A}}/\eta_{\text{B}}$) but rather deviates slightly from it when $(\eta_{\text{A}},\eta_{\text{B}})$ changes. But at least, one general rule is that $\mu_{\text{A}}/\mu_{\text{B}}$ decreases with $x=\eta_{\text{A}}/\eta_{\text{B}}$, or, to put in more simple words, the larger the channel loss, the higher the intensities we should choose to compensate for the loss.

\subsection{Non-smoothness of Key Rate vs Intensities Function}

In the previous section we have shown that the piecewise expression for $Y_{11}^{X,L}$ causes the optimal value to occur on the boundary line ${\mu_{\text{A}} \over \mu_{\text{B}}} = {\nu_{\text{A}} \over \nu_{\text{B}}}$. Here we continue to show that Theorem II is a result of this piecewise function, too.\\

\textit{\textbf{Proof of Theorem II:}} The theorem means that, the key rate does not have a continuous partial derivative with respect to $\mu_{\text{A}}$ or $\mu_{\text{B}}$ at the boundary line. This will cause the boundary line to behave like a sharp ``ridge". To prove this, instead of differentiating $Y_{11}^a$ vs $\mu_{\text{B}}$ and $Y_{11}^b$ vs $\mu_{\text{A}}$, here we perform partial differentiation of both $Y_{11}^a$, $Y_{11}^b$ vs $\mu_{\text{A}}$, and observe this discontinuity of derivative.

First, we rewrite $Y_{11}$ into:

\begin{equation}
\begin{aligned}
Y_{11}^a = {\nu_{\text{A}} \over (\mu_{\text{A}}-\nu_{\text{A}})} [{{1}\over{\nu_{\text{A}}\nu_{\text{B}}}}Q_{\nu\nu}^{M1}-{{1}\over{\mu_{\text{A}}\mu_{\text{B}}}}Q_{\mu\mu}^{M2}] + {{1}\over{\nu_{\text{A}}\nu_{\text{B}}}}Q_{\nu\nu}^{M1}\\
Y_{11}^b = {\nu_{\text{B}} \over (\mu_{\text{B}}-\nu_{\text{B}})} [{{1}\over{\nu_{\text{A}}\nu_{\text{B}}}}Q_{\nu\nu}^{M1}-{{1}\over{\mu_{\text{A}}\mu_{\text{B}}}}Q_{\mu\mu}^{M2}] + {{1}\over{\nu_{\text{A}}\nu_{\text{B}}}}Q_{\nu\nu}^{M1}\\
\end{aligned}
\end{equation}

The last term is not dependent on either $\mu_{\text{A}}$ or $\mu_{\text{B}}$. Note that, here on the boundary of ${\mu_{\text{A}} \over \mu_{\text{B}}} = {\nu_{\text{A}} \over \nu_{\text{B}}}$, the values of $Y_{11}^a$ and $Y_{11}^b$ are equal:

\begin{equation}
Y_{11}^a=Y_{11}^b
\end{equation}

Performing the partial differentiation against $\mu_{\text{A}}$, we can get:

\begin{equation}
\begin{aligned}
{{\partial Y_{11}^a} \over {\partial \mu_{\text{A}}}} =  &-{\nu_{\text{A}} \over{\mu_{\text{A}}-\nu_{\text{A}}}} {{\partial} \over {\partial \mu_{\text{A}}}}({{Q_{\mu\mu}^{M2}}\over {\mu_{\text{A}} \mu_{\text{B}}}}) \\
&+ {\nu_{\text{A}} \over{(\mu_{\text{A}}-\nu_{\text{A}})^2}} ({{Q_{\mu\mu}^{M2}}\over {\mu_{\text{A}} \mu_{\text{B}}}}-{{Q_{\nu\nu}^{M1}}\over {\nu_{\text{A}} \nu_{\text{B}}}})\\
{{\partial Y_{11}^b} \over {\partial \mu_{\text{A}}}} =  &-{\nu_{\text{B}} \over{\mu_{\text{B}}-\nu_{\text{B}}}} {{\partial} \over {\partial \mu_{\text{A}}}}({{Q_{\mu\mu}^{M2}}\over {\mu_{\text{A}} \mu_{\text{B}}}})\\
\end{aligned}
\end{equation}

We can see that, on the boundary of ${\mu_{\text{A}} \over \mu_{\text{B}}} = {\nu_{\text{A}} \over \nu_{\text{B}}}$, the first terms are again equal, however, the second term in ${{\partial Y_{11}^a} \over {\partial \mu_{\text{A}}}}$ is strictly larger than 0 (detailed proof by expanding $Q_{\mu\mu}^{M2}$, $Q_{\nu\nu}^{M1}$ are shown in Appendix C.6.3). Therefore,

\begin{equation}
{{\partial Y_{11}^a} \over {\partial \mu_{\text{A}}}} \neq {{\partial Y_{11}^b} \over {\partial \mu_{\text{A}}}}
\end{equation}

The derivatives of $Y_{11}^{X,L}$ vs $\mu_{\text{A}}$ on the two sides of the ``ridge" are not equal, causing the rate function R to have a non-defined gradient. A similar proof can be applied to $\mu_{\text{B}}$ and it leads to the same result. \qedsymbol\\

An illustration can be seen in main text Fig. 2, which chooses a given set of values $(\nu_{\text{A}}=0.2, \nu_{\text{B}}=0.1)$ and plots the key rate over $(\mu_{\text{A}},\mu_{\text{B}})$. As can be clearly observed, there is a sharp ridge on the line ${\mu_{\text{A}} \over \mu_{\text{B}}}={\nu_{\text{A}} \over \nu_{\text{B}}}=2 $, meaning the key rate function versus intensities is not smooth.

\subsection{Proof of Negativity of Partial Derivatives for Decoy Intensities}

As described above, the expression for the single-photon yield, $Y_{11}^{X,L}$ depends on whether ${\mu_{\text{A}} \over \mu_{\text{B}}} \leq {\nu_{\text{A}} \over \nu_{\text{B}}}$. For case 1, if ${\mu_{\text{A}} \over \mu_{\text{B}}} \leq {\nu_{\text{A}} \over \nu_{\text{B}}}$, we would like to prove that

\textit{\textbf{Lemma I:} ${\partial Y_{11}^a} \over {\partial \mu_{\text{B}}}$ and ${\partial Y_{11}^b} \over {\partial \mu_{\text{A}}}$ are both always negative.\\ }

\textit{\textbf{Proof of Lemma I:}} Here, we use a simplified model of the Gain $Q_{ij}^X$ as in Ref.\cite{mdipractical}, which ignores the dark count rate $Y_0$, and takes a second-order approximation for the modified Bessel function:

\begin{equation}
Q_{\mu\mu}^X={\eta_{\text{B}}^2 \eta_d^2 \over 4} [2x\mu_{\text{A}}\mu_{\text{B}} + (\mu_{\text{B}}^2+x^2\mu_{\text{A}}^2)(2e_d-e_d^2)]
\end{equation}

\noindent where $\eta_{\text{B}}$ is the transmittance in Bob-Charles channel, $x={\eta_{\text{A}} \over \eta_{\text{B}}}$ is the channel mismatch, $\eta_d$ is the detector efficiency, and $e_d$ is the misalignment. Here for convenience we can further define

\begin{equation}
\begin{aligned}
\epsilon &= 2e_d - e_d^2,  \,\,\,\,\,\,\,\,\,\,\,\, T = {\eta_{\text{B}}^2 \eta_d^2 \over 4}
\end{aligned}
\end{equation}

\noindent such that
\begin{equation}
Q_{\mu\mu}^X=T (2x\mu_{\text{A}}\mu_{\text{B}} + \epsilon\mu_{\text{B}}^2 +\epsilon x^2\mu_{\text{A}}^2)
\end{equation}

Now, let us consider  ${\partial Y_{11}^a} \over {\partial \mu_{\text{B}}}$ where

\begin{equation}
Y_{11}^{X,L} = Y_{11}^a = {1 \over (\mu_{\text{A}}-\nu_{\text{A}})} [{{\mu_{\text{A}}}\over{\nu_{\text{A}}\nu_{\text{B}}}}Q_{\nu\nu}^{M1}-{{\nu_{\text{A}}}\over{\mu_{\text{A}}\mu_{\text{B}}}}Q_{\mu\mu}^{M2}]
\end{equation}

To calculate the single-photon gain, the two terms:

\begin{equation}
\begin{aligned}
Q_{\nu\nu}^{M1} &= e^{\nu_{\text{A}}+\nu_{\text{B}}}Q_{\nu\nu}^X - e^{\nu_{\text{A}}}Q_{\nu\omega}^X - e^{\nu_{\text{B}}}Q_{\omega\nu}^X + Q_{\omega\omega}^X\\
Q_{\mu\mu}^{M2} &= e^{\mu_{\text{A}}+\mu_{\text{B}}}Q_{\mu\mu}^X - e^{\mu_{\text{A}}}Q_{\mu\omega}^X - e^{\mu_{\text{B}}}Q_{\omega\mu}^X + Q_{\omega\omega}^X\\
\end{aligned}
\end{equation}

\noindent are linear combinations of the observable Gains $Q_{ij}^Z$.

We can make the observation that, only the term
\begin{equation}
-{\nu_{\text{A}} \over {(\mu_{\text{A}}-\nu_{\text{A}})\mu_{\text{A}}}}\left({Q_{\mu\mu}^{M2}\over \mu_{\text{B}}}\right)
\end{equation}

\noindent contains $\mu_{\text{B}}$, so, we only need to prove the \textit{positivity} of ${\partial \over {\partial \mu_{\text{B}}}} ({Q_{\mu\mu}^{M2}\over \mu_{\text{B}}})$, where

\begin{equation}
\begin{aligned}
Q_{\mu\mu}^{M2} &= e^{\mu_{\text{A}}+\mu_{\text{B}}}Q_{\mu\mu}^X - e^{\mu_{\text{A}}}Q_{\mu\omega}^X - e^{\mu_{\text{B}}}Q_{\omega\mu}^X + Q_{\omega\omega}^X\\
&= e^{\mu_{\text{A}}+\mu_{\text{B}}}Q_{\mu\mu}^X - e^{\mu_{\text{A}}}Q_{\mu\omega}^X - e^{\mu_{\text{B}}}Q_{\omega\mu}^X\\
\end{aligned}
\end{equation}

\noindent substituting with Eq. C.49,

\begin{equation}
\begin{aligned}
{1\over T}Q_{\mu\mu}^{M2} &= (2x\mu_{\text{A}}\mu_{\text{B}}+x^2\epsilon \mu_{\text{A}}^2 + \epsilon \mu_{\text{B}}^2)e^{\mu_{\text{A}}+\mu_{\text{B}}} \\
&- x^2\epsilon \mu_{\text{A}}^2 e^{\mu_{\text{A}}} - \epsilon \mu_{\text{B}}^2 e^{\mu_{\text{B}}}\\
&= 2x\mu_{\text{A}}\mu_{\text{B}} e^{\mu_{\text{A}}+\mu_{\text{B}}} \\
&+ x^2\epsilon \mu_{\text{A}}^2 e^{\mu_{\text{A}}}(e^{\mu_{\text{B}}}-1) + \epsilon \mu_{\text{B}}^2 e^{\mu_{\text{B}}} (e^{\mu_{\text{A}}}-1)\\
\end{aligned}
\end{equation}

\noindent Therefore,

\begin{equation}
\begin{aligned}
{Q_{\mu\mu}^{M2} \over {\mu_{\text{B}}}} &= T \times [2x\mu_{\text{A}} e^{\mu_{\text{A}}+\mu_{\text{B}}} + x^2\epsilon \mu_{\text{A}}^2 e^{\mu_{\text{A}}}{{e^{\mu_{\text{B}}}-1}\over \mu_{\text{B}}} \\
&+ \epsilon(e^{\mu_{\text{A}}}-1) \mu_{\text{B}} e^{\mu_{\text{B}}}]\\
\end{aligned}
\end{equation}

\noindent note that here, as $\mu_{\text{A}}, \mu_{\text{B}}>0$, we have $e^{\mu_{\text{A}}}, e^{\mu_{\text{B}}} > 1$, and each of the three functions satisfy
\begin{equation}
\begin{aligned}
{\partial \over {\partial \mu_{\text{B}}}} (e^{\mu_{\text{A}}+\mu_{\text{B}}}) &>0 \\
{\partial \over {\partial \mu_{\text{B}}}} ({{e^{\mu_{\text{B}}}-1}\over \mu_{\text{B}}}) &>0 \\
{\partial \over {\partial \mu_{\text{B}}}} (\mu_{\text{B}} e^{\mu_{\text{B}}}) &>0 \\
\end{aligned}
\end{equation}

Therefore, we have proven that ${\partial \over {\partial \mu_{\text{B}}}} ({Q_{\mu\mu}^{M2}\over \mu_{\text{B}}}) >0$ and that ${{\partial Y_{11}^a} \over {\partial \mu_{\text{B}}}} <0$. Similarly, we can also prove that  ${{\partial Y_{11}^b} \over {\partial \mu_{\text{A}}}} <0$. Thus, the optimal point $(\mu_{\text{A}}^{opt},\mu_{\text{B}}^{opt})$ must happen on the boundary, i.e.
\begin{equation}
{\mu_{\text{A}}^{opt} \over \mu_{\text{B}}^{opt} } = { \nu_{\text{A}} \over \nu_{\text{B}}}
\end{equation}

\qedsymbol

\subsection{Proof of Discontinuity of Partial Derivatives for Decoy Intensities}

Now, to prove the discontinuity of the first-order derivatives of the key rate function, we need to show that\\

\textit{\textbf{Lemma II:} Partial derivative of $Y_{11}^a$ and $Y_{11}^b$ with respect to $\mu_{\text{A}}$, i.e. ${\partial Y_{11}^a} \over {\partial \mu_{\text{A}}}$ and ${\partial Y_{11}^b} \over {\partial \mu_{\text{A}}}$ are not equal.}

\textit{\textbf{Proof of Lemma II:}}

\begin{equation}
\begin{aligned}
{{\partial Y_{11}^a} \over {\partial \mu_{\text{A}}}} =  &-{\nu_{\text{A}} \over{\mu_{\text{A}}-\nu_{\text{A}}}} {{\partial} \over {\partial \mu_{\text{A}}}}({{Q_{\mu\mu}^{M2}}\over {\mu_{\text{A}} \mu_{\text{B}}}}) \\
&+ {\nu_{\text{A}} \over{(\mu_{\text{A}}-\nu_{\text{A}})^2}} ({{Q_{\mu\mu}^{M2}}\over {\mu_{\text{A}} \mu_{\text{B}}}}-{{Q_{\nu\nu}^{M1}}\over {\nu_{\text{A}} \nu_{\text{B}}}})\\
{{\partial Y_{11}^b} \over {\partial \mu_{\text{A}}}} =  &-{\nu_{\text{B}} \over{\mu_{\text{B}}-\nu_{\text{B}}}} {{\partial} \over {\partial \mu_{\text{A}}}}({{Q_{\mu\mu}^{M2}}\over {\mu_{\text{A}} \mu_{\text{B}}}})\\
\end{aligned}
\end{equation}

On the boundary of ${\mu_{\text{A}} \over \mu_{\text{B}}} = {\nu_{\text{A}} \over \nu_{\text{B}}}$, the first terms are equal, i.e.

\begin{equation}
-{\nu_{\text{A}} \over{\mu_{\text{A}}-\nu_{\text{A}}}} {{\partial} \over {\partial \mu_{\text{A}}}}({{Q_{\mu\mu}^{M2}}\over {\mu_{\text{A}} \mu_{\text{B}}}}) = -{\nu_{\text{B}} \over{\mu_{\text{B}}-\nu_{\text{B}}}} {{\partial} \over {\partial \mu_{\text{A}}}}({{Q_{\mu\mu}^{M2}}\over {\mu_{\text{A}} \mu_{\text{B}}}})
\end{equation}

\noindent However, here we have to show that the second term in ${{\partial Y_{11}^a} \over {\partial \mu_{\text{A}}}}$ is strictly larger than zero:

\begin{equation}
{\nu_{\text{A}} \over{(\mu_{\text{A}}-\nu_{\text{A}})^2}} ({{Q_{\mu\mu}^{M2}}\over {\mu_{\text{A}} \mu_{\text{B}}}}-{{Q_{\nu\nu}^{M1}}\over {\nu_{\text{A}} \nu_{\text{B}}}}) > 0
\end{equation}

\noindent or, since $\mu_{\text{A}} > \nu_{\text{A}}$ and $\mu_{\text{A}}, \nu_{\text{A}} >0$, simply

\begin{equation}
{{Q_{\mu\mu}^{M2}}\over {\mu_{\text{A}} \mu_{\text{B}}}}-{{Q_{\nu\nu}^{M1}}\over {\nu_{\text{A}} \nu_{\text{B}}}} > 0
\end{equation}

Just like in Appendix C.6.3, we can expand the gain $Q_{ij}^X$ using Eq. C.49:

\begin{equation}
\begin{aligned}
{{Q_{\nu\nu}^{M1}}\over {\nu_{\text{A}} \nu_{\text{B}}}} =  2x e^{\nu_{\text{A}}+\nu_{\text{B}}} &+ x^2\epsilon {\nu_{\text{A}}\over \nu_{\text{B}}} e^{\nu_{\text{A}}}(e^{\nu_{\text{B}}}-1) \\
&+ \epsilon {\nu_{\text{B}} \over \nu_{\text{A}}} e^{\nu_{\text{B}}} (e^{\nu_{\text{A}}}-1)\\
{{Q_{\mu\mu}^{M2}}\over {\mu_{\text{A}} \mu_{\text{B}}}} =  2x e^{\mu_{\text{A}}+\mu_{\text{B}}} &+ x^2\epsilon {\mu_{\text{A}}\over \mu_{\text{B}}} e^{\mu_{\text{A}}}(e^{\mu_{\text{B}}}-1) \\
&+ \epsilon {\mu_{\text{B}} \over \mu_{\text{A}}} e^{\mu_{\text{B}}} (e^{\mu_{\text{A}}}-1)\\
\end{aligned}
\end{equation}

\noindent Subtracting them, we can acquire:
\begin{equation}
\begin{aligned}
{{Q_{\mu\mu}^{M2}}\over {\mu_{\text{A}} \mu_{\text{B}}}}-{{Q_{\nu\nu}^{M1}}\over {\nu_{\text{A}} \nu_{\text{B}}}} &=  2x (e^{\mu_{\text{A}}+\mu_{\text{B}}}-e^{\nu_{\text{A}}+\nu_{\text{B}}}) \\
&+ x^2\epsilon (\mu_{\text{A}} e^{\mu_{\text{A}}} {{e^{\mu_{\text{B}}}-1}\over \mu_{\text{B}}} - \nu_{\text{A}} e^{\nu_{\text{A}}} {{e^{\nu_{\text{B}}}-1}\over \nu_{\text{B}}})\\
&+ \epsilon (\mu_{\text{B}} e^{\mu_{\text{B}}} {{e^{\mu_{\text{A}}}-1}\over \mu_{\text{A}}} - \nu_{\text{B}} e^{\nu_{\text{B}}} {{e^{\nu_{\text{A}}}-1}\over \nu_{\text{A}}})\\
\end{aligned}
\end{equation}

\noindent Note that, when a given variable $x>0$, the functions

\begin{equation}
\begin{aligned}
{d \over {d x}} (e^x) &>0 \\
{d \over {d x}} ({{e^{x}-1}\over x}) &>0 \\
{d \over {d x}} (x e^{x}) &>0 \\
\end{aligned}
\end{equation}

\noindent Therefore, these three functions strictly increase with their variable $x$, i.e. for any $x_1>x_2$, $f(x_1)>f(x_2)$. Now, we can use the conditions $\mu_{\text{A}} > \nu_{\text{A}}$, $\mu_{\text{B}} > \nu_{\text{B}}$, and acquire:

\begin{equation}
\begin{aligned}
e^{\mu_{\text{A}}+\mu_{\text{B}}} &> e^{\nu_{\text{A}}+\nu_{\text{B}}}\\
\mu_{\text{A}} e^{\mu_{\text{A}}} &> \nu_{\text{A}} e^{\nu_{\text{A}}}\\
\mu_{\text{B}} e^{\mu_{\text{B}}} &> \nu_{\text{B}} e^{\nu_{\text{B}}}\\
{{e^{\mu_{\text{A}}}-1}\over \mu_{\text{A}}} &> {{e^{\nu_{\text{A}}}-1}\over \nu_{\text{A}}}\\
{{e^{\mu_{\text{B}}}-1}\over \mu_{\text{B}}} &> {{e^{\nu_{\text{B}}}-1}\over \nu_{\text{B}}}\\
\end{aligned}
\end{equation}

Therefore, we have proven that ${{Q_{\mu\mu}^{M2}}\over {\mu_{\text{A}} \mu_{\text{B}}}}-{{Q_{\nu\nu}^{M1}}\over {\nu_{\text{A}} \nu_{\text{B}}}} > 0$, i.e.

\begin{equation}
{{\partial Y_{11}^a} \over {\partial \mu_{\text{A}}}} \neq {{\partial Y_{11}^b} \over {\partial \mu_{\text{A}}}}
\end{equation}

Similarly, one can show that

\begin{equation}
{{\partial Y_{11}^b} \over {\partial \mu_{\text{B}}}} \neq {{\partial Y_{11}^a} \over {\partial \mu_{\text{B}}}}
\end{equation}

Therefore, for any given intensities $(\nu_{\text{A}},\nu_{\text{B}})$, the rate function $R(\mu_{\text{A}},\mu_{\text{B}})$ is not smooth against the two intensities $(\mu_{\text{A}},\mu_{\text{B}})$. \qedsymbol\\

\textit{\textbf{Remark:}} Also, though not explicitly proven here - since $\nu_{\text{A}}, \nu_{\text{B}}$ will affect not only $Y_{11}^{X,L}$, but will affect $e_{11}^{X,U}$ too, their derivatives will be a lot more complex than $\mu_{\text{A}}, \mu_{\text{B}}$ - numerically we observed that for any given $(\mu_{\text{A}},\mu_{\text{B}})$, the rate function $R(\nu_{\text{A}},\nu_{\text{B}})$ is actually not smooth against the two intensities $(\nu_{\text{A}},\nu_{\text{B}})$ either, and the ridge still appears at ${\mu_{\text{A}} \over \mu_{\text{B}}} = {\nu_{\text{A}} \over \nu_{\text{B}}}$.

\section{Local Search Algorithm}

In this section we describe how to perform the optimization for the parameters, which is an indispensable process in obtaining the optimal key rate. In addition, we also discuss the effect of inaccuracies and fluctuations of the intensities and probabilities on the key rate, and show that our method is robust even in the presence of inaccuracies and fluctuations of the parameters.

To provide a good key rate under finite-size effects, the optimal choice of parameters is an extremely important factor in implementing the protocol. However, the 7-intensity protocol has an extremely large parameter space of 12 dimensions, for which a brute-force search is next to impossible. To put into context, a desktop PC (quad-core i7-4790k@4.0GHz) can evaluate the function $R(\vec{v})$ at approximately $10^5$ parameter combinations $\vec{v}$ per second. But searching over a very crude 10-sample resolution for each parameter would take over 4 months, and a 100-sample resolution for each parameter would take $3 \times 10^{11}$ years, a time longer than the age of the universe! Therefore, a local search algorithm must be used to efficiently search the parameters in a reasonable time.

There have been studies to apply convex optimization to QKD e.g. in Ref. \cite{mdiparameter} to find the optimal set of parameters and in Refs. \cite{convex1,convex2,convex3} to bound the information leakage and secure key rate. Here we start by adopting a local search algorithm for parameter optimization, proposed in Ref. \cite{mdiparameter}, called ``coordinate descent" (CD), which requires drastically less time than using an exhaustive search. Instead of performing an exhaustive search over the parameter space, we can descend along each axis at a time, and iterate over each axis in turn. For instance, suppose we currently iterate $s_{\text{A}}$:

\begin{equation}
\begin{aligned}
R^{i+1}=\max_{s_{\text{A}} \in ({s_{\text{A}}}^{min},{s_{\text{A}}}^{max})} &R(s_{\text{A}}, \mu_{\text{A}}^i, \nu_{\text{A}}^i, P_{s_{\text{A}}}^i, P_{\mu_{\text{A}}}^i, P_{\nu_{\text{A}}}^i,\\
&s_{\text{B}}^i, \mu_{\text{B}}^i, \nu_{\text{B}}^i, P_{s_{\text{B}}}^i, P_{\mu_{\text{B}}}^i, P_{\nu_{\text{B}}}^i) \\
= &R(s_{\text{A}}^{i+1}, \mu_{\text{A}}^i, \nu_{\text{A}}^i, P_{s_{\text{A}}}^i, P_{\mu_{\text{A}}}^i, P_{\nu_{\text{A}}}^i, \\
&s_{\text{B}}^i, \mu_{\text{B}}^i, \nu_{\text{B}}^i, P_{s_{\text{B}}}^i, P_{\mu_{\text{B}}}^i, P_{\nu_{\text{B}}}^i)
\end{aligned}
\end{equation}

\noindent which freezes the other coordinates, and replaces $s_{\text{A}}$ with the optimal position on the current coordinate-axis $s_{\text{A}}$. In the next iteration the algorithm will descend along axis $\mu_{\text{A}}$, etc., hence the name coordinate descent. The search space satisfies that: the probabilities lie within $(0,1)$, and while the intensities could be in principle larger than 1, typically that doesn't provide a good key rate, so here we also define the domain for all intensities as $(0,1)$. The decoy intensities also follow two additional constraints $\mu_{\text{A}} > \nu_{\text{A}}$ and $\mu_{\text{B}} > \nu_{\text{B}}$. The CD algorithm is able to reach the same optimal position as a gradient descent algorithm (with descends along the gradient vector), the commonly used approach for parameter optimization.

However, a significant limitation of coordinate descent is that it does not work correctly over functions that have discontinuous first-order derivatives (which cause the gradient to be non-defined). For instance, in the presence of a sharp ``ridge" as in Fig.2 in the main text, any arbitrary point $P$ on the ridge will cause the CD algorithm to terminate incorrectly and fail to find the maximum point. Mathematically, this is caused by the gradient being not clearly defined at a position where derivatives are discontinuous. Therefore, coordinate descent does not work anymore for asymmetric MDI-QKD.

As we discussed above, such discontinuity of derivatives comes from the ``ridge", ${\mu_{\text{A}} \over \mu_{\text{B}}}={\nu_{\text{A}} \over \nu_{\text{B}}}$. Moreover, we know that the optimal parameters must satisfy ${\mu_{\text{A}}^{opt} \over \mu_{\text{B}}^{opt} } = { \nu_{\text{A}}^{opt} \over \nu_{\text{B}}^{opt}}$. Therefore, here we propose to use polar coordinate instead of Cartesian coordinate to perform coordinate descent, and \textit{jointly} search ${\mu_{\text{A}} \over \mu_{\text{B}}}$ and ${\nu_{\text{A}} \over \nu_{\text{B}}}$. In this way, we can make the rate vs parameter function smooth. We redefine $\vec{v}$ as:

\begin{equation}
\vec{v_{polar}}=[s_{\text{A}}, s_{\text{B}}, r_\mu, r_\nu, \theta_{\mu\nu}, P_{s_{\text{A}}}, P_{\mu_{\text{A}}},P_{\nu_{\text{A}}},P_{s_{\text{B}}},P_{\mu_{\text{B}}},P_{\nu_{\text{B}}}]
\end{equation}

\noindent where
\begin{equation}
\begin{aligned}
r_\mu&=\sqrt{\mu_{\text{A}}^2+\mu_{\text{B}}^2}  \,\,\,\,\,\,\,\,\,\,\,\,        r_\nu=\sqrt{\nu_{\text{A}}^2+\nu_{\text{B}}^2} \\
\theta_{\mu\nu}&=tan^{-1}(\mu_{\text{A}}/\mu_{\text{B}})=tan^{-1}(\nu_{\text{A}}/\nu_{\text{B}})\\
\end{aligned}
\end{equation}

\noindent In this way, the expression of $Y_{11}^{L}$ always takes the boundary value (and only has a single expression). Therefore, when other parameters are fixed, $R(\theta_{\mu\nu})$ is actually a smooth function, therefore by searching over the parameters $\vec{v_{polar}}$, we can successfully find the optimal parameters and maximum rate.

After converting to polar coordinates and jointly searching $\theta_{\mu\nu}$, the coordinate descent algorithms becomes:

\begin{equation}
\begin{aligned}
R^{i+1}=\max_{s_{\text{A}} \in ({s_{\text{A}}}^{min},{s_{\text{A}}}^{max})} &R(s_{\text{A}}, s_{\text{B}}^i, r_\mu^i, r_\nu^i, \theta_{\mu\nu}^i,\\
& P_{s_{\text{A}}}^i, P_{\mu_{\text{A}}}^i,P_{\nu_{\text{A}}}^i,P_{s_{\text{B}}}^i,P_{\mu_{\text{B}}}^i,P_{\nu_{\text{B}}}^i) \\
= &R(s_{\text{A}}^{i+1}, s_{\text{B}}^i, r_\mu^i, r_\nu^i, \theta_{\mu\nu}^i,\\
& P_{s_{\text{A}}}^i, P_{\mu_{\text{A}}}^i,P_{\nu_{\text{A}}}^i,P_{s_{\text{B}}}^i,P_{\mu_{\text{B}}}^i,P_{\nu_{\text{B}}}^i) \\
\end{aligned}
\end{equation}

{\color{black}
	Additionally, when searching along each coordinate (for instance, fixing other parameters and searching $s_{\text{A}}$), we employ an iterative searching technique to further accelerate the algorithm, which starts out with a coarse resolution and iteratively narrows the search region while increasing the resolution (this is a similar technique as introduced in Ref. \cite{mdiparameter}, but efficiently parallelized to utilize multi-threading on modern PCs). For instance, we can start out with e.g. 100 samples within the $(0,1)$ region and evaluate them in parallel. After the maximal point is found, we can then choose two neighbouring samples on the left and right of the maximal point, and start a finer search among 10 more samples between them. This process can be iterated until maximum value no longer changes significantly, or until the maximum depth is reached. Such technique allows a search resolution that dynamically changes as needed (from $10^{-2}$ down to even $10^{-5}$, although in practice often $10^{-3}$ is sufficient), and it efficiently uses e.g. the 8 threads on a quad-core CPU, enabling fast and accurate optimization below 0.1s.
	
	One more note is that, the key rate obtained by our method is in fact robust against small inaccuracies in the parameters. For instance, for Point A3 (10km, 60km) in Table 5.3, if we round all parameters to an accuracy of 0.001 (as shown in Table 5.4) and use it for simulation, we can still get $99.5\%$ of the optimal key rate $3.106\times 10^{-5}$, while rounding the parameters to 0.01 will still give us $93.0\%$ of the optimal key rate. In fact, even if we just keep one significant digit of each parameter, we can still get $47.6\%$ of the optimal key rate. This would make it much easier for an experimental implementation of our method, as the key rate is very forgiving of inaccuracies in the parameters, which makes a much less stringent requirement on the intensity modulators and random number generators.
	
	Note that, the above ``accuracy" discusses how strict the requirement is for us to generate an intensity/probability with its mean value close to the desired optimal value (e.g. limited by bits in the random number generator or the accuracy of the intensity modulator), but we are still assuming we have perfect knowledge of the variables we generate. In addition, here we would like to point out that our conclusions remain unchanged, even in the presence of intensity fluctuations, or imprecision in the intensity probabilities. 
	
	Firstly, the system is not very sensitive to the probabilities (since the partial derivatives with respect to them are zero at the optimal points), so even if all signal and decoy probabilities are simultaneously set 5\% away from the optimal value (and we take the global worst-case key rate value among all possible combinations of positive/negative deviation for each variable), the key rate will not significantly drop - for instance for the (10km, 60km) case, one can still obtain 92.3\% the ideal key rate ($2.869\times 10^{-5}$ versus $3.106\times 10^{-5}$) even with a 5\% deviation for the probabilities. 
	
	Similarly, for intensity fluctuations, even if we add a 5\% deviation to \textit{all} intensities (again, taking the (10km, 60km) case as an example) we can still get 73.1\% the ideal key rate ($2.270\times 10^{-5}$ versus $3.106\times 10^{-5}$). Moreover, one important point to note is that, intensity fluctuation is not a problem unique to asymmetric MDI-QKD (or the new asymmetric protocol that we propose in this work). Even if one uses prior protocols (such as the 4-intensity protocol), one would still obtain a significantly lower key rate if taking intensity fluctuation into consideration, such as 39.9\% the key rate ($3.671\times 10^{-5}$ versus $9.206\times 10^{-6}$ with no fluctuation) at (0km, 50km), and zero key rate (versus $3.891\times 10^{-7}$ with no fluctuation) at (10km, 60km). Therefore, the advantage of our method remains unchanged, even if intensity fluctuations are considered.
	
}
\section{Finite Size analysis}

In this section we describe the finite-key analysis used in our simulations.

The analytical proofs in Appendix C.6 are shown for the asymptotic case. Numerically we show that 7-intensity protocol works effectively in the finite-key regime too, as can be observed in the main text Fig. \ref{fig:2d_Results}.

To account for finite-size effects, we perform a standard error analysis\cite{mdifourintensity,mdiparameter}, and estimate the expected value $\langle n \rangle$ of an observable $n$ by

\begin{equation}
\begin{aligned}
\underline{n} = n - \gamma \sqrt{n} \leq \langle n \rangle \leq  n + \gamma \sqrt{n} = \overline{n}
\end{aligned}
\end{equation} 

\noindent where we define the upper and lower bound for an observable $n$ as $\overline{n}$ and $\underline{n}$. Here, $\gamma$ is the number of standard deviations the confidence interval of the observed value is from the expected value (for instance, for a required failure probability of no more than $\epsilon=10^{-7}$, we should set $\gamma=5.3$).

We can denote the observed counts as $n_{\mu_i,\mu_j}^X$, and error counts as $m_{\mu_i,\mu_j}^X$, where $\mu_i \in \{\mu_{\text{A}}, \nu_{\text{A}}, \omega\}$, $\mu_j \in \{\mu_{\text{B}}, \nu_{\text{B}}, \omega\}$. Then, the observed gain and error can be acquired from:

\begin{equation}
\begin{aligned}
Q_{\mu_i,\mu_j}^X &= {{n_{\mu_i,\mu_j}^X} \over {NP_{\mu_i}P_{\mu_j}}} \\
T_{\mu_i,\mu_j}^X &= {{m_{\mu_i,\mu_j}^X} \over {NP_{\mu_i}P_{\mu_j}}} \\
E_{\mu_i,\mu_j}^X &= {{T_{\mu_i,\mu_j}^X} \over {Q_{\mu_i,\mu_j}^X}} \\
\end{aligned}
\end{equation}

\noindent where $N$ is the total number of signals sent, and $P_{\mu_i},P_{\mu_j}$ are the probabilities for Alice and Bob to send the respective intensities. Note that here we define the QBER in terms of error-gains: 

\begin{equation}
T_{\mu_i,\mu_j}^X = Q_{\mu_i,\mu_j}^X E_{\mu_i,\mu_j}^X
\end{equation}

As described in Appendix C.6, we can define the key rate expression as \cite{mdiqkd,mdifourintensity}:
\begin{equation}
\begin{aligned}
R=P_s P_s \{s_{\text{A}} s_{\text{B}} e^{-(s_{\text{A}}+s_{\text{B}})}Y_{11}^{X,L}[1-h_2(e_{11}^{X,U})]\\
-f_e Q_{ss}^Z h_2(E_{ss}^Z)\}
\end{aligned}
\end{equation} 

\noindent and the single-photon gain and error estimated by \cite{mdipractical}:

\begin{equation}
\begin{aligned}
Y_{11}^{X,L}= {1\over {\mu_{\text{A}}-\nu_{\text{A}}}} ({{\mu_{\text{A}}}\over {\nu_{\text{A}}\nu_{\text{B}}}} Q_{\nu\nu}^{M1} - {{\nu_{\text{A}}}\over {\mu_{\text{A}}\mu_{\text{B}}}} Q_{\mu\mu}^{M2})\\
e_{11}^{X,U} = {1\over {\nu_{\text{A}}\nu_{\text{B}}Y_{11}^{X,L}}}(e^{\nu_{\text{A}}+\nu_{\text{B}}}T_{\nu\nu}-e^{\nu_{\text{A}}}T_{\nu\omega}\\
-e^{\nu_{\text{B}}}T_{\omega\nu}+T_{\omega\omega})\\
\end{aligned}
\end{equation}

\noindent where $Q_{\nu\nu}^{M1}, Q_{\mu\mu}^{M2}$ are linear combination terms of the observables

\begin{equation}
\begin{aligned}
Q_{\nu\nu}^{M1} = e^{\nu_{\text{A}}+\nu_{\text{B}}}Q_{\nu\nu}^X - e^{\nu_{\text{A}}}Q_{\nu\omega}^X - e^{\nu_{\text{B}}}Q_{\omega\nu}^X + Q_{\omega \omega}^X\\
Q_{\mu\mu}^{M2} = e^{\mu_{\text{A}}+\mu_{\text{B}}}Q_{\mu\mu}^X - e^{\mu_{\text{A}}}Q_{\mu\omega}^X - e^{\mu_{\text{B}}}Q_{\omega\mu}^X + Q_{\omega \omega}^X\\
\end{aligned}
\end{equation}

Now, with standard error analysis, we can define the upper and lower bounds for the gain and error-gain:

\begin{equation}
\begin{aligned}
\overline{Q_{\mu_i\mu_j}^X}=Q_{\mu_i\mu_j}^X + \gamma \sqrt{Q_{\mu_i\mu_j}^X \over {N P_{\mu_i} P_{\mu_j}}} \\
\underline{Q_{\mu_i\mu_j}^X}=Q_{\mu_i\mu_j}^X - \gamma \sqrt{Q_{\mu_i\mu_j}^X \over {N P_{\mu_i} P_{\mu_j}}} \\
\overline{T_{\mu_i\mu_j}^X}=T_{\mu_i\mu_j}^X + \gamma \sqrt{T_{\mu_i\mu_j}^X \over {N P_{\mu_i} P_{\mu_j}}} \\
\underline{T_{\mu_i\mu_j}^X}=T_{\mu_i\mu_j}^X - \gamma \sqrt{T_{\mu_i\mu_j}^X \over {N P_{\mu_i} P_{\mu_j}}} \\
\end{aligned}
\end{equation}

\noindent 

\noindent Therefore, we have

\begin{equation}
\begin{aligned}
\underline{Q_{\nu\nu}^{M1}} = e^{\nu_{\text{A}}+\nu_{\text{B}}}\underline{Q_{\nu\nu}^X} - e^{\nu_{\text{A}}}\overline{Q_{\nu\omega}^X} - e^{\nu_{\text{B}}}\overline{Q_{\omega\nu}^X} + \underline{Q_{\omega \omega}^X}\\
\overline{Q_{\mu\mu}^{M2}} = e^{\mu_{\text{A}}+\mu_{\text{B}}}\overline{Q_{\mu\mu}^X} - e^{\mu_{\text{A}}}\underline{Q_{\mu\omega}^X} - e^{\mu_{\text{B}}}\underline{Q_{\omega\mu}^X} + \underline{Q_{\omega \omega}^X}\\
Y_{11}^{X,L}= {1\over {\mu_{\text{A}}-\nu_{\text{A}}}} ({{\mu_{\text{A}}}\over {\nu_{\text{A}}\nu_{\text{B}}}} \underline{Q_{\nu\nu}^{M1}} - {{\nu_{\text{A}}}\over {\mu_{\text{A}}\mu_{\text{B}}}} \overline{Q_{\mu\mu}^{M2}}) \\
e_{11}^{X,U} = {1\over {\nu_{\text{A}}\nu_{\text{B}}Y_{11}^{X,L}}}(e^{\nu_{\text{A}}+\nu_{\text{B}}}\overline{T_{\nu\nu}}-e^{\nu_{\text{A}}}\underline{T_{\nu\omega}}\\
-e^{\nu_{\text{B}}}\underline{T_{\omega\nu}}+\overline{T_{\omega\omega}})\\
\end{aligned}
\end{equation}

\noindent which we can use to substitute into Eq. C.75 to obtain the key rate under finite-size effects. (Note that here $Q_{\omega \omega}^X$ takes the lower bound in both $\underline{Q_{\nu\nu}^{M1}}$ and $\overline{Q_{\mu\mu}^{M2}}$, because its overall coefficient is positive in $Y_{11}^{X,L}$).

\begin{table*}[t]
	\caption{{\color{black}Simulation results of key rate estimated with independent-bounds versus joint-bounds, using parameters in Table 5.2. The data points for independent-bounds correspond to the solid red curve in Fig.\ref{fig:2d_Results} (d). As can be seen, using joint-bounds for finite-size estimation can improve the key rate significantly. However, this will result in multiple maxima and cause instabilities in simulations. Therefore, we have used independent-bounds throughout the main text.}}
	\begin{center}
		{\color{black}
			\begin{tabular}{cccc}            
				\hline \hline
				$L_{\text{A}}$ & $L_{\text{B}}$ & $R_{\text{independent}}$ & $R_{\text{joint}}$\\
				\hline
				$60km$ & $10km$ & $3.106\times 10^{-5}$ & $6.714\times 10^{-5}$ \\
				$100km$ & $50km$ & $4.677\times 10^{-11}$ & $7.568\times 10^{-8}$ \\
				$113km$ & $63km$ & $0$ & $7.311\times 10^{-10}$ \\
				\hline \hline
				
			\end{tabular}
		}
	\end{center}
\end{table*}

\begin{figure}[h]
	\includegraphics[scale=0.4]{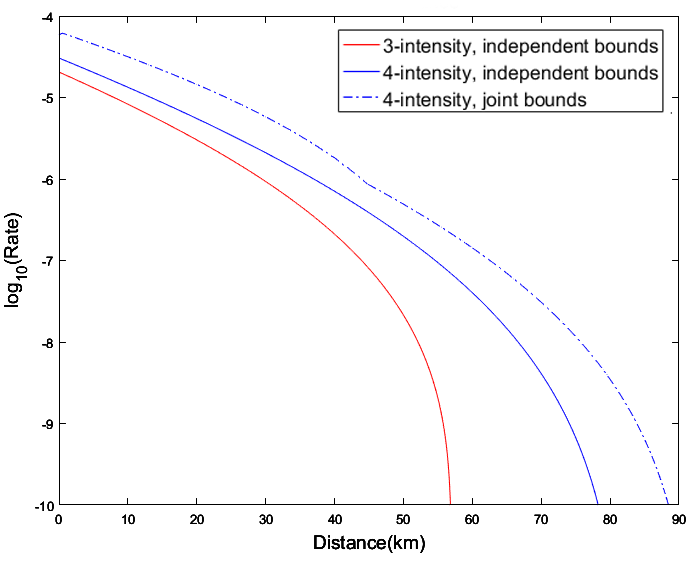}
	\caption{Rate vs distance (Alice to Bob) for symmetric case, for $N=10^{11}$ using parameters $Y_0=6.02\times 10^{-6},\eta_d=14.5\%,e_d=1.5\%$, a parameter set in Zhou et al.'s paper \cite{mdifourintensity}. Here we compare the traditional 3-intensity protocol as proposed in Ref.\cite{mdiparameter} (red solid line), and the 4-intensity protocol\cite{mdifourintensity} with independent-bound (blue solid line) and joint-bound analysis (blue dot-dash line). Reproduced from \cite{this_asymMDI} @2019 APS.}
	\label{fig:rateN11}
\end{figure}

Note that, in Ref. \cite{mdifourintensity}, in addition to proposing the 4-intensity protocol, Zhou et al. have proposed a ``joint-bounds" finite-key analysis, which jointly considers the statistical fluctuations of observable Gain and QBER. It is a tighter bound and can provide a higher rate than considering each observable's fluctuation independently as we've discussed above in this section (i.e. using ``independent-bounds"). To illustrate this, we perform a simple simulation of key rate versus distance plot, using independent-bounds and joint-bounds (as well as using traditional 3-intensity protocol \cite{mdiparameter} for comparison). As can be seen in Fig.\ref{fig:rateN11}, 4-intensity protocol with joint-bounds analysis provides a higher rate than independent-bounds (and both have higher rates than the 3-intensity protocol). 

However, joint-bound analysis is based on linear optimization and sometimes brings multiple maxima for $R(\vec{v})$, which is undesirable for local search, and will result in unpredictable behaviours (such as sudden ``jitters" in the resulting rate versus distance plot, as can be observed in the joint-bound plot in Fig.\ref{fig:rateN11}. Similar behavior is observed in Ref.\cite{mdifourintensity} too). 

{\color{black}
	Here just for comparison, we list in Table C.1 some example data points where we apply both independent-bound and joint-bound analysis. As can be seen, using joint-bounds, we can indeed gain a further improved key rate. However, this comes at the expense of not knowing whether we are indeed at the global maximum or not, due to the existence of multiple maxima (and is not ideal for comparing asymmetric/symmetric protocols, as the key rate estimated could be just local maxima for both of them). Therefore, as the purpose of this work is to study asymmetric MDI-QKD, we focus on independent-bounds throughout the main text. 
}

Also, note that although we have used standard error-analysis for simplicity, our method here can in principle be applied to finite-key analysis with composable security, too, such as using Chernoff bound \cite{mdiChernoff}. The key point is that (as explicitly demonstrated in Appendices B and C), the scaling of asymmetric MDI-QKD key rate versus distances depends on the signal states (which performs a trade-off between error-correction and single photon probability). The decoy states need to maintain balanced arriving intensities at Charles, but only serve to estimate the single-photon contributions as accurately as possible, whose asymptotic bounds are given by the infinite-data, infinite-decoy case. Adopting different finite-key analysis (or no analysis at all, as in asymptotic case) affects the bounds on single photon gain and QBER $Y_{11}^L$ and $e_{11}^U$. The finite-size case can be seen as the asymptotic case with correction terms (i.e. imperfections) added to the privacy amplification, but its key rate will have a similar scaling property as the asymptotic case. This means that the advantage of our method is independent of the finite-size analysis model used (or lack thereof, in the asymptotic case).

\section{Single-Arm MDI-QKD}
\begin{table*}[t]
	\caption{Simulation results of key rate between each pair of nodes in a MDI-QKD network, using parameters from Table 5.2, $N=10^{11}$, and channels in main text Fig. 1(a). As can be seen, using 7-intensity protocol always provides higher rate than either using 4-intensity directly (which fails to establish some connections) or using 4-intensity after adding fibre to each channel to accommodate the longest channel (which results in an identical low rate for every connection - since every channel equals the longest channel after adding fibre). 7-intensity protocol therefore enables high scalability and reconfigurability because each link is independent of other links and no added fibre is needed.}
	\begin{center}
		\begin{tabular}{ccccccc}            
			\hline \hline
			Method & $A_1$-$A_3$ & $A_1$-$A_4$ & $A_1$-$A_5$ \\
			\hline
			4-intensity, add fibre & $1.28\times 10^{-10}$& $1.28\times 10^{-10}$& $1.28\times 10^{-10}$\\
			4-intensity, direct & 0 & 0 & 0\\
			7-intensity, direct &$1.97\times 10^{-7}$& $2.42\times 10^{-7}$& $2.77\times 10^{-7}$ \\
			\hline \hline
			Method & $A_3$-$A_4$ & $A_3$-$A_5$ & $A_4$-$A_5$\\
			\hline
			4-intensity, add fibre & $1.28\times 10^{-10}$& $1.28\times 10^{-10}$& $1.28\times 10^{-10}$\\
			4-intensity, direct & $2.41\times 10^{-4}$& $3.22\times 10^{-4}$& $5.77\times 10^{-4}$\\
			7-intensity, direct & $2.48\times 10^{-4}$& $3.53\times 10^{-4}$& $5.87\times 10^{-4}$\\
			\hline \hline
			
		\end{tabular}
	\end{center}
\end{table*}

\begin{figure}[h]
	\includegraphics[scale=0.3]{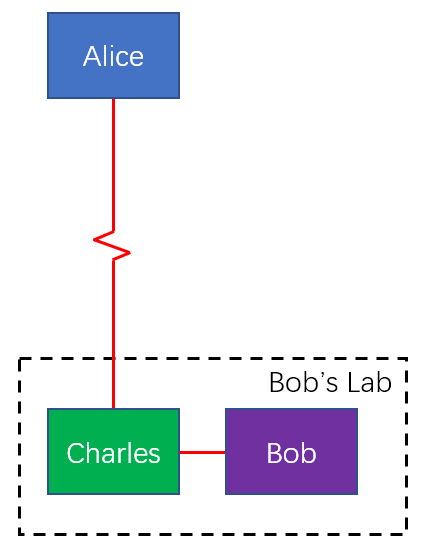}
	\caption{"Single-arm" MDI-QKD where Bob and Charles are both in the same lab, with Bob's channel having as little loss as possible. By optimizing intensities, we can achieve maximum distance (loss) in the single channel between Alice and Charles, while enjoying the security of MDI-QKD. Reproduced from \cite{this_asymMDI} @2019 APS.}
	\label{fig:zero_arm}
\end{figure}

\begin{figure}[h]
	\includegraphics[scale=0.4]{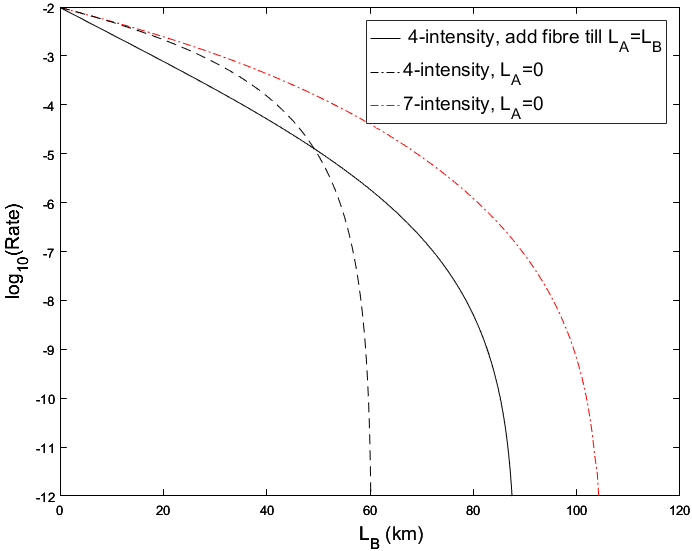}
	\caption{Simulations of ``single-arm" MDI-QKD. We use parameters from Table 5.2, and set $N=10^{11}$. The three lines are generated using 4-intensity protocol and adding fibre until $L_{\text{A}}=L_{\text{B}}$ (black solid line), using 4-intensity  protocol but without being able to add fibre (black dashed line), and using 7-intensity protocol directly (red dot-dash line). As can be seen, using 7-intensity protocol tremendously increases the key rate and maximum distance for the longer single-arm. At $R=10^{-7}$, using 7-intensity protocol (having maximum distance at 90km) increases maximum distance by 17.5 or 33.2km (or, 3.5 to 6.6dB of loss) compared to 4-intensity with/without fibre, respectively. Reproduced from \cite{this_asymMDI} @2019 APS.}
	\label{fig:singlearm}
\end{figure}

In the main text we have proposed a new type of ``single-arm" MDI-QKD setup, which is the extremely asymmetric case where one channel has high loss while the other channel has close to zero loss. In this section we will describe it in more detail and outline its potential applications.

Suppose we have one crucial channel (e.g. a free-space channel, say in a satellite-ground connection, or a ship-to-ship connection) through which we would like to send quantum signals. We would like to prevent all attacks on the detector and improve the security with MDI-QKD, but cannot add a third party in the middle of the free-space channel. In this case, it is possible to add another source Bob in the laboratory (alongside Charles' detectors, with as a small loss as possible in Bob-Charles channel), and use it to interfere with the signals coming from Alice over the longer free-space channel, as shown in Fig.\ref{fig:zero_arm}. With 7-intensity protocol, a high key rate can be generated from this extremely asymmetric case, providing the security of MDI-QKD to a single channel where relays cannot be added while still maintaining good performance.

If one uses 4-intensity protocol, Bob has to add a fibre similar in loss to that of the free-space channel (to maintain the symmetry), while as we've shown with 7-intensity protocol, Bob can simply choose as small a loss as possible, and obtain maximum acceptable loss in Alice's channel. Not only does 7-intensity protocol make such a highly asymmetric MDI-QKD possible, it actually provides a higher rate compared to the symmetric case (if Bob adds a fibre). Moreover, since Alice's channel loss might be constantly changing, it can be very difficult to adjust an added fibre and maintain the symmetry; thus, the convenience of not having to add any loss with 7-intensity protocol is a significant factor, too.

As we can observe in main text Fig. \ref{fig:2d_Results}(a)(b), for the same required minimum rate, rather than performing an experiment at $(L_{max},L_{max})$, if we are free to adjust one channel (and want maximum distance in the other channel), we can set the shorter channel to zero, and obtain a longer distance in the other channel, e.g. $(L_{max}',0)$ with $L_{max}'>L_{max}$. For instance, in main text Fig. \ref{fig:2d_Results}(a)(b), choosing point $B(102km,0km)$ can extend the longer arm from 85km to 102km, from the symmetric point $A(85km,85km)$ for the same $R=10^{-10}$.

Here we list the simulations results for single-arm MDI-QKD. To demonstrate the advantage, here we study three-cases: using 4-intensity (but being able to add fibre until channels are symmetric), using 4-intensity (however, due to being e.g. in a free-space channel or a dynamic network, without the luxury to add fibres and compensate for the channels), and using 7-intensity directly on the asymmetric channels. As can be seen in Fig.\ref{fig:singlearm}, 7-intensity protocol provides better performance than both strategies using 4-intensity, and increases the maximum distance from 56.8km and 72.5km (respectively for adding/not adding fibre) to 90km. Thus, our new protocol can enable a unique new application of providing the security of MDI-QKD to a single channel where relays cannot be added (e.g. a free-space link), while still maintaining high key rate.\\

\section{MDI-QKD Network Numerical Results}

In this section we consider the channels from a real quantum network setup in Vienna, reported in Ref.\cite{quantumnetwork1}, and numerically show that using 7-intensity protocol can provide high-rate communication between each pair of users, while previous protocols either fail to establish some connections in the network, or suffer from a low key rate for all connections.

Here, we focus here on the high-asymmetry nodes in Ref.\cite{quantumnetwork1}, $A_1,A_2,A_3,A_4,A_5$, plotted in main text Fig. 1(a), and consider the case where an untrusted relay is placed at $A_2$. The topology here is a commonly studied model of a star-type network, which is considered for QKD networks in \cite{starnetwork1,starnetwork2}, and is also the model for the MDI-QKD network experiment in Ref.\cite{mdinetwork}. Such a network can provide a complete graph of connections between any two users, but only requires one physical connection from each user. We show the simulation results in Table C.2, where using 7-intensity protocol consistently provides high-rate connections even for nodes with very high asymmetry, and maintains the same (in fact moderately higher) key rate for nodes that are near-symmetric, i.e. including a long channel doesn't affect the rate between pairs of existing shorter channels.

Being able to establish connections with arbitrarily placed new nodes without affecting existing nodes is a very important property for a protocol to be used in a scalable and reconfigurable network, whose links will obviously be, more often than not, asymmetric. For the 4-intensity protocol, to accommodate the highest-loss channel, all connections will suffer from non-optimal key rate. Moreover, since new users might be added/deleted dynamically, such adding-fibre strategy will have poor scalability, since each new node affects the performance of all existing nodes, and also causes interruption of service when users update their fibres. With 7-intensity protocol, we are completely free of the worries of asymmetry, and can directly use the protocol on any channel combination optimally, so each node can be added/deleted without affecting the rest. This greatly improves not only the key rate, but also the scalability of a MDI-QKD network.


\chapter{Supplemental Information on Asymmetric TF-QKD}

This Appendix contains supplemental information for Chapter 6.

\section{Numerically Estimating Photon-Number Yields with Linear Programs}

In this section we briefly describe the linear programming approach we used to estimate the upper bounds for the photon-number yields $p_{ZZ}(k_c,k_d|n_A,n_B)$ - which for simplicity here we will denote as $Y_{nm}$ - which is the probability of obtaining a set of detection events $k_c,k_d$ given that Alice and Bob respectively sent $n_A,n_B$ (or, $n,m$) photons. Such an approach has been widely discussed in literature as in Refs.\cite{MDIAnalytical,mdiparameter,mdiChernoff}, and is also described in the simple TF-QKD proof paper \cite{simpleTFQKD}. We also used a similar linear programming approach for some of the results in Ref. \cite{this_asymMDI} Appendix E, but it was not described in detail in that paper.

For simplicity, in this section we denote the observable gain in Z basis $p_{ZZ}(k_c,k_d|\beta_A,\beta_B)$ as $Q_{\mu_i,\mu_j}$ where $\mu_i=\beta_A^2$ and $\mu_j=\beta_B^2$, and $k_c,k_d$ are omitted (since the same expressions hold true for $k_c,k_d=(0,1)$ or $k_c,k_d=(1,0)$, and we can substitute the observable data for each $k_c,k_d$ respectively to obtain the corresponding $p_{ZZ}(k_c,k_d|n_A,n_B)$). Also, as mentioned above, we denote the yields $p_{ZZ}(k_c,k_d|n_A,n_B)$ as $Y_{nm}$.

\subsection{Linear Program Model}

Following Ref.\cite{mdiqkd}, the yields $Y_{nm}$ where Alice sends n photons and Bob sends m photons, satisfy the constraints:

\begin{equation}
\begin{aligned}
\sum_{n}\sum_{m}P^{\mu_i}_{n}P^{\mu_j}_{m}Y_{nm} &= Q_{\mu_i\mu_j}^Z \\
\end{aligned}
\end{equation}

\noindent where the photon number distributions are Poissonian:

\begin{equation}
\begin{aligned}
P^{\mu_i}_{n} &= e^{\mu_i} {{\mu_i^n}\over {n!}}\\
P^{\mu_j}_{m} &= e^{\mu_j} {{\mu_j^m}\over {m!}}\\
\end{aligned}
\end{equation}

Here, the right-hand-side constants $Q_{\mu_i\mu_j}^Z$ are the ``observables", i.e. the gain and error-gain respectively for the intensity combination $\mu_i,\mu_j$ (which can be any intensity among the set of decoy intensities). For the case of 3 decoys each for Alice and Bob, Equation (A.1) corresponds to 9 sets of constraints. Using Equation (A.1) as linear constraints, and $\{Y_{nm}\}$ as variables, we can apply linear programming, to maximize or minimize any linear combination of any of the variables (called an objective function) - for instance, here we can run the linear program multiple times, each time acquiring the upper bound for a given $Y_{nm}$ where $(n,m)$ ca\textsf{}n be $(0,0),(2,0),(0,2),(1,1),(2,2)$.

Note that, since there are infinitely many photon number states, to solve the linear program on an actual computer, we have to perform a cut-off and discard higher-order terms with large photon numbers. In practice we choose $S_{cut}=10$, such that a term is only discarded when both $n\geq 10$ and $m\geq 10$. For the discarded terms, we can either set them to zero (for lower bounds) or 1 (for upper bounds). 

\begin{equation}
\begin{aligned}
\sum_{n}\sum_{m}P^{\mu_i}_{n}P^{\mu_j}_{m}Y_{nm} &\geq \sum_{n < 10}\sum_{m < 10}P^{\mu_i}_{n}P^{\mu_j}_{m}Y_{nm} \\
\sum_{n}\sum_{m}P^{\mu_i}_{n}P^{\mu_j}_{m}Y_{nm} &\leq \sum_{n < 10}\sum_{m < 10}P^{\mu_i}_{n}P^{\mu_j}_{m}Y_{nm} + \left(1-\sum_{n < 10}\sum_{m < 10}P^{\mu_i}_{n}P^{\mu_j}_{m}\right)\\
\end{aligned}
\end{equation}

Therefore, in practice, the linear constraints can be written as:

\begin{equation}
\begin{aligned}
&Q_{\mu_i\mu_j}^Z - \left(1-\sum_{n < 10}\sum_{m < 10}P^{\mu_i}_{n}P^{\mu_j}_{m}\right) \leq \sum_{n < 10}\sum_{m < 10}P^{\mu_i}_{n}P^{\mu_j}_{m}Y_{nm} \leq Q_{\mu_i\mu_j}^Z\\
\end{aligned}
\end{equation}

\noindent with the additional constraint on variables:

\begin{equation}
\begin{aligned}
0 \leq Y_{nm} \leq 1 \\
\end{aligned}
\end{equation}

The linear program is run multiple times, each time maximizing a given $Y_{mn}$, where $(n,m)$ can be $(0,0),(2,0),(0,2),(1,1),(2,2)$.

\subsection{Finite-Size Effects}

In this subsection we consider finite-size effects for the privacy amplification process. Because of the statistical fluctuations, the observables (gains) we obtain in the Z basis might deviate from their respective expected values, which will lie within a certain ``confidence interval" around the observed values. Here we will perform a standard error analysis, similar to that in \cite{mdiparameter,mdifourintensity,this_asymMDI}, which is meant to be a straightforward estimation of the performance of TF-QKD under asymmetry and with practical data size, but not as a rigorous proof for composable security. 

Consider a random variable, whose observed value is $n$; we can bound its expected value $\langle n \rangle$ with the upper and lower bounds

\begin{equation}
\begin{aligned}
\underline{n} = n - \gamma \sqrt{n} \leq \langle n \rangle \leq  n + \gamma \sqrt{n} = \overline{n}
\end{aligned}
\end{equation} 

\noindent with a confidence (success probability) of $erf({\gamma/\sqrt{2}})$, where $\gamma$ is the number of standard deviations the confidence interval lies above and below the observed value, and $erf$ is the error function. In the simulations we consider a security failure probability of $\epsilon=10^{-7}$, which means we should set $\gamma \approx 5.3$.

In the Z basis, let us denote the observed counts for a given intensity setting $\{\mu_i,\mu_j\}$ as $n_{\mu_i,\mu_j}^Z$, which satisfies

\begin{equation}
\begin{aligned}
n_{\mu_i,\mu_j}^Z = Q_{\mu_i,\mu_j}^Z \times (NP_{\mu_i}P_{\mu_j})
\end{aligned}
\end{equation}

\noindent where $N$ is the total number of signals sent and $P_{\mu_i},P_{\mu_j}$ are the probabilities for Alice and Bob to respectively choose intensities $\mu_i$ and $\mu_j$. By applying Equation (A.6), we can acquire the upper and lower bounds to $Q_{\mu_i,\mu_j}^Z$:

\begin{equation}
\begin{aligned}
\overline{Q_{\mu_i\mu_j}^Z}=Q_{\mu_i\mu_j}^Z + \gamma \sqrt{Q_{\mu_i\mu_j}^Z \over {N P_{\mu_i} P_{\mu_j}}} \\
\underline{Q_{\mu_i\mu_j}^Z}=Q_{\mu_i\mu_j}^Z - \gamma \sqrt{Q_{\mu_i\mu_j}^Z \over {N P_{\mu_i} P_{\mu_j}}} \\
\end{aligned}
\end{equation}

Then, we can substitute them into the upper and lower bounds in the linear program when estimating $Y_{nm}$:

\begin{equation}
\begin{aligned}
&\underline{Q_{\mu_i\mu_j}^Z} - \left(1-\sum_{n < 10}\sum_{m < 10}P^{\mu_i}_{n}P^{\mu_j}_{m}\right) \leq \sum_{n < 10}\sum_{m < 10}P^{\mu_i}_{n}P^{\mu_j}_{m}Y_{nm} \leq  \overline{Q_{\mu_i\mu_j}^Z}\\
\end{aligned}
\end{equation}

\noindent which loosens the bounds and will result in a slightly higher upper bound for $Y_{nm}$ (which is understandable, since we expect a lower key rate with finite-size effect considered). Similar linear programs for finite-size decoy-state have also been considered in Ref. \cite{MDIAnalytical}.

Note that, although here we only consider a standard error analysis, in principle our results in this paper are applicable to e.g. composable security using Chernoff's bound \cite{mdiChernoff}. The key point is, the dependence on channel asymmetry, and the compensation for asymmetry using intensities, are only relevant in the X basis (signal states). The asymptotic case (with infinite decoys, where only signal states are relevant) therefore defines the fundamental scaling of key rate versus asymmetric channels, and all types of finite size analysis on the decoy states (e.g. using standard error analysis, using Chernoff's bound \cite{mdiChernoff}, or adding a ``joint bounds" analysis to tighten the bounds on statistical fluctuation and obtain a higher key rate \cite{mdifourintensity}, versus not considering finite-size effects at all and assuming the asymptotic key rate \cite{TFQKD,TFQKD02,TFQKD03,simpleTFQKD,TFQKD04}, i.e. assuming expected values of the gain and QBER to be identical to observed values in experiment) can be viewed of as correction terms (imperfections) on the yields and the key rate in the asymptotic limit. Our method is only related to the signal states and their intensities in the X basis, and is in principle always applicable regardless of the type of decoy-state analysis (e.g. the number of decoys) and the finite-size analysis used, as long as the Z basis is decoupled from the X basis.

With finite-size effect considered, the optimizable parameters for TF-QKD now include 

\begin{equation}
\begin{aligned}
[&s_A,\mu_A,\nu_A,P_{s_A},P_{\mu_A},P_{\nu_A},\\
&s_B,\mu_B,\nu_B,P_{s_B},P_{\mu_B},P_{\nu_B},]
\end{aligned}
\end{equation}

\noindent where the implicit parameters are $\omega_A,\omega_B$ (which for simplicity we assume to be zero), and $P_{\omega_A}=1-P_{s_A}-P_{\mu_A}-P_{\nu_A}$ and similarly $P_{\omega_B}=1-P_{s_B}-P_{\mu_B}-P_{\nu_B}$, and the choice of signal states $s_A,s_B$ versus the decoy states automatically implies basis choice, too. The above parameters are optimized using the same coordinate descent algorithm as described in Ref. \cite{this_asymMDI}. In Figure \ref{fig:rate_finite}, the dot-dash line (fully asymmetric) optimizes all 12 parameters, while the dashed line (signal-only asymmetric) optimizes only 7 parameters (where all parameters except $s_A,s_B$ are identical for Alice and Bob):

\begin{equation}
\begin{aligned}
[&s_A,\mu,\nu,P_{s},P_{\mu},P_{\nu}, s_B,\mu,\nu,P_{s},P_{\mu},P_{\nu}]
\end{aligned}
\end{equation}

Performing coordinate descent on key rate versus parameters while estimating the yields with linear programming is rather CPU-intensive. We have used a 40-core (80-thread) machine (a single compute node in the Niagara supercomputer \cite{SciNet}, each node with dual 20-core Intel Skylake CPUs) to generate Figure \ref{fig:rate_finite}, where the OpenMP multithreading library is used to parallelize the coordinate descent algorithm (to accelerate the search along each coordinate). The details of the algorithm can be found in Ref. \cite{this_asymMDI,mdiparameter}. Also, we used Gurobi \cite{gurobi}, a commercial linear program solver, to solve the linear programming models. Linear programs sometimes introduce multiple maxima, which means a local search on parameters sometimes might get trapped in a local maximum. To alleviate this, we can start a local search from multiple random starting points, and pick the largest search result, which can be viewed as a form of global search. (In principle, we can permutate the search results and perform multiple iterations of random search using e.g. an evolution algorithm \cite{evolution}, but here using one iteration with multiple random starting points is usually sufficient in finding a good key rate).


\chapter{Supplemental Information for Machine Learning in QKD}

This Appendix contains supplemental information for Chapter 7.

\section{Lookup Table}

\begin{table*}[t]
	\caption{Time benchmarking of using local search algorithm versus using neural network (NN) inference and using a pre-generated lookup table, for the 4-intensity MDI-QKD protocol. The desktop PC has an Intel i7-4790k quad-core CPU (with 16GB of RAM) and an Nvidia Titan Xp GPU. The single-board computers are a Raspberry Pi 3 with a quad-core CPU (with 1GB of RAM), and a Raspberry Pi Zero W with a single-core CPU (with 500MB of RAM). As can be seen, on the single-board computers, using a pre-generated lookup table is slower than directly using a neural network for inference, but it is still significantly faster than performing local search on CPU. By pre-generating a lookup table offline (e.g. on a desktop PC with GPU) and storing them on devices, we can still gain 15-25 times faster speed over local search on low-power devices, making the method suitable for devices where directly running neural networks is not feasible.}
	\begin{center}
		\begin{tabular}{cccc}            
			Device & Local search & NN &  Lookup table\\
			\hline\hline
			Desktop PC with GPU & 0.1s & 0.5-1.0ms  & 0.05s \\
			Raspberry Pi 3 & 3-5s & 2-3ms  & 0.2s \\
			Raspberry Pi 0W & 11-14s & N/A & 0.5s \\
		\end{tabular}
	\end{center}
\end{table*}

As an alternative solution for devices with hardware limitations (very little CPU power and no GPU/AI-chip) or software limitations (libraries unsupported on the platform) that prevent them from directly running a neural network, it is still possible to get a speedup, by using a \textit{pre-generated lookup table} of optimal parameters. For instance, for 4-intensity MDI-QKD, we can set a $100$ point resolution to $Y_0$, $e_d$, and $N$, and $100$ points from $L_{BC}=0-200km$. This will result in a total of $1\times 10^{8}$ data points that need to be calculated. Such a task is only possible with the parallelizable nature of the neural network, and the immense parallel processing power of the GPU. Predicting all the data points with a neural network on a desktop GPU would take an estimated time of 25 minutes. On the other hand, the local search algorithm takes 6 hours to generate the $4\times 10^{5}$ training data alone, and would take as many as two months to sample all $1\times 10^{8}$ input sets. The fast generation of such a lookup table is possible because we only take a small random sample ($4\times 10^{5}$) in the 4-dimensional input space, and use the neural network to learn the overall function shape with these data. Afterwards, once we have ``learned" the function, we can predict (or, intuitively, interpolate) all the $1\times 10^{8}$ points over the entire input parameter space with ease.

In Table E.1 we show a simple time benchmarking of the neural network inference and pre-generated lookup table versus local search algorithm on different devices including a powerful desktop PC and two models of low-power single board computers, for the 4-intensity MDI-QKD protocol. We can see that although using a lookup table is slower than directly running a neural network, it still has a significant advantage over local search on a CPU. This means that, the lookup table method is ideal for systems with extremely limited computing power or with software/hardware restrictions that prevent them from running a neural network (for instance, the Raspberry Pi 0W system has an older armv6 architecture, and neither Intel compute stick nor tensorflow are officially supported on the platform). The lookup table can be generated using a neural network running on a GPU on a desktop computer first, and stored on a mobile system to check when needed. Note that, this does not contradict a neural network's necessity, but rather is one of its applications, since only with a neural network can we possibly generate a lookup table over such a large parameter space.

Nonetheless, such a database would take up more storage resource (generating, for instance, a 100-point resolution lookup table for 4-intensity MDI-QKD would take up roughly 2.4GB of space (assuming single-precision floating point of 4 bytes is used for each output parameter), which we can choose to divide into 10 smaller tables, each taking up 240MB space, to avoid loading the entire table in memory), for such low-power devices, storage space is a lot cheaper than the extremely-limited CPU and memory resource (for instance, Raspberry Pis can read SD cards, which can easily have 64-256GB of storage space), and using small but many databases, they can be quickly loaded in parts into Raspberry Pi's memory, too. 

Therefore, here we show a simple solution to find optimal parameters using a lookup table pre-generated by a neural network offline, such that a speedup of up to 15-25 times can still be gained over running local search on a low-power device, even when directly running a neural network on the device is infeasible due to either hardware or software restrictions.

\addcontentsline{toc}{chapter}{Bibliography}
\bibliographystyle{plain}


\end{document}